\documentclass{article}

\usepackage[margin=1in,footskip=0.25in, a4paper]{geometry}
\usepackage{comment}
\usepackage[english]{babel}
\usepackage{amsmath,amssymb, enumitem}
\usepackage{bbm, dsfont}
\usepackage{tikz}
\usepackage{natbib}
\usepackage{algorithm}
\usepackage{algorithmic}
\usepackage{xspace}
\usepackage{float}
\usepackage{multirow}
\usepackage{arydshln}
\usepackage{rotating}
\usepackage{xcolor}
\usepackage{soul}
\usepackage{minitoc} 
\usepackage{xcolor}
\usepackage[utf8]{inputenc}
\usepackage{hyperref}
\usepackage{graphicx}
\usepackage{ntheorem}
\usepackage{mathrsfs}
\usepackage{subcaption}
\newtheorem{theorem}{Theorem}
\newtheorem{proposition}{Proposition}
\newtheorem{lemma}{Lemma}

\newtheorem{corollary}{Corollary}
\newtheorem{definition}{Definition}

\newtheorem{remark}{Remark}
\newtheorem{assumptions}{Assumption}

\theoremstyle{break}
\usepackage[utf8]{inputenc}
\DeclareMathOperator{\Cov}{Cov}

\DeclareMathOperator{\Var}{Var}

\DeclareMathOperator{\Tr}{Tr}

\newcommand{\E}{\mathbb{E}}
\newcommand{\R}{\mathbb{R}}

\newcommand{\N}{\mathbb{N}}
\newcommand{\Z}{\mathbb{Z}}

\newcommand{\mA}{\mathcal{A}}
\newcommand{\eps}{\varepsilon}
\renewcommand{\tilde}{\widetilde}
\renewcommand{\hat}{\widehat}
\newcommand\independent{\protect\mathpalette{\protect\independenT}{\perp}}
\def\independenT#1#2{\mathrel{\rlap{$#1#2$}\mkern2mu{#1#2}}}

\newcommand{\Ep}{{\mathrm{E}}}

\renewcommand{\Pr}{{\mathrm{P}}}
\newcommand\iid{i.i.d.}

\newcommand{\Xue}{\textbf{Projection}\xspace}
\newcommand{\XY}{\textbf{Max}\xspace}
\newcommand{\Proposed}{\textbf{Proposed}\xspace}

\newcommand{\green}[1]

\title{Simultaneous hypothesis testing for comparing many functional means}

\usepackage{authblk}

\author[1]{Colin Decker}

\author[2]{Dehan Kong}

\author[2]{Stanislav Volgushev}

\affil[1]{
Fakult\"at f\"ur Mathematik, Ruhr Universit\"at Bochum.
}

\affil[2]{
Department of Statistical Sciences, University of Toronto.
}

\begin{document}

\maketitle

\begin{abstract}
Data with multiple functional recordings at each observational unit are increasingly common in various fields including medical imaging and environmental sciences. To conduct inference for such observations, we develop a paired two-sample test that allows to simultaneously compare the means of many functional observations while maintaining family-wise error rate control. We explicitly allow the number of functional recordings to increase, potentially much faster than the sample size. Our test is fully functional and does not rely on dimension reduction or functional PCA type approaches or the choice of tuning parameters. To provide a theoretical justification for the proposed procedure, we develop a number of new anti-concentration and Gaussian approximation results for maxima of $L^2$ statistics which might be of independent interest. The methodology is illustrated on the task-related cortical surface functional magnetic resonance imaging data from Human Connectome Project. 
\end{abstract}
%\keywords{Functional data, two-sample test, Gaussian approximation, multiplier bootstrap}
\section{Introduction}\label{sec:intro}
Advancements in modern technology have enabled the effective representation of numerous real-world datasets as collections of functional objects.   In environmental science, time-series data—including temperature, precipitation, air quality, and other measurements across various geographic locations—can be modeled as multiple functional objects \citep{ZD23}. Similarly, in economics, the trajectories of various stock prices observed over time serve as an example \citep{horvath2014testing}. In neuroscience,  functional magnetic resonance imaging (fMRI), electroencephalography (EEG), and 
magnetoencephalography (MEG) data can often be summarized as multiple functional data observed across different brain regions of interest \citep{tian2010functional}.

In practice, researchers are often interested in determining whether the mean of these functional objects differs under two distinct conditions. For instance, financial practitioners are interested in determining whether intra-day price return curves in different economic regimes are the same on average. Likewise, neuroscientists frequently investigate whether brain signals, represented as a collection of functional curves obtained from an fMRI device, exhibit different patterns when a subject performs various tasks in the scanner. 

Several studies have examined two-sample tests for functional data \citep*{HK07, krzysko2021two, wynne2022kernel, pomann2016two, wang2021two, zhang2010two}. However, these methods typically address cases with only one functional curve per observation. Although methods designed to handle a single functional curve per observation can be adapted to handle many functional curves per observation in ad hoc ways, the resulting tests often have severe limitations. One naive approach is to vectorize multiple functional curves into a single functional curve defined on the union of the domains of the original curves. The global null hypothesis of over-all equality of mean across all functional curves can then be tested using a procedure designed for a single functional curve. In case the global null hypothesis is rejected, this approach does not reveal which of the original functional curves are responsible for the rejection. In addition the test may have low power against alternatives where only a small number of the original functional curves have a non-zero mean. Finally, if the number of curves is growing, classical assumptions that are imposed for theory can fail. Another naive approach is to test each functional curve individually. This approach allows for simultaneous testing of each functional curve, but each individual test is likely to have low power due to multiple testing corrections (e.g., the Bonferroni correction and its variants), especially when the number of functional curves is large. Thus, there is a need for two-sample tests capable of simultaneously handling many functional curves while precisely controlling the family-wise error rate.

In a different direction, because functional data are recorded discretely, it is in principle possible to use high dimensional tests of Euclidean mean vector in a high-dimensional functional setting. Such tests are frequently based either on the $L^{\infty}$ norm or $L^{2}$ norm of the data. The best known $L^{2}$ type tests of high dimensional vector Euclidean mean (see e.g. \cite{CQ10}) are proved to be accurate only under assumptions that do not hold for functional data, such as weak dependence across the marginal distributions and sparsity conditions on the covariance operator. On the other hand, high dimensional tests of Euclidean vectors based on the $L^{\infty}$ norm have been shown to work well in functional settings. The test of \cite{YX} is designed for high dimensional means but can be applied to high dimensional functional data by stacking the many discrete functional recordings into a high dimensional Euclidean vector. In fact, the data analysis considered in \cite{YX} deals with a global test of equality of means for high dimensional functional data. However, the theory in \cite{YX} is not well suited to deal with discretized observations of functional data. \cite{xue2023distribution} extends the test of \cite{YX} to a two-sample test that is explicitly designed to accommodate high dimensional functional data. However, this test requires the choice of a fixed collection of basis functions that the functional observations are projected on. The choice of this basis requires additional tuning and can have a significant impact on power.

The main contribution of the present article is to develop a high dimensional test for the equality of functional means that is based on the full $L^2$ norm of each functional component. By explicitly treating the data as functional,  we prove theoretically that our method has asymptotically correct level and high power even when the number of functional components is much larger than the sample size. Our test is fully functional and does not rely on the projections onto pre-specified basis functions or the choice of cut-off parameters in that are needed in functional PCA type approaches. The proposed test rejects the hypothesis of over-all equality of means if the maximum of the $L^2$ norms over-all functional components is large. Making mild assumptions on the randomness that generates each marginal functional component, and no assumption at all about the dependence structure across functional components, we develop a bootstrap procedure to estimate the quantile of the mixed $L^{\infty}-L^{2}$ norm of normalized sums of independent observations taking values in high dimensional products of the function space $C[0,1]$, but only discretely recorded on a grid of points. 

The quantile of the $L^{\infty}-L^2$ norm also allows to establish family-wise error control for testing many simultaneous functional hypotheses with $L^2$ statistics. In particular, if the global null is rejected we can pin-point which functional components are responsible for this rejection. Our bootstrap quantile converges asymptotically to the smallest possible rejection threshold that controls family-wise error in our setting. Thus the rejection threshold of our test is adaptive, and responds to both the size of the marginal randomness in each functional component and the dependence structure across different components. 

We provide a detailed theoretical justification for the proposed test, with explicit bounds on convergence rate of the family-wise error rate to its nominal level. In order to do so, we prove new Gaussian approximation and Gaussian comparison  bounds for the $L^{\infty}-L^2$ mixed norms of high-dimensional means. A key difference with existing results 
is that we allow both the dimension of each sub-vector for which we compute the $L^2$ norm and also the number of $L^2$ norms to grow quickly with the sample size, and provide an explicit characterization of how this growth affects the approximation error. In contrast, existing results treat the dimension of each  sub-vector for which we compute the $L^2$ norm as fixed, or the number of $L^2$ norms as fixed, or both. We also derive several new anti-concentration bounds for both, $L^\infty$ and $L^{\infty}-L^2$ mixed norms. A key feature of those bounds is that they do not require a lower bound on the smallest variance among components.

The rest of the paper is organized as follows. In Section~\ref{sec:motiv}, we motivate our approach using a functional magnetic resonance imaging as an example, and formally introduce the methodology. Section~\ref{sec:theory} contains theoretical guarantees for our approach. In Section~\ref{sec:sim}, we showcase the finite sample level and power properties of the proposed test in simulations studies and provide comparisons with competing methods. Section~\ref{One_Sample_Real_World} contains a data illustration. Detailed proofs are provided in a series of appendices. 

\section{Motivation and methodology}\label{sec:motiv}

\subsection{Motivation}

\begin{figure}[h]
\begin{center}
\includegraphics[scale=0.45]{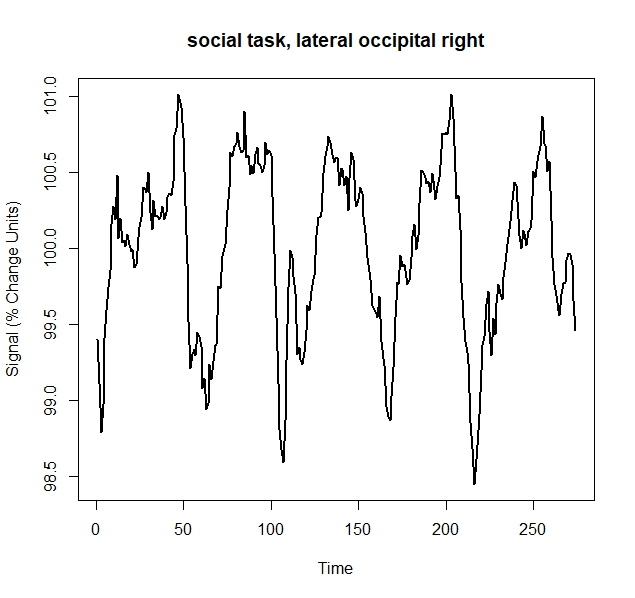}
\includegraphics[scale=0.45]{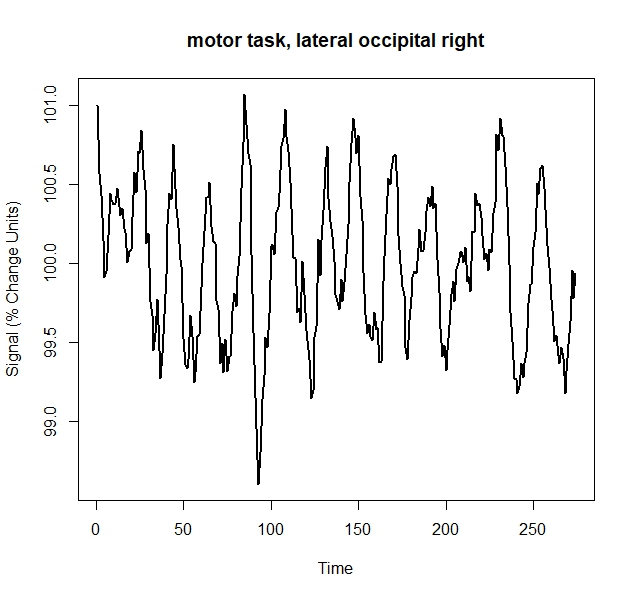}
\end{center}
\caption{A pair of measurements taken on the same arbitrarily chosen participant through an electrode placed over the right lateral occipital region of the brain for two different tasks. Left panel: social task, right panel: motor task.}
\label{figure_signal_plots}
\end{figure}

As a motivation for our approach, consider task-related cortical surface functional magnetic resonance imaging (fMRI) data from Human Connectome Project (https://www.humanconnectome.org/). Here, each participant was recorded performing both a sensory-motor task and social cognition task by an fMRI device. Recordings were taken through electrodes placed above $K=68$ regions of interest (ROI) taken from the Desikan-Killiany atlas \citep{desikan2006automated}. Each functional recording on each participant is associated with one of 68 brain regions, see Figure~\ref{brain_picture} in Section~\ref{One_Sample_Real_World}. In Figure~\ref{figure_signal_plots}, pre-processed recordings from an arbitrarily chosen participant taken through the electrode placed above the lateral occipital right region, a location defined in the Desikan-Killiany atlas, are presented. 

Recordings for each individual for both tasks were performed on a discrete time grid, resulting in paired observations of the form $(g_i^k(t_\ell),h_i^k(t_\ell))$, where $i$ is the index of the individual, $k$ is the channel and $t_\ell$ denotes the observation times. A question of scientific interest is to determine whether there is a difference in brain activity in different brain regions and to identify the regions where activity differs. To this end, we develop a test for equality of functional means which (i) allows to identify deviations in means across brain regions (ii) respects the structure of the data which have different dependence across time than across dimension and (iii) controls the family-wise error rate uniformly over all channels.

\subsection{Proposed methodology}\label{methodology_section}

Assume that for each observational unit $i=1,\dots,n$, we collect a pair of  measurements, where each measurement consists of $K$ functional observations which are recorded on a discrete time grid $0\leq t_1 \leq \dots \leq t_T\leq 1$. Specifically, for $1 \leq k \leq K$,  let $(g_i^k,h_i^k)$ denote the $k$'th underlying pair of functional recordings for subject $i$. Assume that we observe $X_i^{k,\ell} = g_i^k(t_\ell), Y_i^{k,\ell} = h_i^k(t_\ell)$ for $1 \leq \ell \leq T$. The data are thus of the form $(X_i^{k,\ell},Y_i^{k,\ell})_{k=1,\dots,K,\ell = 1,\dots,T,i=1,\dots n}$, where the index $i$ corresponds to the observational unit, the index $k$ is the index of the functional observation for each unit, and $\ell$ indexes the observation time. 

In practice, one is often interested in testing over-all equality of paired means across all of the $K$ functional components of the data-generating distribution. In other words, this corresponds to testing the global null hypothesis 
\begin{equation} \label{eq:h0global}
H^{global}_{0}: \E[h^k(t)] = \E[g^k(t)] \quad \forall~t \in [0,1] \quad \forall~1\leq k \leq K,
\end{equation} 
where $(h_i, g_i)_{i=1,\dots,n}$ are i.i.d. copies of $(h, g)$.

In modern applications, the number of functional components of the data-generating distribution, $K$, can be large compared to the number of independent observational units, $n$. Therefore, we allow for $K \gg n$.

There are many reasonable test statistics that can be used to test $H^{global}_{0}$ against an arbitrary alternative. We consider a test statistic that is designed to have high power against alternatives for which at least one of the $K$ functional components of the data-generating distribution has a dense mean. In the fMRI example, this is well motivated because brain signals are continuous functions of time and can be expected to differ over extended time periods if they differ at all. Similar comments apply to other types of functional data.

Consider the differences $Z_i^{k,\ell} := Y_i^{k,\ell} - X_i^{k,\ell}$. For each $1 \leq k \leq K$, let 
\begin{equation}\label{def:test_stat_channel_k}
T_n^k :=  \frac{1}{T}\sum_{\ell=1}^T \Big(\frac{\sum_{i=1}^n Z_i^{k,\ell}}{\sqrt{n}} \Big)^2 = \frac{n}{T}\sum_{\ell=1}^T \Big(\bar Z^{k,\ell} \Big)^2,
\end{equation}
where $\bar Z^{k,\ell} = n^{-1}\sum_{i=1}^n Z_i^{k,\ell}$. Note that the statistic $T_n^k$ can be interpreted as a Riemann approximation 
\[
T_n^k \approx \int_0^1 \Big(\frac{1}{\sqrt{n}}\sum_{i=1}^n \{g_i^k(t) - h_i^k(t) \}\Big)^2 dt = n \int_0^1 \Big(\bar g_{\cdot}^k(t) - \bar h_{\cdot}^k(t) \Big)^2 dt,
\]
where $g_{\cdot}^k(t) = n^{-1}\sum_{i=1}^n g_i^k(t), h_{\cdot}^k(t) = n^{-1}\sum_{i=1}^n h_i^k(t)$. To test $H^{global}_{0}$, we propose the test statistic
\begin{equation}\label{global_hypothesis_test_stat}
T_n=\max_k T_n^k.
\end{equation}
This statistic acts as (the square of) an $L^{2}$ norm over each functional component of the data, and acts as an $L^{\infty}$ norm across the different functional components. To use this test statistic for level-$\gamma$ testing with $\gamma \in (0,1)$, it is necessary to determine its $(1-\gamma)$ quantile. The dependence structure of a $K$-dimensional functional distribution can be complex, and we place no restriction on the dependence structure across the different functional components of the data-generating distribution. Further, to handle the high dimensional case of $K \gg n$, we consider a model that allows $K$ to grow with $n$. Under this level of generality, classical asymptotic theory is not fully informative. Therefore, to estimate the distribution of $T_n$ under the null hypothesis, we implement a Gaussian multiplier bootstrap procedure. For a vector $e \in \R^n$ with i.i.d. $N(0,1)$ entries $e_i$, define 
\begin{equation}\label{bootstrap_statistic}
T_n(e) = \max_{k=1}^K \frac{1}{T} \sum_{\ell=1}^T \Big(\frac{ \sum_{i=1}^n e_i[Z_i^{k,\ell} - \bar Z^{k,\ell}_\cdot]}{\sqrt{n}} \Big)^2.
\end{equation} 
For an independent collection of $N(0,I_n)$ distributed random vectors $e^1,\dots,e^{N}$, let 
\begin{equation}\label{test:ecdf}
\hat F(t) := \frac{1}{N} \sum_{j=1}^N I\{T_n(e^{j}) \le t\}
\end{equation}
denote the empirical cdf of $\{ T_n(e^{1}),\dots,T_n(e^{N}) \}$. We will show in the next section (see Theoreom \ref{main:thm}) that, under suitable technical assumptions, the rejection rule
\begin{equation}\label{eq:multtest}
	\mbox{Reject  } H_0^{global} \mbox{  iff  } 1 - \hat F(T_n) < \gamma
\end{equation} 
gives an asymptotically level $\gamma$ test. Note that $H^{global}_{0}$ is rejected according to this rule if and only if $T_n:=\max_k T_n^k > \hat q_{1-\gamma}$ where $\hat q_{1-\gamma}$ denotes the $1-\gamma$ sample quantile of the collection $\{ T_n(e^{1}),\dots,T_n(e^{N}) \}$, equal to the $\lceil (1-\gamma)N\rceil$ order statistic of the collection. 

If $H^{global}_{0}$ is rejected, it is of interest to determine which of the $K$ functional components are responsible for the rejection, i.e., for every $1 \leq k \leq K$, we are interested in testing the null hypotheses 
\begin{equation} \label{eq:h0local}
H_0^k: \E[h_i^k(t)] = \E[g_i^k(t)], \quad \forall~t \in [0,1]
\end{equation}
in such a way that the family-wise error rate is precisely controlled. We will show in the next section that, under suitable technical assumptions, the following rejection rule
\begin{equation}\label{eq:multtest_k}
	\mbox{Reject  } H_0^k \mbox{  iff  } 1 - \hat F(T_n^k) < \gamma
\end{equation} 
is a test with asymptotic family-wise type-I error equal to $\gamma$ under the global null. Note that the numbers $1 - \hat F(T_n^k)$ ($1 \leq k \leq K$) can be interpreted as bootstrap p-values for $H_0^k$ that are calibrated for precise family-wise error rate control. The Gaussian multiplier bootstrap statistic $T_{n}(e)$ not only provides information about the quantiles of $T_{n}$ but also the optimal level $\gamma$ rejection threshold for simultaneously testing $H^{k}_{0}$.  In general, the asymptotic level of each individual test of each marginal hypothesis $H_{0}^{k}$ will be less than or equal to $\gamma$. Our procedure is adaptive and uses a data-driven asymptotically consistent estimate of the least conservative possible rejection threshold required to control the probability of at least one false rejection at $\gamma$.

\begin{remark}[Connections to comparisons of high-dimensional means] \label{rem:highdimmeans}
One possibility to analyze data of the form described above would be to treat them as i.i.d. copies of observations of high-dimensional vectors. In this case, tests for comparing very high-dimensional means could be applied. Two classical approaches are to use $L^2$ type statistics \citep{CQ10} or on max type statistics \citep{CCK17,chernozhuokov2022improved, YX}. 
From this perspective, our test can be viewed as utilizing a mixed max-$L^2$ type norm where the $L^2$ norm is taken over time $\ell$ while the max is taken over locations $k$. This approach naturally preserves the functional structure of the data, recognizing that the dependence across time differs in nature from the dependence across locations. The theoretical and practical merits of treating time and dimension differently will be demonstrated in theory,  and through a data illustration later in the manuscript. 
\end{remark}

\begin{remark} [Intuition behind the bootstrap procedure]
Observe that in the definition of $T_{n}(e)$, we use the same univariate $e_i$ across all values of $k$. This is crucial to properly account for the dependence between the test statistics at different values of $k$ and (asymptotically) control the family-wise error rate. The idea of using a collection of independent random variables to create bootstrap replications via multiplication is classical in the literature and known as multiplier bootstrap or wild bootstrap. It was recently shown to work well for approximating the distribution of maxima of high-dimensional means in a line of seminal works by \cite{chernozhukov2013gaussian,CCK17,chernozhuokov2022improved}. In the sections below, we will show that their results extend to our type of max-$L^2$ statistics. To gain some intuition for why this type of bootstrap approach works, observe that, conditional on the data, the vector $\Big( \frac{1}{\sqrt{n}} \sum_{i=1}^n e_i[Z_i^{k,\ell} - \bar Z^{k,\ell}] \Big)_{k,\ell}$ is centered normal with covariance matrix equal to the sample covariance matrix of $\{(Z_1^{k,\ell})_{k,\ell},\dots,(Z_1^{k,\ell})_{k,\ell}\}$. Hence the Gaussian multiplier bootstrap can be interpreted as simulating from a multivariate normal with the estimated covariance matrix.  
\end{remark}

\begin{remark}[Extension to asynchronous observations]
The test statistic can be easily extended to the case where the observations corresponding to $X$ and $Y$ are not taken at the same time points but still on a dense grid. Specifically, assume that $Y_{i}^{k,\ell} = h_{i,k}^{Y}(t_\ell), \ell = 1,\dots, T_Y$ and $X_{i}^{k,\ell} = h_{i,k}^{X}(s_\ell), \ell = 1,\dots, T_X$, whereas the $t_\ell, s_\ell$ are time points in the interval $[0,1]$. Assume without loss of generality that $0 = t_1 < t_2 < \dots < t_{T_Y} = 1$ and similarly for $s_\ell$. Define the functions $Y_i^k(\cdot): [0,1] \to \R$ by linear interpolation of the points $(t_\ell,Y_{i}^{k,\ell})$ and similar for $X_i^k(\cdot)$. Then we could define 
\[
\tilde T_n^k := \int_0^1 \Big(\frac{1}{\sqrt{n}}\sum_{i=1}^n \{Y_{i}^k(t) - X_{i}^k(t)\} \Big)^2 dt
\]   
or consider a Riemann approximation of this integral. Note that when $T_X = T_Y$and $s_\ell = t_\ell~\forall \ell$, the original test statistic $T_n^k$ is exactly a Riemann approximation to $\tilde T_n^k$. Similarly, the bootstrap statistic could be defined as
\[
\tilde T_n(e) := \int_0^1 \Big(\frac{1}{\sqrt{n}}\sum_{i=1}^n e_i \{Y_{i}^k(t) - X_{i}^k(t) - \bar Y_\cdot^k(t) + \bar X_\cdot^k(t)\} \Big)^2 dt,
\]
or a Riemann approximation thereof. When the time points for the measurements of $X,Y$ are aligned (as is the case in our data illustration), our original formulation is substantially more efficient computationally. Below we present theoretical guarantees only for the case of evenly spaced and aligned grid points. In those results, the error is a monotone increasing function of $1/T$. Similar bounds hold in the asynchronous case, but the dependence on the grid points can be replaced with the same monotone function of $\max \{|t_{i+1}-t_{i}|, 1 \leq i \leq T_{Y}-1\}\vee  \max\{|s_{i+1}-s_{i}|, 1 \leq i \leq T_{X}-1\}).$ Thus a dense observation grid is needed in both samples for this extension to work.
\end{remark}

\begin{remark}[Fully functional nature of the proposed test]
Hypothesis testing is a well-explored topic in the functional data analysis literature. A common approach involves initially applying dimension reduction techniques, such as functional principal component analysis, followed by conducting the hypothesis test \citep{fremdt2013testing, kong2016classical, pomann2016two}. In contrast to these classical approaches, our test statistic does not include any dimension reduction via functional principal component analysis. This approach is in line with recent work on ``fully functional" tests, see e.g. \cite{ARS18,DKA20}. As argued by \cite{ARS18}, the fully functional approach can have substantially better power when the signal is not aligned with the first few functional principal components. Additionally, it simplifies the statistical analysis since it removes additional pre-processing steps and arbitrary choices of tuning parameters. 
\end{remark}

\section{Theoretical guarantees} \label{sec:theory}

\textit{Notation} Let $I = [0,1]$ and denote by $C[I]$ the space of continuous, real-valued functions on $I$. For $f \in C[I]$, let $\|f\|_\infty := \sup_{x \in I}|f(x)|$. Let $(C[I])^K$ stand for the $K$-fold Cartesian product of $C[I]$ and for $f = (f_1,\dots,f_K) \in (C[I])^K$ consider the norm $\|f\|_{(C[I])^K} := \max_{k=1}^K \|f_k\|_\infty$. Let $\|\cdot\|_{\Psi_{1}}$ denote the sub-exponential norm. Further, for any $x\in \R$, let $x^{+} = \max\{x,0\}$. For vectors $v_i \in \R^{d_i}, i=1,\dots,m$ let $\oplus_{i=1}^m v_i = (v_1^\top,\dots,v_m^\top)^\top \in \R^{\sum_{i=1}^m d_i}$ denote their concatenation, i.e., the vector obtained by stacking all vectors into one long vector. 

\smallskip

We now state the assumptions that will allow us to control the family-wise error rate of the rejection rule~\eqref{eq:multtest}.

\begin{assumptions}\label{assumptions}

There exist centered, random\footnote{I.e. Borel measurable with respect to the Borel sigma Algebra generated by the norm $\|\cdot\|_{(C[I]^K)}$ on $C[I]^K$. It is straightforward to check (see Lemma~ \ref{measurability}) that $X(\omega, t):=f(\omega)(t)$ is a measurable $K$ dimensional stochastic process and we will use the terms random element, random function, and stochastic process interchangeably.} elements $\tilde f = (\tilde{f}^1,\dots,\tilde{f}^K) \in (C[I])^K$ and functions $\mu \in (C[I])^K$ such that for $f_i^{k}(t_\ell):=\tilde f_i^{k}(t_\ell) + \mu^k(t_\ell)$ the data is generated as $Z_i^{k,\ell} = f_i^{k}(t_\ell)$, where $t_\ell = \ell/T$ and $(\tilde f_i^{1},\dots,\tilde f_i^{K})_{i=1,\dots,n}$ are i.i.d. copies of $(\tilde f^{1},\dots,\tilde f^{K})$ and such that

\begin{itemize}
\item [(A1)] There exist $0 < q \leq 1, C_{H}> 0$, such that for all $k$ and all $(t_{1}, t_{2})\in I^2$,  $\|\tilde{f} _k(t_{1}) - \tilde{f} _{k}(t_{2})\|_{\Psi_{1}} \leq C_{H}|t_{1}-t_{2}|^{q}$.  Further, for each $k$ there exists $x_{k} \in I$ such that $\Var(\tilde{f} _{k}(x_{k}))< \infty$.\label{ass:sub_exp}
\item  [(A2)] Let $\phi^{\tilde{f}_k}_{j}$ denote the $j^{th}$ Karhunen–Loève score of $\tilde{f} _{k}$. There exist $\alpha>1$ and $C_{\phi}>0$, such that for all $k$ and $j$,  $\|\phi^{\tilde{f} _k}_{j}\|_{\Psi_{1}}\leq C_{\phi}j^{-\alpha}$. 

\label{ass:decay}
\item [(A3)] There exists a positive real number $v_{-}$ such that $\max_{k}\max_{j}\Var(\phi^{\tilde{f} _k}_{j}) > v_{-}$.\label{ass:lower_bound}
\item [(A4)] For $\phi^{\tilde{f}_k}_{j}$ from (A2), there exists a constant $C_2 > 0$ and $\beta > \alpha> 1/2$ such that $\min_k \Var(\phi^{\tilde{f}_k}_{j}) \geq C_2 j^{-\beta}$ for all $j \geq 1$. \label{ass:lower_bound_var}
\end{itemize}
\end{assumptions}

Assumption (A1) assumes that the paths of the functional observations $\tilde f_k(\cdot)$ are sufficiently regular in the sense of H\"older increments in the $\|\cdot\|_{\Psi_{1}}$ norm. Note that we do not assume smoothness or differentiability since $q<1$ is allowed. (A1) allows us to control the error incurred by approximating $T_n^k$ by an integral over $[0,1]$. This assumption covers the case of Brownian bridge and Brownian motion on $[0,1]$ which satisfy (A1) with $q =1/2$. Bounds on the increments of a stochastic process do not imply bounds on the variance at any individual time points, so this assumption needs to be added in (A1). Together, (A1) implies that the  $\tilde f_k$ are second order stochastic processes on $[0,1]$ and allows us to apply the Karhunen–Loève Theorem; see Theorem~\ref{kl-theorem} in the Appendix for the specific version we use. This makes the Karhunen–Loève (KL) scores in (A2) well-defined. For part of the theory below we will need to assume that $\alpha > 1$. This barely excludes the Brownian bridge and Brownian motion which satisfy (A2) with $\alpha = 1$. Such processes can be included by assuming the additional lower bound in (A4) which both processes satisfy with $\beta = 1$. Assumption (A3) rules out the completely degenerate case where all functions $\tilde f^k$ are zero almost surely.

We note that (A1)-(A4) are purely marginal assumptions on the functional components of the data-generating distribution (i.e., they are assumed to hold for each $1 \leq k\leq K$ separately), and do not  place any restriction on the joint dependence across different marginal functional distributions whatsoever.

\begin{remark}[On the sampling scheme]
To keep the presentation simple and focus on the main points, we have opted for a simple sampling scheme where observation times are equidistant and there is no noise added on top of the functional observations. Relaxing the equidistant sampling assumption would only require minor modifications of the argument and would not change the results if assumptions are placed on the maximal spacing between consecutive observation times. If the functional observations are corrupted by additional i.i.d. noise at each measurement point, local smoothing techniques such as kernel smoothing can serve as an initial step \citep{zhang2007statistical}. This would induce additional technicalities into our assumptions and arguments and we do not pursue this direction here. 
\end{remark}

Our first main result concerns the asymptotic family-wise error rate control of our procedure. For $\alpha$ from (A2) and $\beta$ from (A4) define 

%\cd{6.5.25: changed $K+1$ to $K+2$ in some places so that all log factors are greater than 1. }
\begin{align}
d_n = d(n,K,T,N) &:= \log^{3}(K+2)\log^{3}(n)\log(n(K+1))[n^{\frac{1-\alpha}{6\alpha-1}}+n^{\frac{1}{12\alpha-2}}T^{-q}] + \sqrt{\frac{\log(N)}{N}}, \label{eq:defdn}
\\
\tilde d_n = \tilde d(n,K,T,N) &:= \log^{3+\frac{3\beta}{2\alpha-1}}(n)\log^{\frac 72+\frac{3\beta}{2\alpha-1}}(K+2)\Big[n^{\frac{\frac{1}{2}-\alpha}{6\alpha-1}}+T^{-q}\Big]+\sqrt{\frac{\log(N)}{N}}. \label{eq:deftildedn}
\end{align}
Further we recall the definition of the  the family-wise error rate when testing at level $\gamma$,
\[
FWER_{\gamma} := \Pr\Big(\exists k \in \{1, \dots, K\} \textrm{ such that } \mu^k \equiv 0,1-\hat F(T^k_n)<\gamma\big)
\]
i.e., the probability of making at least one false rejection using the rejection rule in~\eqref{eq:multtest}, 
where $\hat F$ is defined in~\eqref{test:ecdf}
and $T^k_n$ in~\eqref{def:test_stat_channel_k}.

\begin{theorem}(Control of family-wise error rate)\label{main:thm}
Suppose (A1)--(A3) hold with $\alpha$ from (A2) satisfying $\alpha > 1$. There exists a constant $C_{\gamma}$ that is independent of $K$, $T$, $n$, $N$, and depends on $\gamma$ and on the distribution of $\tilde f=f-\mu$ only through the model parameters appearing in (A1)-(A3), such that for all $n \geq 3$, $K \geq 1$, $T \geq 1$, $N \geq 1$,
\begin{equation}\label{eq_FWER_general}
FWER_\gamma \leq \gamma + C_{\gamma} d(n,K,T,N).
\end{equation}
Moreover, under the global null $H^{global}_0$, we have 
\begin{equation}\label{eq_FWER_global_null}
|FWER_\gamma - \gamma| \leq C_{\gamma} d(n,K,T,N).
\end{equation}
If in addition (A4) holds and $\alpha$ from (A2) instead satisfies $\alpha>\frac 12$, the same conclusions hold with $\tilde C_\gamma \tilde d(n,K,T,N)$, where $\tilde C_\gamma$ denotes a constant that is independent of $K$, $T$, $n$, $N$, and depends on the distribution of $\tilde{f}=f-\mu$ only through the model parameters appearing in (A1)-(A4).
\end{theorem}

The bounds in Theorem~\ref{main:thm} hold for fixed $n \geq 3,K \geq 1, T \geq 1, N \geq 1$ and give explicit dependence on all of those quantities. This can be easily turned into an asymptotic statement in a triangular array framework where $N = N_n,T = T_n, K = K_n$ are allowed to change with $n$. For instance, if $K_n = O(n^m)$ for some fixed $m < \infty$, the test has asymptotic level control under (A3) provided that $N_n \to \infty, T_n \gtrsim n^{\zeta}$ for $\zeta > \{q(12\alpha-2)\}^{-1}$ and under (A4) as soon as $T_n \gtrsim n^\xi$ for an arbitrarily small $\xi$. It is also possible to let $K_n$ grow like $\exp(n^a)$ for a sufficiently small $a$ which depends on $\alpha,\beta$ at the cost of further strengthening the assumption on $T$. 

The error term $d(n, K, T, N)$ consists of three pieces that have different sources. The part $\sqrt{N^{-1}\log(N)}$ stems from the Monte-Carlo error when approximating the distribution of $T_n(e)$ by simulations. This error can be made arbitrarily small by increasing the number of bootstrap replications $N$. The role of the other two terms is discussed in more detail below.   

\begin{remark}[Related work on high-dimensional functional data]
High-dimensional Gaussian approximation techniques for functional data are also mentioned in \cite{lopes2020bootstrapping,YX,lin2023high} and utilized in \cite{ZD23}. Our work differs from all of those approaches in several important ways.

\cite{YX} mainly consider two-sample comparisons of high-dimensional means and only mention functional data in their application. All of their theory is for the vector setting. In their application section, \cite{YX} consider testing for functional data through vectorization as described in Remark~\ref{rem:highdimmeans}. The resulting test statistic is of a max-max type, as opposed to our max-$L^2$ type statistic. Specifically, it takes the form (see page 1304, \cite{YX})
\begin{equation}\label{Xue_yao_test_stat}
T^{\text{Max}}_n:=\sqrt{n}\max_{k=1}^{K}\max_{\ell=1}^{T}\big|\bar{Z}^{k,\ell}\big|.
\end{equation}
Additionally, a direct application of high-dimensional Gaussian approximation results to vectorized functional data without taking into account regularity of functional observations along time direction leads to bounds that become worse as the number of time points increases. In contrast, our bounds improve as the observation grid gets denser.

\cite{lopes2020bootstrapping} show that the rates for Gaussian approximation of maxima in high dimension can be close to parametric and dimension-free if the variances of the mean vector can be ordered to decay along dimension. Such a variance decay is always present in the KL scores of functional observations. However, their approach does not apply to our setting since we have variance decay along sequences of KL scores for each $k$, but also a growing number $K$ of different functional observations. And we do not put any assumptions on variance decay in the first KL score across different $k$. \cite{lin2023high} extend the findings in \cite{lopes2020bootstrapping} to MANOVA-type pair-wise testing, but they only allow the number of pair-wise comparisons go grow at a logarithmic rate. In contrast, $K$ can grow like $\exp(n^a)$ for a small $a$ in our setting.

\cite{xue2023distribution} considers high-dimensional functional data. Their test statistic looks at the supremum over dimension, and in the time direction the supremum over projections of the data onto each member of a pre-specified collection of unit vectors. Specifically, denoting by $\mathcal{X}$ a finite and pre-specified collection of unit vectors in $\R^T$ \footnote{We note that in \cite{xue2023distribution}, the test statistic is implemented by smoothing the discrete observations and using the $L^2$ inner-product, whereas in the formulation of \eqref{Xue_test_stat} we use un-smoothed data and the Euclidean inner-product; for densely observed functional data, the difference is small.}, the text statistic in \cite{xue2023distribution} takes the form
(see page 2338, \cite{xue2023distribution})
\begin{equation}\label{Xue_test_stat}
{T^{\text{Projection}}_n}:=\frac{\sqrt{n}}{\sqrt{T}}\max_{k=1}^{K}\max_{v \in \mathcal{X}}\big|\sum_{\ell=1}^{T}v_{\ell}^\top\bar{Z}^{k,\ell}\big|.
\end{equation}
This can be thought of as a finite dimensional approximation of our test statistic which can be written as (see \eqref{global_hypothesis_test_stat})
\begin{equation}\label{proposed_test_stat}
T_n:=\frac{n}{T}\max_{k=1}^{K}\sum_{l=1}^{T}\big(\bar{Z}^{k,\ell}\big)^2 = \Big(\frac{\sqrt{n}}{\sqrt{T}}\max_{k=1}^{K}\sup_{v \in \R^T, \|v\|_2=1}\big|\sum_{l=1}^{T}v_{\ell}^\top\bar{Z}^{k,\ell}\big|\Big)^2.
\end{equation}
The error rate in their theoretical guarantees grows as a power of the cardinality of the collection of unit vectors, and among collections of fixed size, the error blows up when unit vectors in directions of low variance are included in the collection.  In contrast, our test considers a supremum over dimension, and a supremum over all unit vectors in the time dimension, while preserving very weak dependence on dimension. Passing from a finite collection to the entire unit ball poses several technical challenges that require taking the intrinsic functional structure of the data into account.

\end{remark}

Before proceeding to the power analysis, we briefly comment on the main proof ideas for Theorem~\ref{main:thm}. A detailed proof is provided in the supplement.
Let $\{G_{k,j}: k=1,\dots,K,j \geq 1\}$ denote a collection of centered normal random variables with the same covariance structure as the collection of KL scores $\{\phi_{j}^{\tilde f_k}: k=1,\dots,K,j \geq 1\}$. The main idea of the proof consists in carefully bounding the errors in the following sequence of approximations under the global null $H^{over-all}_{0}$ in \eqref{eq:h0global}
\begin{align}
T_n 
&\approx \max_{k=1}^K \int_0^1 \Big(\frac{1}{\sqrt{n}}\sum_{i=1}^n \tilde f_i^k(t) \Big)^2 dt 
= \max_{k=1}^K \sum_{j=1}^\infty \Big(\frac{1}{\sqrt{n}}\sum_{i=1}^n \phi^{\tilde{f}_{k,i}}_{j} \Big)^2 \label{eq:approx1}
\\
&\approx 
\max_{k=1}^K \sum_{j=1}^M \Big(\frac{1}{\sqrt{n}}\sum_{i=1}^n \phi^{\tilde{f}_{k}}_{j,i} \Big)^2 \label{eq:approx2}
\\
& \stackrel{\mathcal{D}}{\approx} \max_{k=1}^K \sum_{j=1}^M G_{k,j}^2 \label{eq:approx3}
\\
& \stackrel{\mathcal{D}}{\approx} 
\max_{k=1}^K \sum_{j=1}^M \Big(\frac{1}{\sqrt{n}}\sum_{i=1}^n e_i [\phi^{\tilde{f}_{k}}_{j,i} - \bar \phi^{\tilde{f}_{k}}_{j,\cdot}] \Big)^2 \label{eq:approx4}
\\ 
& \approx
\max_{k=1}^K \sum_{j=1}^\infty \Big(\frac{1}{\sqrt{n}}\sum_{i=1}^n e_i [\phi^{\tilde{f}_{k}}_{j,i} - \bar \phi^{\tilde{f}_{k}}_{j,\cdot}] \Big)^2 \label{eq:approx5}
\\
& = \int_0^1 \Big(\frac{1}{\sqrt{n}}\sum_{i=1}^n e_i [\tilde f_i^k(t) - \bar f_\cdot^k(t)] \Big)^2 \approx T_n(e). \label{eq:approx6}
\end{align}
Here, $\stackrel{\mathcal{D}}{\approx}$ indicates approximate equality in distribution and $\phi^{\tilde{f}^{k}}_{j,i}$ denotes the $j$'th KL score of $\tilde{f}^k_{i}$. 

The approximations in~\eqref{eq:approx1} and \eqref{eq:approx6} utilize assumptions (A1) and result in a term involving $T^{-q}$ in the rates $d(n,K,T,N), \tilde d(n,K,T,N)$. If full functional samples are available and the corresponding integrals can be computed directly, those terms could be dropped from the final bound. If a different approximation than the simple Riemann approximation we utilize is used, $T^{-q}$ in the bounds should be replaced by the corresponding approximation errors. We note that the factors in front of $T^{-q}$ are due to Gaussian anti-concentration and are independent of the method used for approximating the integral. 

The approximations in~\eqref{eq:approx2} and \eqref{eq:approx5} are from truncating infinite sums. Those truncations result in an error that depends on $M$ and the coefficient decay of the KL scores from Assumption (A2). We note that larger $M$ leads to smaller errors and that $M$ appears as a technical tool in the proof and is not a tuning parameter  of the statistical procedure itself. To implement the test, choosing $M$ is not required.

In~\eqref{eq:approx3} we use new high-dimensional Gaussian approximation results for max-$L^2$ norms which we develop in the Appendix. Those results are one of the major challenges in the proof and do not follow from existing findings. While some key steps in the proofs rely on ideas from~\cite{CCK17}, several important steps are different. First,~\cite{CCK17} only consider sparsely convex sets which allow to derive Gaussian approximation of maxima of $L^2$ statistics where the dimension of each $L^2$ statistic is fixed while we need $M$ to grow in order to control the approximation errors in~\eqref{eq:approx2},~\eqref{eq:approx5}. Second, the results in~\cite{CCK17}, and especially their anti-concentration results, assume a fixed lower bound on the variances of all terms in the $L^2$ statistics, which in our case correspond to the KL scores $\phi_{j,i}^{\tilde f_k}.$ To remove this dependence, we derive new anti-concentration results for maxima of high-dimensional Gaussian vectors and for maxima of $L^2$ statistics of such vectors when the variance is allowed to decay to zero. Assumption (A4) allows us to derive a sharper anti-concentration result, leading to a relaxed assumption on $\alpha$ under (A4). 

Finally, the approximation in~\eqref{eq:approx4} compares a mixed $L^2$-max norm of the Gaussian vector $\oplus_{k=1}^K \oplus_{j=1}^M G_{k,j}$ to that of $\oplus_{k=1}^K \oplus_{j=1}^M \Big(\frac{1}{\sqrt{n}}\sum_{i=1}^n e_i [\phi^{\tilde{f}_{k}}_{j,i} - \bar \phi^{\tilde{f}_{k}}_{j,\cdot}] \Big)$ which is Gaussian conditional on the data. Here, we extend the Gaussian comparison results in \cite{CCK17} to the case of $L^2$ statistics with increasing dimensions and without variance lower bounds.

The terms in $d, \tilde d$ which involve $n^{\frac{1-\alpha}{6\alpha-1}}$ and $n^{\frac{1/2-\alpha}{6\alpha-1}}$, respectively, result from~\eqref{eq:approx3} and~\eqref{eq:approx4}. The bounds in~\eqref{eq:approx3} and~\eqref{eq:approx4} are increasing in $M$ while~\eqref{eq:approx2} and \eqref{eq:approx5} result in bounds that are decreasing in $M$. Balancing those errors by making the optimal choice of $M$ leads to the final bound on the convergence rate. 

We next discuss the power properties of the proposed bootstrap test. 

\begin{theorem}[Power for detecting deviations at individual channels]\label{main:thm-power}
	$\newline$
Let $T^G := \max_{k=1}^K \frac{1}{T}\sum_{\ell=1}^T [G_n^{k,\ell}]^2$, where $G_n \in (\R^T)^K$ is a centered Gaussian with the same covariance as $\oplus_{k=1}^K \oplus_{\ell=1}^T f^k_1(t_\ell)$. Let $q_G$ be the quantile function of $T^G$. Let 
\[
\mu^{k}= \sqrt{\frac{1}{T}\sum_{l=1}^{T}\big(\E[f^k_1(t_{l})-g^k_1(t_{l})]\big)^2}.
\]
Fix $\gamma \in (0,1)$ and $\eps \in (0, \gamma)$. There exists a constant $B:=B(\eps, \gamma)$ (that is independent of the distribution of $f^{d}_1$ and all other parameters) and a constant $C_{\gamma,\eps}$ (that depends only on the distribution of $\tilde f$ through $\tilde f^k$ and only the distribution of $\tilde f^k$ only through the model parameters of (A1)-(A3))  such that  if $n\geq 3$,
\begin{equation}\label{th_power:main_equation}
	\Pr\Big(T^k_n \geq \hat q_{1-\gamma}\Big) \geq 1-2\exp\Big(-\frac{1}{B}(\sqrt{n}\mu^{k}-\sqrt{q_{G}(1-\gamma+\eps)})_{+}^{2} \Big)-C_{\gamma,\eps}d_n.
\end{equation}
If instead (A1)-(A4) holds the same statement holds with  possibly different values of $B$ and $C_{\gamma, \eps}$ and $d$ replaced by $\tilde d_n$
\end{theorem}

Under (A1)--(A3) we always have $q_G = O(\sqrt{\log (K+1)})$ by Lemma~ \ref{lem:tail-bound-from-trace-bound}. Assuming that $T$ is sufficiently large, the term $\mu^k$ can be replaced by the simpler expression $\sqrt{\int_0^1 \mu_k^2(t) dt}$. The asymptotic power of our test is thus governed by an $L^2$-type deviation of the functional mean from zero. This can be contrasted with the power of max-type tests for which the power is governed by $\sup_{t \in [0,1]} |\mu_k(t)|$. If the functional mean deviates from zero over an extended time interval, as is the case in many applications, the proposed max-$L^2$ test can thus expected to have better power. 

The leading term in the power of our test does not have an explicit dependence on dimension. Rather, it depends on the quantile function $q_G$ which adapts to the dependence structure of the data. To illustrate this, consider the extreme case where the functional data are perfectly dependent at different $k$. In this case all vectors $\oplus_{\ell=1}^T G_n^{k,\ell}$ are also perfectly dependent, and $T^G$ behaves like a univariate quantity with quantile that is independent of $K$. In contrast, simple multiple testing corrections such as the Bonferroni procedure are not able to capture this kind of dependence in the data and will not adapt to the dependence across different values of the dimension $K$. While such an extreme scenario is unrealistic in practice, the quantiles of $T^G$ will also adapt to more subtle dependence structures across $k$.

\section{Simulation Study}\label{Simulation_Study} \label{sec:sim}

In this section we evaluate the performance of the proposed method with a simulation study. We also compare the proposed method to paired two-sample versions of the methods of \cite{xue2023distribution} and \cite{YX}. We refer to these three methods as \Proposed for the proposed method, \Xue for \cite{xue2023distribution}, and \XY for \cite{YX}.  

For fair comparison, we largely follow \cite{xue2023distribution} for the basic simulation set-up. For each $1 \leq k \leq K$ we generate $n$ independent differenced functional realizations, denoted by $\{d^k_i\}_{i=1}^{n}$, where each $d^k_i$ is a function defined on $I=[0,1]$. Let $\phi_1 = 1$, and for $2 \leq v \leq 25$, let $\phi_{2v-1}(t):= \sqrt{2}\sin((v-1)\pi(2t-1))$, and for $1 \leq v \leq 25$, let $\phi_{2v}(t):=\sqrt{2}\cos(v\pi(2t-1))$, for $ t \in [0,1]$. The functions $\{\phi_v\}_{v=1}^{50}$ form an orthonormal collection. Define a $K\times K$ matrix $\Sigma$ by $[\Sigma_{\rho}]_{ij}: = \rho +(1-\rho)\delta_{ij}$, where $\delta_{ij}=1$ if $i =j $ and $0$ if $i \neq j$. We set $K=80$ in all simulations.

Under the global null hypothesis $H_0^{global}$, we generate the realizations $\{d^k_i\}_{i=1}^{n}$ in two stages. First we define 
\begin{equation}\label{def:independent_noise}
v^k_i=\sum_{j=1}^{50}j^{-\alpha}\phi_{\sigma(j)}g_{ikj}.
\end{equation}
Here, $\sigma$ is a  random permutation of the set $\{1, \dots, 50\}$, which is fixed for all Monte Carlo runs. The collection ${g_{ikj}}_{i=1, k=1, j=1}^{n, K, 50}$ consists of independent and identically distributed random variables generated from one of the following distributions: (i) standard normal; (ii) standardized $\chi^2(1)$, with mean zero and variance one. We then define 
\begin{equation}
d^k_i=\Big[\Sigma^{1/2}_{\rho}[v^1_i, \dots, v^K_i]^{\top}\Big]_k,
\end{equation}
for $\rho \in [0,1)$. The correlation parameter $\rho$ dictates the degree of dependence across the functional components of the $K$ dimensional functional realization $\big[d^1_i, \dots d^K_i\big]^{\top}$.

We consider alternatives from the parametrized family 
\begin{equation}\label{sine_definition}
\mu^k(s, \delta) = \delta\mathbf{1}\{k \leq  \lfloor 80s \rfloor\}.
\end{equation}
The sparsity parameter $s$ controls the proportion of the $K$ functional components that have non-zero mean. The signal size parameter $\delta$ controls the size of the mean on components where it is non-zero. Under an alternative (i.e., for a fixed set of nonzero $s$ and $\delta$), we generate each $d^k_i$ as

\begin{equation}\label{simulated_distribtuion_alternatives}
d^k_i=\Big[\Sigma^{1/2}_{\rho}[v^1_i, \dots, v^K_i]^{\top}\Big]_k+\mu^k(s, \delta).
\end{equation}

For the actual observations, we assume them to be realizations of $d^k_i(\cdot)$ at $T=300$ equally spaced time points on $I=[0,1]$, i.e., \(
Z_i^{k,\ell} = d^k_i(\frac \ell T).
\) for $ l=1, 2, \ldots, T$. 

In all simulations, we set $\alpha=0.55$ in equation~\eqref{def:independent_noise}. We consider $\rho =0, 0.3, 0.6, 0.9$ and $n=100, 150$. We use $5000$ Monte-Carlo runs and $N=500$ bootstrap resamples when estimating the quantile of the global test-statistic of \eqref{global_hypothesis_test_stat} as described in \eqref{test:ecdf}. We consider sparsity parameter $s=0.1, 0.3, 0.6, 0.9$ and signal sizes $\delta= 0, .05, .1, .15, .2, .25, .3, .35, .4$.

\subsection{Results}
\subsubsection{Type I Error}
Table \ref{sim_results_level} presents the empirical Type I Error of \Proposed, \Xue, and \XY  
when used to test the hypothesis $H^{global}_0$ of \eqref{eq:h0global} at nominal level $\gamma=.05$.  Higher values of $\rho$ correspond to more dependent marginal functional observations ($\rho=0$ corresponds to the fully independent case). As a general trend, the empirical Type I Error becomes closer to the nominal Type I Error of $5\%$ as $\rho$ increases, reflecting the fact that when the marginal functional observations are more dependent, the bootstrap statistic used to approximate the rejection has weaker dependence on the dimension $K$.  

\Xue has empirical Type I error rates that are closer to the nominal levels than \Proposed  across most settings presented in Table 4.1.1.   This is expected, since \Xue  can be thought of as a finite dimensional approximation to \Proposed based on projection, as can be seen by comparing \eqref{proposed_test_stat} and \eqref{Xue_test_stat}; the test statistic used in \Xue is effectively lower dimenisional and therefore the Gaussian multiplier bootstrap approximation errors that result in the gap between the nominal and empirical Type I Error are smaller.

\begin{table}
\centering

\begin{tabular}{|l|ccc|ccc|}
\hline
\textbf{Gaussian Noise} & \multicolumn{3}{c|}{$n=100$} & \multicolumn{3}{c|}{$n=150$} \\
\hline
$\underline{\rho}$ & Max & Proposed & Projection & Max & Proposed & Projection \\ 
\hline
0.0 & 3.54 & 3.60 & 4.36 & 3.54 & 4.76 & 5.46 \\
0.3 & 3.50 & 4.32 & 5.00 & 4.12 & 4.46 & 4.54 \\
0.6 & 3.78 & 4.50 & 4.76 & 3.96 & 4.98 & 5.24 \\
0.9 & 4.26 & 4.94 & 5.10 & 5.40 & 5.40 & 5.32 \\
\hline \hline
\textbf{$\chi^2(1)$ Noise} & \multicolumn{3}{c|}{$n=100$} & \multicolumn{3}{c|}{$n=150$} \\
\hline
$\underline{\rho}$ & Max & Proposed & Projection & Max & Proposed & Projection \\
\hline
0.0 & 2.68 & 2.44 & 2.86 & 3.80 & 3.38 & 3.70 \\
0.3 & 2.94 & 3.04 & 3.04 & 3.40 & 3.58 & 3.86 \\
0.6 & 3.40 & 3.56 & 3.70 & 4.44 & 3.68 & 3.66 \\
0.9 & 4.68 & 4.56 & 4.92 & 4.32 & 4.52 & 4.86 \\
\hline
\end{tabular}
\caption{Empirical level of \Proposed, \Xue, and \XY at nominal level $5\%$, when the data is generated according to \eqref{simulated_distribtuion_alternatives} with $\delta = 0$ for a range of correlation parameters $\rho$. Both Gaussian and Sub-Exponential noises are considered. }\label{sim_results_level}
\end{table}

\subsubsection{Power}\label{sec:simspower}
\begin{figure}[H]
    \centering

    \begin{minipage}{0.45\textwidth}
        \centering
        \includegraphics[width=\linewidth]{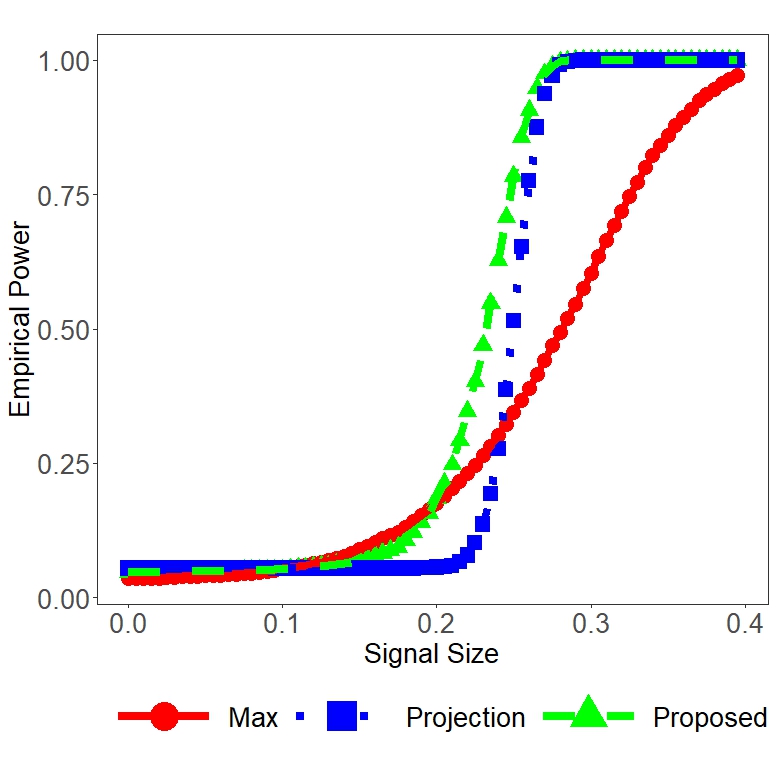}
        \\ \small Parameters $s=0.1$, $\rho=0$
    \end{minipage}
   \hspace{1cm} %\hfill
    \begin{minipage}{0.45\textwidth}
        \centering
        \includegraphics[width=\linewidth]{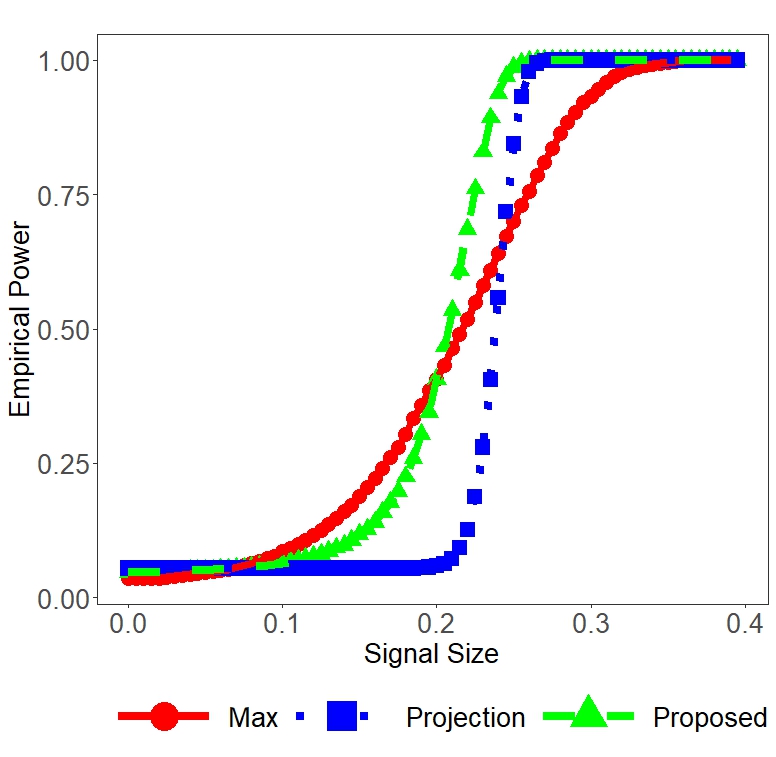}
        \\ \small Parameters $s=0.3$, $\rho=0$
    \end{minipage}
%\hfill
    \begin{minipage}{0.45\textwidth}
        \centering
        \includegraphics[width=\linewidth]{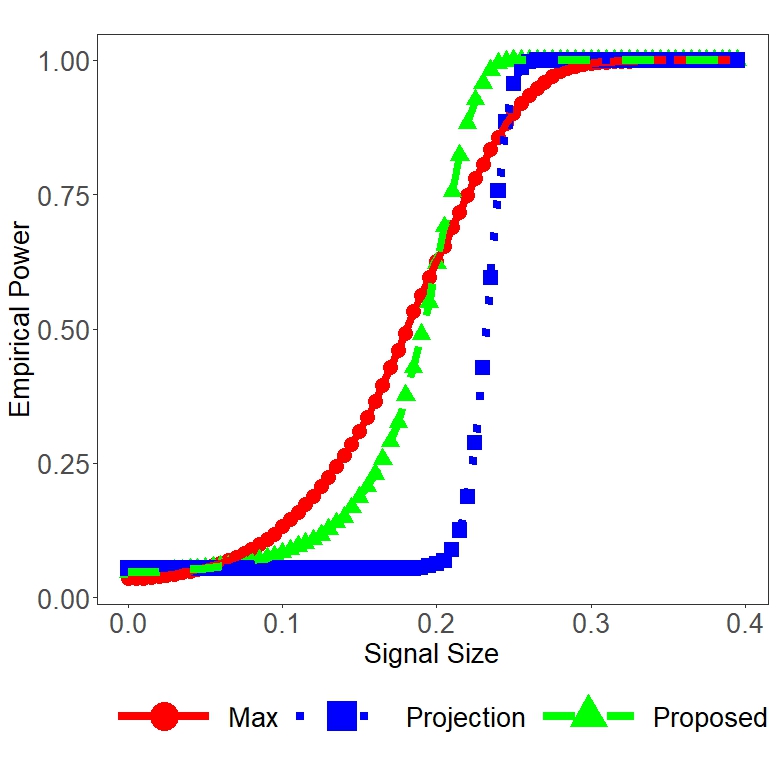}
        \\ \small Parameters $s=0.6$, $\rho=0$
    \end{minipage}
\hspace{1cm} % \hfill
      \begin{minipage}{0.45\textwidth}
        \centering
        \includegraphics[width=\linewidth]{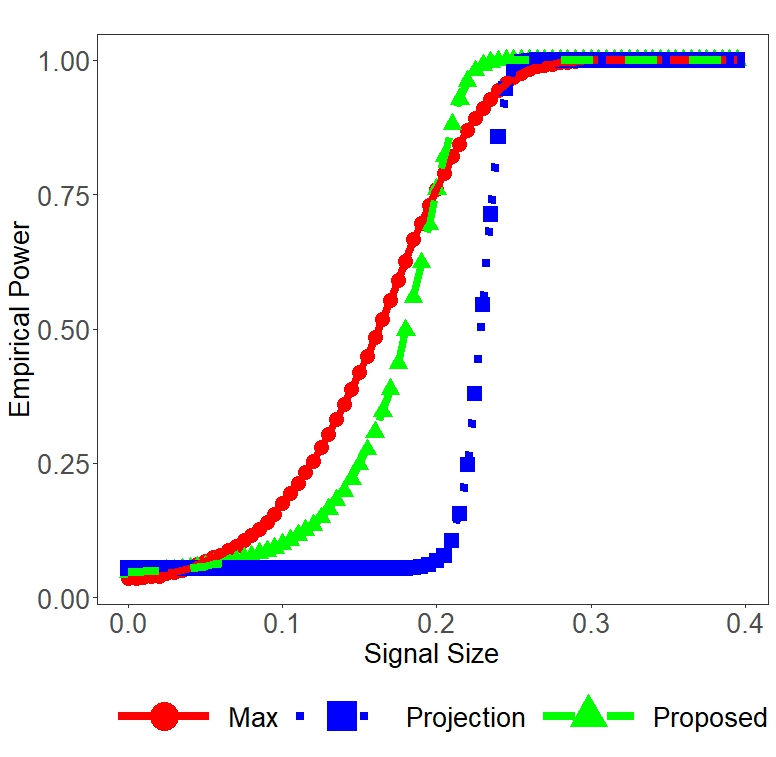}
        \\ \small Parameters $s=0.9$, $\rho=0$
    \end{minipage}

    \caption{
    {Empirical power of \Proposed, \Xue, and \XY, with data generated according to \eqref{simulated_distribtuion_alternatives} against alternatives from the family \eqref{sine_definition} with $s, \rho$ fixed to the particular value indicated in the sub-caption, and $\delta$, the signal size, shown on the horizontal axis.} }
    \label{fig:side_by_side}
    \end{figure}

Figure \ref{fig:side_by_side} shows empirical power curves of \Proposed, \XY and \Xue against the global null as a function of the signal-size parameter $\delta$ specified in \eqref{sine_definition} when the correlation parameter is set at $\rho=0$, the sample size is set at $n=150$, and the functional noise is generated with centered and unit variance Gaussian noise in the basis expansion, i.e, the scalars  $g_{ikj}$ in \eqref{def:independent_noise} are standard Gaussian variates and $\alpha = 0.55$. The sparsity parameter $s$ specified in \eqref{sine_definition} varies from sub-plot to sub-plot as indicated in the respective captions. We consider two settings for $\rho$: $\rho = 0$, which corresponds to complete independence across channels, see Figure~\ref{fig:side_by_side}, and $\rho = 0.9$, corresponding to a substantial dependence. The intermediate cases $\rho = 0.3, 0.6$ are presented in the Supplement; see Section~\ref{sec:additional_sims}, Figures~\ref{fig:side_by_side_.3}, \ref{fig:side_by_side_.6}.

First consider the case of complete independence, Figure~\ref{fig:side_by_side}. The three power curves show different rates of increase as the signal gets stronger. \XY has the slowest increase, but starts to have a non-trivial power earlier than the other tests. It has higher power than \Proposed for smaller signal strength, but lower power for larger signals due to a steeper increase in the power of \Proposed. \Xue has the highest rate of increase, but starts to have non-trivial power for larger signals and its power is dominated by \Proposed for all settings we considered. Increasing the sparsity parameter, which corresponds to signals at more channels, increases the power of all methods. The power of \XY is increased by the largest amount, leading to a higher advantage for small signal strengths. For larger signal strengths \Proposed still has the highest power.

\begin{figure}[h]
    \centering

    \begin{minipage}{0.45\textwidth}
        \centering
        \includegraphics[width=\linewidth]{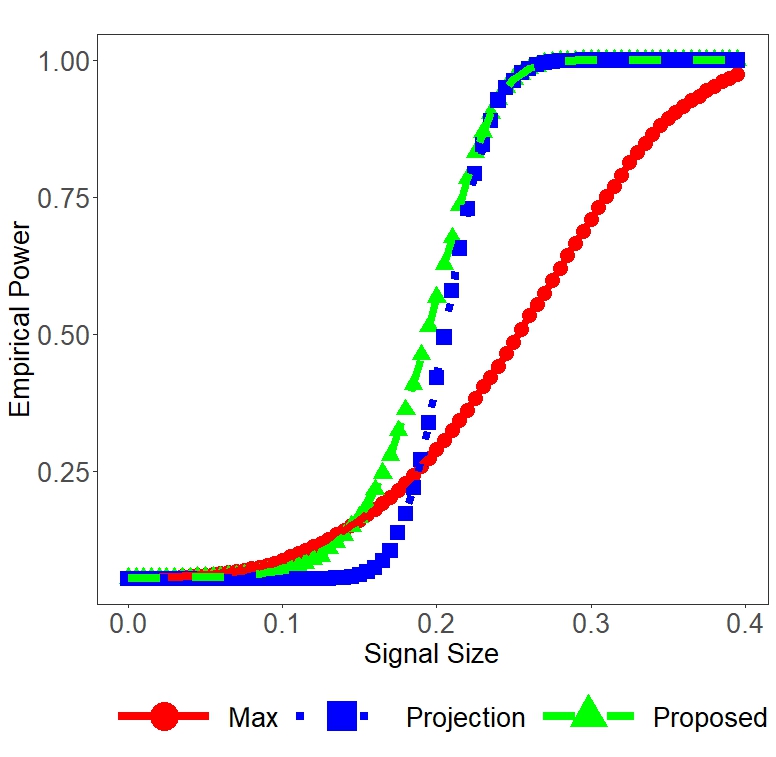}
        \\ \small Parameters $s=0.1$, $\rho=.9$
    \end{minipage}
    \hspace{1cm} %\hfill
    \begin{minipage}{0.45\textwidth}
        \centering
        \includegraphics[width=\linewidth]{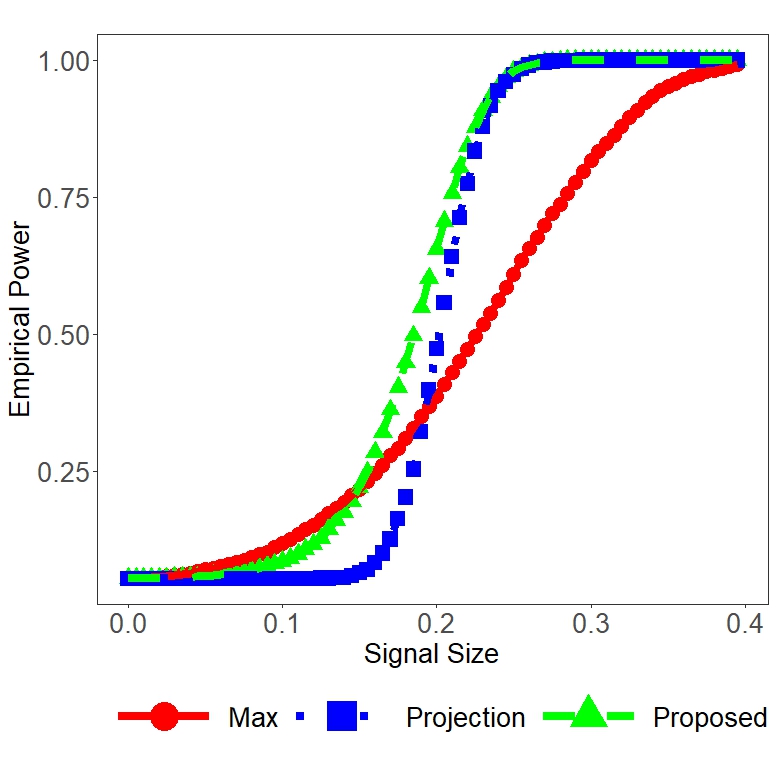}
        \\ \small Parameters $s=0.3$, $\rho=0.9$
    \end{minipage}
    %\hfill
    \begin{minipage}{0.45\textwidth}
        \centering
        \includegraphics[width=\linewidth]{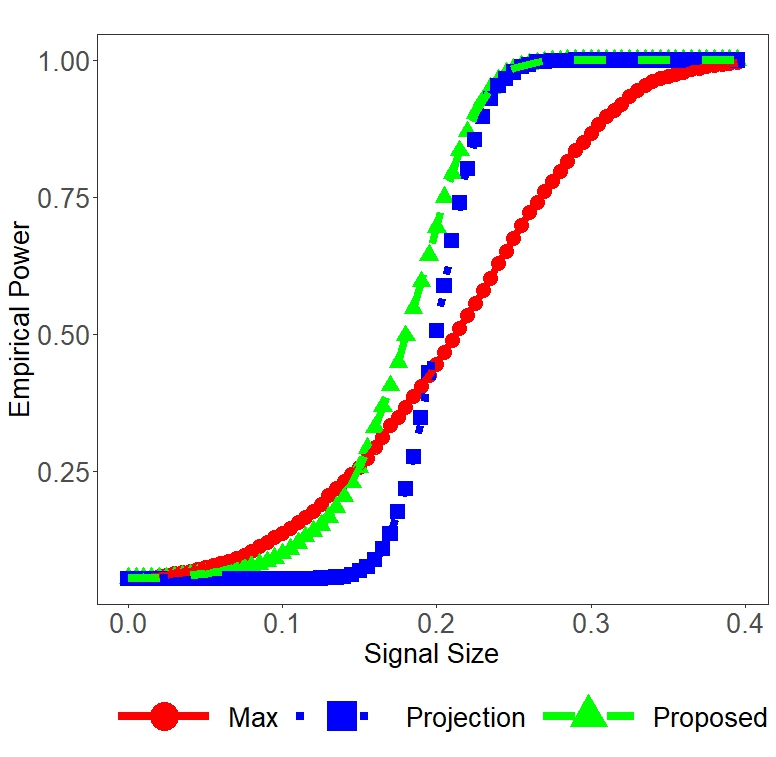}
        \\ \small Parameters $s=0.6$, $\rho=0.9$
    \end{minipage}
    \hspace{1cm}  %\hfill
      \begin{minipage}{0.45\textwidth}
        \centering
        \includegraphics[width=\linewidth]{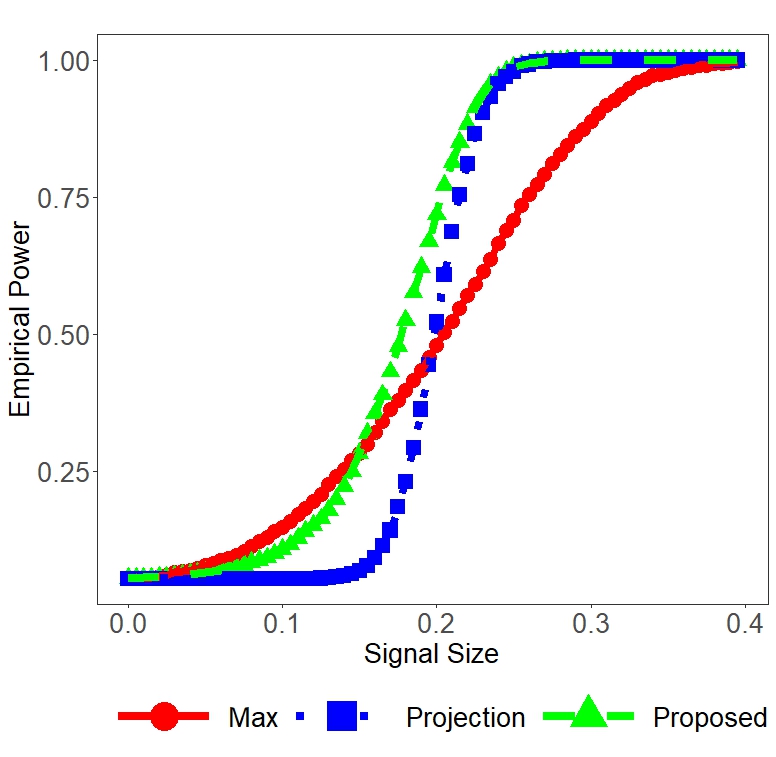}
        \\ \small Parameters $s=0.9$, $\rho=0.9$
    \end{minipage}
    \caption{  {Empirical power of \Proposed, \Xue, and \XY, with data generated according to \eqref{simulated_distribtuion_alternatives} against alternatives from the family \eqref{sine_definition} with $s, \rho$ fixed to the particular value indicated in the sub-caption, and $\delta$, the signal size, shown on the horizontal axis.} 
    }
   \label{fig:side_by_side2}
\end{figure}

For the case of stronger dependence in Figure~\ref{fig:side_by_side2}, all methods exhibit higher power compared to the independence setting. This is intuitive, since stronger dependence makes the maximum of the noise smaller. This demonstrates the adaptivity of the proposed bootstrap procedure to the underlying dependence structure. The gains are largest for \Proposed, and it dominates the other two methods for all but very small signal strengths where \XY has a very slight edge. The results for intermediate values of $\rho = 0.3, 0.6$ in Section~\ref{sec:additional_sims}; Figures~\ref{fig:side_by_side_.3},~ \ref{fig:side_by_side_.6} show a similar trend.

\section{Real Data Application}  \label{One_Sample_Real_World}
\subsection{Background and test on the full sample} 

We apply the proposed method to the task-related cortical surface functional magnetic resonance imaging (fMRI) data from Human Connectome Project (https://www.humanconnectome.org/). We include the data usage acknowledgment in Section \ref{datausage} of the supplementary material. We use the 900 Subjects release that includes behavioral and 3T MR imaging data of healthy adult participants collected between 2012 and 2015. Our analysis includes 841 participants with both motor task and social task evoked fMRI data available. 

Sensory-motor processing is assessed using a task in which participants are presented with visual cues instructing the movement of either the right hand, left hand, right foot, or left foot. Movements are paced with a visual cue, which is presented in a blocked design. This motor task was adapted from the one developed by \citet{buckner2011organization, yeo2011organization}. Participants are presented with visual cues that ask them to either tap their left or right index and thumb fingers or squeeze their left or right toes. Each block of a movement type lasts 12 seconds (10 movements), and is preceded by a 3 second cue. In the run, there are 32 blocks, with 16 of hand movements (8 right and 8 left), and 16 of foot movements (8 right and 8 left). In addition, there are nine 15-second fixation blocks. The motor task evoked fMRI data consist of 284 volumes, i.e., observed at 284 equally spaced time points \citep{HCP}.

Social cognition is assessed using a task in which participants are presented with short (20-second) video clips of objects (squares, circles, triangles) that either interacted in some way, or moved randomly on the screen. These videos were developed either by Castelli and colleagues \citep{castelli2013movement} or Martin and colleagues \citep{wheatley2007understanding}. After each video clip, participants judge whether the objects had a mental interaction (an interaction in which the objects are taking into account each others' thoughts and feelings) or not. Each of the two task runs has five video blocks (two mental, three random in one run, three mental and two random in the other run), and five fixation blocks of 15 seconds each. The social cognition task evoked fMRI data consist of 274 volumes, i.e., observations at 274 evenly spaced time points. 

We utilized the Desikan-Killiany atlas to segment the brain into 68 distinct regions of interest (ROIs), and for each of $n=841$ participants collected 68 functional curves, one at each ROI, representing the blood-oxygen-level dependent (BOLD) signal.  The task-evoked fMRI data for both the social cognition and motor tasks were sampled at the same frequency. To facilitate a direct comparison, we truncated the length of the motor task data from $284$ to $274$ time points, matching the length of social cognition task data. Therefore,  for each participant, we collected $68$ functional curves representing the blood-oxygen-level-dependent (BOLD) signals across the 68 ROIs, and each consisting of 274 time points. We normalized these signals for each subject using the percent signal change method, as recommended by Rainer Goebel 
\footnote{\url{https://www.brainvoyager.com/bv/doc/UsersGuide/StatisticalAnalysis/TimeCourseNormalization.html}}.  

After preprocessing, the datasets for the two tasks are represented as $\{X_i^{k,l}, Y_i^{k,l}\}_{k=1, \ldots, K, l=1, \ldots, T, i=1, \ldots, n}$, with $K=68$ regions, $T=274   $ time points, and $n=841$ subjects, where 
$X_i^{k,\ell} = g_i^k(t_\ell), Y_i^{k,\ell} = h_i^k(t_\ell)$ for $1 \leq \ell \leq T$,  denote the BOLD signal values from the $i^{th}$ subject in region $k$ at time point $\ell$ during the motor and emotion tasks, respectively. {Here, we set {$t_\ell = \ell/T$}.} To visualize the data, we plot in Figure \ref{figure_signal_plots} the post-processed social cognition evoked fMRI time series and the post-processed motor evoked fMRI time series from a randomly selected subject and a randomly selected brain region (the left medial orbitofrontal cortex). In our study, we are exploring whether tasks related to motor functions and social cognition elicit distinct patterns of brain activity across all 68 ROIs. We frame this inquiry by testing the null hypothesis $H^{global}_{0}: \E[h_i^k(t)] = \E[g_i^k(t)] \quad \forall~t \in [0,1] \quad \forall~1\leq k \leq K.$ 

Recall from the previous section that we refer to the proposed method as \Proposed, and the paired two-sample versions of the methods of \cite{xue2023distribution} and \cite{YX} as \Xue and \XY respectively. We also simultaneously test each ($1 \leq k \leq K$) marginal null hypothesis of the form  $H_0^k: \E[h_i^k(t)] = \E[g_i^k(t)], \quad \forall~t \in [0,1]$, controlling the family-wise error using our multiplier bootstrap procedure (see  \eqref{eq:multtest}). To implement \Proposed we use the rejection rules of \eqref{eq:multtest} (to test $H^{global}_0$) and  \eqref{eq:multtest_k} (to test $H^{k}_0$). We use $N=1000$ bootstrap multiplier draws when estimating the distribution of the the bootstrap statistic of \eqref{bootstrap_statistic} according to \eqref{test:ecdf}.

We also implement \XY and \Xue. These tests use test statistics that are functions of the differenced sample mean, see (\eqref{Xue_test_stat}, \eqref{Xue_yao_test_stat}, \eqref{proposed_test_stat}) and a multiplier bootstrap procedure similar to the one used in the proposed method, and in these cases we use $N=1000$ multiplier draws as well. All tests reject the global null with very small p-values $< 0.001$. \Proposed rejects the marginal null hypothesis $H^{k}_{0}$ (and so detects a signal at the $k^{th}$ ROI) for $67$ values of $k$, \XY rejects the marginal null hypothesis $H^{k}_{0}$ for $63$ values of $k$, and \Xue rejects the marginal null hypothesis $H^{k}_{0}$ for $60$ values of $k$. Every  marginal null hypothesis rejected by \Xue or \XY is rejected by \Proposed. In the next section, we thus take a look at power for smaller sample sizes.

\subsection{Power on smaller samples}\label{sec:global-null-subsample}

Since all three tests from the previous section provide overwhelming evidence of differences in brain activity in at least one region of the brain (i.e., the global null hypothesis is rejected with overwhelming evidence) we treat these results as the ground truth, indicating a significant distinction in brain activity between the two tasks.  

\begin{figure}[H]
\centering
\includegraphics[scale=.4]{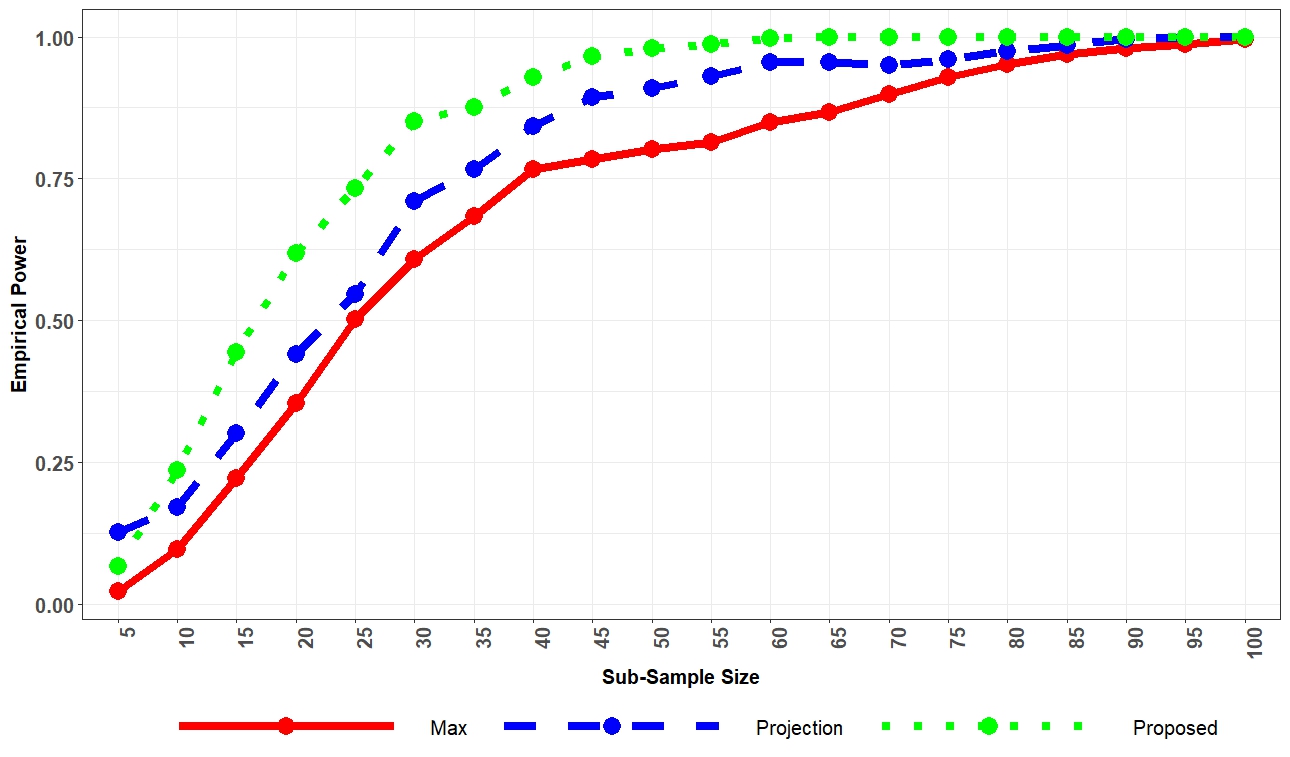}
\caption{Empirical power as described in the beginning of section~\ref{sec:global-null-subsample} for \Proposed (green dotted line), \Xue (blue dashed line) and \XY (red solid line) against sub-sample size $m\in \{5k\}_{k=1}^{20}$.} 
\label{realdata_power_global}
\end{figure}

To evaluate the power for testing against the global null on smaller samples, 
we consider various sub-sample sizes $m\in \{5j: j=1,\dots,20\}$. For each $m$, we draw a sample of size $m$ for $1000$ times and report the proportion of times that the global null hypothesis is rejected by \Proposed as well as by \XY and  \Xue, on each of the sub-samples. The results of these comparisons are depicted in the power curves shown in Figure \ref{realdata_power_global}. The proposed method appears to be more powerful than \Xue and \XY at all considered $ m $ except $m=5$, showing the advantage of the proposed method on a real world data set. We also note that the power of \Xue dominates \XY uniformly across $m$. 

\begin{figure}[H]
\centering
%\ContinuedFloat
\begin{subfigure}[b]{1.0\textwidth}
\includegraphics[scale=0.72]{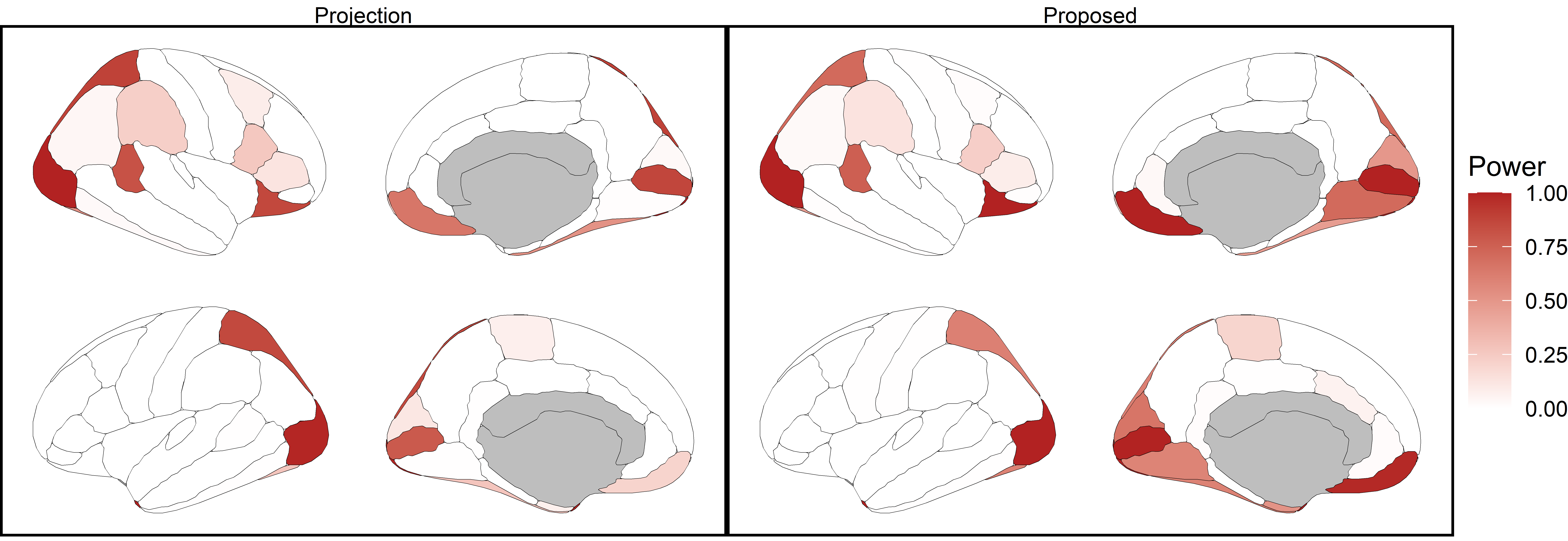} 
\caption{$m=100$. }
\end{subfigure}
\begin{subfigure}[b]{1.0\textwidth}
\includegraphics[scale=0.72]{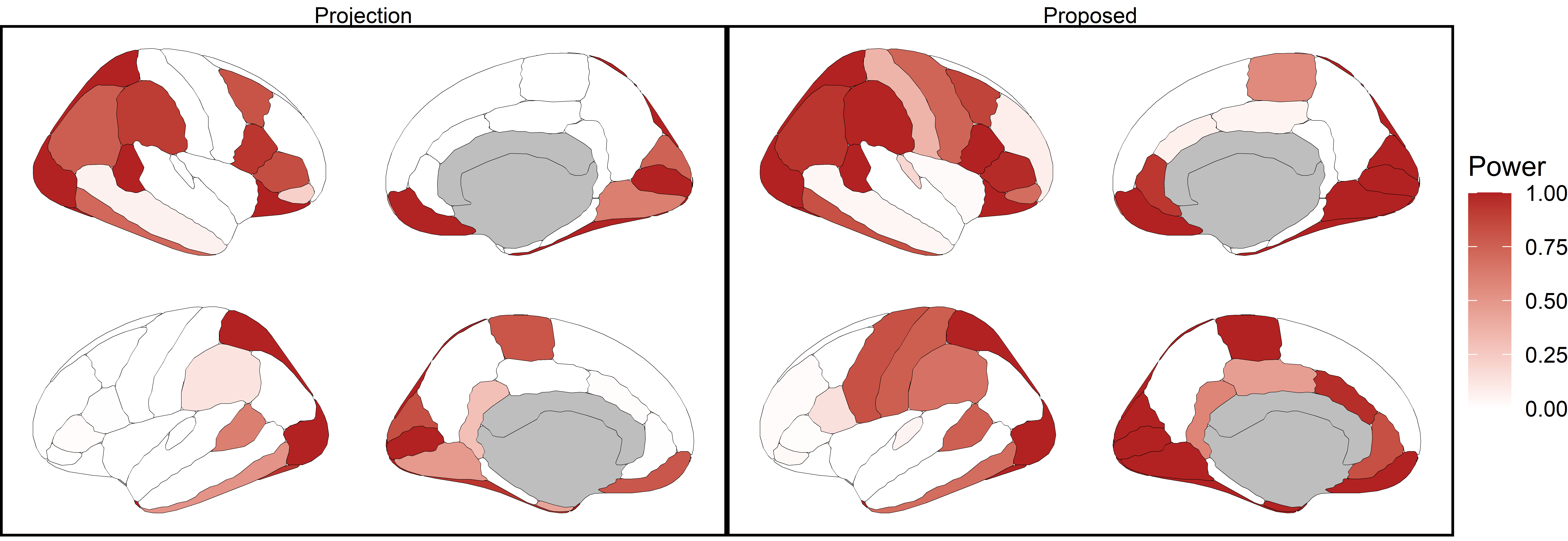} 
\caption{$m=200$. }
\end{subfigure}
\begin{subfigure}[b]{1.0\textwidth}
\includegraphics[scale=0.72]{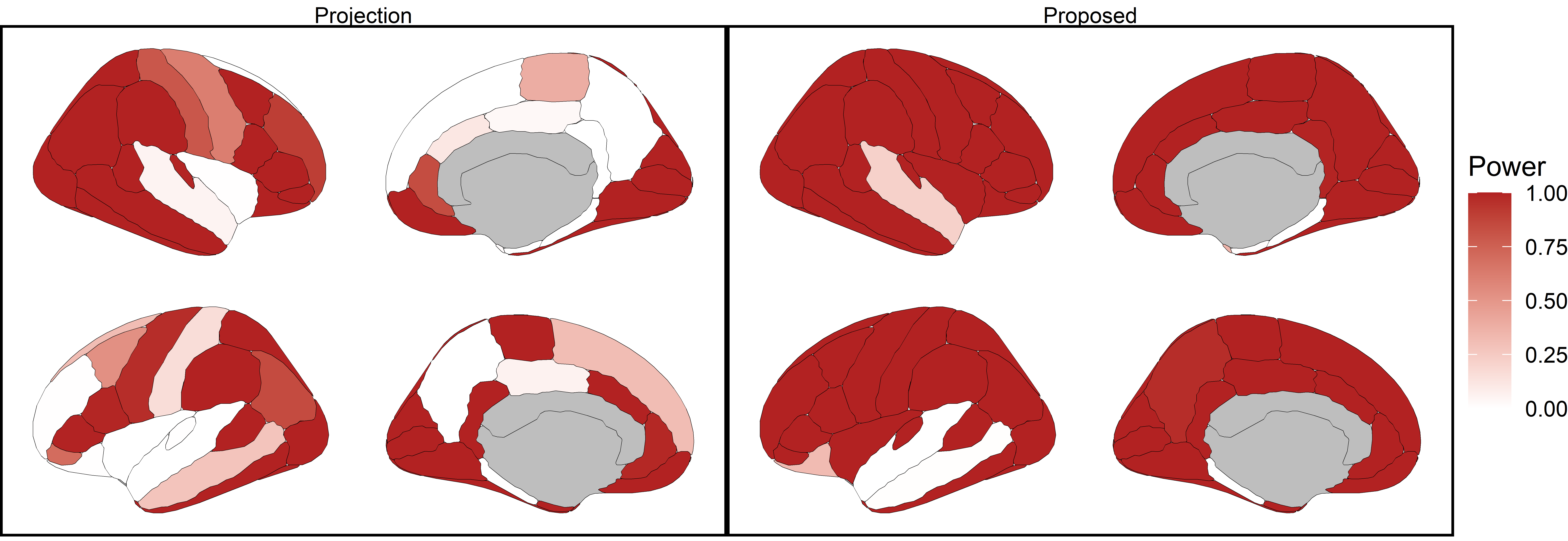} 
\caption{$m=500$. }
\end{subfigure}
\caption{
Empirical power of marginal tests for the null of no difference at every of the $68$ regions of interest with FWER control for $m = 100$ (Panel (a)), $m = 200$ (Panel (b)), $m = 500$ (Panel (c)).
Power is computed as described in the beginning of Section~\ref{sec:global-null-subsample} with nominal level $\gamma = 0.05$. For each subfigure, the right panel corresponds to \Proposed, while the left panel corresponds to \Xue. Gray-shaded regions indicate areas not included in the Desikan-Killiany atlas. }

\label{brain_picture}
\end{figure}

Using all three of \Proposed, \XY, and \Xue  we also simultaneously tested all marginal hypotheses, i.e., we tested for significant differences in mean signals at each of the 68 regions of interest. Due to the increased difficulty of detecting deviations from the null for each individual channel, we consider larger sub-sample-sizes $m \in \{100, 200, 500\}$ in this comparison. The powers of \Proposed and \Xue are shown in Figure \ref{brain_picture}.\footnote{We used the R package ggseg \cite{ggseg} to make the brain plots.} Darker shades of red in Figure \ref{brain_picture} correspond to higher empirical power. The results indicate that \Proposed has higher power at most regions of interest (ROI), but \Xue has higher empirical power on a small percentage of the ROI at each considered sub-sample size. We note that \Xue can be thought of as an approximation to \Proposed, in the sense that the data are first projected onto a collection of pre-defined basis vectors before a supremum is taken.  When the signal happens to align well with one of those basis vectors, \Xue may have higher power, see \eqref{Xue_test_stat}. 

The power of \XY was clearly inferior to the power of the proposed test across all brain regions and sample sizes, and hence we do not display the corresponding results here. For completeness, the corresponding side by side comparison of \XY and the proposed method can be found in Section~\ref{sec:additional_real_data}, Figure~\ref{brain_pic_supp} in the supplement. Overall, the findings regarding power at individual channels largely mirror the power against the global null.

\section*{Acknowledgments}
Dehan Kong and Stanislav Volgushev were partially supported by a discovery grant from NSERC of Canada (Grant RGPIN-2022-04646 to DK and Grant RGPIN-2024-05528
to SV).

\bibliographystyle{Chicago}
\bibliography{comparing_many_functional_means}

\begin{thebibliography}{}

\bibitem[\protect\citeauthoryear{Aue, Rice, and S{\"o}nmez}{Aue
  et~al.}{2018}]{ARS18}
Aue, A., G.~Rice, and O.~S{\"o}nmez (2018).
\newblock Detecting and dating structural breaks in functional data without
  dimension reduction.
\newblock {\em Journal of the Royal Statistical Society Series B: Statistical
  Methodology\/}~{\em 80\/}(3), 509--529.

\bibitem[\protect\citeauthoryear{Ball et~al.}{Ball et~al.}{1997}]{B97}
Ball, K. et~al. (1997).
\newblock An elementary introduction to modern convex geometry.
\newblock {\em Flavors of Geometry\/}~{\em 31\/}(1--58), 26.

\bibitem[\protect\citeauthoryear{Belloni and Oliveira}{Belloni and
  Oliveira}{2018}]{Bel2018}
Belloni, A. and R.~I. Oliveira (2018).
\newblock A high dimensional central limit theorem for martingales, with
  applications to context tree models.
\newblock {\em arXiv preprint arXiv:1809.02741\/}.

\bibitem[\protect\citeauthoryear{Buckner, Krienen, Castellanos, Diaz, and
  Yeo}{Buckner et~al.}{2011}]{buckner2011organization}
Buckner, R.~L., F.~M. Krienen, A.~Castellanos, J.~C. Diaz, and B.~T. Yeo
  (2011).
\newblock The organization of the human cerebellum estimated by intrinsic
  functional connectivity.
\newblock {\em Journal of Neurophysiology\/}~{\em 106\/}(5), 2322--2345.

\bibitem[\protect\citeauthoryear{Castelli, Happ{\'e}, Frith, and
  Frith}{Castelli et~al.}{2013}]{castelli2013movement}
Castelli, F., F.~Happ{\'e}, U.~Frith, and C.~Frith (2013).
\newblock Movement and mind: a functional imaging study of perception and
  interpretation of complex intentional movement patterns.
\newblock In {\em Social Neuroscience}, pp.\  155--169. Psychology Press.

\bibitem[\protect\citeauthoryear{Chen and Qin}{Chen and Qin}{2010}]{CQ10}
Chen, S.~X. and Y.-L. Qin (2010).
\newblock A two-sample test for high-dimensional data with applications to
  gene-set testing1.
\newblock {\em The Annals of Statistics\/}~{\em 38\/}(2), 808--835.

\bibitem[\protect\citeauthoryear{Chernozhukov, Chetverikov, and
  Kato}{Chernozhukov et~al.}{2013}]{chernozhukov2013gaussian}
Chernozhukov, V., D.~Chetverikov, and K.~Kato (2013).
\newblock Gaussian approximations and multiplier bootstrap for maxima of sums
  of high-dimensional random vectors.
\newblock {\em Journal of the American Statistical Association\/}~{\em
  41\/}(6), 2786--2819.

\bibitem[\protect\citeauthoryear{Chernozhukov, Chetverikov, and
  Kato}{Chernozhukov et~al.}{2015}]{CCK12c}
Chernozhukov, V., D.~Chetverikov, and K.~Kato (2015).
\newblock Comparison and anti-concentration bounds for maxima of gaussian
  random vectors.
\newblock {\em Probability Theory and Related Fields\/}~{\em 162}, 47--70.

\bibitem[\protect\citeauthoryear{Chernozhukov, Chetverikov, and
  Kato}{Chernozhukov et~al.}{2017a}]{CCK17}
Chernozhukov, V., D.~Chetverikov, and K.~Kato (2017a).
\newblock Central limit theorems and bootstrap in high dimensions.
\newblock {\em The Annals of Probability\/}~{\em 45\/}(4), 2309--2352.

\bibitem[\protect\citeauthoryear{Chernozhukov, Chetverikov, and
  Kato}{Chernozhukov et~al.}{2017b}]{CCKNaz}
Chernozhukov, V., D.~Chetverikov, and K.~Kato (2017b).
\newblock Detailed proof of {N}azarov's inequality.
\newblock {\em arXiv preprint arXiv:1711.10696\/}.

\bibitem[\protect\citeauthoryear{Chernozhuokov, Chetverikov, Kato, and
  Koike}{Chernozhuokov et~al.}{2022}]{chernozhuokov2022improved}
Chernozhuokov, V., D.~Chetverikov, K.~Kato, and Y.~Koike (2022).
\newblock Improved central limit theorem and bootstrap approximations in high
  dimensions.
\newblock {\em The Annals of Statistics\/}~{\em 50\/}(5), 2562--2586.

\bibitem[\protect\citeauthoryear{Decker}{Decker}{2024}]{Decker24}
Decker, D.~C. (2024).
\newblock {\em Hypothesis Tests for High-Dimensional Functional Data}.
\newblock Ph.\ D. thesis, University of Toronto.

\bibitem[\protect\citeauthoryear{Desikan, S{\'e}gonne, Fischl, Quinn,
  Dickerson, Blacker, Buckner, Dale, Maguire, Hyman, et~al.}{Desikan
  et~al.}{2006}]{desikan2006automated}
Desikan, R.~S., F.~S{\'e}gonne, B.~Fischl, B.~T. Quinn, B.~C. Dickerson,
  D.~Blacker, R.~L. Buckner, A.~M. Dale, R.~P. Maguire, B.~T. Hyman, et~al.
  (2006).
\newblock An automated labeling system for subdividing the human cerebral
  cortex on {MRI} scans into gyral based regions of interest.
\newblock {\em NeuroImage\/}~{\em 31\/}(3), 968--980.

\bibitem[\protect\citeauthoryear{Dette, Kokot, and Aue}{Dette
  et~al.}{2020}]{DKA20}
Dette, H., K.~Kokot, and A.~Aue (2020).
\newblock Functional data analysis in the banach space of continuous functions.
\newblock {\em The Annals of Statistics\/}~{\em 48\/}(2).

\bibitem[\protect\citeauthoryear{Fremdt, Steinebach, Horv{\'a}th, and
  Kokoszka}{Fremdt et~al.}{2013}]{fremdt2013testing}
Fremdt, S., J.~G. Steinebach, L.~Horv{\'a}th, and P.~Kokoszka (2013).
\newblock Testing the equality of covariance operators in functional samples.
\newblock {\em Scandinavian Journal of Statistics\/}~{\em 40\/}(1), 138--152.

\bibitem[\protect\citeauthoryear{Guide}{Guide}{2006}]{CAB}
Guide, A.~H. (2006).
\newblock {\em Infinite dimensional analysis}.
\newblock Springer.

\bibitem[\protect\citeauthoryear{Hall and Van~Keilegom}{Hall and
  Van~Keilegom}{2007}]{HK07}
Hall, P. and I.~Van~Keilegom (2007).
\newblock Two-sample tests in functional data analysis starting from discrete
  data.
\newblock {\em Statistica Sinica\/}, 1511--1531.

\bibitem[\protect\citeauthoryear{Horv{\'a}th, Kokoszka, and Rice}{Horv{\'a}th
  et~al.}{2014}]{horvath2014testing}
Horv{\'a}th, L., P.~Kokoszka, and G.~Rice (2014).
\newblock Testing stationarity of functional time series.
\newblock {\em Journal of Econometrics\/}~{\em 179\/}(1), 66--82.

\bibitem[\protect\citeauthoryear{Hsing and Eubank}{Hsing and
  Eubank}{2015}]{TE15}
Hsing, T. and R.~Eubank (2015).
\newblock {\em Theoretical foundations of functional data analysis, with an
  introduction to linear operators}, Volume 997.
\newblock John Wiley \& Sons.

\bibitem[\protect\citeauthoryear{Kong, Staicu, and Maity}{Kong
  et~al.}{2016}]{kong2016classical}
Kong, D., A.-M. Staicu, and A.~Maity (2016).
\newblock Classical testing in functional linear models.
\newblock {\em Journal of Nonparametric Statistics\/}~{\em 28\/}(4), 813--838.

\bibitem[\protect\citeauthoryear{Krzy{\'s}ko and Smaga}{Krzy{\'s}ko and
  Smaga}{2021}]{krzysko2021two}
Krzy{\'s}ko, M. and {\L}.~Smaga (2021).
\newblock Two-sample tests for functional data using characteristic functions.
\newblock {\em Austrian Journal of Statistics\/}~{\em 50\/}(4), 53--64.

\bibitem[\protect\citeauthoryear{Ledoux and Talagrand}{Ledoux and
  Talagrand}{2013}]{LT19}
Ledoux, M. and M.~Talagrand (2013).
\newblock {\em Probability in Banach Spaces: isoperimetry and processes}.
\newblock Springer Science \& Business Media.

\bibitem[\protect\citeauthoryear{Lin, Lopes, and M{\"u}ller}{Lin
  et~al.}{2023}]{lin2023high}
Lin, Z., M.~E. Lopes, and H.-G. M{\"u}ller (2023).
\newblock High-dimensional {MANOVA} via bootstrapping and its application to
  functional and sparse count data.
\newblock {\em Journal of the American Statistical Association\/}~{\em
  118\/}(541), 177--191.

\bibitem[\protect\citeauthoryear{Lopes, Lin, and M{\"u}ller}{Lopes
  et~al.}{2020}]{lopes2020bootstrapping}
Lopes, M.~E., Z.~Lin, and H.-G. M{\"u}ller (2020).
\newblock Bootstrapping max statistics in high dimensions: Near-parametric
  rates under weak variance decay and application to functional and multinomial
  data.
\newblock {\em The Annals of Statistics\/}.

\bibitem[\protect\citeauthoryear{Massart}{Massart}{1990}]{M90}
Massart, P. (1990).
\newblock The tight constant in the {D}voretzky-{K}iefer-{W}olfowitz
  inequality.
\newblock {\em The Annals of Probability\/}, 1269--1283.

\bibitem[\protect\citeauthoryear{Mowinckel and Vidal-Piñeiro}{Mowinckel and
  Vidal-Piñeiro}{2019}]{ggseg}
Mowinckel, A.~M. and D.~Vidal-Piñeiro (2019).
\newblock Visualisation of brain statistics with {R}-packages ggseg and
  ggseg3d.

\bibitem[\protect\citeauthoryear{Oksendal}{Oksendal}{2013}]{oksendal2013stochastic}
Oksendal, B. (2013).
\newblock {\em Stochastic differential equations: an introduction with
  applications}.
\newblock Springer Science \& Business Media.

\bibitem[\protect\citeauthoryear{Pomann, Staicu, and Ghosh}{Pomann
  et~al.}{2016}]{pomann2016two}
Pomann, G.-M., A.-M. Staicu, and S.~Ghosh (2016).
\newblock A two-sample distribution-free test for functional data with
  application to a diffusion tensor imaging study of multiple sclerosis.
\newblock {\em Journal of the Royal Statistical Society Series C: Applied
  Statistics\/}~{\em 65\/}(3), 395--414.

\bibitem[\protect\citeauthoryear{Spivak}{Spivak}{2006}]{Sp94}
Spivak, M. (2006).
\newblock {\em Calculus}.
\newblock Cambridge University Press.

\bibitem[\protect\citeauthoryear{Tian}{Tian}{2010}]{tian2010functional}
Tian, T.~S. (2010).
\newblock Functional data analysis in brain imaging studies.
\newblock {\em Frontiers in Psychology\/}~{\em 1}, 35.

\bibitem[\protect\citeauthoryear{Van Der~Vaart and Wellner}{Van Der~Vaart and
  Wellner}{2023}]{VW96}
Van Der~Vaart, A.~W. and J.~A. Wellner (2023).
\newblock {\em Weak convergence and empirical processes}.
\newblock Springer Science \& Business Media.

\bibitem[\protect\citeauthoryear{Vershynin}{Vershynin}{2018}]{V18}
Vershynin, R. (2018).
\newblock {\em High-dimensional probability: An introduction with applications
  in data science}, Volume~47.
\newblock Cambridge university press.

\bibitem[\protect\citeauthoryear{Wang}{Wang}{2021}]{wang2021two}
Wang, Q. (2021).
\newblock Two-sample inference for sparse functional data.
\newblock {\em Electronic Journal of Statistics\/}, 1395–1423.

\bibitem[\protect\citeauthoryear{Wheatley, Milleville, and Martin}{Wheatley
  et~al.}{2007}]{wheatley2007understanding}
Wheatley, T., S.~C. Milleville, and A.~Martin (2007).
\newblock Understanding animate agents: distinct roles for the social network
  and mirror system.
\newblock {\em Psychological Science\/}~{\em 18\/}(6), 469--474.

\bibitem[\protect\citeauthoryear{WU-Minn}{WU-Minn}{2015}]{HCP}
WU-Minn, H. (2015).
\newblock 900 subjects data release reference manual.
\newblock {\em URL https://www. humanconnectome. org\/}~{\em 565}.

\bibitem[\protect\citeauthoryear{Wynne and Duncan}{Wynne and
  Duncan}{2022}]{wynne2022kernel}
Wynne, G. and A.~B. Duncan (2022).
\newblock A kernel two-sample test for functional data.
\newblock {\em Journal of Machine Learning Research\/}~{\em 23\/}(73), 1--51.

\bibitem[\protect\citeauthoryear{Xue}{Xue}{2023}]{xue2023distribution}
Xue, K. (2023).
\newblock Distribution/correlation-free test for two-sample means in
  high-dimensional functional data with eigenvalue decay relaxed.
\newblock {\em Science China Mathematics\/}~{\em 66\/}(10), 2337--2346.

\bibitem[\protect\citeauthoryear{Xue and Yao}{Xue and Yao}{2020}]{YX}
Xue, K. and F.~Yao (2020).
\newblock Distribution and correlation-free two-sample test of high dimensional
  means.
\newblock {\em The Annals of Statistics\/}~{\em 48\/}(3), 1304--1328.

\bibitem[\protect\citeauthoryear{Yeo, Krienen, Sepulcre, Sabuncu, Lashkari,
  Hollinshead, Roffman, Smoller, Z{\"o}llei, Polimeni, et~al.}{Yeo
  et~al.}{2011}]{yeo2011organization}
Yeo, B.~T., F.~M. Krienen, J.~Sepulcre, M.~R. Sabuncu, D.~Lashkari,
  M.~Hollinshead, J.~L. Roffman, J.~W. Smoller, L.~Z{\"o}llei, J.~R. Polimeni,
  et~al. (2011).
\newblock The organization of the human cerebral cortex estimated by intrinsic
  functional connectivity.
\newblock {\em Journal of Neurophysiology\/}~{\em 106\/}(3), 1125--1165.

\bibitem[\protect\citeauthoryear{Zhang, Peng, and Zhang}{Zhang
  et~al.}{2010}]{zhang2010two}
Zhang, C., H.~Peng, and J.-T. Zhang (2010).
\newblock Two samples tests for functional data.
\newblock {\em Communications in Statistics—Theory and Methods\/}~{\em
  39\/}(4), 559--578.

\bibitem[\protect\citeauthoryear{Zhang and Chen}{Zhang and
  Chen}{2007}]{zhang2007statistical}
Zhang, J.-T. and J.~Chen (2007).
\newblock Statistical inferences for functional data.
\newblock {\em The Annals of Statistics\/}~{\em 35\/}(3), 1052--1079.

\bibitem[\protect\citeauthoryear{Zhou and Dette}{Zhou and Dette}{2023}]{ZD23}
Zhou, Z. and H.~Dette (2023).
\newblock Statistical inference for high-dimensional panel functional time
  series.
\newblock {\em Journal of the Royal Statistical Society Series B: Statistical
  Methodology\/}~{\em 85\/}(2), 523--549.

\end{thebibliography}

\newpage
\section*{Supplementary Material}

\setcounter{subsection}{0}
\renewcommand{\thesubsection}{S.\arabic{subsection}}

\subsection{Additional Simulation Results}\label{sec:additional_sims}

\begin{figure}[H]
    \centering

    \begin{minipage}{0.45\textwidth}
        \centering
        \includegraphics[width=\linewidth]{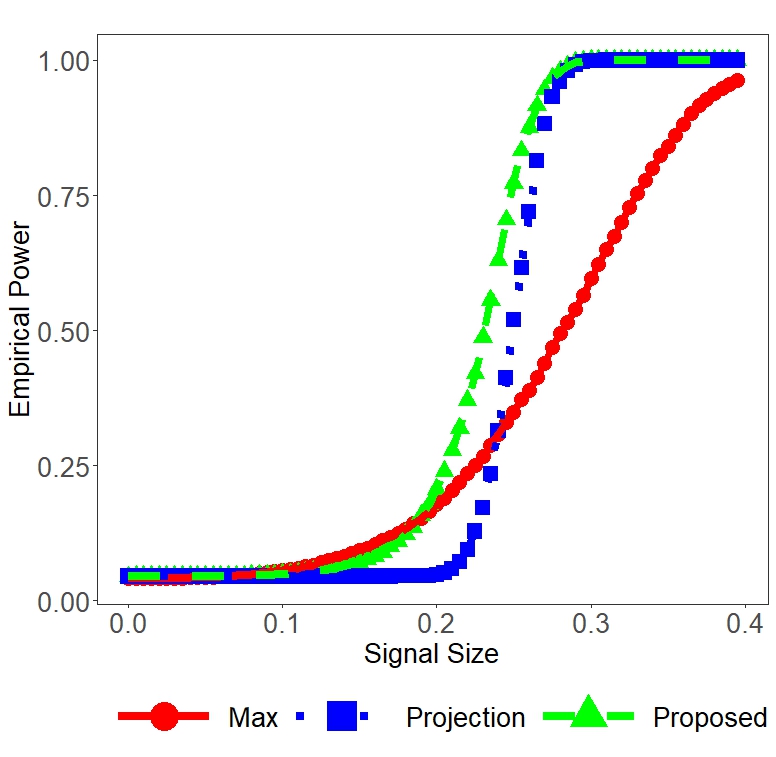}
        \\ \small Parameters $s=0.1$, $\rho=.3$
    \end{minipage}
    \hspace{1cm} % \hfill
    \begin{minipage}{0.45\textwidth}
        \centering
        \includegraphics[width=\linewidth]{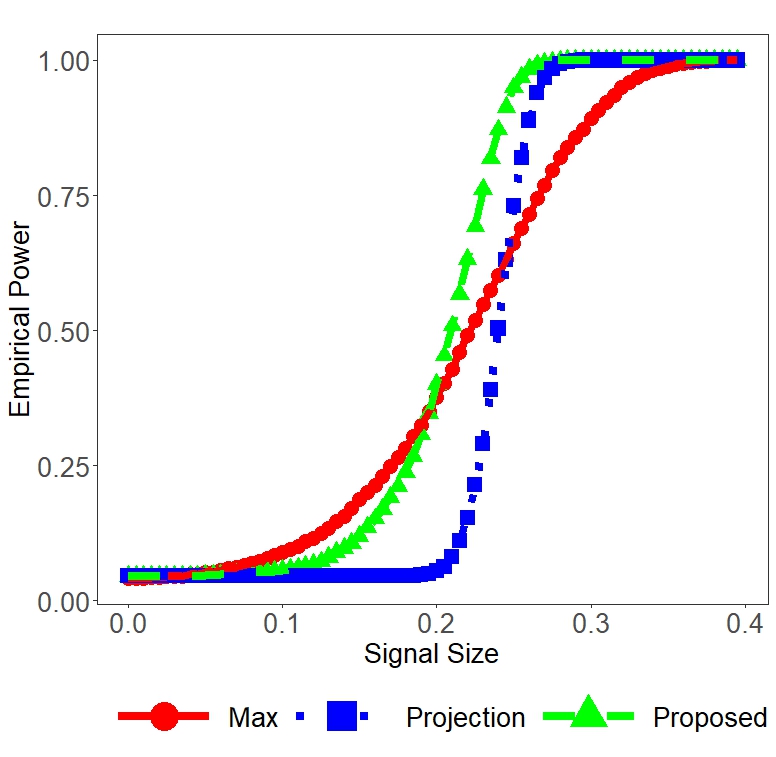}
        \\ \small Parameters $s=0.3$, $\rho=0.3$
    \end{minipage}
   \hspace{1cm} % \hfill
    \begin{minipage}{0.45\textwidth}
        \centering
        \includegraphics[width=\linewidth]{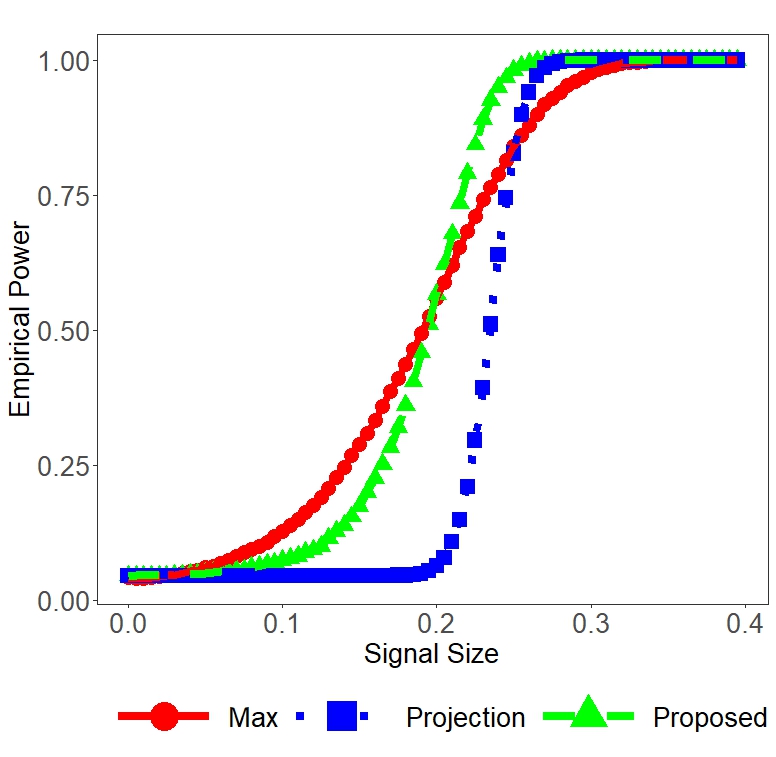}
        \\ \small Parameters $s=0.6$, $\rho=0.3$
    \end{minipage}
    \hspace{1cm} % \hfill
      \begin{minipage}{0.45\textwidth}
        \centering
        \includegraphics[width=\linewidth]{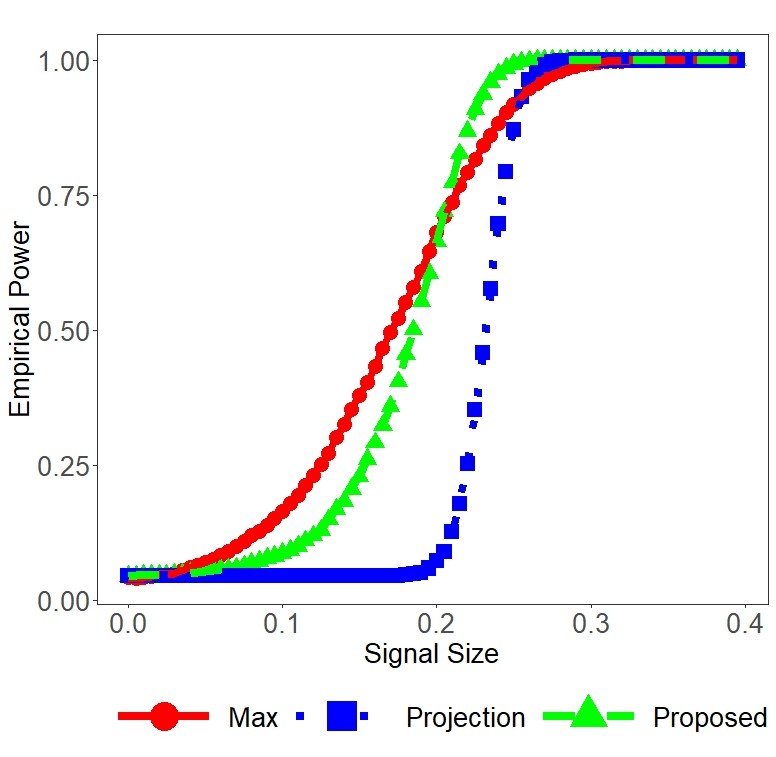}
        \\ \small Parameters $s=0.9$, $\rho=0.3$
    \end{minipage}
    \caption{Empirical power of \Proposed, \Xue, and \XY, with data generated according to \eqref{simulated_distribtuion_alternatives} against alternatives from the family \eqref{sine_definition} with $s, \rho$ fixed to the particular value indicated in the sub-caption, and $\delta$, the signal size, shown on the horizontal axis. Compare with Figure \ref{fig:side_by_side}. }
    
    \label{fig:side_by_side_.3}
   
\end{figure}

\begin{figure}[H]
    \centering

    \begin{minipage}{0.45\textwidth}
        \centering
        \includegraphics[width=\linewidth]{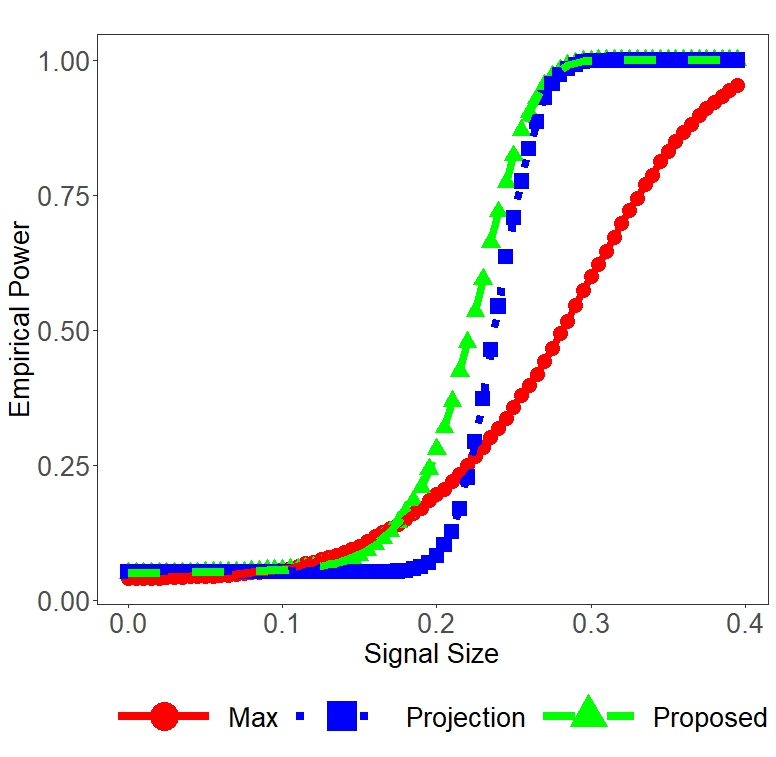}
        \\ \small Parameters $s=0.1$, $\rho=.6$
    \end{minipage}
    \hspace{1cm} % \hfill
    \begin{minipage}{0.45\textwidth}
        \centering
        \includegraphics[width=\linewidth]{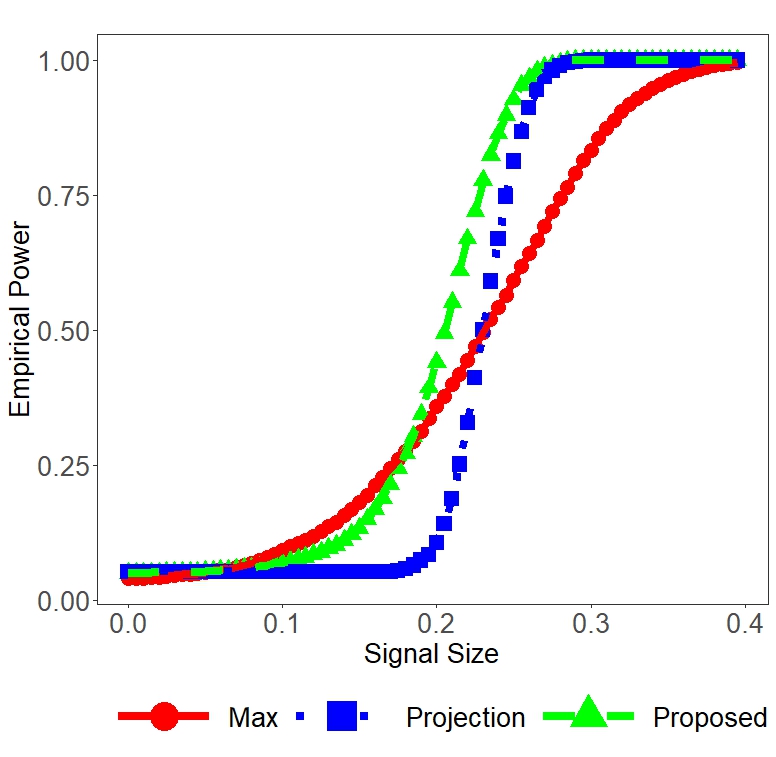}
        \\ \small Parameters $s=0.3$, $\rho=0.6$
    \end{minipage}
    \hspace{1cm} % \hfill
    \begin{minipage}{0.45\textwidth}
        \centering
        \includegraphics[width=\linewidth]{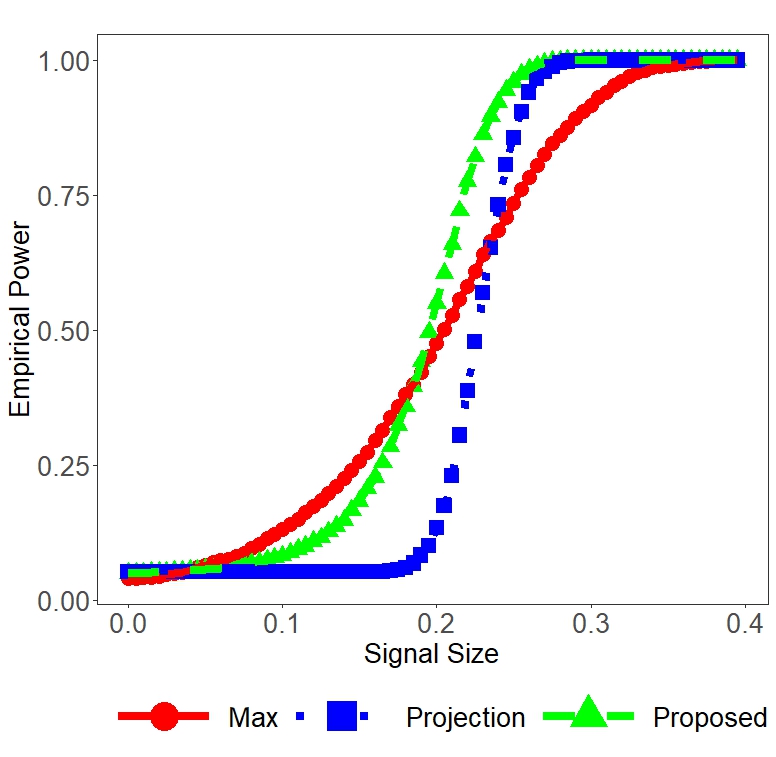}
        \\ \small Parameters $s=0.6$, $\rho=0.6$
    \end{minipage}
    \hspace{1cm} % \hfill
      \begin{minipage}{0.45\textwidth}
        \centering
        \includegraphics[width=\linewidth]{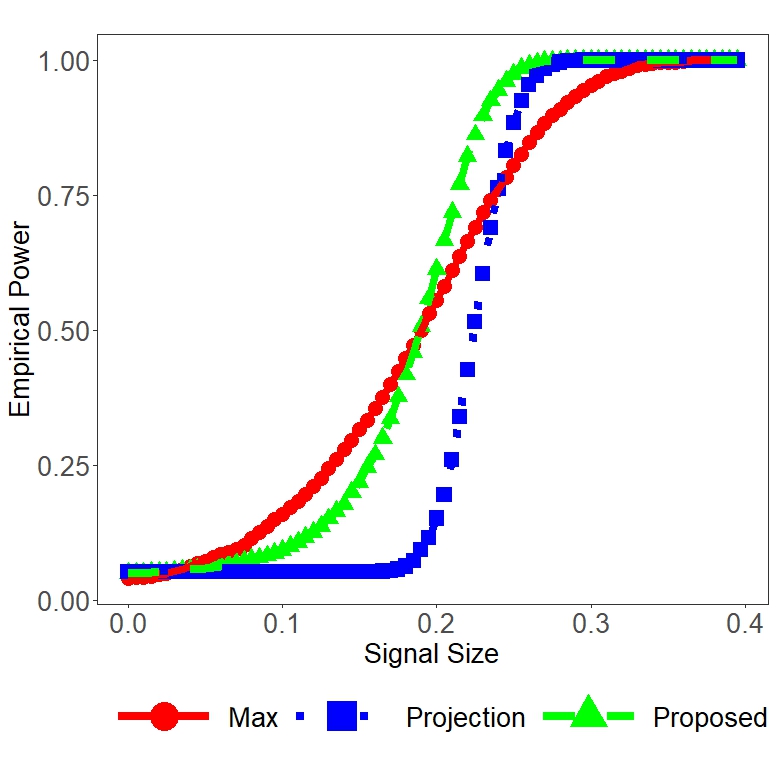}
        \\ \small Parameters $s=0.9$, $\rho=0.6$
    \end{minipage}

     \caption{Empirical power of \Proposed, \Xue, and \XY, with data generated according to \eqref{simulated_distribtuion_alternatives} against alternatives from the family \eqref{sine_definition} with $s, \rho$ fixed to the particular value indicated in the sub-caption, and $\delta$, the signal size, shown on the horizontal axis. Compare with Figure \ref{fig:side_by_side}}
    \label{fig:side_by_side_.6}
\end{figure}

In this section we provide additional simulation results to compare the power of \Proposed, \XY and \Xue against the global null. The setting is as described in the beginning of Section~\ref{sec:simspower} in the main text: powers are plotted against the signal-size parameter $\delta$ specified in \eqref{sine_definition} for samples of size $n=150$ where the functional noise is generated with centered and unit variance Gaussian noise in the basis expansion, i.e, the scalars  $g_{ikj}$ in \eqref{def:independent_noise} are standard Gaussian variates and $\alpha = 0.55$. The sparsity parameter $s$ specified in \eqref{sine_definition} varies from sub-plot to sub-plot as indicated in the respective captions and we consider two different values for the parameter $\rho$: $\rho = 0.3$ in Figure~\ref{fig:side_by_side_.3} and $\rho = 0.6$ in Figure~\ref{fig:side_by_side_.6}.

\subsection{Data Usage Acknowledgement}\label{datausage}
Data used in the preparation of this work were obtained from the Human Connectome Project (HCP) database (https://ida.loni.usc.edu/login.jsp). The HCP project (Principal Investigators: Bruce Rosen, M.D., Ph.D., Martinos
Center at Massachusetts General Hospital; Arthur W. Toga, Ph.D., University of Southern California, Van J. Weeden, MD,
Martinos Center at Massachusetts General Hospital) is supported by the National Institute of Dental and Craniofacial
Research (NIDCR), the National Institute of Mental Health (NIMH) and the National Institute of Neurological Disorders and
Stroke (NINDS). HCP is the result of efforts of co-investigators from the University of Southern California, Martinos Center
for Biomedical Imaging at Massachusetts General Hospital (MGH), Washington University, and the University of
Minnesota. 

\newpage 

\subsection{Additional Real Power Analysis}\label{sec:additional_real_data}
In this section we complement Figure \ref{brain_picture}, which compares the power of \Xue and \Proposed, by including a similar figure that compares the power of \XY and \Proposed. See discussion near Figure \ref{brain_picture} for additional context.

\begin{figure}[H]
\centering
%\ContinuedFloat
\begin{subfigure}[b]{1.0\textwidth}
\includegraphics[scale=0.72]{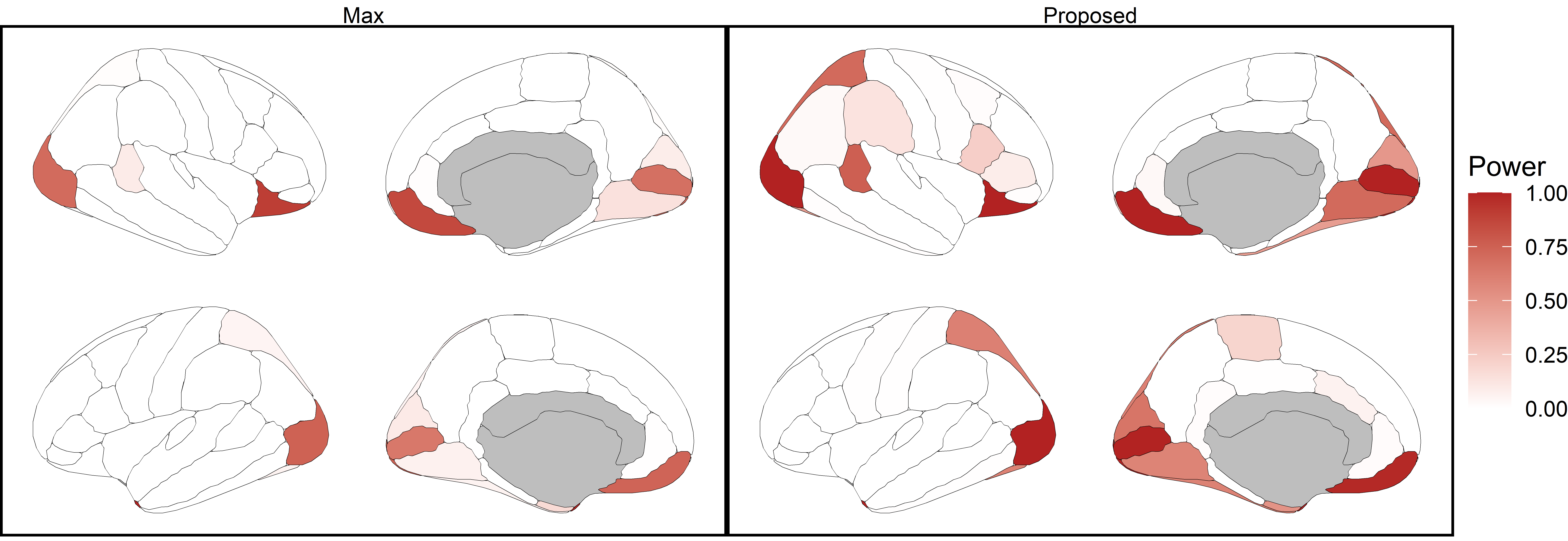} 
\caption{$m=100$. }
\end{subfigure}
\begin{subfigure}[b]{1.0\textwidth}
\includegraphics[scale=0.72]{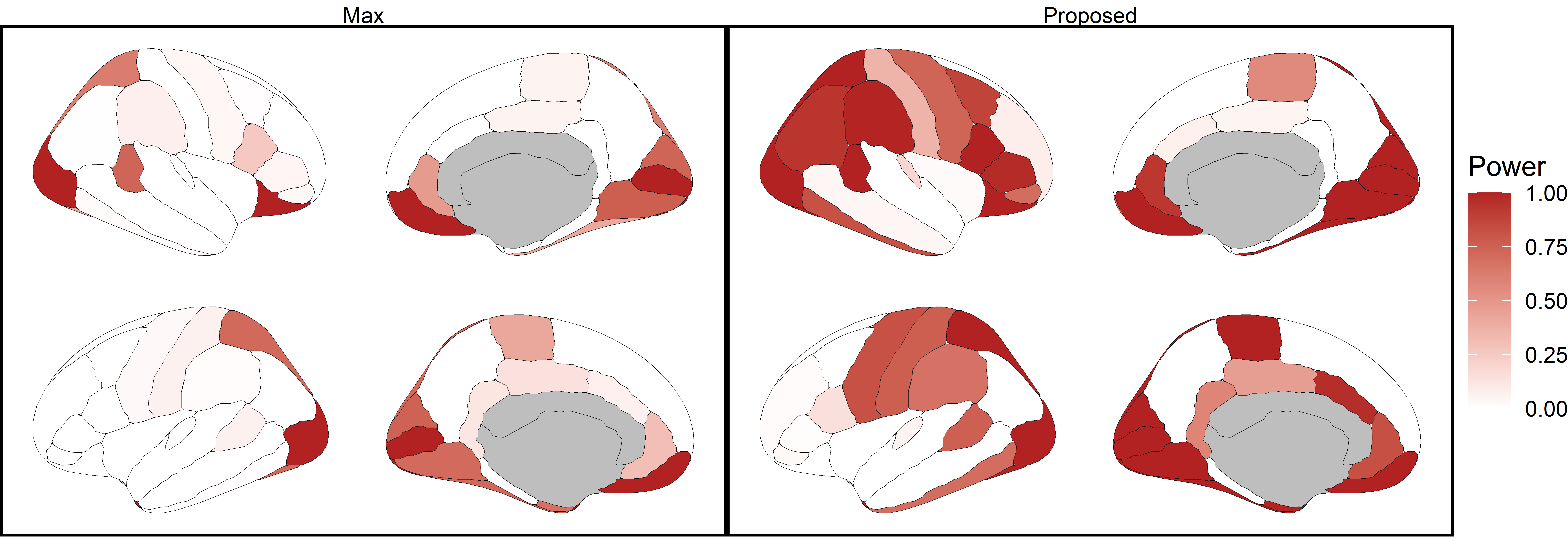} 
\caption{$m=200$. }
\end{subfigure}
\begin{subfigure}[b]{1.0\textwidth}
\includegraphics[scale=0.72]{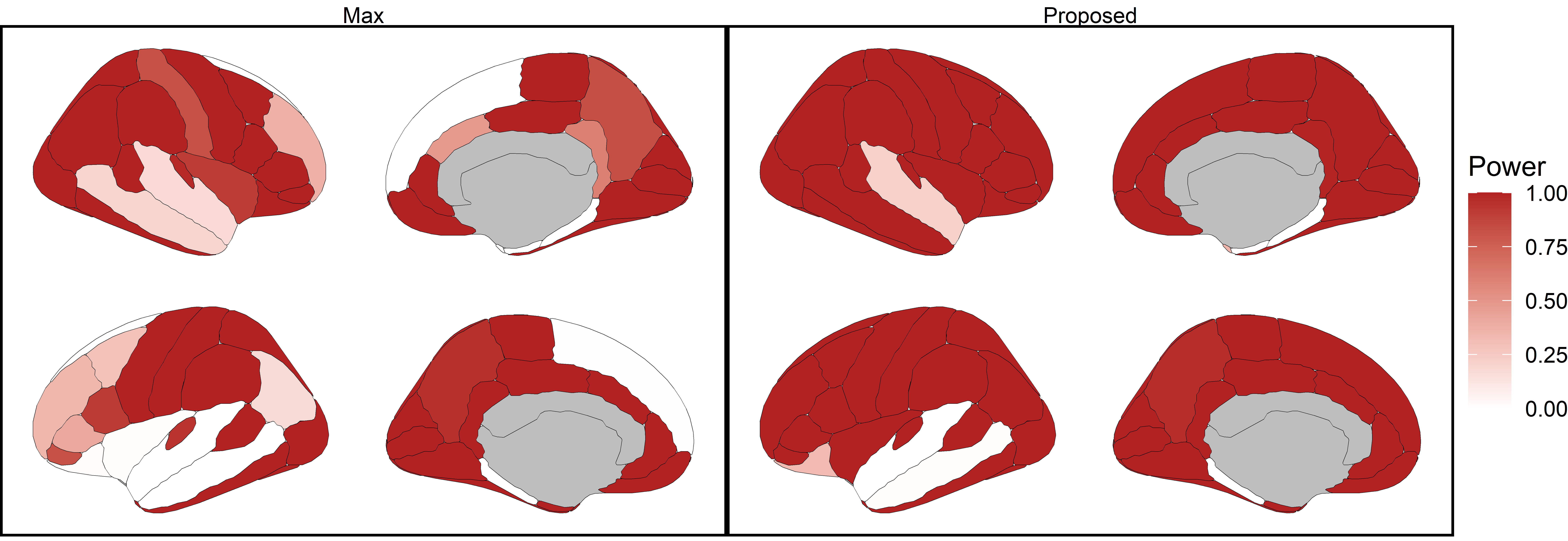} 
\caption{$m=500$. }
\end{subfigure}
\caption{
Empirical power of marginal tests for the null of no difference at every of the $68$ regions of interest with FWER control for $m = 100$ (Panel (a)), $m = 200$ (Panel (b)), $m = 500$ (Panel (c)).
Power is computed as described in the beginning of Section~\ref{sec:global-null-subsample} with nominal level $\gamma = 0.05$. For each subfigure, the right panel corresponds to the proposed test, while the left panel corresponds to the \XY based test. Gray-shaded regions indicate areas not included in the Desikan-Killiany atlas. }
\label{brain_pic_supp}
\end{figure}

\subsection{Additional notation used in the proofs}\label{sub-section:additional_notation}
Given $a \in \R^{n_1}$ and $b \in \R^{n_2}$, we use direct product notation for their concatenation, so that $a \oplus b:= [a^{\top}, b^{\top}]^{\top} \in \R^{n_1+n_2}$, and we will frequently use this operation iteratively; for example if $\{x_1,x_2, x_3\}$ is a collection of 3 elements of $\R^{M}$,  $\oplus_{k=1}^{3}x_{k} = (x_1 \oplus x_2)\oplus x_3 \in \R^{3M}$. If for all $1 \leq k \leq K$, $x_{k}\in \R^M$, we define $\pi_{\ell}[\oplus_{k=1}^{K}x_{k}]:=x_{\ell} \in \R^{M}$, so that $\pi_{\ell}$ stands for projection of a vector $y \in (\R^{M})^K$  onto the $\ell^{th}$ copy of $\R^{M}$. Following \cite{oksendal2013stochastic} we frequently use the notation $(\R^{M})^K$ to mean $KM$ dimensional Euclidean space,  rather than $\R^{MK}$, to emphasize a particular decomposition of the vector space. Given independent random variables $X$ and $Y$, and a measurable real valued function $h$, we define $\Pr_X(h(X,Y) \leq t):= g(Y)$, where $g(y) = \E[\Pr(h(X,y) \leq t)]:= \E[\mathbf{1}\{h(X,y) \leq t\}]$. Given random elements $X$ and $Y$ taking values in the same measure  space, $X \overset{d}{=} Y$ means that $\Pr(X \in B) = \Pr(Y \in B)$ for all measurable subsets  $B$ of the space, and for general random elements $X$ and $Y$ the notation $X \independent Y$ means that $X$ and $Y$ are stochasticallly independent. If $\mathcal{F}$ and $\mathcal{G}$ are $\sigma$-fields, the notation $\mathcal{F} \independent \mathcal{G}$ means that $\mathcal{F}$ and $\mathcal{G}$ are independent $\sigma$-fields. By $\sigma(X)$ we mean the $\sigma$-field generated by $X$. If $S$ is a topological space, the notation $\mathcal{B}_{S}$ refers to the Borel $\sigma$-field on $S$. For a real number $x$ we define $x_{+}:=\max\{x,0\}$. Given real numbers $a_{1}, \dots a_{n}$, $\textrm{diag}[a_1, \dots a_n] \in M_{n \times n}$ is defined entry-wise by the equation $[\textrm{diag}[a_1, \dots a_n]]_{ij} = \delta_{ij}a_{i}$, where $\delta_{ij}=1$ if $i =j $ and $0$ if $i \neq j$. Given $a \in \R$, we define $\lfloor a \rfloor  = \sup \{z \in \Z, z \leq a\}$, and $\lceil a \rceil = \inf\{z \in \Z, z \geq a\}$, i.e. the floor and ceiling functions.  If $X_1, \dots, X_n$ are independent random elements satisfying $X_i\overset{d}{=}X_j$ for all $1\ \leq i \leq n$ and all $1 \leq j \leq n$, we say that $X_1, \dots, X_n$ are an $\iid$ collection or are $\iid$. If $a$ and $b$ are real numbers we define $a \vee b:\max\{a,b\}$, and $a \wedge b:= \min\{a,b\}$. If $v \in \R^{n}$, $\|v\|_{2} := \sqrt{\sum_{j=1}^{n}v_{j}^{2}}$, $\|v\|_{1} := \sum_{j=1}^{n}|v_j|$, and $\|v\|_{\infty} := \max_{j=1}^{n}|v_j|$, and these definitions are extended to matrices by identifying the matrix with a vector. We define $\exp$ to be the exponential function. For $f$ and $g$ real valued functions (with domain $\R$, $\N$, etc.) the notation $f \lesssim g$ means that there is a universal constant $C$ such that $f \leq Cg$. Let $I = [0,1] \subset \R$, unless otherwise stated. If $A$ is a finite set, we denote by $|A|$ the number of elements contained in $A$. If $A \subset \R$ is an interval, we denote by $|A|$ the number $\sup A - \inf A$.

We now introduce notation that is used in section \ref{sub-section:proofs}, which contains proofs of Theorem \ref{main:thm} and \ref{main:thm-power}. The notation also appears in section \ref{sub-section:technical_results} when specializing our general technical results into a form that directly applies to the proof arguments made in section \ref{sub-section:proofs}. Recall the definition of the test statistic $T_n^k$ (defined for all $1 \leq k \leq K$), the global test statistic $T_n = \max_{k=1}^{K}T_n^{k}$, the bootstrap statistic $T_{n}(e)$, and the empirical c.d.f $\hat F(t)$, all defined in Section \ref{methodology_section}. Recall the notation introduced in Assumptions \ref{assumptions}. Let $\Phi_{i,k} := \oplus_{j=1}^M \phi_j^{\tilde f_{k,i}} \in \R^M$ and $\Phi_i := \oplus_{k=1}^K \Phi_{i,k} \in \R^{MK}$ and $\Phi:=\oplus_{i=1}^{n}\Phi_i$. Let  $\Sigma^{\Phi}$ denote the covariance matrix of $n^{-1/2}\sum_{i=1}^{n}\Phi_i$. Let $\hat \Sigma^{\Phi}$ denote the empirical covariance of  $\Phi_1, \dots,\Phi_n$ and let $\hat \Sigma^\Phi_{k}$ denote the empirical covariance of $\pi_k\Phi_1, \dots, \pi_k\Phi_n$. Let $g$ be a Gaussian in $(\R^{M})^K$ with $\Cov(g) = \Sigma^{\Phi}$ and let $e_i, i=1,\dots,n$ denote $\iid$ standard Gaussians with $e = (e_1,\dots,e_n)^\top \in \R^n$ independent of $\Phi$.  Define
\begin{align}
{T^d(f_1,\dots,f_n)} &:= \max_{k=1}^K \frac{1}{T}\sum_{\ell=1}^T \Big(\frac{1}{\sqrt{n}}\sum_{i=1}^n f^k_i(t_\ell)\Big)^2, \label{eq:defT^d}
\\
T^{M}(\Phi) &:= \max_{k=1}^K \Big\|\frac{1}{\sqrt{n}} \sum_{i=1}^n \Phi_{i,k} \Big\|_2^2, \label{eq:defT^M}
\\
T^{M,G}(g) &:= \max_{k=1}^K \|\pi_k g\|_{2}^2, \label{eq:defT^MG}
\\
T^{M,e}(\Phi,e) &:= \max_{k=1}^K \Big\|\frac{1}{\sqrt{n}} \sum_{i=1}^n e_i(\Phi_{i,k} - \overline{\Phi}_{\cdot,k}) \Big\|_{2}^2, \label{eq:defT^Me}
\\
{T^{d,e}(f_1,\dots,f_n,e)} &:= \max_{k=1}^K \frac{1}{T}\sum_{\ell=1}^T \Big(\frac{1}{\sqrt{n}}\sum_{i=1}^n e_i[ f^k_i(t_\ell) - \bar f^k_{\cdot}(t_\ell) ]\Big)^2. \label{eq:defT^de}
\end{align}
In many cases we will lighten notation by omitting the arguments of ${T^d}, T^M,T^{M,G},T^{M,e}, {T^{d,e}}$ above. Let $E = \oplus_{i=1}^N e^i$ for i.i.d. $N(0,I_n)$ random vectors $e^i \in \R^n$, independent of $\Phi,g$, and let $t \mapsto \hat F(t;f,E)$ denote the empirical cdf of the collection $\{ T_n(e^{1}),\dots,T_n(e^{N}) \}$, and let $\hat q_{1-\gamma}$ be the $1-\gamma$ empirical quantile of the collection. Sometimes we will write $\hat F(t)$ instead of $\hat F(t;f,E)$ to lighten notation. With this notation 
\[
\hat q_{1-\gamma} := \inf\{s \in \mathbb{R}: 1-\gamma \leq \hat F(s)\},
\] 
which is equal to the $\lceil (1-\gamma)N \rceil$ order statistic of the collection $\{ T_n(e^{1}),\dots,T_n(e^{N}) \}$.  Let $\sigma(f)$ be the $\sigma$-field generated by the random element $f \in ((C[I])^K)^n$.

\subsection{Proofs}\label{sub-section:proofs}

\subsubsection{Proof of Theorem~\ref{main:thm}}

Note that 
\begin{align*}
\mbox{FWER}_{\gamma} &= \Pr\Big(\exists k \in \{1, \dots, K\} \mbox{ such that } \mu^k \equiv 0,1-\hat F(T^k_n)<\gamma\Big) \\
&\leq \Pr\Big(\exists k \in \{1, \dots, K\}~:~ 1-\hat F(T^k_n)<\gamma \mbox{ when } \mu^j \equiv 0~ \forall j=1,\dots,K  \Big),
\end{align*}
where the final term in the above display is the probability of one false rejection under the global null hypothesis $H^{global}_0$. Thus~\eqref{eq_FWER_general} is implied by~\eqref{eq_FWER_global_null}. Since it suffices to prove~\eqref{eq_FWER_global_null} we assume throughout the proof that $H^{global}_0$ holds. In particular we assume that  for all $1 \leq k \leq K$ we have $\mu^k\equiv0$, i.e that the continuous function $\mu^k$ is identically equal to 0 and so for all $1 \leq i \leq n$ and all $1 \leq k \leq K$, $f^k_i$ is centered and therefore  $\tilde f \overset{d}{=} f$. Note that
\begin{equation}\label{eq:fwer_ub}
\Big|\mbox{FWER}_{\gamma}-\gamma\Big| = \Big|1-\gamma -\Pr\Big(T_n \leq \hat q_{1-\gamma}\Big)\Big|.
\end{equation}

Observe that $T^d(f)= T_n$ (see \eqref{global_hypothesis_test_stat}) and $T^{d,e}(f,e) = T_n(e)$ (see \eqref{bootstrap_statistic}) almost surely. We will thus work with $T^d, T^{d,e}$ in what follows (for example, the empirical quantile $\hat q_{1-\gamma}$ may be derived from the collection $\{ T^{d,e}(f,e^{1}),\dots,T^{d,e}(f,e^{N})\}$). 
By \eqref{eq:fwer_ub} it suffices to show that
under (A1) -- (A3),

\begin{equation}
|\Pr(T^{d} \leq  \hat q_{1-\gamma})-(1-\gamma)|\leq C_{1-\gamma}d(n,K,T,n),
\end{equation}
and under (A1)---(A4),
\begin{equation}
|\Pr(T^{d} \leq \hat q_{1-\gamma})-(1-\gamma)|\leq \tilde{C}_{1-\gamma}\tilde{d}(n,K,T,n),
\end{equation}
for  constants $C_{1-\gamma}$ and $\tilde{C}_{1-\gamma}$ that depend on the distribution of the data-generating distribution only through the model parameters appearing in (A1)-(A3) and (A1)-(A4) respectively. 
We start by collecting several preliminary results. Let $q_{X}(\cdot)$ denote the quantile function of a generic random variable $X$, and recall that for any $\gamma$, $F_{X}(q_{X}(\gamma))= \gamma$ if $F_{X}(\cdot)$ is a continuous function.

Recall that under Assumption (A3) there exist $k^*, j^*$ such that $\Var(g^{f_{k^*}}_{j^*}) \geq v_{-} > 0$. Then we have almost surely  
\[
T^{M,G} = \max_{k=1}^{K}\sum_{j=1}^{M}(g^{f_{k}}_{j})^{2} \geq (g^{f_{k^*}}_{j^*})^{2} 
\]
and thus for all $\eta \in (0,1)$,
\[
q_{T^{M,G}}(\eta) \geq v_{-} q_{\chi_1^2}(\eta),
\]
where $q_{\chi_1^2}$ is the quantile function of a $\chi^2$ random variable with one degree of freedom. Fix
\begin{equation}\label{eq:defsgamma}
s_\gamma := v_{-} q_{\chi_1^2}( (1-\gamma)/2).
\end{equation}
By Lemma~ \ref{lem:tail_bound_TMG}, $F_{T^{M,G}}(\cdot)$ is a continuous function and so
\begin{equation}\label{key_property_s}
	F_{T^{M,G}}(s_\gamma) \le (1-\gamma)/2,
\end{equation}
which by the definition of the quantile function implies that
\begin{equation}\label{key_property_s_2}
q_{T^{M,G}}((1-\gamma)/2) \geq s_\gamma.
\end{equation}
If (A1)-(A3) holds, let $C=[80\big(1\vee C_{4,1} \vee C_{5,1}\vee C_{7,1}\vee C_{7,3}\big)]^{-1}$ for the constants $C_{4,1}$ and $C_{5,1}$, $C_{7,1}$ and $C_{7,3}$ introduced in the proof below; it can be seen in the proof that these constants depend only on the model parameters that appear in (A1)-(A3). If in addition (A4) holds  let $\tilde{C} = [50\big(C_{4,1} \vee C_{5,1} \vee C_{8,3} \vee C_{8,1} \vee 1\big)]^{-1}$, where $C_{8,1}$ and $C_{8,3}$ are likewise defined in the proof below and depend only on the model parameters of (A1)-(A4). Let $D_{\gamma}=[s_{\gamma} \wedge (1-\gamma)]/2$. 

If $d_n:=d(n,K,T,N)$ or $\tilde d_n:=\tilde{d}(n,K,T,N)$ satisfy $d_n \geq CD_{\gamma}, \tilde{d}_n \geq \tilde{C}D_{\gamma}$, then 
the choice $C_{1-\gamma} = [D_\gamma C]^{-1}$ implies that \eqref{eq_FWER_global_null} holds under (A1)-(A3) and likewise for $\tilde{C}_{1-\gamma} = [D_\gamma\tilde{C}]^{-1}$ if (A4) holds additionally.  We therefore assume in the remainder of the proof that
\begin{equation}\label{eq_def_C}
d_n< CD_\gamma \mbox{ when working under (A1)-(A3) and } \tilde d_n< \tilde{C}D_\gamma \mbox{ under (A1)-(A4)}.
\end{equation}

 The remaining proof relies on many intermediate technical results which we will derive in the supplement below. Here, we provide the main steps and explain the high level ideas on the way. The following bound will be used repeatedly: for two generic random variables $X,Y$ we have for any $\delta > 0, s \in \R$
\begin{equation}\label{eq:diffcdf}
	\sup_{t \geq s} |\Pr(X \leq t) - \Pr(Y \leq t)| \leq \Pr(|X-Y|>\delta) + \sup_{t \geq s} \Pr(Y \in [t-\delta,t+\delta]).	
\end{equation}
It follows by elementary computations and the proof is omitted. A crucial role will be played by the following quantity related to anti-concentration of $T^{M,G}$
\begin{equation}\label{def:ac-function}
\mA(s,\delta) := \sup_{t \geq s} \Pr(|T^{M,G}-t|\leq \delta).
\end{equation}
We first provide a general approach linking $\Pr(T^d \leq  \hat q_{1-\gamma})$ to various approximation errors given next. We will give explicit bounds on those errors under the different sets of assumptions made in Theorem~\ref{main:thm} at the end of the proof.  Define
\begin{align*}
\Delta_{de,MG}(s;f) &:= \sup_{t \geq s} \Big|\Pr(T^{M,G} \le t) - \Pr_e(T^{d,e}(f,e)\le t) \Big|,
\\
\Delta_{d,MG}(s) &:= \sup_{t \geq s} \Big|\Pr(T^{M,G} \le t) - \Pr(T^d\leq t) \Big|, 
\\
\Delta_{M,MG}(s) & := \sup_{t \geq s} \Big|\Pr(T^M \le t) - \Pr(T^{M,G} \le t)\Big|
\\
\Delta_{Me,MG}(s;f) & := \sup_{t \geq s} \Big|\Pr_e(T^{M,e}(\Phi,e) \leq t) - \Pr(T^{M,G}\leq t) \Big|.
\end{align*}
We assume the existence of sequences $a_n, b_n, v_n, w_n$ with the property that
\begin{align}\label{eq:defanbn}
\Pr(\Delta_{de,MG}(f;s_{\gamma}/2) \geq a_n) &\leq b_n,
\\
\label{eq:defcndn}
\Pr(\Delta_{Me,MG}(f;s_{\gamma}/2) \geq v_n) &\leq w_n.
\end{align}
and also
\begin{equation}\label{eq_bound_an_bn_v_n_w_n}
a_n \leq \frac{d_n}{2C}
\end{equation}
if (A1)-(A3) holds, and the same bound with $d_n$ replaced by $\tilde d_n$ and $C$ replaced by $\tilde C$ if in addition (A4) holds.
Later on we will show that such sequences exist under (A1)-(A3) and also (A1)-(A4)  for sufficiently small value of $C$ and $\tilde C$; the sequences $a_n$ and $v_n$ will differ under the two different sets of conditions.  

Over the course of the proof we will calculate various bounds in terms of the dimensional parameter $M$. We set $M = 3\lceil n^{\frac{1}{12\alpha-2}}\rceil$ and note that $3 \leq M \leq 3(1+n^{\frac{1}{12\alpha-2}})\leq 6n^{\frac{1}{12\alpha-2}}$.  However we use the dimensional parameter $M$ as notation until the end of the proof when we confirm the rates in $n$, $T$, $K$, and $N$ that are stated in Theorem \ref{main:thm} by plugging in the definition of $M$ in terms of the above sequence in $n$. This facilitates readability and allows us to directly apply technical results proved in the supplement that apply to $L^\infty-L^2$ mixed norms of random vectors defined on $(\R^M)^K$.
\medskip

\noindent\textbf{Step 1: $\hat F$ and the conditional distribution of $T^{d,e}$ given data are close}

Recall that $e \in \R^n$ is an isometric Gaussian independent of $(f,E)$. Define
\begin{equation}
\Delta_1(f,E) := \sup_{t\in \R}\Big|\hat F(t;f,E) - \Pr_e(T^{d,e}(f,e)\leq t)\Big|.
\end{equation}
Let $\tau :=\sqrt{\frac{2\log(N)}{N}}$ and define $A_\tau := (\tau,1]$. Since $f$ and $E$ are independent and $\Delta_1(f,E) \leq 1$ a.s the Fubini-Tonelli theorem gives, for $\mathbf{1}_{A_\tau}$ denoting the indicator fo the set $A_\tau$, 
\begin{equation}\label{step1-eq1}
\E[\mathbf{1}_{A_\tau}\circ \Delta_1(f,E)] = \int_{((\R^T)^K)^{n}}\E[\mathbf{1}_{A_\tau}\circ \Delta_1(x,E)]d\mu^{1}_x,
\end{equation}
where the real Borel probability measure $\mu^{1}$ corresponds to the distribution of the collection $\{f_{k,i}(t_{\ell}) -\bar f_{k,\cdot}(t_{\ell})\}_{k=1,i=1, \ell=1}^{K,n, T}$.

The Dvoretzky–Kiefer–Wolfowitz inequality (see \cite{M90}) implies the inequality
\[
\E[\mathbf{1}_{A_\tau}\circ \Delta_1(x,E)] \leq 2\exp(-2N\tau^2) \leq \frac{1}{N}
\]
for every $x$ since 
\[
\E[\mathbf{1}_{A_\tau}\circ \Delta_1(x,E)] =  \Pr\Big(\sup_{t\in \R}|\hat F(t;x,E) - \Pr(T^{d,e}(x,e)\leq t)]|>\tau\Big)
\]
and since, for fixed $x$, $\hat F(t;x,E)$ is the empirical cdf of $T^{d,e}(x,e^1),\dots,T^{d,e}(x,e^N)$, an i.i.d. sample from the distribution of $T^{d,e}(x,e)$. Combined with~\eqref{step1-eq1} this shows
\begin{equation}\label{eq:step1fin}
\Pr\Big(\Delta_1(f,E) >  \sqrt{\frac{2\log(N)}{N}}\Big) \leq \frac{1}{N}
\end{equation}
and completes Step 1.

\medskip

\noindent\textbf{Step 2: relating $\hat q_{1-\gamma}$ and $q_{T^{M,G}}(\cdot)$}

Define
\begin{equation}\label{def:kappa}
\kappa_n := \sqrt{\frac{2\log(N)}{N}} + a_n 
\end{equation}
for $a_n$ from~\eqref{eq:defanbn} and also define the event 
\[
\Omega_1 := \Big\{  \sup_{t \geq s_\gamma/2} \big|\hat F(t) - F_{T^{M,G}}(t)\big| \leq \kappa_n \Big\}. 
\]

As seen in \eqref{key_property_s}, $F_{T^{M,G}}(s_\gamma/2) \leq (1-\gamma)/2$. Also by~\eqref{eq_bound_an_bn_v_n_w_n} and the definition of $D_\gamma$, we have $\kappa_n \leq \sqrt{2}d_n +d_n/(2C)<(1-\gamma)/2$, and also $\kappa_n <s_\gamma/2$ under the conditions (A1)-(A3), and we obtain the same inequalities if (A1)-(A4) hold. We apply  Lemma~\ref{lem:koltoquant} on the event $\Omega_1$  with $\alpha=1-\gamma$, $G = F_{T^{M,G}}$, $F = \hat F$, $\delta = \kappa_n$ and $t_* = s_\gamma/2$. Since $G(t_*)+\kappa_n <F_{T^{M,G}}(s_\gamma/2)+(1-\gamma)/2 \leq (1-\gamma)=\alpha$ we obtain that on $\Omega_1$
\[
q_{T^{M,G}}(1-\gamma-\kappa_n) \le \hat q_{1-\gamma} \le q_{T^{M,G}}(1-\gamma+\kappa_n).
\]

By a combination of~\eqref{eq:step1fin} from Step 1 and the definition of $b_n$ in~\eqref{eq:defanbn} the union bound gives $\Pr(\Omega_1^C) \leq b_n + N^{-1}$ and thus
\begin{equation}\label{eq:step2fin}
	\Pr\Big(q_{T^{M,G}}(1-\gamma-\kappa_n) \le \hat q_{1-\gamma} \le q_{T^{M,G}}(1-\gamma+\kappa_n)\Big) \geq 1 - b_n - \frac{1}{N}.
\end{equation}
This completes Step 2. 

\medskip 

\noindent\textbf{Step 3: deriving a general bound on $|\Pr(T^d \leq \hat q_{1-\gamma}) - (1-\gamma)|$} 

In the previous step we showed that $\kappa_n < (1-\gamma)/2$. Therefore by~\eqref{key_property_s} 
\begin{equation}\label{eq:lowboundq_new}
q_{T^{M,G}}(1-\gamma-\kappa_n) \geq q_{T^{M,G}}((1-\gamma)/2) \geq s_\gamma.
\end{equation} 
Then, since on $\Omega_1$ we have $\hat q_{1-\gamma} \le q_{T^{M,G}}(1-\gamma+\kappa_n)$ and by~\eqref{eq:lowboundq_new}
\begin{align*}
	&\Pr(T^d \leq \hat q_{1-\gamma})
	\\
	&= \Pr( \{T^d \leq \hat q_{1-\gamma}\}\cap \Omega_1 ) + \Pr( \{T^d \leq \hat q_{1-\gamma}\}\cap \Omega_1^C)
	\\
	&\le \Pr(T^d \leq q_{T^{M,G}}(1-\gamma+\kappa_n)) + \Pr(\Omega_1^C)
	\\
	&\le \Pr(T^{M,G} \leq q_{T^{M,G}}(1-\gamma+\kappa_n)) + \Delta_{d,MG}(s_\gamma) + b_n + \frac{1}{N}
	\\
	&\le 1-\gamma+\kappa_n + \Delta_{d,MG}(s_\gamma) + b_n + \frac{1}{N},
\end{align*} 
where to get the last inequality we used absolute continuity of $T^{M,G}$, shown in Lemma~ \ref{lem:tail_bound_TMG}.
Similarly, since on $\Omega_1$ we have $\hat q_{1-\gamma} \ge q_{T^{M,G}}(1-\gamma-\kappa_n) \ge s_\gamma$ by~\eqref{eq:lowboundq_new}, it follows that
\begin{align*}
	&\Pr(T^d \leq \hat q_{1-\gamma})
	\\
	&= \Pr( \{T^d \leq \hat q_{1-\gamma}\}\cap \Omega_1 ) + \Pr( \{T^d \leq \hat q_{1-\gamma}\}\cap \Omega_1^C)
	\\
	&\ge \Pr(\{T^d \leq q_{T^{M,G}}(1-\gamma-\kappa_n)\}\cap \Omega_1)
	\\
	& = \Pr(T^d \leq q_{T^{M,G}}(1-\gamma-\kappa_n)) - \Pr(\{T^d \leq q_{T^{M,G}}(1-\gamma-\kappa_n)\}\cap \Omega_1^C)
	\\
	&\geq\Pr(T^{M,G} \leq q_{T^{M,G}}(1-\gamma-\kappa_n)) - \Delta_{d,MG}(s_\gamma) - \Pr(\Omega_1^C) 
	\\
	&\geq 1-\gamma-\kappa_n - \Delta_{d,MG}(s_\gamma) - b_n - \frac{1}{N}.
\end{align*}
where to get the last inequality we used absolute continuity of $T^{M,G}$. In summary, we find
\begin{equation}\label{eq:step3fin}
\Big|\Pr(T^d \leq \hat q_{1-\gamma}) - (1-\gamma) \Big| \leq \kappa_n + \Delta_{d,MG}(s_\gamma) + b_n + \frac{1}{N}.
\end{equation}

\medskip

\noindent\textbf{Step 4: bounding $\Delta_{d,MG}$ through $\Delta_{M,MG}$}
By Lemma~\ref{lem:|T^d-T^M|}, there exists a constant $C_{4,1}\geq 1$ (we choose this constant to be greater than one for technical reasons that will become apparent later) such that
\[
\|T^d - T^M\|_{\Psi_{\frac{1}{2}}}\leq C_{4,1}(M^{1-2\alpha} + T^{-q})[\log(K+1)]^2.
\]
Let $\delta_{1,n}: = 4C_{4,1}(M^{1-2\alpha} + T^{-q})[\log(K+1)]^2[\log(n)]^2 \geq C_{4,1}(M^{1-2\alpha} + T^{-q})[\log(K+1)]^2[\log(2n)]^2$, so that by Lemma ~\ref{lim_max_sum}(iv), 
\begin{equation}\label{eq:step4-1}
\Pr(|T^d - T^M|>\delta_{1,n}) \leq \frac{1}{n}.
\end{equation}
We verify in Remark~\ref{remark:sequence_bounds} that under (A1)-(A3) or (A1)-(A4) we have $\delta_{1,n} \leq s_{\gamma}/2$, as the constants $C$ and $\tilde C$ were chosen large enough (see \eqref{eq_def_C}). Then by definition of $\Delta_{M,MG}$ we have
\[
\sup_{t \geq s_\gamma} \Pr(|T^M-t|\leq \delta_{1,n}) \leq 2\Delta_{M,MG}(s_\gamma/2) + \mA(s_\gamma,\delta_{1,n}).
\]
Combined with~\eqref{eq:diffcdf} and~\eqref{eq:step4-1} this yields
\begin{equation*}
\sup_{t \geq s_\gamma} \Big|\Pr(T^d \le t) - \Pr(T^{M} \le t)\Big| \leq 2\Delta_{M,MG}(s_\gamma/2) + \mA(s_\gamma,\delta_{1,n}) + \frac{1}{n}. 
\end{equation*}
The final bound of Step 4 is
\begin{equation}\label{eq:step4fin}
\Delta_{d,MG}(s_\gamma)=\sup_{t \geq s_\gamma} \Big|\Pr(T^d \le t) - \Pr(T^{M,G} \le t)\Big| \leq 3\Delta_{M,MG}(s_\gamma/2) + \mA(s_\gamma,\delta_{1,n}) + \frac{1}{n},
\end{equation}
where we have used that $\Delta_{M,MG}(\cdot)$ is a non-increasing function. 
\medskip

\noindent\textbf{Step 5: bounding $\Delta_{de,MG}$ through $\Delta_{Me,MG}$}
Define
\[
\Delta_5(f;s) := \sup_{t \geq s} \Big|\Pr_e(T^{M,e}(\Phi,e) \le t) - \Pr_e(T^{d,e}(f,e)\leq t) \Big|. 
\]
Lemma ~\ref{lem:|T^{d,e}-T^{M,e}|} shows that there exists a random variable $W$ such that almost surely
\[
\|T^{M,e}(\Phi,e) - T^{d,e}(f,e)\|_{\Psi_{1}|\sigma(f)} \leq W 
\] 
where $W$ is a random variable with
\[
\|W\|_{\Psi_{\frac{1}{2}}} \le C_{5,1} (M^{1-2\alpha} + T^{-q}) [\log(K+1)]^3,
\] 
and $C_{5,1}$ is a constant that depends only on the fixed model parameters of (A1)-(A3). Thus by a conditional version of~\eqref{eq:diffcdf} we have a.s., for any $\delta_{n} < s_\gamma/2$
\begin{align*}
\Delta_5(f;s_\gamma) &\leq 2\exp(-\delta_{n}/W) + \sup_{t\geq s_\gamma} \Pr_e(|T^{M,e}(f,e)-t| \leq \delta_n).
\\
&\leq 2\exp(-\delta_{n}/W) + 2\Delta_{Me,MG}(f;s_\gamma/2) + \mA(s_\gamma,\delta_{n}),
\end{align*}
xwhere we have applied Lemma~\ref{lim_max_sum} (iv) conditionally, as demonstrated in Remark \ref{remark:conditional_tail_bound}. Let
\[
 \delta_{2,n}:=8(C_{5,1}\vee C_{4,1})(M^{1-2\alpha} + T^{-q}) [\log(K+1)]^3[\log n]^3 \geq \|W\|_{\Psi_{\frac{1}{2}}} [\log (2n)]^3.
\]

We verify in Remark \ref{remark:sequence_bounds} that $\delta_{2,n}< s_{\gamma}/2$. The constant in the definition of $\delta_{2,n}$ was chosen so that $\delta_{2,n} \geq \delta_{1,n}$. Then for $n \geq 2$
\[
\Pr(\exp(-\delta_{2,n}/W) \ge 1/n) = \Pr(\delta_{2,n}/W \le  \log n ) \le \Pr( W/\|W\|_{\Psi_{\frac{1}{2}}} \geq  [\log 2n]^2) \le 1/n
\]
where the last line follows from Lemma~\ref{lim_max_sum}(iv) since $\big\|W/\|W\|_{\Psi_{\frac{1}{2}}}\big\|_{\Psi_{\frac{1}{2}}} = 1$. Combining the bounds above we see that 
\begin{equation}\label{eq:step5fin}
\Pr\Big(\Delta_{de,MG}(f;s_\gamma) > 2/n + 3\Delta_{Me,MG}(f;s_\gamma/2) + \mA(s_\gamma,\delta_{2,n})\Big) \le \frac{1}{n}.
\end{equation} 
We used that under (A1)-(A3) or (A1)-(A4) we have $\delta_{2,n} \leq s_{\gamma}/2$, as the constants $C$ and $\tilde C$ were chosen large enough (see \eqref{eq_def_C}) which is verified in Remark~\ref{remark:sequence_bounds}.
This completes Step 5.

\medskip 

\noindent\textbf{Step 6: combining the bounds so far}
 Using the definition of $w_n$, $v_n$ from \eqref{eq:defcndn} and the union bound we have
\begin{equation}\label{eq:df}
	\Pr\Big(\Delta_{de,MG}(f;s_\gamma) > 2/n + 3v_n + \mA(s_\gamma,\delta_{2,n})\Big) \le \frac{1}{n}+w_n.
\end{equation} 
Equation \eqref{eq:df} shows that we may set
\begin{equation}\label{eq:bound_an}
b_n=1/n+w_n, \hspace{.5cm} a_n =2/n+3 v_n+\mathcal{A}(s_\gamma, \delta_{2,n}).
\end{equation}
Indeed, we show in Remark \ref{remark:sequence_bounds} that under (A1)-(A3) we have, $a_n \leq \frac{d_n}{2C}$ and under (A1)-(A4) the same inequalities hold with $d_n$ replaced by $\tilde{d}_n$ and $C$ replaced by $\tilde C$ so that \eqref{eq:defanbn} and also \eqref{eq_bound_an_bn_v_n_w_n} are satisfied.
Noting that $\mA(s, \delta)$  is non-decreasing in $\delta$ we therefore have after 
\begin{comment}

\end{comment}
plugging in the values of $a_n$ and $b_n$, by Equations \eqref{eq:step3fin}, \eqref{eq:step4fin} (which controls the term $\Delta_{d,MG}(s_\gamma)$ that appears in \eqref{eq:step3fin}) we have by the union bound
\begin{align}\label{eq:pfthm1step6}
\Big|\Pr(T^d \leq \hat q_{1-\gamma}) - (1-\gamma) \Big|\leq C_{6,1}\{v_n+\mA(s_\gamma,\delta_{2,n})+w_{n}+\sqrt{\frac{\log(N)}{N}}+\Delta_{M,MG}(s_\gamma/2)
	+ \frac{1}{n}\},
\end{align}
for a constant $C_{6,1}$ that depends only on the fixed model parameters of (A1)-(A3) if (A4) does not hold, and may depend additionally on (A4) if (A4) holds. We also note for use in a later proof that the arguments so far imply
\begin{equation}\label{eq:finalrelatequantile}
\Pr\Big(q_{T^{M,G}}(1-\gamma-\theta_n-3v_n) \le \hat q(1-\gamma) \le q_{T^{M,G}}(1-\gamma+\theta_n+3v_n)\Big) \geq 1 - \frac{2}{n} - \frac{1}{N} - w_n
\end{equation}
for 
\begin{equation}\label{eq:defthetan}
\theta_n := \frac{2}{n} + A(s_\gamma,\delta_{2,n}) + \sqrt{\frac{2\log N}{N}}.
\end{equation} 
This bound is obtained from \eqref{eq:step2fin} and \eqref{eq:df}, the union bound, and plugging in the previously stated expressions for $\kappa_n$ and $a_n$.
\medskip 

\noindent\textbf{Step 7: finalizing the result under (A1)--(A3)}
It remains to control all terms on the right-hand side of~\eqref{eq:pfthm1step6}. By Lemma~ \ref{lem:ac_mainthm_A1_A3}, there is a constant $C_{7,1}$ that depends only on fixed model parameters in (A1)-(A3) such that for all $n \geq 3$ and all $K \geq 1$,
\begin{equation}\label{eq: bound mA (A1)-(A3)}
\mA(s_\gamma/2,\delta_{2,n}) \leq C_{7,1}(M^{1-2\alpha}+T^{-q})[\log(K+1)]^{3}[\log(n)]^3M\log(nK).
\end{equation}
We set the free parameter $m$ in the bound of Lemma~\ref{lem:ac_mainthm_A1_A3} as $m=n$, and by our choice of $\delta_{1,n}$ (and in particular our choice $C_{4,1} \geq 1$) it is easily seen that $\delta_{2,n} \geq \delta_{1,n} \geq \frac 1n$. By Lemma~\ref{lem:A1_A3_Delta_M_MG}, for a constant $C_{7,2}$ that depends only on fixed model parameters in (A1)-(A3) and all $n \geq 3$ and all $K \geq 1$ (recall that we previously set $M$ as a function of $n$ in such a way that $n \geq 3$ implies $M \geq 3$),
\[
\Delta_{M,MG}(s_\gamma/2) \leq C_{7,2}\Big(\frac{M^{10}(\log(Kn))^{10}}{n}\Big)^{1/6}.
\]

By Lemma \ref{lem:bootstrap} for a constant $C_{7,3}$ that depends only on fixed model parameters in (A1)-(A3) and all $n \geq 3$ and all $K \geq 1$, (again we use $M \geq 3$),
\begin{equation}\label{eq:mainlemgauscomp_mainthm}
	\Pr\Big(\Delta_{Me,MG}(f;s_{\gamma}/2) > C_{7,3}\Big[M^{4/3} \log(Kn)\Big( \frac{\log(KMn)\log^{2}(n)}{n}\Big)^{1/6}\Big]\Big) \leq 1/n.
\end{equation}
Thus we take $v_n =  C_{7,3}\Big[M^{4/3} \log(Kn)\Big( \frac{\log(KMn)\log^{2}(n)}{n}\Big)^{1/6}\Big]$ and $w_n=\frac{1}{n}$, and note that they satisfy \eqref{eq:defcndn}.  Recall that  $M = 3 \lceil n^{\frac{1}{12\alpha-2}}\rceil$ and so $M \leq 6n^{\frac{1}{12\alpha-2}}$. This upper bound on $M$  balances the terms $[M^{10}n^{-1}]^{1/6}$ and $M^{2-2\alpha}$ and they are both of order $n^{\frac{1-\alpha}{6\alpha-1}}$. Noting that the leading $\log$ factor is $[\log(K+1)]^3[\log(n)]^3\log(nK)$ completes the proof under (A1)-(A3).

\medskip

\noindent\textbf{Step 8: finalizing the result under (A1)--(A4)}
It remains to control all terms on the right-hand side of~\eqref{eq:pfthm1step6}. By Lemma~\ref{lem:ac_mainthm_A1_A4}, there is a constant $C_{8,1}$ such that depends only on fixed model parameters in (A1)-(A4) such that for all $n \geq 3$ and all $K \geq 1$, and all $M\geq 1$.
\begin{equation}\label{eq: bound mA (A1)-(A4)}
\mA(s_\gamma/2,\delta_{2,n}) \leq C_{8,1}(M^{1-2\alpha}+T^{-q})[\log(K+1)\log(n)]^3[\log(K+1)]^{\frac{1}{2}+\frac{3\beta}{2\alpha-1}}[\log(n)]^{\frac{3\beta}{2\alpha-1}}.
\end{equation}
We set the free parameter $m$ in the bound of Lemma~\ref{lem:ac_mainthm_A1_A4} as $m=n$, and by our choice of $\delta_{1,n}$ (and in particular our choice $C_{4,1} \geq 1$) it is easily seen that $\delta_{2,n} \geq \delta_{1,n} \geq \frac 1n$.
By Lemma~\ref{lem:A1_A4_Delta_M_MG}, for a constant $C_{8,2}$ that depends only on fixed models parameters in (A1)-(A4), and all $n \geq 3$, and all $K \geq 1$, (recall that we defined $M$ as a function of $M$ so that $n \geq 3$ satisfies $M \geq 3$), 
\[
\Delta_{M,MG}(s_\gamma/2) \leq C_{8,2}\Big[\frac{M^{4}[\log(nK)]^{4}}{n}\Big]^{1/6}\log^{1+\frac{3\beta}{2\alpha-1}}(K+1)\log^{\frac{1}{2}+\frac{3\beta}{2\alpha-1}}(n).
\]
By Lemma~\ref{lem:bootstrap-2}, for a constant $C_{8,3}$ that depends only on fixed model parameters, we may set 
\[
v_{n} =C_{8,3}\Big[M^{2/3} \log^{1/3}(Kn)\Big( \frac{\log(KMn)\log^{2}(n)}{n}\Big)^{1/6}\Big]\log^{\frac12+\frac{3\beta}{2\alpha-1}}(K+1)\log^{\frac{3\beta}{2\alpha-1}}(n)
\] 
and $w_{n}=\frac{1}{n}$. Following Step 7, let $M = 3\lceil n^{\frac{1}{12\alpha-2}}\rceil$ (it turns out the same choice of $M$ works in both cases), and note that $M \lesssim  n^{\frac{1}{12\alpha-2}}$. This choice balances the terms $[M^{4}n^{-1}]^{1/6}$ and $M^{1-2\alpha}$ and they are both of order $n^{\frac{\frac{1}{2}-\alpha}{6\alpha-1}}$. Noting that the leading $\log$ factor is $[\log(K+1)]^{\frac{7}{2} +\frac{3\beta}{2\alpha-1}} [\log n]^{3 + \frac{3\beta}{2\alpha-1}}$ completes the proof. 
\hfill $\Box$

\subsubsection{Proof of Theorem~\ref{main:thm-power}}

Fix $\gamma \in (0,1)$ and fix $\eps \in (0, \gamma)$. Recall the constants $C$ and $\tilde C$ and $D_{\gamma}$ defined in the proof of Theorem \ref{main:thm} above \eqref{eq_def_C}. Let $D_{\gamma, \eps}= D_{\gamma}\wedge \eps$, and assume without loss of generality as in \eqref{eq_def_C} (but this time with $D_{\gamma}$ replaced by $D_{\gamma, \eps}$) that 
\begin{align}\label{eq_D_gamma_eps}
d_n &<\big(C\wedge C_{3,2}\wedge C_{4,4}\big)D_{\gamma, \eps} \mbox{ when working under (A1)-(A3) and }
\\
\tilde d_n &< \big(\tilde{C}\wedge C_{3,3} \wedge C_{4,4}\big)D_{\gamma, \eps} \mbox{ under (A1)-(A4)}.
\end{align}
where the constants $C_{4,4}, C_{3,2}$ and $C_{3,3}$ are defined in the proof below. For $1 \leq k \leq K$, let 
\begin{align*}
	v^k := \sqrt{\frac{1}{T}\sum_{i=1}^{T}\Big(\frac{\sum_{i =1}^{n}\tilde f^k_{i}(t_{\ell})}{\sqrt{n}}\Big)^2},\hspace{.5cm} \mu^k:=  \sqrt{\frac{1}{T}\sum_{i=1}^{T}\Big(\mu_k(t_{\ell})\Big)^2}.
\end{align*}
Recall the definition of $s_\gamma$ in ~\eqref{eq:defsgamma}. Set $B_1:=\{(s_\gamma \vee 1)[\log(2)]^{-1}\}\vee\{D_1^2\}$ for $D_1$ defined later in the proof. If $\sqrt{n}\mu^{k}-\sqrt{q_{G}(1-\gamma+\eps)} \leq \sqrt{s_\gamma}$ 
the right-hand side of \eqref{th_power:main_equation} is non-positive for $B=B_1$, and the inequality holds trivially. From now on we assume that  
\[
\sqrt{n}\mu^{k}-\sqrt{q_{G}(1-\gamma+\eps)} > \sqrt{s_\gamma}.
\]
By the reverse triangle inequality for the Euclidean norm on $\R^T$,
\[
\sqrt{T_n^k} 
= \sqrt{\frac{1}{T}\sum_{i=1}^{T}\Big(\frac{\sum_{i=1}^{n}\tilde f^k_{i}(t_{\ell})+\mu_k(t_{\ell})}{\sqrt{n}}\Big)^2} 
\geq 
\Bigg|\sqrt{\frac{1}{T}\sum_{i=1}^{T}\Big(\frac{\sum_{i =1}^{n}\tilde f^k_{i}(t_{\ell})}{\sqrt{n}}\Big)^2} - \sqrt{\frac{1}{T}\sum_{i=1}^{T}\Big( \sqrt{n}\mu_k(t_{\ell})\Big)^2}\Bigg|,
\]
which implies $\sqrt{T_n^k}  \geq \sqrt{n}\mu^k - v^k$ and so
\begin{equation}\label{power_bound_1}
\Pr\Big(\sqrt{T_n^k} \geq \sqrt{\hat q_{1-\gamma}}\Big) \geq \Pr\Big(\sqrt{n}\mu^k \geq v^k+ \sqrt{\hat q_{1-\gamma}}\Big).
\end{equation}
In the remainder of the proof we give a lower bound to the right hand side of \eqref{power_bound_1}.

\noindent\textbf{Step 1: relating $\hat q_{1-\gamma}$ and $q_{T^{G}}(\cdot)$}

By Lemma~\ref{lem:gaussian_process}, there exists a $K$ dimensional Gaussian process $\tilde G = \tilde{G}_n$ on $(C[I])^K$ (to lighten notation we suppress dependence of $\tilde G$ on $n$). With the property that for all $1 \leq k_1 \leq k_2 \leq K$, and all $t_1, t_2 \in I$ and all $n$,
\[
\Cov(\tilde{G}^{k_1}(t_1),\tilde{G}^{k_2}(t_2)) = \Cov\Big(\frac{\sum_{i=1}^{n}\tilde f_{i}^{k_1}}{\sqrt{n}}(t_1),\frac{\sum_{i=1}^{n}\tilde f_{i}^{k_2}}{\sqrt{n}}(t_2)\Big). 
\]  
We now show that that the process $\tilde G$ satisfies (A1) and (A2) (with the constants $C_H$ and $C_\phi$ possibly enlarged by a numeric multiplicative factor) and (A3), (A4) whenever $\tilde f$ does. 

With respect to (A1), denoting by $\tilde G^k$ the $k^{th}$ marginal process of $\tilde G$, we have that  
for all $1 \leq k \leq K$, and all $t_{1}, t_{2} \in I$,
\begin{align*}
&\|{\tilde G^{k}}(t_{1})-{\tilde G^{k}}(t_{2})\|_{\Psi_{1}} = \sqrt{\frac 83}(\Var({\tilde G^{k}}(t_{1}) - {\tilde G^{k}}(t_{2})) )^{1/2} = \sqrt{\frac 83}(\Var(\tilde f{^{k}}(t_{1}) - \tilde f{^{k}}(t_{2})) )^{1/2}
\\ &\leq \sqrt{\frac 83}C_{2,1}\|\tilde f{^{k}}(t_{1})-\tilde f{^{k}}(t_{2})\|_{\Psi_{1}} \leq \sqrt{\frac 83}C_{2,1}C_{H}|s-t|^{q},
\end{align*}
where in the first equality we used that the increments of each $G_k$ are Gaussian and $\sqrt{\frac 83}$ is equal to the $\Psi_1$ norm of a standard Gaussian and in the first inequality we utilize Lemma \ref{relationship_norms} and the constant $C_{2,1}$ defined in its statement, and in the final inequality we use assumption (A1). Therefore $\tilde G$ satisfies (A1) (with constant$\sqrt{\frac 83}C_{2,1}C_{H}$ instead of $C_H$) 

With respect to (A2), by Lemma~\ref{lem:gaussian_process}, and using the notation introduced in there, the covariance function of the KL scores of $\tilde G$ equals the covariance function of the KL scores of $\tilde f$ so that $\tilde G$ also satisfies (A2): 
\[
\|\zeta_i^k\|_{\psi_1} = \sqrt{\frac 83}\sqrt{\Var(\zeta_i^k)} = \sqrt{\frac 83}\sqrt{\Var(\phi_i^{f_k})} \leq \sqrt{\frac 83}C_{2,1} \|\phi_i^{f_k}\|_{\psi_1}
\]
where we use Lemma \ref{relationship_norms} and the constant $C_{2,1}$ defined in its statement in the last inequality. Therefore $\tilde G$ satisfies (A2) (with constant$\sqrt{\frac 83}C_{2,1}C_{\phi}$ instead of $C_\phi$) whenever $\tilde f$ does and also (A3), (A4) whenever $\tilde f$ does (with the same constant), because their Karhunen–Loève scores have the same second order properties. Define for each $1 \leq k \leq K$

\begin{align}
&T^{\tilde G,k}(\tilde G_1^k,\dots,\tilde G_n^k)=\frac{1}{T}\sum_{\ell=1}^T \Big(\frac{1}{\sqrt{n}}\sum_{i=1}^n \tilde G^k_i(t_\ell)\Big)^2
\label{eq:defT^Gk}
\\
&T^{\tilde G}(\tilde G_1,\dots,\tilde G_n) :=  \max_{k=1}^KT^{\tilde G,k}, %\label{eq:defT_tilde G}
\\
&T^{M,\tilde G}(\tilde G_1,\dots,\tilde G_n) :=  \max_{k=1}^K\Big\|\frac{\sum_{i=1}^{n}\oplus_{j=1}^{M}\zeta^k_i}{\sqrt{n}}\Big\|^2_2, %\label{eq:defT^\tilde G}
\end{align} 
where $\tilde G_1,\dots,\tilde G_n$ are i.i.d copies of $\tilde G$. Noting the definitions of $T^d$ and $T^{M}$ in~\eqref{eq:defT^d}, ~\eqref{eq:defT^M} it is seen that $T^{\tilde G}$ and $T^{M,\tilde G}$ are derived from $\tilde G$ in the same way that $T^d$ and $T^M$ are derived from $\tilde f$, and we previously showed that $\tilde G$ satisfies (A1) and (A2).  Therefore we may apply Lemma~\ref{lem:|T^d-T^M|} with $f_1,\dots,f_n$ replaced by $\tilde G_1,\dots,\tilde G_n$ to obtain
\[
\|T^{\tilde G} -T^{M,\tilde G}\|_{\psi_{1/2}} \leq C_{3,1}[\log(K+1)]^2(M^{1-2\alpha}+T^{-q})
\]
for a constant $C_{3,1}\geq 1$ that depends only on the fixed model parameters in (A1)-(A3). Let $\delta_{3,n} = 4 C_{3,1} (M^{1-2\alpha} + T^{-q})[\log(K+1)]^2 [\log(n)]^2 \geq C_{3,1} (M^{1-2\alpha} + T^{-q})[\log(K+1)]^2 [\log(2n)]^2$ and set $M=3\lceil n^{\frac{1}{12\alpha-2}} \rceil$, as in the proof of Theorem \ref{main:thm}. By~\eqref{eq:diffcdf} and the definition of $\mA(\cdot)$ we obtain for any $s>0$
\begin{equation}\label{eq: T^G-T^MG}
\sup_{t \geq s} \Big| \Pr(T^{\tilde{G}}\leq t) - \Pr(T^{M,\tilde{G}}\leq t)\Big| \leq \Pr(|T^{M,\tilde{G}} -T^{\tilde G}| \geq \delta_{3,n} )+\mA(s, \delta_{3,n})
\leq \frac{1}{n}+\mA(s, \delta_{3,n}),
\end{equation} 
where we applied Lemma \ref{lim_max_sum} (iv) to bound $\Pr(|T^{M,\tilde{G}} -T^{\tilde G}| \geq \delta_{3,n} )$ in the last line. We also used that $T^{M,\tilde{G}}\overset {d}{=} T^{M,G}$. This is true since $n^{-1/2}\sum_{i=1}\tilde G_i$ has the same covariance structure as $n^{-1/2}\sum_{i=1}\tilde f_i$ and their KL scores, which are determined by only by second order information, have the same covariance structure. Moreover, they are both multivariate Gaussian (see Lemma~\ref{lem:gaussian_process}) and centered, thus they have the same law. Likewise, we have $T^{\tilde{G}} \overset{d}{=}T^G$, which follows from the definition of $T^G$ in the statement of the Lemma and the fact that the Gaussian vectors described there is equal in distribution to the process $\tilde G$ after it has been evaluated on the points $\{t_{\ell}\}_{\ell=1}^{T}$. Recall the sequences $\theta_n$ and $v_n$ defined in the proof of Theorem~\ref{main:thm}. Note that 
\[
q_{T^{M,G}}(1-\gamma +\theta_n+3v_n) \geq q_{T^{M,G}}((1-\gamma)/2)\geq s_{\gamma}>s_{\gamma}/2.
\]
The first inequality follows from positivity of $\theta_n$ and $v_n$ and the second from \eqref{key_property_s_2}. It follows from \eqref{eq: T^G-T^MG} with $s=s_\gamma/2$ and Lemma ~\ref{lem:koltoquant} (with $t_{*} = s_{\gamma}/2$, $F$ the cumulative distribution function of $T^{M,G}$, $G$ the cumulative distribution function of $T^G$, $\delta=\frac 1n+\mA(s_\gamma/2, \delta_{3,n})$, $\alpha=1-\gamma+\theta_n + 3v_n$) that 
\[
q_{T^G}\big(1-\gamma+\theta_n + 3v_n + n^{-1} +\mA(s_\gamma/2, \delta_{3,n})\big) \geq  q_{T^{M,G}}\big(1-\gamma+\theta_n + 3v_n\big).
\]
It then follows from ~\eqref{eq:finalrelatequantile} that 
\begin{align*}
&\Pr\Big( \hat q_{1-\gamma} \leq q_{T^G}\big(1-\gamma+\theta_n + 3v_n + n^{-1} +\mA(s_\gamma/2, \delta_{3,n})\big) \Big)
\\
&\geq \Pr\Big(\hat q_{1-\gamma} \leq q_{T^{M,G}}\big(1-\gamma+\theta_n+3v_n\big) \Big) \geq 1 - \frac{2}{n} - \frac{1}{N} - w_n
\end{align*}
We note that in Steps 7 and 8 of the proof of Theorem \ref{main:thm}, $w_n$ is set as $w_n=\frac 1n$. 
We also note that there there is a constant $C_{\delta,3,1}\geq 1$ such that $\delta_{3,n} \leq C_{\delta,3,1}\delta_{1,n}$ where $\delta_{1,n}$ is defined in the proof of Theorem~\ref{main:thm}. Under (A1)-(A3) this implies that there is a constant $C_{3,2}$ such that 
\begin{equation}\label{eq:pfThm2boundAsdelta3}
\mA(s_\gamma/2, \delta_{3,n}) \leq C_{3,2}(M^{1-2\alpha}+T^{-q})[\log(K+1)]^{3}[\log(n)]^3M\log(nK).    
\end{equation}
where we have argued as we did to obtain \eqref{eq: bound mA (A1)-(A3)} and used that (when applying  Lemma~\ref{lem:ac_mainthm_A1_A3}) $\delta_{3,n} \geq \frac 1n$. Likewise under (A1)-(A4),
\begin{equation}\label{eq:pfThm2boundAstildedelta3}
\mA(s_\gamma/2,\delta_{3,n}) \leq C_{3,3}(M^{1-2\alpha}+T^{-q})[\log(K+1)\log(n)]^3[\log(K+1)]^{\frac{1}{2}+\frac{3\beta}{2\alpha-1}}[\log(n)]^{\frac{3\beta}{2\alpha-1}}
\end{equation}
where we argue as we did leading up to ~\eqref{eq: bound mA (A1)-(A4)}.

We further note that $\theta_n+3v_n+\frac 1n+\mA(s_\gamma/2, \delta_{3,n}) \leq \eps$ under both sets of assumptions that we consider, as shown in Remark \ref{remark:sequence_bounds}. Therefore, 
\begin{equation}\label{eq:power_proof_step_1_final_bound}
\Pr\Big(\hat q_{1-\gamma} \leq q_{T^{G}}\big(1-\gamma+\eps\big) \Big) \geq 1 - \frac{3}{n} - \frac{1}{N}.
\end{equation}

\noindent\textbf{Step 2: completion of Proof}
Combining \eqref{power_bound_1} and \eqref{eq:power_proof_step_1_final_bound} we have
\begin{align}
	&\Pr\Big(T^{k}_n \geq \hat q_{1-\gamma}\Big) = \Pr\Big(\sqrt{T^{k}_n}\geq \sqrt{\hat q_{1-\gamma}}\Big) \geq \Pr\Big(\sqrt{n}\mu^k \geq v^k+ \sqrt{\hat q_{1-\gamma}}\Big) \notag
    \\
    &\geq \Pr\Big(\sqrt{n}\mu^k \geq v^k+ \sqrt{ q_{T^G}\big(1-\gamma+\eps)\big)}\Big)-\frac3n-\frac1N \label{eq:pfthm2lowbound1}
\end{align}
We now make the following definitions for all $1 \leq k \leq K$:
\begin{align}
&T^{M,k}(\Phi_{1,k},\dots,\Phi_{n,k}) :=  \Big\|\frac{1}{\sqrt{n}} \sum_{i=1}^n \Phi_{i,k} \Big\|_2^2, \label{eq:defT^Mk}
\\ 
&T^{M,G,k}(g) := \|\pi_k g\|_2^2,
\label{eq:defT^MGk}
\\
&\Delta_{M,MG,k}(s)  := \sup_{t \geq s} \Big|\Pr(T^{M,k} \le t) - \Pr(T^{M,G,k} \le t)\Big|,
\\
&\mA_k(s,\delta) := \sup_{t \geq s} \Pr(|T^{M,G,k}-t|\leq \delta).
\end{align} 
We note that $T^{M,k}$ is a version of $T^M$ (compare~\eqref{eq:defT^M}) where we only consider one dimension, and similar comments link $T^{M,G,k}$ to $T^{M,G}$ (c.f.~\eqref{eq:defT^MG}) and $[v^k]^2$ to $T^d$; see~\eqref{eq:defT^d}. Thus we may argue as was done in Step 4 of the proof of Theorem~\ref{main:thm} to arrive at~\eqref{eq:step4fin} and obtain
\[
\sup_{t \geq s_\gamma} \Big|\Pr([v^k]^2 \leq t) - \Pr(T^{G,M,k} \leq t)\Big| \leq 3\Delta_{M,MG,k}(s_\gamma/2)+\mA_k(s_\gamma,\delta_{4,n})+\frac 1n
\]
where $\delta_{4,n}:=C_{4,4}(M^{1-2\alpha}+T^{-q})[\log(n)]^2$. We show in Remark \ref{remark:sequence_bounds} that $\delta_{4,n} <  s_\gamma/2$. Similarly,
\[
 \sup_{t \geq s_\gamma} \Big|\Pr(T^{G,k}\leq t) - \Pr(T^{G,M,k} \leq t)\Big|  \leq 3\Delta_{M,MG,k}(s_\gamma/2)+\mA_k(s_\gamma,\delta_{4,n})+\frac 1n
\]
where we have replaced $\tilde f_1^k,\dots,\tilde f_n^k$ with $\tilde G_1^k,\dots,\tilde G_n^k$ similarly to the argument in Step 1 of the present proof. Combining the results above, we find 
\begin{align*}
&\sup_{t \geq s_\gamma} \Big|\Pr([v^k]^2 \leq t) - \Pr(T^{G,k} \leq t)\Big|
\\
&\leq \sup_{t \geq s_\gamma} \Big|\Pr([v^k]^2 \leq t) - \Pr(T^{G,M,k} \leq t)\Big|+\sup_{t \geq s_\gamma} \Big|\Pr(T^{G,k}\leq t) - \Pr(T^{G,M,k} \leq t)\Big| 
\\ 
&\leq 2\Big(3\Delta_{M,MG,k}(s_\gamma/2)+\mA_k(s_\gamma,\delta_{4,n})+\frac 1n\Big)
\\
&\leq C_{5,1}d_n,
\end{align*}
where $C_{5,1}$ is a sufficiently large constant that depends only on the model parameters in (A1)-(A3). If instead (A1)-(A4) holds, a similar bound with $d_n$ replaced by $\tilde d_n$ and $C_{5,1}$ replaced by a possibly different constant $C_{5,1,1}$ that depends on the model parameters of (A1)-(A4) holds. Here, the last inequality follows from a direct computation after noting that $\mA_k$ is a one-dimensional version of $\mA$ and can be bounded by the same upper bounds as in~\eqref{eq: bound mA (A1)-(A3)} under conditions (A1)-(A3) and~\eqref{eq: bound mA (A1)-(A4)} under (A1)-(A4). 

Finally, recall that we assumed in the beginning of the proof that $\sqrt{n}\mu^k - \sqrt{ q_{T^G}\big(1-\gamma+\eps)} \geq \sqrt{s_\gamma} > 0$ and thus
\begin{align*}
&\Pr\Big(\sqrt{n}\mu^k \geq v^k+ \sqrt{ q_{T^G}\big(1-\gamma+\eps)}\Big) = \Pr\Big(v^k \leq \sqrt{n}\mu^k - \sqrt{ q_{T^G}\big(1-\gamma+\eps)} \Big) 
\\
&\geq \Pr\Big([v^k]^2 \leq \big(\sqrt{n}\mu^k - \sqrt{ q_{T^G}\big(1-\gamma+\eps)}\big)_+^2 \Big) 
\\
&\geq \Pr\Big(T^{G,k} \leq \big(\sqrt{n}\mu^k - \sqrt{ q_{T^G}\big(1-\gamma+\eps)}\big)_+^2 \Big) - C_{5,1} d_n.
\end{align*}

Combined with~\eqref{eq:pfthm2lowbound1} we obtain under (A1)-(A3)
\begin{align}
\Pr\Big(T^{k}_n \geq \hat q_{1-\gamma}\Big) &\geq \Pr\Big(T^{G,k} \leq \big(\sqrt{n}\mu^k - \sqrt{ q_{T^G}\big(1-\gamma+\eps)}\big)_+^2 \Big) - C_{5,1} d_n - \frac3n - \frac1N \notag
\\
&\geq \Pr\Big(T^{G,k} \leq \big(\sqrt{n}\mu^k - \sqrt{ q_{T^G}\big(1-\gamma+\eps)}\big)_+^2 \Big) - C_{6,1} d_n. \label{eq:pfthm2lowbound2}
\end{align}
Under (A1)-(A4), a similar bound holds with $d_n$ replaced by $\tilde d_n$ and $C_{6,1}$ replaced by a possibly different constant $\tilde C_{6,1}$.

To give a further lower bound, it is now useful to develop a tail bound for $\sqrt{T^{G,k}}$. Note that
\begin{align*}
&\|\sqrt{T^{G,k}}\|_{\Psi_{2}} =\|T^{G,k}\|^{\frac12}_{\Psi_{1}} =\Big(\Big\|\frac{1}{T}\sum_{\ell=1}^T [G_n^{k,\ell}]^2\Big\|_{\Psi_{1}}\Big)^{\frac12} 
\leq \Big(\frac{1}{T}\sum_{\ell=1}^{T}\|[G_n^{k,\ell}]^2\|_{\Psi_{1}}\Big)^{\frac12} =\Big(\frac{1}{T}\sum_{\ell=1}^{T}\|G_n^{k,\ell}\|_{\Psi_2}^2\Big)^{\frac12} 
\\
&\leq\big(\max_{\ell=1}^{T}\|G_n^{k,\ell}\|_{\Psi_2}^2\big)^{\frac12} =\max_{\ell=1}^{T}\|G_n^{k,\ell}\|_{\Psi_2} \leq \sqrt{\frac83\max_{\ell=1}^{T}\Var(f_{k,1}(t_{\ell}))} \leq D_1,
\end{align*}
for a constant $D_1$ that depends only on $\alpha$, $C_{\phi}$, $C_{H}$ and $q$ under either set of conditions we consider. We have applied Lemma~\ref{relationship_norms} in the first equality, the triangle inequality for the $\psi_1$ norm in the first inequality, Lemma~\ref{relationship_norms} in the third and fourth equality. The penultimate inequality uses the fact that if $Z$ is centered Gaussian, $\|Z\|_{\Psi_2}=\sqrt{\tfrac83\Var(Z)}$, and the fact that $G_n$ defined in the statement of the Lemma we are proving satisfies $\Var(G_n^{k,\ell}) = \Var(f_{k,1}(t_\ell))$ for all $1 \leq k \leq K$, and all $1 \leq \ell \leq T$. The final inequality uses Lemma~\ref{lem:sup_process} (note that we do not explicitly assume that $\sup_{t}\Var(f_{k,1}(t))$ is finite in Assumptions \ref{assumptions}) and the fact that $f_{k,1}$ is centered which implies $\sqrt{\Var(f_{k,1}(t_\ell))} = \E[(f_{k,1}(t_\ell))^2]^{1/2} \leq \E[\sup_t (f_{k,1}(t))^2]^{1/2} \leq C_{1,2} \big\|\sup_t |f_{k,1}(t)|\big\|_{\psi_1}$ where the last inequality utilizes Lemma~\ref{relationship_norms}. It follows then from Lemma ~\eqref{lim_max_sum}(iv) combined with~\eqref{eq:pfthm2lowbound2} that
\[
\Pr\Big(T^{k}_n \geq \hat q_{1-\gamma}\Big) \geq 1-2\exp\big(-(\sqrt{n}\mu^k -\sqrt{q_{T^{G}}(1-\gamma+\eps)})^2_{+}/D_1^2\big)-C_{6,1} d_n
\]
under (A1)-(A3) and similarly with $\tilde d_n$ in place of $d_n$ and a possibly different constant $\tilde C_{6,1}$ under (A1)-(A4). This completes the proof.

\hfill $\Box$

\bigskip
\bigskip

\begin{remark}\label{remark:sequence_bounds}
We demonstrate the bounds on $\delta_{1,n}$, $\delta_{2,n}$ claimed in Step 4 and Step 5 of the proof of Theorem \ref{main:thm} and verify that our choices of $a_n$, $b_n$, $v_n$ and $w_n$ satisfy \eqref{eq:defanbn} and \eqref{eq:defcndn}. Recall that $M \leq 6n^{\frac{1}{12\alpha-2}}$ and

\[
C = [80\big(C_{4,1} \vee C_{5,1}\vee C_{7,3}\vee C_{7,1} \vee 1\big)]^{-1}, \hspace{.5cm}\tilde{C} = [50\big(C_{4,1} \vee  C_{5,1} \vee C_{8,3} \vee C_{8,1} \vee1 \big)]^{-1}
\]
and 
\[d_n = \log^{3}(K+2)\log^{3}(n)\log(n(K+1))[n^{\frac{1-\alpha}{6\alpha-1}}+n^{\frac{1}{12\alpha-2}}T^{-q}] + \sqrt{\frac{\log(N)}{N}}
\]
and 
\[
\tilde d_n= \log^{3+\frac{3\beta}{2\alpha-1}}(n)\log^{\frac 72+\frac{3\beta}{2\alpha-1}}(K+2)\Big[n^{\frac{\frac{1}{2}-\alpha}{6\alpha-1}}+T^{-q}\Big]+\sqrt{\frac{\log(N)}{N}},
\]
and we have assumed without loss of generality that

\[d_n \leq Cs_{\gamma}/2, \hspace{.5cm} \tilde d_n \leq \tilde{C}s_{\gamma}/2.\]
In many cases we use a calculator to determine the size of numeric constants.
\newline
\newline
\textbf{\em It holds that $\delta_{1,n}< s_{\gamma}/2$ under (A1)-(A3) and also under (A1)-(A4)}
Since $\delta_{1,n} \leq \delta_{2,n}$ this follows from the claim for $\delta_{2,n}$ which we show next.
\newline
\newline
\textbf{\em It holds that $\delta_{2,n}< s_{\gamma}/2$ under (A1)-(A3) and also under (A1)-(A4)}
\newline
\textbf{\em Case (A1)-(A3):}
\begin{align*}
&\delta_{2,n}=8(C_{5,1}\vee C_{4,1})(M^{1-2\alpha} + T^{-q}) [\log(K+1)]^3[\log (n)]^3\\
&< 48(C_{5,1}\vee C_{4,1})[\log(K+1)]^3[\log(n)]^3(\big(n^{\frac{1-\alpha}{6\alpha-1}}n^{\frac{-\alpha}{6\alpha-1}}\big)^{\frac 12}+T^{-q})\leq 27(C_{5,1}\vee C_{4,1})d_n < s_{\gamma}/2 \\
\end{align*}
\newline
\textbf{\em Case (A1)-(A4):}
Arguing as in the previous case for the first inequality,
\begin{align*}
&\delta_{2,n}<48(C_{5,1}\vee C_{4,1})[\log(K+1)]^3[\log(n)]^3(\big(n^{\frac{1-\alpha}{6\alpha-1}}n^{\frac{-\alpha}{6\alpha-1}}\big)^{\frac 12}+T^{-q}) <48(C_{5,1}\vee C_{4,1})\tilde d_n < s_{\gamma}/2
\end{align*}
\newline
\textbf{\em It holds that
\begin{equation}\label{eq:boundvnbydntildedn}
v_n \leq 10C_{7,3} d_n \mbox{ under (A1)-(A3) and also } v_n \leq 7 C_{8,3} \tilde d_n \mbox{ under (A1)-(A4)}
\end{equation}
}
\newline
Note that $\frac{\log(K+1)+\log(n)}{\log(K+1)\log(n)} \leq \frac{1}{\log(2)}+\frac{1}{\log(3)} \leq 3$ and also $Mn \leq 6 n^{1+\frac{1}{12\alpha-2}} \leq n^3$ since $n\geq 3, \alpha \geq 1$
\newline
\medskip

\noindent
\textbf{\em Case (A1)-(A3):}
\begin{align*}
v_n&=C_{7,3}M^{4/3} \log(Kn)\Big( \frac{\log(KMn)\log^{2}(n)}{n}\Big)^{1/6} 
\\
&\leq C_{7,3}\times 6^{4/3}3^{1/6}\big[n^{\frac{\frac 56-\alpha}{6\alpha-1}}\big][\log(Kn)]^{1/6}[\log(n)]^{1/3}\log(Kn)\\
&<16C_{7,3}\big[n^{\frac{\frac 56-\alpha}{6\alpha-1}}\big][\log(K+1)\log(n)]^{1/6}[\log(n)]^{1/3}\log(Kn)\\
&< 10 C_{7,3}\big[n^{\frac{\frac 56-\alpha}{6\alpha-1}}\big][\log(K+2)]^3[\log(n)]^{3}\log(n(K+1)) \leq 10 C_{7,3}d_n <  s_\gamma/2
\end{align*}

\noindent
\textbf{\em Case (A1)-(A4):}
Arguing as in the previous case for the first inequality, 
\begin{align*}
&v_n =C_{8,3}\Big[M^{2/3} \log^{1/3}(Kn)\Big( \frac{\log(KMn)\log^{2}(n)}{n}\Big)^{1/6}\Big]\log^{\frac12+\frac{3\beta}{2\alpha-1}}(K+1)\log^{\frac{3\beta}{2\alpha-1}}(n)\\
&<6^{2/3}3^{1/6} C_{8,3}\big[n^{\frac{1}{3(6\alpha-1)}-\frac 16}\big][\log(Kn)]^{1/2}[\log(n)]^{1/3}\log^{\frac12+\frac{3\beta}{2\alpha-1}}(K+1)\log^{\frac{3\beta}{2\alpha-1}}(n)\\
&<7C_{8,3}\big[n^{\frac{1}{3(6\alpha-1)}-\frac 16}\big][\log(K+1)\log(n)]^{1/2}[\log(n)]^{1/3}\log^{\frac12+\frac{3\beta}{2\alpha-1}}(K+1)\log^{\frac{3\beta}{2\alpha-1}}(n)\\
&<7C_{8,3}\times\tilde d_n < s_\gamma/2
\end{align*}
\newline
\textbf{\em It holds that $a_n \leq d_n/(2C)$ under (A1)-(A3) and $a_n \leq \tilde{d}_n/(2\tilde{C})$ under (A1)-(A4)}
Recall that in  \eqref{eq:bound_an} we set $a_n =2/n+3v_n+\mathcal{A}(s_\gamma, \delta_{2,n})$.
\newline

\textbf{\em Case (A1)-(A3):}
We note that the bound in~\eqref{eq: bound mA (A1)-(A3)} implies $\mathcal{A}(s_\gamma, \delta_{2,n}) \leq 6 C_{7,1} d_n$ and thus by~\eqref{eq:boundvnbydntildedn}
\begin{align*}
&a_n =  2/n+3v_n+\mathcal{A}(s_\gamma, \delta_{2,n})\\
&<2d_n+30C_{7,3}d_n+ 6 C_{7,1} d_n
< 38 (1 \vee C_{7,1} \vee C_{7,3})d_n< d_n/(2C)
\end{align*}

\textbf{\em Case (A1)-(A4):}
Using the bound of \eqref{eq: bound mA (A1)-(A4)} we obtain $\mathcal{A}(s_\gamma, \delta_{2,n}) \leq C_{8,1}\tilde d_n$ and thus by~\eqref{eq:boundvnbydntildedn}
\begin{align*}
&a_n =  2/n+3v_n+\mathcal{A}(s_\gamma, \delta_{2,n}) \leq 24(1 \vee C_{8,1} \vee C_{8,3})\tilde d_n < \frac{\tilde d_n}{2\tilde C}
\end{align*}
\newline 
We now demonstrate the bound on $\theta_n+3v_n+\frac 1n+\mA(s_\gamma/2, \delta_{3,n})$ indicated in the proof of Theorem~\ref{main:thm-power}. Recall that 
\begin{align*}
\theta_n &= \frac{2}{n} + A(s_\gamma,\delta_{2,n}) + \sqrt{\frac{2\log N}{N}}
\end{align*}
and that equations \eqref{eq:pfThm2boundAsdelta3} and \eqref{eq:pfThm2boundAstildedelta3} in the body of the proof of Theorem \ref{main:thm-power} imply $\mA(s_\gamma/2, \delta_{3,n}) \leq C_{3,2} d_n$ under (A1)-(A3) and $\mA(s_\gamma/2, \delta_{3,n}) \leq C_{3,3} \tilde d_n$ under (A1)-(A4). We have 
\begin{align*}
&\theta_n+3v_n+\frac 1n+\mA(s_\gamma/2, \delta_{3,n}) = \frac 3n+\mA(s_\gamma,\delta_{2,n})+\mA(s_\gamma/2,\delta_{3,n})+3v_n + \sqrt{\frac{2\log N}{N}}\\
&=a_n+\frac 1n+\mA(s_\gamma/2,\delta_{3,n})+ \sqrt{\frac{2\log N}{N}}
\end{align*}
If (A1)-(A3) holds, we have 
\begin{align*}
&a_n+\frac 1n+\mA(s_\gamma/2,\delta_{3,n})+ \sqrt{\frac{2\log N}{N}} \leq \big({38}(1 \vee C_{7,1} \vee C_{7,3}) +1 +\sqrt{2}+C_{3,2})\big)d_n\\
&\leq {42}(1 \vee C_{7,1} \vee C_{7,3}\vee C_{3,2})d_n <\eps.
\end{align*}
\medskip
If (A1)-(A4) holds, we have
\begin{align*}
&a_n+\frac 1n+\mA(s_\gamma,\delta_{3,n})+ \sqrt{\frac{2\log N}{N}} \leq \big(24(1 \vee C_{7,1} \vee C_{7,3}) +1 +\sqrt{2}+C_{3,3})\big)d_n\\
&\leq 28(1 \vee C_{7,1} \vee C_{7,3}\vee C_{3,3})\tilde d_n <\eps.
\end{align*}
\textbf{\em It holds that $\delta_{4,n}< s_{\gamma}/2$ under (A1)-(A3) and also under (A1)-(A4)}
Noting that by similar calculations as performed for $\delta_{2,n}$ above we have under (A1)-(A3) noting that   $6/(\log(6) \times (\log 3)^4 ) \leq 3$,
\[
\delta_{4,n}:=C_{4,4}(M^{1-2\alpha}+T^{-q})[\log(n)]^2 \leq 3(C_{5,1} \vee C_{4,1} \vee C_{4,4})d_n  < s_\gamma/2.
\]
Under (A1)-(A4) we have similarly
\[
\delta_{4,n}:=C_{4,4}(M^{1-2\alpha}+T^{-q})[\log(n)]^2 \leq 6(C_{5,1} \vee C_{4,1} \vee C_{4,4})\tilde d_n < s_\gamma/2.
\]
\end{remark}

\subsection{Technical results}\label{sub-section:technical_results}

%\newpage

\subsubsection{Approximations of Convex Sets by Polytopes}\label{Appendix:Polytopes}

Let $\bar{B}^{M}_{t} \subset \R^{M}$ be the closed unit ball of radius $t$ for $\|\cdot\|_{2}$, and let $S^{M-1}_{t} = \partial \bar{B}^{M}_{t} = \{x \in \R^{M}, \|x\|_{2}=t\}$.  For $A \subset \R^m$ and $B\subset \R^n$ let $A \times B \subset \R^{m+n}$ be the cartesian product of $A$ and $B$. Sometimes we will write $(\R^{M})^{K}$ instead of $\R^{MK}$ to emphasize the identity $\R^{MK} = \times_{k=1}^{K}\R^{M} = \oplus_{k=1}^{K}\R^{M}$. Sometimes we will drop superscripts and subscripts to lighten notation where no confusion will arise.

Given $x \in (\R^{M})^{K}$ we approximate each level set (in a sense defined below) of the Euclidean $L^{\infty}-L^{2}$ mixed norm $x \rightarrow \max_{k=1}^{K}\|\pi_{k}x\|_{2}$ (recall the definition of $\pi_k$ in \ref{sub-section:additional_notation}) by a level set of a  polytope (defined below) that has special properties. 

\begin{definition}\label{def:E_t}
	$\newline$
	Let 
	\[
	E_{t} := \Big\{x \in (\R^{M})^{K}: \max_{k=1}^K \|\pi_{k}(x)\|_{2}\leq t \Big\} = \times_{k=1}^K \bar{B}^{M}_{t}.
	\] 
\end{definition}
Our general strategy for working with a set $E_{t}$ is to approximate it by half-spaces.  Given a finite collection of unit vectors $\mathcal{V} \subset \R^{d}$ and a positive real number $s$, we define 
\[
A(\mathcal{V},s) = \{w \in \R^d: \forall v \in \mathcal{V},  v^{\top}w \leq s\}.
\]
We will refer to sets $A(\mathcal{V},s)$ as polytopes. Note that for any $\mathcal{V}$, $\bar{B}^{d}_{s} \subset A(\mathcal{V},s)$.

\begin{lemma}\label{lem:product_spheres_approximation}
	For each pair of real numbers $t$ and $\eps$ satisfying $t \geq \eps>0$, and each $K \geq 1$, $M \geq 3$, there exists a collection of unit vectors $\mathcal{V}_{\eps,t}$ such that
	
	\begin{enumerate}
		\item $|\mathcal{V}_{\eps,t}| \leq K(1 + t\eps^{-1})^{\frac{M-1}{2}} 4^{M-1}$ so that $\log(|\mathcal{V}_{\eps,t}|)\leq 3M\log(K(1 + t\eps^{-1}))$.
		
		\item $E_{t-\eps}\subset A(\mathcal{V}_{\eps,t},t-\eps) \subset E_t \subset A(\mathcal{V}_{\eps,t},t) \subset E_{t+\eps}$
		\item If $v \in \mathcal{V}_{\eps,t}$ there is a unique $k$ such that  $\pi_{k}v \neq 0$. 
	\end{enumerate} 
\end{lemma}

\begin{definition}\label{def:collection_V}
	For $t \geq \eps$, $K \geq 1$,  $M \geq 3$ let  $\mathcal{V}_{\eps, t} \subset (\R^{M})^{K}$ \footnote{Note that applying a fixed non-identity orthogonal transformation of the form $O=\oplus^{K}_{k=1}O_k$, $O_{k}\in O(M, \R)$, to each element of a collection $\mathcal{V}^{1}_{\eps, t}$ of Lemma \ref{lem:product_spheres_approximation} produces a distinct collection $\mathcal{V}^{2}_{\eps, t}$ satisfying the same properties.} be a collection of unit vectors satisfying conditions (a)--(c)of Lemma~\ref{lem:product_spheres_approximation}.
\end{definition}

To prove Lemma \ref{lem:product_spheres_approximation} we use Lemma \ref{lem:sphericalapproximation} whose proof relies on the estimate from Lemma \ref{lem:rad}. Lemma \ref{lem:rad} is a restatement of Lemma 2.3  on page 11 of \cite{B97}. Lemma \ref{lem:sphericalapproximation} is a straightforward generalization of an argument contained in \cite{B97}.

\begin{lemma}[Lower bound on surface area of spherical cap]\label{lem:rad}
	Let $S$ be the boundary of the $M$ dimensional unit ball $B$. Let $\mu$ be normalized surface measure on $S$, so that $\mu(S)=1$. Let $A_{r,v} = \{x \in S, \|x-v\|_2 \leq r\}$. Then for $0 \leq r \leq 2$, $\mu(A_{r,v})\geq \frac{1}{2}(r/2)^{M-1}$. 
\end{lemma}	

\begin{lemma}[Approximation of Ball by Polytopes]\label{lem:sphericalapproximation}
	Let $B_t \subset \R^{M}$ be the closed unit ball of radius $t$. Suppose $M \geq 3$.   For each $1 \geq \eps>0$ and every $t\geq \eps$  there exists a collection of unit vectors $\mathcal{V}$ such that
	\begin{enumerate}
		\item $|\mathcal{V}| \leq (1 + \eps^{-1})^{\frac{M-1}{2}} 4^{M-1}$
		\item $A(\mathcal{V},\frac{t}{1+\epsilon}) \subset B_t \subset A(\mathcal{V},t)$
	\end{enumerate}
\end{lemma}
\bigskip

\subsubsection{Proofs for Section~\ref{Appendix:Polytopes}}

\noindent \textbf{Proof of Lemma \ref{lem:product_spheres_approximation}}

Fix $t$ and $\eps$ satisfying $t\geq \eps>0$. We first prove the claim in case $K=1$. By Lemma~ \ref{lem:sphericalapproximation} there exists a collection $\mathcal{V}$ such that 
\begin{equation}\label{first_equation_lem:product_spheres_approximation}
A(\mathcal{V},\tfrac{t}{1+t^{-1}\eps}) \subset B_t \subset A(\mathcal{V},t) \subset(1+\tfrac{\eps}{t})B_t = B_{t+\eps},
\end{equation}
and $|\mathcal{V}|\leq (1 + t\eps^{-1})^{\frac{M-1}{2}} 4^{M-1}$. Since 
\[
\frac{t}{1+t^{-1}\eps} = \frac{t^2}{t+\eps} > \frac{t^2}{t+\eps} - \frac{\eps^2}{t+\eps} =  t-\eps,
\]
we have
\[
A(\mathcal{V},t-\eps) \subset A(\mathcal{V},\tfrac{t}{1+t^{-1}\eps}),
\]
and so,
\[
A(\mathcal{V},t-\eps) \subset B_t.
\]
The inclusions $B_{t-\eps} \subset A(\mathcal{V},t-\eps)$ and $B_{t}\subset A(\mathcal{V},t)$ hold trivially by the remark following Definition \ref{def:E_t}. Note that

\begin{equation}\label{bound on log|V|}
\begin{split}
&\log\Big(|\mathcal{V}|\Big)\leq 2(M-1)\log(2)+\Big((M-1)/2\Big)\log(1+t\eps^{-1})	\\
&\leq 2(M-1)\log(1+t\eps^{-1})+\Big((M-1)/2\Big)\log(1+t\eps^{-1})	\\
&\leq 3M\log(1+t\eps^{-1}).	\\
\end{split}
\end{equation}
This proves the claim in case $K=1$ and we now prove the general case. To each $v \in \mathcal{V} \subset \R^{M}$ we now assign a collection of $K$ vectors $\{v_{1}, \dots, v_{k}\}\subset (\R^{M})^K$. Let $v_1(v) =  v \oplus_{i=2}^{K}\mathbf{0}_{\R^M}$, let $v_{K}(v) =  \oplus_{i=1}^{K-1}\mathbf{0}_{\R^M} \oplus v$, and for $2 \leq k \leq K-1$, let $v_k(v) = \oplus_{i=1}^{k-1}\mathbf{0}_{\R^M}\oplus v \oplus_{i=k+1}^{K}\mathbf{0}_{\R^M}$. Now let $\mathcal{V}_{K} = \cup_{i=1}^{K} \cup_{v \in\mathcal{V}} \{v_k(v)\}$. It is straightforward to see that  
\[
E_{t-\eps} \subset A(\mathcal{V}_K,t-\eps) \subset E_t \subset A(\mathcal{V}_K,t) \subset E_{t+\eps},
\]
and that if $w \in \mathcal{V_{K}}$, $\pi_{k}w\neq 0$ for a unique $k$. Then for every $K \geq 1$ %(in particular for $K \geq 1$)

\begin{align*}
&\log\Big(|\mathcal{V}_K|\Big) = \log\Big(K|\mathcal{V}|\Big) = \log\Big(K\Big)+\log\Big(|\mathcal{V}|\Big) \leq 3M\log\Big(K\Big)+3M\log\Big(1+t\eps^{-1}\Big)\\ 
&= 3M\log\Big(K(1+t\eps^{-1})\Big),
\end{align*}
where in the first inequality we used the bound of \eqref{bound on log|V|}.

\hfill $\Box$

\textbf{Proof of Lemma \ref{lem:sphericalapproximation}}
It suffices to find a collection $\mathcal{V}$ of asserted size such that
\[
B_1 \subset A(\mathcal{V},1) \subset (1+\eps)B_1.
\]
Scaling by $\frac{1}{1+\eps}$ all sets in the chain of inclusions then gives
\[
A(\mathcal{V},\tfrac{1}{1+\epsilon}) \subset B_1 \subset A(\mathcal{V},1).
\]
Scaling next by $t$ gives
\[
A(\mathcal{V},\tfrac{t}{1+\epsilon}) \subset B_t \subset A(\mathcal{V},t).
\]
The remainder of the proof is a modification of the argument presented on page 11 of \cite{B97}, which deals with the case $\eps=1$. We modify it to cover the case of arbitrary $1\geq \eps>0$ and add some details. 

$\newline$
We will show that $\mathcal{V}$ can be chosen so that   $|\mathcal{V}| \leq (1 + \eps^{-1})^{\frac{M-1}{2}} 4^{M-1}$ while 
\[
B_1 \subset A(\mathcal{V}, 1) \subset (1+\eps)B_1.
\]
Recalling that $\mathcal{V}$ is a collection of unit vectors $B_1 \subset A(\mathcal{V}, 1)$ is automatic. The inclusion $A(\mathcal{V}, 1) \subset (1+\eps)B_1$ holds if and only if  $\forall \theta \in S_1$, $\exists v \in \mathcal{V}$ such that $\langle \theta, v \rangle \geq \frac{1}{1+\eps}$, however we only require the `if' direction of this statement, which is straightforward to check using the definitions of $A(\mathcal{V}, 1)$ and $(1+\eps)B_1 = B_{1+\eps}$.

We will now identify a collection $\mathcal{V}$ with property that $\forall \theta \in S_1$, $\exists v \in \mathcal{V}$ such that $\langle \theta, v \rangle \geq \frac{1}{1+\eps}$. To this end, let $u$ be a generic unit vector, and consider the set
\[
C(\eps, u) = \{y \in S_1,\langle u, y \rangle \geq \tfrac{1}{1+\eps}\}.
\] 
Elementary computations shows that for $\tilde{r}=\sqrt{\frac{2\eps}{1+\eps}}$.
\[
C(\eps, u) = \{y \in S_1, \|y-u\|_{2} \leq \tilde{r}\}.
\]
Let $\{p_{i}\}_{i=1}^{m}$ be a maximal $\tilde{r}$ packing of $S_1$ with respect to the Euclidean metric inherited from $\R^{M}$. By the triangle inequality, the sets $\{y \in S_1, \|y-p_{i}\|_2 \leq \frac{\tilde{r}}{2}\}$ are disjoint. Let $\mu$ be normalized surface measure, and note that $0 \leq \tilde{r} \leq 2$. By Lemma \ref{lem:rad},
\[
m \frac{1}{2}(\tilde{r}/4)^{M-1} \leq \sum_{i=1}^{m}\mu(\{y \in S_1, \|y-p_{i}\|_{2} \leq \frac{\tilde{r}}{2}\}) \leq 1.
\]
Rearranging this equation shows that for $M \geq 3$, $m \leq 4^{M-1}(1+\eps^{-1})^{\frac{M-1}{2}}$. \hfill $\Box$

\subsubsection{Anti-Concentration Bounds}
\label{Appendix:Anti-Concentration}

We begin the section by stating two Lemmas that are used in the proof of Theorem~\ref{main:thm}. Recall the definition of $T^{M,G}$ in Section \ref{sub-section:additional_notation}, and that

\[
	\mA(s,\delta) := \sup_{t \geq s} \Pr(|T^{M,G}-t|\leq \delta),
\]
 defined above in Equation~\eqref{def:ac-function}.

\begin{lemma}\label{lem:ac_mainthm_A1_A3}
	Suppose conditions (A1) - (A3) hold. There exists a constant $A = A(\alpha, C_{\phi})$ such that that if $K \geq 1$, $M \geq 3$, $m\geq 3$, and $\delta \geq \frac{1}{m}$,
	\[
	\mA(s,\delta) \leq \frac{A}{s \wedge 1}\delta M\log\Big(Km\Big).
	\]
\end{lemma}

\begin{lemma}\label{lem:ac_mainthm_A1_A4}
	Suppose conditions (A1) - (A4) hold. There exists a constant $A=A(\alpha, \beta, C_{\phi}, C_2)$ that does not depend on $K$, $n$, $M$, $s$ or $\delta$such that if $K \geq 1$, $M \geq 1$, $m \geq 3$,  $\delta \geq \frac{1}{m}$, 
	\[
	\mA(s,\delta) \leq \frac{A}{(s \wedge 1)^{\frac{1}{2}+\frac{3\beta}{2\alpha-1}}}\delta\log^{\frac{1}{2}+\frac{3\beta}{2\alpha-1}}(K+1)\log^{\frac{3\beta}{2\alpha-1}}(m).
	\]
\end{lemma}

In the rest of this section we state  anti-concentration bounds for Gaussian maxima, $L^{\infty}-L^{2}$ mixed norms and $\eps$-nets for the sets $E_t$.
Recall Nazarov's inequality (see \cite{CCKNaz}): 
\begin{lemma}\label{lem:naz_ineq}
	Let $X \in \R^{p}$ ($p\geq 1$), be a centered gaussian with positive semi-definite covariance matrix $\Sigma$. If $\min_{j}\sqrt{\Sigma_{jj}} \geq \sigma_{\min}$,
	\begin{equation}\label{eq:naz_ineq}
		\sup_{t \in \mathbb{R}}\Pr\Big(\max_{j=1}^{p}X_{j} \in [t, t+\delta]\Big) \leq \delta\frac{\sqrt{2\log(p)}+2}{\sigma_{\min}}.
	\end{equation}
\end{lemma}
Our proof of the following bound is mainly from \cite{Bel2018}.

\begin{lemma}\label{lem: Belloni anti-concentration}
	
	Let $X \in \R^{p}$( $p\geq 2)$, be a centered gaussian with positive semi-definite covariance matrix $\Sigma$. For any $s>0$ and any $\delta >0$,
    \[
	\sup_{t \geq s } \Pr\Big(\Big|\max_{j=1}^{p}X_{j}-t\Big| \leq \delta\Big) \leq \frac{84}{5} \frac{\delta }{s}\log(p).
	\]
\end{lemma}

We will also use the following bound which appears in \cite{CCK12c} as Theorem 3 (ii). Dependence of the constant on $\sigma_{\max}$ and $\sigma_{\min}$ is easily determined by inspecting the proof in \cite{CCK12c}; note that the bound behaves correctly under rescaling of $X$.
\begin{lemma}[Theorem~3~(ii) in \cite{CCK12c}] \label{dimension_free_extension}
	Let $X \in \R^{p}$( $p\geq 2)$, be a centered Gaussian with covariance matrix $\Sigma$. Let $\sigma_{\min}=\min_{j}\sqrt{\Sigma_{jj}} \leq \max_{j}\sqrt{\Sigma_{jj}} = \sigma_{\max}$. Then for every $\delta>0$,
	\[
	\sup_{t\in \R}\Pr\Big(|\max_{j=1}^{p}X_j-t| \leq \delta\Big)\leq C\frac{\sigma_{\max}}{\sigma_{\min}^{2}}\delta\Big\{\E\Big[\max_{j=1}^{p}\Big(\frac{X_j}{\sqrt{\Sigma_{jj}}}\Big)\Big]+\sqrt{1\vee \log\Big(\frac{\sigma_{\min}}{\delta}\Big)}\Big\},
	\]
	for a universal constant $C$.
\end{lemma}

We note that if $W \in (\R^{M})^K$ is a centered Gaussian, the variable $S = \max_{k=1}^{K}\|W_k\|_2$ is absolutely continuous.

\begin{lemma}[Anti-Concentration for Mixed Norms of Gaussian Variables]\label{lem:ac_mixed_norm}
	$\newline$
	Let $W \in (\R^M)^K$ be a centered gaussian with positive semi-definite covariance matrix $\Sigma$. Let $S = \max_{k=1}^{K}\|W_k\|_2$.  Suppose $M \geq 3$ and $K \geq 1$. Then, for all $\delta, \eps, t >0$ satisfying $\delta+\eps <t$,
	\begin{equation}\label{ac:mn_general_bound}
		\Pr\Big(S \in [t-\delta, t]\Big) 
		\leq \frac{51}{2}\frac{\delta+\eps}{t}M\log(K(1 + (t-\delta)\eps^{-1})).
	\end{equation}	
	For each $k$ let $\Sigma_{k}:= \Var(\pi_{k}W)$, and suppose
	\begin{equation}\label{def:U_for_ac}
	\max_{k=1}^{K}\sqrt{\sum_{j=1}^{M}(\Sigma_k)_{jj}}\leq U.
	\end{equation}
	If in addition to the conditions above we assume $\eps<2/\exp(1)$ then
	\begin{equation}\label{ac:smn_tail_bound}
		\Pr\Big(S \in [t-\delta,t]\Big)  
		\leq  51\Big\{\frac{\delta+\eps}{t}M\log\Big(\frac{4(K_{\Psi_{1}}\vee1)^{\frac{1}{2}}K(U\vee 1)}{\eps}\Big) \Big\}\vee \eps,
	\end{equation}
where $K_{\Psi_{1}}$ is the constant from Lemma ~\ref{lim_max_sum} (ii). 
\end{lemma}

\begin{corollary}[Anti-Concentration for Squared Mixed Norms of Gaussian Variables]\label{lem:ac_smixed_norm}
Under the conditions of Lemma~\ref{lem:ac_mixed_norm} and with the additional condition that $0<\eps < \frac{2t-3\delta}{2\sqrt{t-\delta}}$,
\[
\Pr\Big(S^2 \in [t-\delta,t]\Big)  
		\leq  102\Big\{\frac{\delta+\eps}{t\wedge 1}M\log\Big(\frac{4(K_{\Psi_{1}}\vee1)^{\frac{1}{2}}K(U\vee 1)}{\eps}\Big) \Big\}\vee \eps
        \]
\end{corollary}

\begin{lemma}[Improved Anti-Concentration for Mixed Norms of Gaussian Variables]\label{lem:acnew}
	Let $\alpha$, $\beta$, $C_1$, $C_2$, $\rho$ be positive real numbers satisfying $\beta\geq\alpha>\frac{1}{2}$,$C_1 \geq 1$ and $C_1 \geq C_2$ and $\rho\leq 1$. Let $\{\sigma_{k,j}\}_{k=1,j=1}^{K,M}$ be a collection of real numbers such that for all $1\leq k \leq K$, and all $1 \leq j \leq M$,
	\begin{equation}\label{ac_variance_envelope}
	C_2j^{-\beta}\leq \sigma_{k,j}\leq C_1j^{-\alpha}.
	\end{equation}
	Let $W \in (\R^{M})^{K}$ be a centered Gaussian such that for each $1 \leq k \leq K$,
	\[
	\Var(\pi_k W)=\textrm{diag}[\sigma^2_{k,1}, \cdots, \sigma^2_{k,M}].
	\] 
	Let $S = \max_{k=1}^{K}\| W_k\|_{2}$. Suppose $m \in \N$ satisfies $m\geq 3$, $M \geq 1$, and $K\geq 1$. There exists a constant $D$ that may depend on $\alpha$ and $\beta$ and is universal for fixed values of $\alpha$ and $\beta$ such that 
	\begin{equation}\label{eq:1_lemma}
		\sup_{t\geq \rho}\Pr(S\in[t, t+\delta]) \leq \frac{DC_1^{\frac{6\beta}{2\alpha-1}+2}}{C_2^3\rho ^{\frac{6\beta}{2\alpha-1}}}\Big(\delta\vee \frac{1}{{m}[\log({m})]^{\frac{\beta}{2\alpha-1}}} \Big)\log^{\frac{1}{2}+\frac{3\beta}{2\alpha-1}}(K+1)\log^{\frac{3\beta}{2\alpha-1}}({m}).
	\end{equation}
\end{lemma}

\begin{corollary}[Improved Anti-Concentration for Squared Mixed Norms of Gaussian Variables]\label{lem:acsnew}
Under the conditions of Lemma \ref{lem:acnew} there is a universal constant $D$ that depends only on $\alpha$ and $\beta$ such that 
\begin{equation}\sup_{t\geq \rho}\Pr(S^2\in[t-\delta, t+\delta]) 
\leq 
\frac{DC_1^{\frac{6\beta}{2\alpha-1}+2}}{C_2^3\rho^{\frac{3\beta}{2\alpha-1}+\frac{1}{2}}} \Big(\delta\vee \frac{1}{{m}[\log(m)]^{\frac{\beta}{2\alpha-1}}} \Big)\log^{\frac{1}{2}+\frac{3\beta}{2\alpha-1}}(K+1)\log^{\frac{3\beta}{2\alpha-1}}({m}).
\end{equation}

\end{corollary}

\begin{lemma}\label{lem:ac_ep_net}
Assume the conditions on $M$, ${m}$, $K$, $q$, and $W$ as in Lemma \ref{lem:acnew}. Let $\mathcal{V}_{\eps,t}$ be a collection from Lemma~\ref{lem:product_spheres_approximation}.
There is a constant $D_{\alpha, \beta, q, C_1, C_2}$ that may depend only on $\alpha$, $\beta$, $q$, $C_1$, and $C_2$ such that for any  $0 < \eps <\min\{t, \delta\}$ and any $0<q \leq 1$, 
\begin{equation}\label{ac_eps_new_from_ac_new_1}
\sup_{t \geq q}\Pr\Big(\max_{v \in \mathcal{V}_{\eps,t}}v^{\top}W \in [t-\delta, t+\delta]\Big) \leq D_{\alpha, \beta, q, C_1, C_2}(\delta\vee {m}^{-1})\log^{\frac{1}{2}+{\frac{3\beta}{2\alpha-1}}}(K+1)\log^{\frac{3\beta}{2\alpha-1}}({m}).
\end{equation}
Furthermore, there is a constant $F_{\alpha, \beta, q,C_1, C_2}$ such that 
\begin{equation}\label{ac_eps_new_from_ac_new_2}
\sup_{t \geq q}\sup_{w \geq q/2}\Pr\Big(\max_{v \in \mathcal{V}_{\eps,t}}v^{\top}W \in [w-\delta, w+\delta]\Big) \leq F_{\alpha, \beta, q, C_1, C_2}(\delta\vee {m^{-1}})\log^{1+{\frac{3\beta}{2\alpha-1}}}(K+1)\log^{\frac{1}{2}+\frac{3\beta}{2\alpha-1}}({m}).
\end{equation}
\end{lemma}

\begin{lemma}\label{lem:V_approx}
	Let $X$ and $Y$ be random vectors taking values in $(\R^{M})^{K}$ $(M \geq 3, K \geq 1)$. For any $t>0$ and any $\eps$ satisfying $\min(\eps, t-\eps)>0$,
	\begin{equation}
		|\Pr(X \in E_t)-\Pr(Y \in E_t)| \leq I_{1}(t,\eps)+I_{2}(t,\eps)
	\end{equation}
	where
	\begin{align*}
		& I_{1}(t,\eps) = \max_{w \in \{t-\eps,t\}}\Big|\Pr\Big(\max_{v \in \mathcal{V}_{\eps, t}}v^{\top}X \leq w\Big)- \Pr\Big(\max_{v \in \mathcal{V}_{\eps, t}}v^{\top}Y \leq w\Big)\Big|,
		\\
		&I_{2}(t,\eps) = \Big|\Pr\Big(\max_{v \in \mathcal{V}_{\eps, t}}v^{\top}Y \leq t\Big)-\Pr\Big(\max_{v \in \mathcal{V}_{\eps, t}}v^{\top}Y \leq t-\eps\Big)\Big|,
		\\
	\end{align*}
	and $\mathcal{V}_{\eps, t}$ is a collection from Lemma~\ref{lem:product_spheres_approximation}.
\end{lemma}

\medskip

\subsubsection{Proofs for section~\ref{Appendix:Anti-Concentration}}

\noindent\textbf{Proof of Lemma~\ref{lem:ac_mainthm_A1_A3}} 
Recall the notation introduced in Section \ref{sub-section:additional_notation}.
Let $g \in \R^{MK}$ be a centered Gaussian with $\Var(g) = \Sigma^{\Phi}$, so that $T^{M,G} \overset{d}{=} \max_{k=1}^{K}\|\pi_{k}g\|_{2}^{2}$. Recalling Definition \ref{def:E_t} we see that 
\[
\mA(s,\delta) = \sup_{t \geq s}\Pr(g \in E_{\sqrt{t+\delta}}/E_{\sqrt{t-\delta}}) = \sup_{t \geq s} \Pr(S^2 \in [t-\delta,t+\delta])
\]
where $S := \max_{k=1}^K \|\pi_k g\|_2$.
    
If $\delta \geq (s \wedge 1)/10$, we have, 
\[
\mA(s,\delta) \leq 10\frac{\delta}{s \wedge 1},
\]
and so the bound holds with $A=10$, noting that $M\log\big(Kn\big)\geq 1$. If $\delta <(s \wedge 1)/10$, let $\eps=\frac{1}{m}$ and note that 

    \[
    \frac{2(s\wedge 1)-3\delta}{2\sqrt{(s\wedge 1)-\delta}} \geq \frac{2(s\wedge 1)-(3/10)(s\wedge 1)}{2\sqrt{(s\wedge 1)}} \geq \frac{17}{20}\sqrt{s\wedge 1} > \frac{17}{20}\sqrt{10}\sqrt{\delta} \geq \frac 1m = \eps, 
    \]
where we used the assumption $\delta \geq \frac{1}{m}$.  
We will apply Corollary~\ref{lem:ac_smixed_norm}. Note that the sequence $U_m:=C_{2,1}C_{\phi}\sqrt{\frac{2\alpha}{2\alpha-1}}$ satisfies \eqref{def:U_for_ac}, where $C_{2,1}$ is the constant from Lemma~\ref{relationship_norms} (ii): under (A2), for all $1 \leq k \leq K$, 
	
\begin{align*}
&\sum_{j=1}^{M}[\Var(\pi_k g)]_{jj}=\sum_{j=1}^{M}\Big[\sum_{i=1}^{n}\Var(\Phi_{i,k})/n\Big]_{jj} \leq C_{2,1}^2\sum_{j=1}^{M}\|\phi^{f_k}_j\|_{\Psi_{1}}^2 \leq C_{2,1}^2C_{\phi}^2\frac{2\alpha}{2\alpha-1}.	
\end{align*}
Plugging in the lower bounds assumed on $\delta$ (i.e., $\delta \geq 1/m=:\eps$)gives:
\begin{align*}
&\mA(s,\delta) \leq \Big[102\frac{4\delta}{s \wedge 1}M\log\Big(C_2Kn\Big\{\Big(C_{2,1}C_{\phi}\sqrt{\frac{2\alpha}{2\alpha-1}}\Big)\vee 1\Big\}\Big)\Big] \vee \frac{1}{m}
\\
&\leq C_1 \frac{\delta}{s\wedge 1}M\log\Big(Km\Big).\\
\end{align*}
where $C_1$ is a constant that depends only on $U_n$, and therefore only on $\alpha$ and $C_{\phi}$. To pull the constant out of the argument of the logarithm and absorb it into $C_1$, we used $Km \geq 3$, assumed in the statement of the lemma. $C_2$ is the universal constant given by Corollary~\ref{lem:ac_smixed_norm}, i.e. $C_2:=4(K_{\Psi_{1}}\vee1)^{\frac{1}{2}}$. Set $A = C_1$. \hfill $\Box$

\bigskip

\noindent\textbf{Proof of Lemma~\ref{lem:ac_mainthm_A1_A4}} 
If (A1)-(A4) hold, the random variable  $g \in \R^{MK}$, a centered Gaussian with $\Var(g) = \Sigma^{\Phi}$, satisfies the assumptions of Lemma~ \ref{lem:acnew}. The lower bound assumption holds directly by (A4) and the upper bound holds by (A2) and Lemma~\ref{relationship_norms} (ii) applied with $p=1$ and $q=2$. Since $T^{M,G} \overset{d}{=} \max_{k=1}^{K}\|\pi_{k}g\|_{2}^{2}$, 
    \[
    \mA(s,\delta) = \sup_{t \geq s} \Pr(S^2 \in [t-\delta,t+\delta])
    \]
    where $S := \max_{k=1}^K \|\pi_k g\|_2$. The result follows from the bound of Corollary ~\ref{lem:acsnew}. \hfill $\Box$

\bigskip

\noindent
\textbf{Proof of Lemma~\ref{lem: Belloni anti-concentration}} 
We follow the proof of Theorem 3.2 in \cite{Bel2018}, and make only one essential modification. Observe that $2^2 > e$, so $\log 2 > 1/2$. This implies $(\log p)^{-1/2} < \sqrt{2}$ for all $p\geq 2$.
    
First, consider the case that $\delta \geq s/(14\log(2)-1)$. Then $\delta > s/(14 \log p)$ and so 
    \begin{equation}\label{belloni_anti_concentration_case_1}
    \Pr\Big(\Big|\max_{j=1}^{p}-t\Big| \leq \delta\Big)\leq 1 \leq 14\frac{\delta}{s}\log(p).
    \end{equation}

    Now consider the case $\delta< s/(14\log(2)-1)$, and let $c=(s-\delta)/\sqrt{2}$. Divide the index set of the random variables $X_{j}$ into two groups. Let $A_{1} = \{j|\sigma_{j}\geq c \log^{-1/2}(p)\}$, and let $A_{2} = \{j|\sigma_{j}< c\log^{-1/2}(p)\}$. By the union bound, 
	\begin{equation}\label{belloni_anti_concentration_union_bound}
	\Pr\Big(\Big|\max_{j=1}^{p}-t\Big| \leq \delta\Big) \leq \Pr\Big(\Big|\max_{j \in A_{1}}X_{j}-t\Big| \leq \delta\Big)+ \Pr\Big(\Big|\max_{j \in A_{2}}X_{j}-t\Big| \leq \delta\Big).
	\end{equation}
	We will bound each term on the right hand side separately, starting with $A_{2}$. Observe that the density function $f_{\sigma^2}(t)$ of $N(0,\sigma^2)$ is non-decreasing in $\sigma$ on $[0,t]$ and is decreasing in $t$ on $[0,\infty)$. Noting that $t-\delta \geq s-\delta \geq s/(14\log(2)-1) -\delta>0$,
	\begin{align*}
		& \Pr\Big(\Big|\max_{j \in A_{2}}X_{j}-t\Big| \leq \delta\Big) 
		\\
		&\leq \sum_{j \in A_{2}}\Pr(|X_{j}-t| \leq \delta)
		\\
		&\leq 2\delta\sum_{j \in A_{2}}\frac{1}{\sqrt{2\pi\sigma_{j}^2}}\exp\Big(-\frac{(t-\delta)^{2}}{2\sigma_{j}^{2}}\Big)
		\\
		&\leq 2\delta\sum_{j \in A_{2}}\frac{1}{\sqrt{2\pi\sigma_{j}^2}}\exp\Big(-\frac{(s-\delta)^{2}}{2\sigma_{j}^{2}}\Big)
		\\
		&= 2\delta\sum_{j \in A_{2}}\frac{1}{\sqrt{2\pi\sigma_{j}^2}}\exp\Big(-\frac{(\sqrt{2}c)^{2}}{2\sigma_{j}^{2}}\Big)
		\\
		&\leq 2\delta\sum_{j \in A_{2}}\frac{1}{\sqrt{2\pi(c\log^{-1/2}(p))^2}}\exp\Big(-\frac{(\sqrt{2}c)^{2}}{2(c\log^{-1/2}(p))^{2}}\Big)
		\\
		&=2\delta\sum_{j \in A_{2}}\frac{1}{\sqrt{2\pi}c}\log^{1/2}(p)\exp(-\log(p))
		\\
		&\leq \frac{2}{\sqrt{2\pi}}\delta \frac{1}{c} \log^{1/2}(p) = \frac{2}{\sqrt{\pi}}\frac{\delta}{s-\delta}\log^{1/2}(p) \leq \frac{2\sqrt{2}}{\sqrt{\pi}}\frac{\delta}{s-\delta}\log(p)\leq 2\frac{\delta}{s-\delta}\log(p).
	\end{align*}
	since $|A_{2}|\leq p$. The fourth line follows from the third because $s-\delta \leq t-\delta$ for all $t$. The fifth line follows from the fourth by solving for $s-\delta$ in terms of $c$. The sixth line follows from the fifth due to our observation that $\sigma_{j}<\sqrt{2}c$, that therefore increasing $\sigma_{j}$ to some value less than $\sqrt{2}c$ causes the density to increase, and that finally by definition $c\log^{-1/2}(p) <\sqrt{2}c$. The second last inequality follows from $\frac{1}{\sqrt{\log(p)}} \leq \sqrt{2}$. The constant could be improved by allowing only $p\geq 3$. 
	
	The term involving $A_{1}$ is bounded as follows
	\[
	\Pr\Big(|\max_{j \in A_{1}}X_{j}-t| \leq \delta\Big) \leq 2\delta\frac{\log^{1/2}(p)}{c}(\sqrt{2}\log^{1/2}(|A_{1}|)+2),
	\]
	this is a direct consequence of Nazarov's inequality (Lemma \ref{lem:naz_ineq}) and the fact that $\min_{j\in A_1}\sigma_j \geq c \log^{-1/2}(p)$. Since $|A_{1}|\leq p$, we have: 
	\begin{align*} 
		&\Pr\Big(|\max_{j \in A_{1}}X_{j}-t| \leq \delta\Big) \leq 2\frac{\delta}{s-\delta}\sqrt{2}\log^{1/2}(p)(\sqrt{2}\log^{1/2}(p)+2) 
        \\
        & = \frac{\delta}{s-\delta}(4\log(p)+4\sqrt{2}\log(p)\log^{-1/2}(p)) \leq  \frac{\delta}{s-\delta}(4\log(p)+8\log(p))\leq 12\frac{\delta}{s-\delta}\log(p)\\
	\end{align*}
Therefore by Equation \eqref{belloni_anti_concentration_union_bound} we have shown that
\[
\Pr\Big(\Big|\max_{j=1}^{p}-t\Big| \leq \delta\Big) \leq 14\frac{\delta}{s-\delta}\log(p).
\]
To complete the argument, note that $14\log(2)-1 \geq 6$. Thus $\delta \leq s/(14\log(2)-1) \leq s/6$ and $1/(s-\delta) \leq \tfrac{6}{5s}$. This implies
\[
\Pr\Big(\Big|\max_{j=1}^{p}-t\Big| \leq \delta\Big) \leq 14\frac{\delta}{s-\delta}\log(p) \leq \frac{84}{5}\frac{\delta}{s}\log(p).
\]
This concludes the proof of Lemma \ref{lem: Belloni anti-concentration}.  \hfill $\Box$

\bigskip
    
\textbf{Proof of Lemma~\ref{lem:ac_mixed_norm}} 
$\newline$

First note that
\begin{equation*}
\Pr\Big(S \in [t-\delta, t]\Big)=\Pr\Big(W \in E_{t}/E_{t-\delta }\Big).
\end{equation*}
Since $t-\delta>\eps>0$, for any collection $\mathcal{V}_{\eps, t-\delta} \subset (\R^{M})^{K}$ of Lemma~ \ref{lem:product_spheres_approximation},
\[
A(\mathcal{V}_{\eps,t-\delta},t-\delta-\eps) \subset E_{t-\delta} \subset A(\mathcal{V}_{\eps,t-\delta },t-\delta).
\] 
Thus,
\[
E_t = \frac{t}{t-\delta}E_{t-\delta} \subset \frac{t}{t-\delta}A(\mathcal{V}_{\eps,t-\delta },t-\delta) = A(\mathcal{V}_{\eps,t-\delta },t),
\]
so that
\[
A(\mathcal{V}_{\eps, t-\delta },t-\delta-\eps) \subset E_{t-\delta} \subset E_{t} \subset A(\mathcal{V}_{\eps, t-\delta },t).
\]
Hence
\begin{align*}
	& \Pr(W \in E_{t})-P(W\in E_{t-\delta})&\\  
	&\leq \Pr(W \in A(\mathcal{V}_{\eps, t-\delta },t)) - \Pr(W \in A(\mathcal{V}_{\eps, t-\delta },t-\delta -\eps))&\\
	& =\Pr\Big(\max_{v \in \mathcal{V}_{\eps, t-\delta }} v^{\top}W \leq t\Big) - \Pr\Big(\max_{v \in \mathcal{V}_{\eps, t-\delta }} v^{\top}W \leq t-\delta-\eps\Big)&\\ 
	&= \Pr\Big(\max_{v \in \mathcal{V}_{\eps, t-\delta }} v^{\top}W  \in  [t-\delta-\eps, t]\Big).
\end{align*}
By Lemma \ref{lem: Belloni anti-concentration} and the bound $\log(|\mathcal{V}_{\eps,t-\delta}|)\leq 3M\log(K(1 + (t-\delta)\eps^{-1}))$ (see Lemma ~ \ref{lem:product_spheres_approximation}) the last line is bounded by 
\begin{align*}
\frac{17}{2}\frac{\delta+\eps}{t}\log(|\mathcal{V}_{\eps, t-\delta }|) \leq  \frac{51}{2} \frac{\delta+\eps}{t}M\log(K(1 + (t-\delta)\eps^{-1})).
\end{align*}
This completes the proof of the first bound in the Lemma.

To prove the second statement of the Lemma, note that 
Lemma~ \ref{lem:tail-bound-from-trace-bound} shows that for $C_1 = \sqrt{8/3}K_{\psi_{1}}^{1/2}$( where $K_{\psi_{1}}$ is the constant of Lemma~ \ref{lim_max_sum} (ii)) if $t-\delta\geq C_1\sqrt{\log(K+1)}U\sqrt{\log(\frac{2}{\eps})}$,
\[
\Pr(W \in E_{t}/E_{t-\delta})=\Pr(W \in [t-\delta,t])\leq \Pr(W \geq t-\delta)\leq \eps.
\] 
If $t-\delta \leq C_1\sqrt{\log(K+1)}U\sqrt{\log(\frac{2}{\eps})}$, use the general bound proved in the first part which gives

\begin{align*}
	&\Pr(W \in E_{t}/E_{t-\delta})=\Pr(W \in [t-\delta,t]) \leq \frac{51}{2}\frac{\delta+\eps}{t}M\log(K(1 + (t-\delta)\eps^{-1}))
    \\
	& \leq \frac{51}{2}\frac{\delta+\eps}{t}M\log\Big[K\Big(1 + C_1 U \eps^{-1} \sqrt{\log(K+1)}\sqrt{\log(2/\eps)}\Big)\Big]
    \\
& \leq \frac{51}{2}\frac{\delta+\eps}{t}M\log(K(1 + ( 2C_1\sqrt{\log(K+1)}U\eps^{-2}))
    \\
& \leq \frac{51}{2}\frac{\delta+\eps}{t}M\log(K(1 + ( 2C_1KU\eps^{-2}))
\\
& \leq \frac{51}{2}\frac{\delta+\eps}{t} M\log\Big(\Big[\frac{(2(C_1 \vee1))(U\vee1)K}{\eps \wedge 1}\Big]^2\Big)
\\
& = 51\frac{\delta+\eps}{t}M\log\Big(\frac{2(C_1 \vee1)(U\vee1)K}{\eps \wedge 1}\Big)
\\
&\leq 51\frac{\delta+\eps}{t} M\log\Big(\frac{4( K_{\psi_1}^{1/2}\vee1)(U\vee1)K}{\eps \wedge 1}\Big)
\end{align*}
This completes the proof of Lemma~\ref{lem:ac_mixed_norm}. \hfill $\Box$

\bigskip

\textbf{Proof of Lemma~\ref{lem:acnew}}

We may re-order the components of $W_k$ for every $1 \leq k \leq K$ without affecting the distribution of $S$; without loss of generality we assume that for all $1 \leq k \leq K$ and all $1 \leq j \leq M$, $\sigma_{k,j+1}\leq \sigma_{k,j}$. We define $W_k:=\pi_{k}W$, so that $W_k \in \R^{M}$. We further assume without loss of generality that there is no pair $W_{k_1}$, $W_{k_2}$ with $k_1 \neq k_2$  such that $W_{k_1} = OW_{k_2}$ almost surely for an orthogonal transformation $O$, as in this case $\|W_{k_2}\|_{2} = \|W_{k_1}\|_{2}$ almost surely.

Let $S^{M-1} = \{v \in \R^{M} : \|v\|_{2}=1\}$. Then $S= \max_{k=1}^{K}\sup_{v\in S^{M-1}}v^{\top}W_k$.  We will now define a family (indexed by $\tau \in \R$) of subsets of $S^{M-1}$ called $V_{M}(\tau)$ and a proxy for $S$ called $\tilde{S}$ that will depend on $\tau$. Let
\[
V_{M}(\tau) = \Big\{v \in S^{M-1} : \forall  k ,\sum_{j=1}^{M}v_{j}^2\sigma_{k,j}^{2} \geq \tau^2 \Big\} 
\]
Let $\tilde{S}(\tau) = \max_{k=1}^{K}\sup_{v \in V_{M}(\tau)}v^{\top}W_{k}$. The inequality $S \geq \tilde{S}$ holds (to lighten notation we do not always include the argument $\tau$). The left hand side of \eqref{eq:1_lemma} can be broken into two pieces as follows: 
\begin{align*}
	&\sup_{t\geq \rho}\Pr(S\in[t, t+\delta])
	\\
	=&\sup_{t\geq \rho}\Pr(S\in[t, t+\delta], S\geq \rho)
	\\
	=&\sup_{t\geq \rho}\Big\{\Pr(S\in[t, t+\delta] , S\geq \rho ,S = \tilde{S}) +\Pr(S\in[t, t+\delta] ,S\geq \rho , S > \tilde{S})\Big\}
	\\
	\leq& \sup_{t\geq \rho}\Pr(\tilde{S}\in[t, t+\delta])+\Pr(S > \tilde{S},S\geq \rho).
\end{align*}
We will now bound 
\begin{align*}
	\sup_{t\geq \rho}\Pr(\tilde{S}\in[t, t+\delta]) 
	=  \sup_{t\geq \rho}\Pr\Big(\max_{k=1}^{K}\sup_{v \in V_{M}(\tau)}v^{\top}W_k\in[t, t+\delta]\Big)
\end{align*}
in terms of the  unspecified constant $\tau$. As a subset of the compact metric space $S^{M-1}$, $V_{M}(\tau)$ is totally bounded. Let $\eps_\ell \downarrow 0$, and let $\mathcal{N}_{\eps_\ell} \subset V_{M}(\tau)$ be a finite $\eps_\ell$-net for $V_{M}(\tau)$. Let $\tilde{S}_{\ell} = \max_{k=1}^{K} \sup_{v \in \mathcal{N}_{\eps_\ell}}v^{\top}W_k$. It is straightforward to see that $\tilde{S}_{\ell} \uparrow \tilde{S}$ as $\ell \to \infty$:
\begin{align*}
	0 &\leq \tilde{S}-\tilde{S}_{\ell}
	\\
	&\leq \max_{k=1}^{K}\Big\{\sup_{v \in\mathcal{N}_{\eps_\ell}}\sup_{w: \|w-v\|\leq \eps_\ell}w^{\top}W_k - \sup_{v\in \mathcal{N}_{\eps_\ell}}v^{\top}W_k\Big\}
	\\
	&\leq \max_{k=1}^{K}\Big\{ \sup_{v \in\mathcal{N}_{\eps_\ell}}\Big( \sup_{w: \|w-v\|\leq \eps_\ell}w^{\top}W_k - v^{\top}W_k \Big) \Big\}
	\\
	&\leq \eps_\ell\max_{k=1}^{K}\|W_k\|_{2},
\end{align*}
and by choice of the sequence $\eps_\ell$, $\eps_\ell \max_{k=1}^{K}\|W_k\|_{2}\downarrow 0$. Thus for each fixed $t$, and each fixed $\delta$, by the bounded convergence theorem applied to appropriate indicator functions, $\lim_{\ell \rightarrow \infty}\Pr(\tilde{S}_\ell \in [t, t+\delta])=\Pr(\tilde{S}\in [t, t+\delta])$ and so $\Pr(\tilde{S} \in [t, t+\delta]) \leq  \sup_{\ell}\Pr(\tilde{S}_{\ell} \in [t, t+\delta])$. 

For any $v \in V_M(\tau) \subset S^{M-1}$ and all $k$ we have $\Var(v^\top W_k) = \Var(\sum_{j=1}^{M}v_{j}[W_{k}]_j) = \sum_{j=1}^{M}v_{j}^2\Var([W_{k}]_{j}) \leq C_1^2$, and also $\Var(v^\top W_k) \geq \tau^2$. By Lemma~\ref{dimension_free_extension}, for a universal constant $D_1$ and any $\ell$,

\[
\sup_{t \in \R}\Pr(\tilde{S}_{\ell} \in [t, t+\delta])\leq D_1\frac{C_1}{\tau^2}\delta\Big\{ \E\Big[\max_{k=1}^K\max_{v \in V_{M}(\tau)} \frac{v^\top W_k}{\sqrt{\Var(v^\top W_k)}} \Big]+\sqrt{1\vee \log(\frac{\tau}{\delta})}\Big\}.
\]
A straightforward computation using Lemma~ \ref{relationship_norms}(ii) and Lemma~\ref{lim_max_sum}(ii), gives 
\begin{align*}
&\E\Big[\max_{k=1}^K\max_{v \in V_{M}(\tau)} \frac{v^\top W_k}{\sqrt{\Var(v^\top W_k)}} \Big]
\leq \frac{1}{\tau}\E\Big[\max_{k=1}^K\max_{v \in V_{M}(\tau)} v^\top W_k \Big] 
\leq \frac{1}{\tau}\E\Big[\max_{k=1}^K\max_{v \in S^{M-1}} v^\top W_k \Big]\\
&=\frac{1}{\tau}\E\Big[\max_{k=1}^K\|W_k\|_{2} \Big] \leq C_{2,1}\frac{1}{\tau}\|\max_{k=1}^{K}\|W_k\|_{2}\|_{\psi_{2}} \leq C_{2,1}K^{1/2}_{\Psi_{1}}\frac{\sqrt{\log(K+1)}}{\tau}\max_{k=1}^{K}\|\|W_k\|_{2}\|_{\Psi_2}\\
& =C_{2,1}K^{1/2}_{\Psi_{1}}\frac{\sqrt{\log(K+1)}}{\tau}\max_{k=1}^{K}(\|\|W_k\|^2_{2}\|_{\Psi_1})^{1/2} =C_{2,1}K^{1/2}_{\Psi_{1}}\frac{\sqrt{\log(K+1)}}{\tau}\max_{k=1}^{K}(\|\sum_{j=1}^{M}[W_{k}]^2_j\|_{\Psi_1})^{1/2}  \\
&\leq C_{2,1}K^{1/2}_{\Psi_{1}}\frac{\sqrt{\log(K+1)}}{\tau}\max_{k=1}^{K}(\sum_{j=1}^{M}\|[W_{k}]^2_j\|_{\Psi_1})^{1/2} = C_{2,1}K^{1/2}_{\Psi_{1}}\frac{\sqrt{\log(K+1)}}{\tau}\max_{k=1}^{K}(\sum_{j=1}^{M}\|[W_{k}]_j\|^2_{\Psi_2})^{1/2}\\
& \leq \sqrt{8/3}C_{2,1}K^{1/2}_{\Psi_{1}}C_{1}\frac{\sqrt{\log(K+1)}}{\tau}(\sum_{j=1}^{M}j^{-2\alpha})^{1/2} \leq
\sqrt{8/3}\sqrt{\frac{2\alpha}{2\alpha-1}}C_{2,1}K^{1/2}_{\Psi_{1}}C_{1}\frac{\sqrt{\log(K+1)}}{\tau}.
\end{align*}
 We used that if $X$ is centered Gaussian with standard deviation $\sigma$ that $\|X\|_{\psi_{2}} = \sqrt{8/3}\sigma$ and the $L^{\psi_{1}}$ triangle inequality, which follows from Lemma \ref{orlicz-banach}, and bounded the finite sum in the last line with an infinite sum, and the constants $C_{2,1}$ and $K_{\psi_1}$ are defined in Lemmas \ref{relationship_norms} and \ref{lim_max_sum}. 
Defining $D_2=\sqrt{8/3}\sqrt{(2\alpha)/(2\alpha-1)}[C_{2,1}\vee 1][K_{\Psi_1}^{1/2}\vee 1]$ we have
\begin{equation}\label{eq:upboundexpect}
\E\Big[\max_{k=1}^K\max_{v \in V_{M}(\tau)} \frac{v^\top W_k}{\sqrt{\Var(v^\top W_k)}} \Big]
\leq  C_1 D_2 \frac{\sqrt{\log(K+1)}}{\tau}
 \end{equation}
and thus for any $\delta>0$,
\[
\sup_{t \in \R}\Pr(\tilde{S}_{\ell} \in [t, t+\delta])\leq D_1 C_1 \frac{\delta}{\tau^2}\Big\{ \frac{D_2C_1}{\tau}\sqrt{\log(K+1)}+\sqrt{1\vee \log(\frac{\tau}{\delta})}\Big\}.
\]
Set  
\[
C_\tau^{-1} = \Big([K_{\psi_1}\vee1](8/3)\frac{2\alpha}{2\alpha-1}C^2_1\Big)^{\beta/(2\alpha-1)}2^{(1+2\beta)/2+\beta/(2\alpha-1)}(1/C_2) \rho^{-(2\beta)/(2\alpha-1)}
\]
and let $\tau = C_\tau(\log(K+1)\log(2{m}))^\frac{-\beta}{2\alpha-1}$.
This choice of $C_{\tau}$ is justified by Equation~ \eqref{justify_C_tau} below. Let 
\[
D_3 = \Big((K_{\psi_1}\vee 1)(8/3)\frac{2\alpha}{2\alpha-1}\Big)^{\beta/(2\alpha-1)}2^{(1+2\beta)/2+\beta/(2\alpha-1)},
\]
and let $\gamma  = \beta/(2\alpha-1)$. Note for later that 
\begin{equation}\label{C_tau_interms of D_3}
C^{-1}_{\tau} = D_3(C_1/\rho)^{2\beta/(2\alpha-1)}(1/C_2).
\end{equation}
With this choice of $\tau$ we obtain,
\begin{align*}
&\sup_{t \in \R}\Pr(\tilde S \in [t, t+\delta]) \leq D_1D_3^2\frac{C_1^{4\gamma+1}}{C_2^{2}\rho^{4\gamma}}\delta[\log(K+1)]^{2\gamma}[\log(2{m})]^{2\gamma}\\
&\times \Big\{D_2D_3\frac{C_1^{2\gamma+1}}{C_2\rho^{2\gamma}}[\log(K+1)]^{\gamma+\frac{1}{2}}[\log(2{m}))]^{\gamma}+\sqrt{1 \vee \log(\tau/\delta)}\Big\}
\end{align*}
Now note that the term $\sqrt{\log(\tau/\delta)}$ is less than or equal to the first term in the same bracket if and only if
\begin{align*}
&\delta\geq \frac{\frac{C_2\rho^{2\gamma}}{C_1^{2\gamma}D_3}}{[\log(K+1)]^{\gamma}[\log(2{m})]^{\gamma}\exp\Big(D_2^2D_3^2\frac{C_1^{4\gamma+2}}{C_2^2\rho^{4\gamma}}[\log(K+1)]^{2\gamma+1}[\log(2{m})]^{2\gamma}\Big)}
\end{align*}
The left hand side of the preceding equation can be bounded as follows 
\begin{align*}
&\frac{\frac{C_2\rho^{2\gamma}}{C_1^{2\gamma}D_3}}{[\log(K+1)]^{\gamma}[\log(2{m})]^{\gamma}\exp\Big(D_2^2D_3^2\frac{C_1^{4\gamma+2}}{C_2^2\rho^{4\gamma}}[\log(K+1)]^{2\gamma+1}[\log(2{m})]^{2\gamma}\Big)}\\
&\leq \frac{2^{\gamma}\frac{C_2\rho^{2\gamma}}{C_1^{2\gamma}D_3}}{[\log(2{m})]^{\gamma}\exp\Big((1/2)^{\gamma+1}D_2^2D_3^2\frac{C_1^{4\gamma+2}}{C_2^2\rho^{4\gamma}}[\log(2{m})]^{2\gamma}\Big)}\\
&\leq \frac{C_2(\rho/C_1)^{2\gamma}}{[\log(2{m})]^{\gamma}\exp\Big(\frac{C_1^{4\gamma+2}}{C_2^2\rho^{4\gamma}}[\log(2{m})]^{2\gamma}\Big)}\\
&\leq \frac{1}{[\log({m})]^{\gamma}{m}}.
\end{align*}
In the first inequality we used $\log(K+1) \geq \frac{1}{2}$ both inside and outside of the argument of the exponential. In the second inequality we used $D_2 \geq 1$ and $\frac{D^2_3}{2^{\gamma+1}} \geq \frac{(8/3)^{2\gamma}}{2^{\gamma+1}}>\frac{2^{2\gamma}} {2^{\gamma+1}}  \geq \frac{2^{\gamma+1}}{2^{\gamma+1}}=1$. 
In the final inequality we used that $\rho \leq 1$ and that $C_1\geq C_2$, that $C_1 \geq 1$ and everywhere use used $2\gamma>1$. It follows that for $\delta \geq \frac{1}{{m}[\log({m})]^{\gamma}}$,
\begin{align*}
&\sup_{t \in \R}\Pr(\tilde S \in [t, t+\delta]) \leq 2D_1D_2D_3^3\frac{C_1^{6\gamma+2}}{C_2^{3}\rho^{6\gamma}}\delta[\log(K+1)]^{3\gamma+\frac{1}{2}}[\log(2{m})]^{3\gamma},
\end{align*}
and plugging in the formula for $\gamma$ in terms of $\alpha$ and $\beta$ given above yields the first summand in the right-hand side of Equation \eqref{eq:1_lemma}, after noting that the left-hand side of Equation \eqref{eq:1_lemma} is non-decreasing in $\delta$.
Now we will show with this choice of $\tau$ that 
\begin{equation}\label{eq:StildeSlarge}
\Pr(S > \tilde{S},S\geq \rho) \leq \frac{1}{{m}}.
\end{equation}
With this result established, the proof will be complete upon setting the constant $D$ in the formula of the Lemma we are proving equal to $4D_1D_2D_3^3/\log(2)^{3\gamma+\frac{1}{2}}$. We note that by Lemma \ref{maximizer_is_unique} the set $\arg \max_{k=1}^{K}\|W_k\|_{2}$ is almost surely a singleton. Let $\hat k = \arg \max_{k=1}^{K}\|W_k\|_{2}$, and for $k=1,\dots, K$ define  $\hat v_{k} = \frac{W_{k}}{\|W_{k}\|_{2}}$,  and
\[
s_k(v) := \sum_{j=1}^{M}v_{j}^{2}\sigma_{k,j}^{2} = \Var(v^{\top}W_k).
\] 
With this notation we have $S= \hat v_{\hat k}^{\top}W_{\hat k}$, that for all $1 \leq k \leq K$,$\hat v_k = \arg\max_{v \in S^{M-1}} v^\top W_k$, and that
\begin{equation}
\{S > \tilde{S}\} \subset \{\hat v_{\hat k} \notin V_{M}(\tau)\} =\{\exists k, s_{k}(\hat v_{\hat k})<\tau^{2}\},
\end{equation}
which implies
\begin{equation}
\{S > \tilde S, S\geq \rho\} \subset \{\hat v_{\hat k} \notin V_{M}(\tau),S\geq \rho\}\subset \{\exists k, s_{k}(\hat v_{\hat k})<\tau^{2},S\geq\rho\}.
\end{equation}
We will now construct an event $\Omega_1$ with the property that $\Omega_1\subset \{\forall k, s_k(\hat v_{\hat k}) \geq \tau^2, S \geq \rho\}$, and also, $\Pr(\Omega_1^{c}, S \geq \rho) \leq \frac{1}{{m}}.$ Since 
\[
\Omega_1^{c}\cap \{S \geq \rho\} 
\supset 
(\{\exists k, s_{k}(\hat v_{\hat k})<\tau^{2}\} \cup \{S<\rho\})\cap \{S\geq\rho\} 
=
\{\exists k, s_{k}(\hat v_{\hat k}) \tau^{2},S\geq\rho\},
\]
this implies~\eqref{eq:StildeSlarge}. Let
\[
L = \lfloor 2^{-1/(2\beta)}C_2^{1/\beta}\tau^{-1/\beta} \rfloor
\]
The inequality $\frac{x}{2} \leq \lfloor x \rfloor \leq x$ holds whenever $x \geq 2$, and so whenever
\begin{equation}
2^{-1/2\beta}C_2^{1/\beta}\tau^{-1/\beta} \geq 2,
\end{equation}
we have
\begin{equation}\label{condition_on_n_new_anti_concentration}
2^{-1/(2\beta)-1}C_2^{1/\beta}\tau^{-1/\beta}\leq  L \leq 2^{-1/(2\beta)}C_2^{1/\beta}\tau^{-1/\beta}.
\end{equation}
Note that $D_3 \geq 2^{(1+2\beta)/2}2^{\beta/(2\alpha-1)}$ and so
\begin{equation}\label{new_ac_proof_condition_on_n}
\begin{split}
&2^{-1/2\beta}C_2^{1/\beta}\tau^{-1/\beta} =2^{-1/2\beta}C_2^{1/\beta}[D_3(C_1/\rho)^{2\beta/(2\alpha-1)}(1/C_2)]^{1/\beta}[\log(K+1)\log(2{m})]^{\frac{1}{2\alpha-1}}
\\
&\geq 2^{-1/2\beta}[2^{(1+2\beta)/2}2^{\beta/(2\alpha-1)}]^{1/\beta}[\log(K+1)\log(2{m})]^{\frac{1}{2\alpha-1}}
\\
&\geq 2[\log(2{m})]^{\frac{1}{2\alpha-1}},
\end{split}
\end{equation}
where in the last line we used $\log(K+1) \geq 1/2$. Therefore it suffices that $\log(2{m}) \geq 1$ and so it suffices that ${m} \geq 2$ for \eqref{condition_on_n_new_anti_concentration} to hold. 
Let 
\[
\Omega_1:=\Big \{\sum_{j=L+1}^{M}[W_{\hat k}]_j^{2} \leq \rho^2/2, S^2 \geq \rho^2\Big\}.
\]
Since $S^2 = \sum_{j=1}^{M}[W_{\hat k}]_j^{2}$, on the event $\Omega_1$, $\sum_{j=1}^{L}[W_{\hat k}]_j^{2} \geq S^{2}/2.$ Thus on the event $\Omega_1$, for all $1 \leq k \leq K$,
\[
s_k(\hat v _{\hat k})
= \frac{\sum_{j=1}^{M}\sigma_{k,j}^{2}[W_{ \hat k}]^2_j}{S^2} 
\geq \frac{\sum_{j=1}^{L}\sigma_{k,j}^{2}[W_{ \hat k}]^2_j}{2\sum_{j=1}^{L}[W_{\hat k}]_j^{2}} 
\geq \frac{\sigma_{k,L}^{2}}{2} 
\ge C_2^2 L^{-2\beta}/2 \geq \tau^{2}.
\]
For all $1 \leq j \leq M$ the variable $[W_k]_j$ is Gaussian. A straightforward computation using Lemma~ \ref{relationship_norms} and Lemma~\ref{lim_max_sum}, similar to the derivation of~\eqref{eq:upboundexpect} earlier in this proof, noting that $2\alpha \geq 1$, gives 
\begin{align*}
	&\Big\| \max_{k=1}^{K}\sum_{j=L+1}^M [W_k]_j^2 \Big\|_{\Psi_{1}}\leq  K_{\psi_1}(8/3)\frac{2\alpha}{2\alpha-1}C^2_1\log(K+1)L^{-2\alpha+1}\\
	& \leq [K_{\psi_1}\vee1](8/3)\frac{2\alpha}{2\alpha-1}C^2_1\log(K+1)2^{(2\alpha-1)(2\beta+1)/(2\beta)}C^{-(2\alpha-1)/(\beta)}_2\tau^{(2\alpha-1)/\beta},
\end{align*}
where in the last line we plugged in the lower bound on $L$ from \eqref{condition_on_n_new_anti_concentration}.
Recall that $\tau = C_\tau(\log(K+1)\log(2{m}))^{\frac{-\beta}{2\alpha-1}}$, where the exact form of the constant $C_{\tau}$ is defined above. Plugging this into the last line gives
\begin{equation}\label{justify_C_tau}
\Big\| \max_{k=1}^{K}\sum_{j=L+1}^M [W_k]_j^2\Big\|_{\Psi_{1}} \leq \rho^2[2\log(2{m})]^{-1},
\end{equation}
and so, using Lemma~\ref{lim_max_sum}(iv)
\begin{align*}
&\Pr(\Omega_1^{c}, S \geq \rho) \leq \Pr\Big(\sum_{j=L+1}^{M}[W_{\hat k}]_j^{2} \geq \rho^{2}/2\Big) 
\leq 
\Pr\Big(\max_{k=1}^{K}\sum_{j=L+1}^{M}[W_{k}]_j^{2} \geq \rho^{2}/2\Big)\\
	&\leq 2\exp(-\log(2{m}))=
	\frac{1}{{m}}.
\end{align*}

\hfill $\Box$

\textbf{Proof of Corollary~\ref{lem:ac_smixed_norm}} 
$\newline$
If $\delta \geq t/2$, we have $\Pr\Big(S^2 \in [t-\delta,t]\Big) \leq 2\delta/t$, from which the bound readily follows. Therefore we may assume that $\delta <t/2$. Then,
\begin{align*}
&\Pr\Big(S^2 \in [t-\delta,t]\Big) = \Pr\Big(S \in [\sqrt{t-\delta},\sqrt{t}]\Big) \leq \Pr\Big(S \in [\sqrt{t}-\frac{\delta}{2\sqrt{t-\delta}},\sqrt{t}]\Big) \\
&  \leq  51\Big\{\frac{\delta/(2\sqrt{t-\delta})+\eps}{\sqrt{t}}M\log\Big(\frac{4(K_{\Psi_{1}}\vee1)^{\frac{1}{2}}K(U\vee 1)}{\eps}\Big) \Big\}\vee \eps\\
&  =  51\Big\{\frac{\delta/2+\eps\sqrt{t-\delta}}{\sqrt{t}\sqrt{t-\delta}}M\log\Big(\frac{4(K_{\Psi_{1}}\vee1)^{\frac{1}{2}}K(U\vee 1)}{\eps}\Big) \Big\}\vee \eps\\
&  \leq   51\Big\{\frac{\delta/2+\eps\sqrt{t-\delta}}{t-\delta}M\log\Big(\frac{4(K_{\Psi_{1}}\vee1)^{\frac{1}{2}}K(U\vee 1)}{\eps}\Big) \Big\}\vee \eps.
\end{align*}
In the first line we used the mean-value theorem, and in the second line we used Lemma~\ref{lem:ac_mixed_norm}. Note that to use Lemma~\ref{lem:ac_mixed_norm} in this setting we require
\begin{align*}
\delta/(2\sqrt{t-\delta})+\eps \leq \sqrt{t}, 
\end{align*}
which is equivalent to 
\begin{align*}
\delta+2\sqrt{t-\delta}\eps \leq 2\sqrt{t-\delta}\sqrt{t}, 
\end{align*}
which is implied by
\begin{align*}
\delta+2\sqrt{t-\delta}\eps \leq 2(t-\delta), 
\end{align*}
which is equivalent to
\begin{align*}
\eps \leq (2t-3\delta)/2\sqrt{t-\delta}, 
\end{align*}
which is what we assumed in the statement of the Corollary. \hfill $\Box$
 
\bigskip

\noindent\textbf{Proof of Corollary~\ref{lem:acsnew}}

We will treat the cases $\delta \geq \rho/2$ and $\delta < \rho/2$ separately. When $\delta \geq \rho/2$,
\begin{align*}
& \frac{C_1^{\frac{6\beta}{2\alpha-1}+2}}{C_2^3(\rho )^{\frac{3\beta}{2\alpha-1}+\frac{1}{2}}} \Big(\delta\vee \frac{1}{{m}[\log({m})]^{\beta/2\alpha-1}} \Big)\log^{\frac{1}{2}+\frac{3\beta}{2\alpha-1}}(K+1)\log^{\frac{3\beta}{2\alpha-1}}({m})
\\
& \geq \frac{\delta}{\rho} \frac{1}{\rho^{\frac{3\beta}{2\alpha-1}-\frac{1}{2}}} \frac{C_1^3}{C_2^3} C_1^{\frac{6\beta}{2\alpha-1}-1}2^{-\frac{1}{2}-\frac{2\beta}{2\alpha-1}}
\\
& \geq 2\cdot 2^{-\frac{1}{2}-\frac{2\beta}{2\alpha-1}}
\end{align*}
where in the last line we used ${m} \geq 3, \rho < 1, C_1/C_2\geq 1, C_1\geq 1, \tfrac{6\beta}{2\alpha-1} > 1$. Thus the conclusion of the proposition holds in this case by picking $D \geq 2^{\frac{2\beta}{2\alpha-1}-\frac{1}{2}}$.

In the case $\delta<\rho/2$, note that
\begin{align*}
\Pr(S^{2}\in [t-\delta, t+\delta]) = \Pr\big(S\in [\sqrt{t-\delta}, \sqrt{t+\delta}]\big)\leq \Pr\Big(S^{2} \in \Big[\sqrt{t-\delta},\sqrt{t-\delta}+\frac{\delta}{\sqrt{t-\delta}}\Big]\Big).
\end{align*}
We used the bound $|\sqrt{t+\delta} - \sqrt{t-\delta}| \leq  \frac{\delta}{\sqrt{t-\delta}}$ which follows from the mean value theorem. 
Since $\delta <\rho /2$ and $t \geq \rho$, $\sqrt{t-\delta} \geq \sqrt{\rho/2}$.
Thus,
\begin{align*}
\sup_{t \geq \rho}\Pr(S^{2}\in [t-\delta, t+\delta]) \leq \sup_{t \geq \sqrt{\rho/2}}\Pr\Big(S\in \Big[t, t+\frac{\delta}{\sqrt{t-\delta}}\Big]\Big)
\end{align*}
Then by Lemma~ \ref{lem:acnew}, denoting the constant $D$ of Lemma~\ref{lem:acnew} by $D_2$,
\begin{align*}
 &\sup_{t\geq \rho}\Pr(S\in [t-\delta, t+\delta])
 \\
 &\leq 
 \frac{D_2C_1^{\frac{6\beta}{2\alpha-1}+2}}{C_2^3(\sqrt{\rho/2} )^{\frac{6\beta}{2\alpha-1}}} \Big(\frac{\delta}{\sqrt{\rho/2}}\vee \frac{1}{{m}[\log({m})]^{\frac{\beta}{2\alpha-1}}} \Big)\log^{\frac{1}{2}+\frac{3\beta}{2\alpha-1}}(K+1)\log^{\frac{3\beta}{2\alpha-1}}({m}) + \frac{1}{{m}}
 \\
  &\leq \frac{2^{\frac{3\beta}{2\alpha-1}+\frac{1}{2}}D_2C_1^{\frac{6\beta}{2\alpha-1}+2}}{C_2^3(\rho \wedge 1)^{\frac{3\beta}{2\alpha-1}+\frac{1}{2}}} \Big(\delta\vee \frac{1}{{m}[\log({m})]^{\frac{\beta}{2\alpha-1}}} \Big)\log^{\frac{1}{2}+\frac{3\beta}{2\alpha-1}}(K+1)\log^{\frac{3\beta}{2\alpha-1}}({m})
  \\
  &\leq \frac{2^{\frac{3\beta}{2\alpha-1}+\frac{1}{2}}D_2C_1^{\frac{6\beta}{2\alpha-1}+2}}{C_2^3(\rho )^{\frac{3\beta}{2\alpha-1}+\frac{1}{2}}} \Big(\delta\vee \frac{1}{{m}[\log({m})]^{\frac{\beta}{2\alpha-1}}} \Big)\log^{\frac{1}{2}+\frac{3\beta}{2\alpha-1}}(K+1)\log^{\frac{3\beta}{2\alpha-1}}({m})
  \\
  &\leq \frac{D C_1^{\frac{6\beta}{2\alpha-1}+2}}{C_2^3(\rho)^{\frac{3\beta}{2\alpha-1}+\frac{1}{2}}} \Big(\delta\vee \frac{1}{{m}[\log({m})]^{\frac{\beta}{2\alpha-1}}} \Big)\log^{\frac{1}{2}+\frac{3\beta}{2\alpha-1}}(K+1)\log^{\frac{3\beta}{2\alpha-1}}({m})
\end{align*}
where  $D$ is the larger of $2^{\frac{3\beta}{2\alpha-1}+\frac{1}{2}}D_2$ and $2^{\frac{2\beta}{2\alpha-1}-\frac{1}{2}}$. This completes the proof. 

\hfill $\Box$

\noindent\textbf{Proof of Lemma~\ref{lem:ac_ep_net}}

Set $D^{(1)}_{\alpha, \beta, q, C_1,C_2} =F^{(1)}_{\alpha, \beta, q, C_1, C_2} = \frac{4}{\log^{1+{\frac{3\beta}{2\alpha-1}}}(2)}\frac{1}{q}$, and note that if $\delta\geq q/4$, it follows that
\begin{equation}
\sup_{t \geq q}\Pr(\max_{v \in \mathcal{V}_{\eps,t}}v^{\top}W \in [t-\delta, t+\delta]\Big) \leq 1 \leq D^{(1)}_{\alpha, \beta, q, C_1,C_2}\delta\log^{\frac{1}{2}+{\frac{3\beta}{2\alpha-1}}}(K+1)\log^{\frac{3\beta}{2\alpha-1}}({m}),
\end{equation}
and also that 
\begin{equation}
\sup_{t \geq q}\sup_{w \geq q/2}\Pr(\max_{v \in \mathcal{V}_{\eps,t}}v^{\top}W \in [w-\delta, w+\delta]\Big) \leq 1 \leq F^{(1)}_{\alpha, \beta,q, C_1, C_2}\delta\log^{1+{\frac{3\beta}{2\alpha-1}}}(K+1)\log^{\frac{1}{2}+\frac{3\beta}{2\alpha-1}}({m}).
\end{equation}
We now consider the case $\delta<q/4$. Let $\mathcal{V}_{\eps, t}$ denote a collection from Lemma~\ref{lem:product_spheres_approximation}. For each $t>0$, $\delta>0$ and $w >0$ , $\eps>0$ satisfying $t \wedge \delta >\eps>0$, 
\[
A(\mathcal{V}_{\eps, t}, w+\delta)=\frac{w+\delta}{t}A(\mathcal{V}_{\eps,t},t)\subset \frac{w+\delta}{t}E_{t+\eps}=E_{w+\delta +w\eps t^{-1}+\eps\delta t^{-1}} \subset E_{w+\delta(2+wt^{-1})},
\]
where we used $\eps \leq \delta$.

Since $\mathcal{V}_{\eps,t}$ is a collection of unit vectors it holds by definition that $E_{w-\delta} \subset A(\mathcal{V}_{\eps,t},w-\delta)$. Therefore,
\begin{align*}
&\Pr\Big(\max_{v \in \mathcal{V}_{\eps,t}}v^{\top}W \in [w-\delta, w+\delta]\Big) 
= \Pr\Big(\max_{v \in \mathcal{V}_{\eps,t}}v^{\top}W \leq w+\delta\Big) - \Pr\Big(\max_{v \in \mathcal{V}_{\eps,t}}v^{\top}W < w-\delta\Big)
\\
& \leq \Pr(W \in E_{w+\delta(2+wt^{-1})}) - \Pr(W \in E_{w-\delta}) \leq \Pr\Big(\max_{k=1}^{K}\|W_k\|_{2} \in [w-\delta,w+\delta(2+wt^{-1})]\Big).
\end{align*}

If $w=t$, by Lemma \ref{lem:acnew}, for any $t \geq q$, there is a constant $A_{\alpha, \beta, C_1, C_2}$ such  that 

\begin{align*}
&\Pr\Big(\max_{k=1}^{K}\|W_k\|_{2} \in [w-\delta,w+\delta(2+wt^{-1})]\Big)=\Pr\Big(\max_{k=1}^{K}\|W_k\|_{2} \in [t-\delta,t+3\delta ]\Big)\\
& \leq \sup_{t \geq q/2}\Pr\Big(\max_{k=1}^{K}\|W_k\|_{2} \in [t,t+4\delta ]\Big)\\
&\leq %2^{\frac{6\beta}{2\alpha-1}}
A_{\alpha, \beta, C_1, C_2}q^{-\frac{6\beta}{2\alpha-1}}\Big((4\delta)\vee(1/{m})\Big)\log^{\frac{1}{2}+\frac{3\beta}{2\alpha-1}}(K+1)\log^{\frac{3\beta}{2\alpha-1}}({m}). 
\end{align*}

Now set $D_{\alpha, \beta, q, C_1,C_2}^{(2)} =4
%\times2^{\frac{6\beta}{2\alpha-1}}
A_{\alpha, \beta, C_1, C_2}q^{-\frac{6\beta}{2\alpha-1}}$, and set $D_{\alpha, \beta, q, C_1,C_2}$ in the formulation of the Lemma equal to $D_{\alpha, \beta, q, C_1,C_2}:=D^{(1)} _{\alpha, \beta, q, C_1,C_2} \vee D^{(2)} _{\alpha, \beta, q, C_1,C_2}$. This proves the first claim by demonstrating the bound in Equation \eqref{ac_eps_new_from_ac_new_1}.

We now consider the case $w \neq t$, on the way to justifying the second claim. By Lemma \ref{lem:tail-bound-from-trace-bound} there is a constant $C_{\alpha, \beta, C_1}$ such that if $w-\delta \geq C_{\alpha, \beta, C_1}\sqrt{\log(K+1)}\sqrt{\log(2{m})}$, we have
\begin{align*}
&\Pr\Big(\max_{k=1}^{K}\|W_k\|_{2} \in [w-\delta,w+\delta(2+wt^{-1})]\Big)\leq \Pr\Big(\max_{k=1}^{K}\|W_k\|_2 \geq w-\delta\Big) \leq \frac{1}{{m}}.
\end{align*}
Thus for any $t>\eps>0$ and any $\delta>0$ and $w \geq \delta + C_{\alpha, \beta, C_1}\sqrt{\log{K+1}}\sqrt{\log(2{m})}$ we have 
\begin{equation}
\Pr\Big(\max_{v \in \mathcal{V}_{\eps,t}}v^{\top}W \in [w-\delta, w+\delta]\Big) \leq \Pr\Big(\max_{k=1}^{K}\|W_k\|_2  \geq w-\delta\Big) \leq \delta \vee (1/{m}).
\end{equation}
Set $F^{(2)}_{\alpha, \beta, q, C_1, C_2} = \frac{1}{\log^{1+{\frac{3\beta}{2\alpha-1}}}(2)}$ so that
\begin{equation}
\Pr\Big(\max_{v \in \mathcal{V}_{\eps,t}}v^{\top}W \in [t-\delta, t+\delta]\Big) \leq F^{(2)}_{\alpha, \beta, q, C_1, C_2}(\delta \vee (1/{m}))\log^{1+{\frac{3\beta}{2\alpha-1}}}(K+1)\log^{\frac{1}{2}+\frac{3\beta}{2\alpha-1}}({m}).
\end{equation}
Now suppose $q/2\leq w < \delta + C_{\alpha, \beta, C_1}\sqrt{\log(K+1)}\sqrt{\log(2{m})}$ and $t \geq q$. Recall that we are in the case $\delta \leq q/4 < 1$. Then,
\begin{align*}
&\Pr\Big(\max_{v \in \mathcal{V}_{\eps,t}}v^{\top}W \in [w-\delta, w+\delta]\Big) \leq \Pr\Big(\max_{k=1}^{K}\|W_k\|_{2} \in [w-\delta,w+\delta(2+wt^{-1})]\Big)\\
&\leq \sup_{x \geq q/4 }\Pr\Big(\max_{k=1}^{K}\|W_k\|_{2} \in [x,x+\delta(3+wt^{-1})]\Big)\\
&\leq \sup_{x \geq q/4}\Pr\Big(\max_{k=1}^{K}\|W_k\|_{2} \in \Big[x,x+\frac{\delta}{q}\Big\{ 4+C_{\alpha, \beta, C_1} \sqrt{\log(K+1)}\sqrt{\log(2{m})}\Big\}\Big]\Big).
\end{align*}
We have used $q \leq 1$, $q\leq t$ and $\delta <1$ in the last line. Set $C^{(2)}_{\alpha, \beta, C_1} = 4\sqrt{2}+C_{\alpha, \beta, C_1}$ so that we have
\begin{align*}
&\Pr\Big(\max_{v \in \mathcal{V}_{\eps,t}}v^{\top}W \in [w-\delta, w+\delta]\Big)
\\
&\leq \sup_{x \geq q/4 }\Pr\Big(\max_{k=1}^{K}\|W_k\|_{2} \in \Big[x,x+\frac{C^{(2)}_{\alpha, \beta, C_1}\delta}{q} \sqrt{\log{(K+1)}}\sqrt{\log(2{m})}\Big]\Big)\\
&\leq F^{(3)}_{\alpha, \beta, q, C_1, C_2}(\delta \vee (1/{m}))\log^{\frac{3\beta}{2\alpha-1}+1}(K+1)\log^{\frac{3\beta}{2\alpha-1}+\frac{1}{2}}({m}),
\end{align*}
where $F^{(3)}_{\alpha, \beta, q, C_1, C_2}$ is a constant that follows from combining the previous constant with the constant from Lemma \ref{lem:acnew}. Then set $F_{\alpha, \beta, q, C_1, C_2}$ in the formulation of the Lemma as $F_{\alpha, \beta, q, C_1, C_2} = F^{(2)}_{\alpha, \beta, q, C_1, C_2} \vee F^{(2)}_{\alpha, \beta, q, C_1, C_2} \vee F^{(3)}_{\alpha, \beta, q, C_1, C_2}$. This completes the proof of Lemma~\ref{lem:ac_ep_net}. \hfill $\Box$

\bigskip

\noindent\textbf{Proof of Lemma~\ref{lem:V_approx}}

	$\newline$
	By Lemma~\ref{lem:product_spheres_approximation},
	
	\[
	A(\mathcal{V}_{\eps, t},t-\eps) \subset E_t \subset A(\mathcal{V}_{\eps, t},t).
	\]
Thus 
	\begin{align*}
		&\Pr(X \in E_t)-\Pr(Y \in E_t) 
		\\
		&\leq \Pr\Big(\max_{v \in \mathcal{V}_{\eps, t}}v^{\top}X \leq t\Big) -\Pr\Big(\max_{v \in \mathcal{V}_{\eps, t}}v^{\top}Y \leq t-\eps\Big)
		\\
		&\leq \Pr\Big(\max_{v \in \mathcal{V}_{\eps, t}}v^{\top}X \leq  t\Big) -\Pr\Big(\max_{v \in \mathcal{V}_{\eps, t}}v^{\top}Y \leq  t\Big)+\Big|\Pr\Big(\max_{v \in \mathcal{V}_{\eps, t}}v^{\top}Y \leq t\Big)-\Pr\Big(\max_{v \in \mathcal{V}_{\eps, t}}v^{\top}Y \leq t-\eps\Big)\Big|,
		\\
	\end{align*}
	and 
	
	\begin{align*}
		\Pr(Y \in E_t)-\Pr(X \in E_t) 
&\leq \Pr\Big(\max_{v \in \mathcal{V}_{\eps, t}}v^{\top}Y \leq t\Big) -\Pr\Big(\max_{v \in \mathcal{V}_{\eps, t}}v^{\top}X \leq t-\eps\Big)
		\\
		&\leq \Pr\Big(\max_{v \in \mathcal{V}_{\eps, t}}v^{\top}Y \leq t-\eps\Big) -\Pr\Big(\max_{v \in \mathcal{V}_{\eps, t}}v^{\top}X \leq t-\eps\Big)
        \\
        &\quad+\Big|\Pr\Big(\max_{v \in \mathcal{V}_{\eps, t}}v^{\top}Y \leq t\Big)-\Pr\Big(\max_{v \in \mathcal{V}_{\eps, t}}v^{\top}Y \leq t-\eps\Big)\Big|.
	\end{align*}
Hence, 
\[
\Big|\Pr(Y \in E_t)-\Pr(X \in E_t)\Big| \leq I_{1}(t,\eps) +I_{2}(t,\eps).
\]
\hfill $\Box$

\subsubsection{Gaussian approximation results revisited}
Let $\{X_{i}\}_{i=1}^n$  be an independent collection of centered variables taking values in $\R^p$. Let $\{Y_{i}\}_{i=1}^n$ be an independent collection of centered Gaussian variables with $\Var(Y_i) = \Var(X_i)$. Let $1 \leq B_n <\infty$ satisfy 
\begin{align}
	\max_{j=1}^{p} \frac{1}{n}\sum_{i=1}^{n}\E[|[X_{i}]_{j}|^{3}]\leq B_{n}, \label{eq:Bn1}
	\\
	\max_{j=1}^{p} \frac{1}{n}\sum_{i=1}^{n}\E[|[X_{i}]_{j}|^{4}]\leq B_{n}^{2}, \label{eq:Bn2}
	\\
	\max_{i=1}^{n}\max_{j=1}^{p}\|[X_{i}]_{j}\|_{\Psi_{1}} \leq B_{n}. \label{eq:Bn3}
\end{align}
Define $S^Y_n:=\frac{1}{\sqrt{n}}\sum_{i=1}^n Y_i$ and $S^X_n:=\frac{1}{\sqrt{n}}\sum_{i=1}^n X_i$. The following is modified version of Proposition~2.1 in \cite{CCK17}:

\begin{theorem}[Gaussian Approximation on Rectangles Revisited -- Abstract Version]\label{max_CLT_abstract}
		$\newline$
	Let 
	\[
	\rho_n(t):=\sup_{x\geq t}\Big|\Pr\Big(\max_{1\leq j \leq p} [S^{X}_{n}]_j \leq x\Big)-\Pr\Big(\max_{1\leq j \leq p}[S^{Y}_{n}]_j \leq x\Big)\Big|.
	\]
	For $s \in \R$, define 
	
	\[\mA(s, \delta):=\sup_{x \geq s}\Pr(|\max_{j=1}^{p}[S_n^Y]_j-x|\leq \delta).
	\]
      
        Let $\phi_{*} = \Big(\frac{\log^{4}(p)B_{n}^2}{n}\Big)^{1/6}$. There is a universal constant $C$ such that for all $p \geq 3$ and all $n \geq 3$,

        \begin{equation}\label{eq_statement:max_CLT_abstract}
	\rho_n(t)\leq C\max\Big\{\phi_{*}\log^{1/2}(n),B_n\mA\big(t-C\log_2(n)\phi_{*},C\phi_{*}\big)\Big\}.
	\end{equation}
\end{theorem}

Note that the anti-concentration function $\mA$ is given a different definition in Equation~\eqref{def:ac-function}, and we use the other definition in the proof of Theorem \ref{main:thm} and Theorem \ref{main:thm-power}.

\bigskip

\noindent \textbf{Proof of Theorem~\ref{max_CLT_abstract}} 

The first part of the proof closely follows the steps in the proof of Proposition~2.1 in \cite{CCK17} and we will only point out the places where modifications need to be made in order to obtain our version of the result. We begin by introducing additional notation which is as close as possible to that in \cite{CCK17} in order to make it easier to compare our and their results. Let
\[
L_n:=\max_{1\leq j\leq p} \frac{1}{n}\sum_{i=1}^{n}\Ep[|X_{ij}|^3]. 
\]
For all $\phi\geq 1$, define

\begin{equation}\label{eq: M definition}
	M_{n,X}(\phi):=\frac{1}{n}\sum_{i=1}^{n}\Ep\left[\max_{1\leq j\leq p}|X_{ij}|^31\left\{\max_{1\leq j\leq p}|X_{i j}|>\sqrt{n}/(4\phi\log p)\right\}\right].
\end{equation}
Similarly, define $M_{n,Y}(\phi)$ with $X_{ij}$ replaced by $Y_{ij}$ in (\ref{eq: M definition}), and let 
\[
M_n(\phi):=M_{n,X}(\phi)+M_{n,Y}(\phi).
\]
Following the analysis of $M_n(\phi)$ in the proof of Proposition~2.1 in \cite{CCK17} in their case (E.1), we see that under~\eqref{eq:Bn1}--\eqref{eq:Bn3} the following bound holds for universal constants $C_2, C_3$
\begin{equation}\label{eq:boundMnphi}
	M_n(\phi) \leq C_3\Big(\sqrt{n}/(\phi\log p)+B_n\log p\Big)^3\exp\Big(-\frac{\sqrt{n}}{4C_{2}\phi B_n\log^{2} p}\Big).
\end{equation}
Define 
\[
\varrho_{n}(c) := \sup_{x\geq c,v\in[0,1]} \Big|\Pr\Big(\max_{1\leq j \leq p} \sqrt{v}S^X_{n}+\sqrt{1-v}S^Y_{n} \leq x \Big)-\Pr(\max_{1\leq j \leq p}S^Y_{n} \leq x)\Big|,
\]
where the random vectors $Y_1,\dots,Y_n$ are assumed to be independent of the random vectors $X_1,\dots,X_n$. We first state an intermediate bound on $\varrho_{n}(c)$ which can be seen as a modification of Lemma~5.1 in \cite{CCK17}.

There exists a universal constant $C_1$ such that for all $\phi \geq 1$ 
\begin{equation}\label{eq:keylemma}
	\varrho_n(c) \leq C_1 \frac{\phi^3 (\log p)^2}{n^{1/2}} \Big\{M_n(\phi) + {L_n\mA}(c - 2/\phi,5/\phi) + L_n\varrho_n(c - 8\phi^{-1}) \Big\} + \mA(c - 2/\phi,1/\phi).
\end{equation}

The derivation of this result relies on arguments from the proof of Lemma~5.1 in \cite{CCK17} with several modifications which will be discussed in detail below. Before providing this proof we complete the proof of our main result taking \eqref{eq:keylemma} as given.

Note that equation~\eqref{eq:keylemma} can be restated as
\[
\varrho_{n}(c) \leq a_n(\phi)\varrho_{n}(c-8/\phi) + b_n(\phi,c)
\] 
where $a_n(\phi) = C_1 \frac{B_n \phi^3 (\log p)^2}{n^{1/2}}$,
\[
b_n(\phi,c) = C_1 \frac{\phi^3 (\log p)^2}{n^{1/2}} \Big\{M_n(\phi) + {L_n\mA}(c - 2/\phi,5/\phi)\Big\} + \mA(c - 2/\phi,1/\phi).
\]

Note that above we have used the bound $L_n \leq B_n$, which holds by definition. Iteratively plugging applying this equation $M$ times and noting that $b_n(\phi,c)$ is non-increasing in $c$ for every $\phi$ (this follows by definition of $\mA(t,\delta)$ as a supremum over $[t, \infty)$) yields
\begin{align*}
	\varrho_{n}(c) &\leq a_n(\phi)\varrho_{n}(c-8/\phi) + b_n(\phi,c) 
	\\
	&\leq a_n(\phi)\{a_n(\phi)\varrho_{n}(c-16/\phi) + b_n(\phi,c-8/\phi)\} + b_n(\phi,c)
	\\
	&\leq a_n(\phi)^M \varrho_{n}(c-8M/\phi) + \sum_{j=0}^{M-1} a_n(\phi)^j b_n(\phi,c-8j/\phi)
	\\
	&\leq a_n(\phi)^M + b_n(\phi,c-8M/\phi) \sum_{j=0}^{M-1} a_n(\phi)^j,
\end{align*}
where we used the bound $\varrho_{n}(c) \leq 1$ for every $c$ in the last line. Now pick 
\[
\phi = \phi^* := \Big(\frac{n}{4 C_1^2 \log^4 (p) B_n^2} \Big)^{1/6}
\] 
resulting in $a_n(\phi) = 1/2$ and let $M = \lceil \log_2(n)\rceil$ where $\log_2$ denotes the logarithm with base $2$. This yields the bound
\[
\varrho_{n}(c) \leq n^{-1} + 2 b_n(\phi^*,c-8\lceil \log_2(n) \rceil/\phi^*).
\]
We may assume without loss of generality that $\log^{1/2}(n)/\phi^* \leq C_b$ for any universal constant $C_b$ by \eqref{eq_statement:max_CLT_abstract}, and we choose the value of $C_b$ below. This implies that $\big[\frac{4C^2_1}{C_b^6}\big]B_n^2\log^2(p) \leq n$, and so $B_n\log(p) \leq (C^6_b n/4C^2_1)^{1/2}$. By \eqref{eq:boundMnphi}, 
\begin{align*}
&M_n(\phi^*) \leq C_3\Big(\sqrt{n}/(\phi^*\log p)+B_n\log p\Big)^3\exp\Big(-\frac{\sqrt{n}}{4C_{2}\phi^* B_n\log^{2} p}\Big)\\
&\leq C_3\Big(\sqrt{n}/(C_b\log^{1/2}(n)\log p))+(C_b^6n/4C^2_1)^{1/2}\Big)^3\exp\Big(-\frac{\sqrt{n}}{4C_{2}\phi^* B_n\log^{2} p}\Big)\\
&\leq C_4n^{3/2}\exp\Big(-\frac{\sqrt{n}}{4C_{2}\phi^* B_n\log^{2} p}\Big)
\end{align*}
for a universal constant $C_4$. Then, 
\begin{align*}
&\exp\Big(-\frac{\sqrt{n}}{4C_{2}\phi^* B_n\log^{2} p}\Big) = \exp\Big(-(4C_1^2)^{1/6}\frac{n^{1/3}}{4C_{2}B_n^{2/3}\log^{4/3} p}\Big)\\
&=\exp\Big(-(\phi^{*})^2\frac{{(4C_1^2)^{1/2}}}{4C_{2}}\Big)\leq \exp(-2\log(n)) \leq n^{-2},
\end{align*}
where we set $C_b$ such that it implies that $(\phi^{*})^2\frac{{(4C_1^2)^{1/2}}}{4C_{2}} \geq 2\log(n)$. It follows that $M_n(\phi^*)\leq C_4n^{-1/2}$. Continuing we have
\begin{align*}
&\varrho_{n}(t) \leq n^{-1} + 2 b_n(\phi^*,t-8\lceil \log_2(n) \rceil/\phi^*)\\
& \leq n^{-1} +2\big\{M_n(\phi^*)+B_n\mA(t-18\log_2 (n)/\phi_*,5/\phi^*)\big\}+\mA(t-18\log_2 (n)/\phi_*,5/\phi^*)\\
& \leq n^{-1} +2C_4n^{-1/2}+2B_n\mA(t-18\log_2 (n)/\phi_*,5/\phi^*),\\
\end{align*}
where we used that $B_n$ is and upper bound on $L_n$ and $B_n>1$. By picking $C$ to be a sufficiently large universal constant we obtain \eqref{eq_statement:max_CLT_abstract}.

It remains to prove~\eqref{eq:keylemma}. To this end introduce the following definitions from \cite{CCK17}: $\beta := \phi\log p$,
\begin{equation}\label{eq:smoothmax}
	F_\beta(w,x):=\beta^{-1} \log\left( {\textstyle \sum}_{j=1}^p\exp \left( \beta (w_{j}-x) \right) \right), \ w \in \R^{p}.
\end{equation}
Note that in \eqref{eq_bound:smoothmax} we give the function $F_\beta$, a different definition, and the definition of \eqref{eq:smoothmax} is used only within the body of the current proof. 
The function $F_\beta(w,x)$ can be considered as a smoothed version of $w \mapsto \max_{1\leq j\leq p}(w_j-x)$ and satisfies (see equation (22) in \cite{CCK17})
\begin{equation}%
	\label{eq: F properties}
	0 \leq F_\beta(w,x)-\max_{1\leq j\leq p}(w_j-x)\leq \beta^{-1}\log p=\phi^{-1}, \ \text{for all} \ w \in \R^{p}.
\end{equation}
Pick a three times continuously differentiable function $g_0:\R \to [0,1]$ whose derivatives up to the third order are all bounded such that $g_{0}(t)=1$ for $t \leq 0$ and $g_{0}(t)=0$ for $t\geq 1$, i.e. a smoothed version of the indicator function $I\{\cdot \leq 0\}$. Define $g(t):=g_{0}(\phi t), t \in \R$ and set 
\[
m(w,x):=g(F_\beta(w,x)), \ w \in \R^{p}.
\]
Further, let 
\begin{align}
	&h_x(w,t) := 1\left\{-\phi^{-1}-t/\beta< \max_{1\leq j\leq p}(w_j-x)\leq \phi^{-1}+t/\beta\right\}, \ w \in \R^{p}, t > 0, \label{eq: def iota} 
	\\
	&\omega(t) :=\frac{1}{\sqrt{t}\wedge\sqrt{1-t}}, \ t \in (0,1). \notag
\end{align}
Let 
\[
\mathcal{I}_n(v,x):=m(\sqrt{v}S^X_{n}+\sqrt{1-v}S^Y_{n},x)-m(S^W_{n},x).
\]
where $S^W_{n}$ is based on independent copies $W_i$ of $Y_i$ define
\[
V_n(v) := \sqrt{v}S^X_{n}+\sqrt{1-v}S^Y_{n}.
\]
Then we have for all $v \in [0,1]$ and $x \in \R$
\begin{align*}
	&\Pr(\max_{j=1}^{p}[V_{n}(v)]_{j}\leq x-\phi^{-1})
	\\
	&\quad \leq \Pr(F_\beta(V_{n}(v),x)\leq 0)
	\\
	&\quad \leq \Ep[m(V_{n}(v),x)] 
	\\
	&\quad \leq \Pr(F_\beta(S_{n}^W,x)\leq\phi^{-1})+(\E[m(V_{n}(v),x)]-\E[m(S_{n}^W,x)]) \\
	&\quad \leq \Pr(\max_{j=1}^{p}([S_{n}^W]_{j} -x)\leq \phi^{-1})+|\E[\mathcal{I}_n(v,x)]| 
	\\
	&\quad =\Pr(\max_{j=1}^{p}[S_{n}^W]_{j} \leq \phi^{-1}+x)+|\E[\mathcal{I}_n(v,x)]| 
	\\
	&\quad \leq \Pr(\max_{j=1}^{p}[S_{n}^W]_{j} \leq x-\phi^{-1}) + \Pr(|\max_{j=1}^{p}[S_{n}^W]_{j} - x| \leq \phi^{-1}) +|\E[\mathcal{I}_n(v,x)]|.
\end{align*}
where the inequality symbol is applied element wise in the final line. We used~\eqref{eq: F properties} for the first inequality, the definition of $m$ in the second and third inequality. The fourth inequality follows from~\eqref{eq: F properties} and the definition of $\mathcal{I}_n$. We note that

\begin{align*}
	&\E[m(V_{n}(v), x-2\phi^{-1})]\leq \Pr \Big(F_\beta(V_{n}(v),x-2\phi^{-1})\leq \phi^{-1} \Big) \\
	& \leq \Pr \Big(\max_{j=1}^{p}V_{n}(v)_j -(x-2\phi^{-1})\leq \phi^{-1}\Big) = \Pr\Big(\max_{j=1}^{p}[V_{n}(v)]_j \leq x -\phi^{-1}\Big)\\
\end{align*}
The inequality on the first line follows from the identity 

\[m(V_{n}(v), x-2\phi^{-1}) = g\circ F_\beta(V_{n}(v),x-2\phi^{-1}) \leq \mathbf{1}\{F_\beta(V_{n}(v),x-2\phi^{-1}) \leq \phi^{-1}\},\] 
after taking expectations. The equality on the second line follow from the inequality 

\[F_\beta(V_{n}(v),x-2\phi^{-1}) \geq \max_{j=1}^{p}[V_{n}(v)]_j -(x-2\phi^{-1}).\]
Therefore,
\begin{align*}
	&\Pr(\max_{j=1}^{p}[S_{n}^W]_{j}\leq x-\phi^{-1})  - \Pr (\max_{j=1}^{p}[V_{n}(v)]_j\leq x-\phi^{-1})
	\\
	&\quad \leq \Pr(\max_{j=1}^{p}[S_{n}^W]_{j}\leq x-\phi^{-1}) - (\E[m(V_{n}(v), x-2\phi^{-1})]-\E[m(S_{n}^W,x-2\phi^{-1})])
    \\
    &\quad \quad -\E[m(S_{n}^W,x-2\phi^{-1})]
	\\
	&\quad \leq |\E[\mathcal{I}_n(v, x-2\phi^{-1})]|+ \Pr(\max_{j=1}^{p}[S_{n}^W]_{j}\leq x-\phi^{-1})-\Pr(\max_{j=1}^{p}[S_{n}^W]_{j}\leq x-3\phi^{-1})
	\\
	&\quad \leq |\E[\mathcal{I}_n(v,x-2\phi^{-1})]| +\Pr(|\max_{j=1}^{p}[S_{n}^W]_{j} - x-2\phi^{-1}| \leq \phi^{-1})
\end{align*}
We note that $\Pr(\max_{j=1}^{p}[S_{n}^W]_{j}\leq x-3\phi^{-1}) \leq \Pr \Big(F_{\beta}(S_{n}^W,  x-2\phi^{-1})\leq 0\Big)  \leq  \E[m(S_{n}^W,x-2\phi^{-1})]$, justifying the inequality between the third last and second last line. 
Combining the inequalities above we find that for every $x \in \R$
\begin{align}\label{eq:approxbound}
	&\Big|\Pr(\max_{j=1}^{p}[V_{n}(v)]_{j}\leq x-\phi^{-1}) - \Pr(\max_{j=1}^{p}[S_{n}^W]_{j} \leq  x-\phi^{-1}) \Big| \notag
    \\
	&\leq \Pr(|\max_{j=1}^{p}[S_{n}^W]_{j}- x-2\phi| \leq \phi^{-1})\vee\Pr(|\max_{j=1}^{p}[S_{n}^W]_{j} - x| \leq \phi^{-1}) 
    \\
    &\quad +|\E[\mathcal{I}_n(v,x)]| \vee |\E[\mathcal{I}_n(v,x-2\phi^{-1})]|. \notag
\end{align}
It remains to bound $|\E[\mathcal{I}_n(v,x)]|$. To this end, a close look at Step 1 in the proof of Lemma~5.1 in \cite{CCK17} reveals that the arguments used to bound $I, II, III_2$ in the notation from that paper remain unchanged in our setting and thus for a universal constant $C_4$
\[
|\E[\mathcal{I}_n(v,x)]| \leq C_4 M_n(\phi) \phi \beta^2/n^{1/2}+ |III_1|.
\]
The bound for $III_1$ takes the form
\[
III_1 \leq C_5 M_n(\phi) \phi\beta^2/n^{1/2} + C_5 \phi\beta^2 \int_0^1  \E[h(Z(t,v),2)] \frac{\omega(t)L_n}{n^{1/2}}dt 
\]
where $C_5$ denotes a universal constant and 
\begin{align*}
	Z(t,v) & =  \frac{1}{\sqrt{n}}\sum_{i=1}^{n}(\sqrt{tv}X_{i}+\sqrt{t(1-v)}Y_{i}+\sqrt{1-t}W_{i})
	\\
	& \stackrel{d}{=}  \frac{1}{\sqrt{n}}\sum_{i=1}^{n}(\sqrt{tv} X_{i}+\sqrt{1-tv}Y_{i}),
\end{align*}
with $\stackrel{d}{=}$ denoting equality in distribution. Now for any $x > \phi^{-1} + 2/\beta$, by definition of $h_x$, 
\begin{align*}
	& \E[h_x(Z(t,v),2)] 
	\\
	=~& \Pr\Big( \Big|\max_{i=1,\dots,p} Z_i(t,v) - x \Big| \leq \phi^{-1} + 2/\beta\Big)
	\\
	\leq~ &  \Pr\Big( \Big|\max_{i=1,\dots,p} Z_i(0,0) - x \Big| \leq \phi^{-1} + 2/\beta\Big)
	\\
	& + 2\sup_{y \geq x - \phi^{-1} - 3/\beta} \Big|\Pr\Big(\max_{i=1,\dots,p} Z_i(0,0) \leq y \Big) - \Pr\Big(\max_{i=1,\dots,p} Z_i(t,v) \leq y \Big)\Big|
	\\
	\leq~&  \mA(x,\phi^{-1}+2/\beta) + 2\varrho_n(x - \phi^{-1} - 3/\beta),
\end{align*}
where we used the fact that $Z(0,0) = n^{-1/2}\sum_{i=1}^n Y_i$. Combining the bounds above we obtain the inequality
\begin{align*}
	|\E[\mathcal{I}_n(v,x)]| \leq~& (C_4 + C_5) M_n(\phi) \phi \beta^2/n^{1/2}
	\\
	& + C_6 \phi \beta^2 L_n n^{-1/2}\Big\{ \mA(x,\phi^{-1}+2/\beta) + \varrho_n(x - \phi^{-1} - 3/\beta) \Big\}  
\end{align*}
where $C_6 = 2\times C_5 \int_0^1 \omega(t) dt$. We note that $\varrho_n(c)$ is a decreasing function of $c$ and so this bound is decreasing in $x$. We may therefore write 

\begin{align*}
	&|\E[\mathcal{I}_n(v,x)]| \vee |\E[\mathcal{I}_n(v,x-2\phi^{-1})]| \\
	&\leq(C_4 + C_5) M_n(\phi) \phi \beta^2/n^{1/2}\\
	&+2C_6 \phi \beta^2 L_n n^{-1/2}\Big\{ (\mA(x - 2/\phi,\phi^{-1}+2/\beta)+\varrho_n(x - 3\phi^{-1} - 3/\beta))\Big\} 
\end{align*}
%	$\mA(x,\phi^{-1}+2/\beta)$
Finally, note that $2/\log p \leq 4, 3/\log p \leq 5$ and recall $\beta = \phi \log p$ so that 
\[
|\E[\mathcal{I}_n(v,x)]| \vee |\E[\mathcal{I}_n(v,x-2\phi^{-1})]| \leq C_7 \frac{\phi^3 (\log p)^2}{n^{1/2}} \Big\{M_n(\phi) + L_n\mA(x - 2/\phi,5/\phi) + L_n\varrho_n(c - 8\phi^{-1}) \Big\}
\] 
for a universal constant $C_7$. Combining this with~\eqref{eq:approxbound}  yields
\[
\varrho_n(c) \leq C_8 \frac{\phi^3 (\log p)^2}{n^{1/2}} \Big\{M_n(\phi) + L_n \mA(c - 2/\phi,5/\phi) + L_n\varrho_n(c - 8\phi^{-1}) \Big\} + \mA(c- 2/\phi,1/\phi).
\]
This proves~\eqref{eq:keylemma} and completes the proof of Theorem~\ref{max_CLT_abstract}.

\hfill $\Box$

\subsubsection{Gaussian Approximation for squared $L^{\infty} - L^{2}$ Mixed Norms}
We begin this section by stating Lemmas that are used in the proof of Theorem \ref{main:thm}. Recall the definitions of $T^M$ and $T^{MG}$ made in Section \ref{sub-section:additional_notation}.
	
\begin{lemma}\label{lem:A1_A3_Delta_M_MG}
Suppose conditions (A1) -- (A3) hold. There exists a constant  $C = C(C_{\phi}, \alpha)$ such that for all $s >0$, all $n \geq 3$, all $K \geq 1$ and all $M \geq 3$,
\begin{equation}\label{eq:lem:A1_A3_Delta_M_MG}
\sup_{t \geq s} \Big|\Pr(T^{MG}\leq t) - \Pr(T^{M}\leq t) \Big| \leq \frac{C}{\sqrt{s \wedge 1}}\Big(\dfrac{M^{10} \log^{10}(Kn)}{n}\Big)^{1/6}.
\end{equation}
\end{lemma}

\begin{lemma}\label{lem:A1_A4_Delta_M_MG}
	Suppose conditions (A1)--(A4) hold.  There exists a constant $C = C(C_{\phi},C_2, \alpha, \beta, s)$ such that for all $n \geq 3$, and all $K \geq 1$ and all $M \geq 3$,
	
	\begin{equation}\label{lem:A1_A4_Delta_M_MG_main_bound}
	\sup_{t \geq s} \Big|\Pr(T^{MG}\leq t) - \Pr(T^{M}\leq t) \Big| \leq C\Big(\dfrac{M^4 \log^{4}(Kn)}{n}\Big)^{1/6}\log^{1+\frac{3\beta}{2\alpha-1}}(K+1)\log^{\frac{1}{2}+\frac{3\beta}{2\alpha-1}}(n).
	\end{equation}
\end{lemma}

Let $\{X_{i}\}_{i=1}^n$  be an independent collection of centered variables taking values in $(\R^M)^K$. Let $\{Y_{i}\}_{i=1}^n$ be an independent collection of centered Gaussian variables with $\Var(Y_i) = \Var(X_i)$. Let $1\leq H_n <\infty$ satisfy 
\begin{equation}
\max_{k=1}^{K} \sqrt{\sum_{j=1}^{M}\max_{i=1}^{n}\|[\pi_{k}X_{i}]_j\|^2_{\Psi_{1}}} \leq H_n. \label{eq:Hn3}
\end{equation}

\begin{theorem}[Gaussian Approximation for Maxima of $L^{2}$ norms -- Abstract Version]\label{clt_l2_abstract}
Let 
\[
\mA^{t}_n(s, \delta):=\sup_{x\geq s}\Pr\Big(\Big|\max_{v \in \mathcal{V}_{\frac{1}{n}, t}}v^\top S_{n}^Y - x\Big|\leq \delta\Big).
\]
There exist universal constants $D$ and $D_1$ such that if $M \geq 3$, and $K \geq 1$, $n >2 \vee (1/s)$,
\[
\sup_{t \geq s} \Big| \Pr(S^{Y}_{n} \in E_{t})- \Pr(S^{X}_{n} \in E_{t}) \Big| \leq D\Big[\Big\{\sup_{t \in [s,t_{*}]}H_n^3\mA^{t}_n(s-D\log(n)\delta_{*}, D\delta_{*}) \Big\}\vee \big\{\delta_{*}\sqrt{\log(n)}\big\}\Big],
\]
for $t_{*} = D_1 H_n\log(K+1) \log(n)$ and  $\delta_{*} = \Big(\dfrac{H_n^6M^4 \log^{4}(H_nKn)}{n}\Big)^{1/6}$.
\end{theorem}
Lemmas~ \ref{lem:A1_A3_Delta_M_MG} and \ref{lem:A1_A4_Delta_M_MG} are specializations of Theorem~\ref{clt_l2_abstract} using bounds on $\mA^{t}_n(w, \delta)$ that are available under (A1)--(A3) and (A1) --(A4) respectively.

$\newline$

\textbf{Proof of Theorem~\ref{clt_l2_abstract}}

We specify the value of $D_1$ in the formulation of the Theorem we are proving at the end of the proof. Let 
	\begin{align*}
		&I_{1}:=\sup_{t > t_{*}} \Big| \Pr(S^{Y}_{n} \in E_{t})- \Pr(S^{X}_{n} \in E_{t}) \Big|,
		\\
		&I_{2}:=\sup_{t \in [s,t_{*}]} \Big| \Pr(S^{Y}_{n} \in E_{t})- \Pr(S^{X}_{n} \in E_{t})\Big|.
	\end{align*}
By the triangle inequality, 
\[
\sup_{t \geq s} \Big| \Pr(S^{Y}_{n} \in E_{t})- \Pr(S^{X}_{n} \in E_{t}) \Big| \leq I_1+I_2.
\]
We show below that (for the chosen value of $D_1$), $t_{*}$ is sufficiently large so that 
\begin{equation}\label{bound_on_I_1}
I_1 \leq \frac{1}{n}.
\end{equation}
Recall the definition of $\mathcal{V}_{\eps,t}$ from Lemma~\ref{lem:product_spheres_approximation}. To bound $I_2$, for any $0 < \eps <s$ let 
\begin{align*}
&I^{\eps}_{2,1}:=\sup_{t \in[s, t_{*}]}\sup_{w \geq t-\eps}\Big|\Pr(\max_{v \in \mathcal{V}_{\eps,t}}v^\top S_n^Y \leq w) - \Pr(\max_{v \in \mathcal{V}_{\eps,t}}v^\top S_n^X \leq w)\Big|,
\\
&I^{\eps}_{2,2}:=\sup_{t \in[s, t_{*}]}\sup_{w \geq t}\Big|\Pr(\max_{v \in \mathcal{V}_{\eps,t}}v^\top S_n^Y \leq w) - \Pr(\max_{v \in \mathcal{V}_{\eps,t}}v^\top S_n^Y \leq w-\eps)\Big|.
\end{align*}
By Lemma \ref{lem:V_approx}, $I_2 \leq I^{\eps}_{2,1}+I^{\eps}_{2,2}$ for any $\eps<s$. Let $\eps_{*} = 1/n$, which requires the constraint $1/n<s$, assumed in the formulation of the theorem we are proving. We show below that there is a universal constant $C_{11} \geq 1$ (this condition is a convenience for keeping track of constants and does not play an essential role in the proof) such that 
\begin{equation}\label{bound_I_{2,1}}
I^{\eps_{*}}_{2,1}\leq C_{11}\Big[H_n^3\sup_{t \in [s, t_{*}]}\mA^{t}_n(s-C_{11}\log(n)\delta_{*},C_{11}\delta_{*})\vee \big\{\delta_{*}\sqrt{\log(n)} \big\}\Big]
\end{equation}
for $t_{*} = D_3 H_n\log(K+1) \log(n)$ as defined above, and $\delta_{*}$  as defined as in the formulation of the theorem we are proving. Now note that
\begin{equation}\label{bound_I_{2,2}}
 I^{\eps_{*}}_{2,2} \leq \sup_{t \in [s,t_{*}]}\mA^{t}_n(s, 1/n) \leq C_{11} \sup_{t \in [s,t_{*}]}\mA^{t}_n(s-C_{11}\log(n)\delta_{*}, C_{11}\delta_{*}).
\end{equation}
The first inequality holds by definition, and the second holds because $\mA^{t}_n$ is monotone non-increasing in its first argument, and monotone non-decreasing in its second argument. It follows that 
    \[
    \sup_{t \geq s} \Big| \Pr(S^{Y}_{n} \in E_{t})- \Pr(S^{X}_{n} \in E_{t}) \Big| \leq 2C_{11}\Big[H_n^3\sup_{t \in [s, t_{*}]}\mA^{t}_n(s-C_{11}\log(n)\delta_{*},C_{11}\delta_{*})\vee \big\{ \delta_{*}\sqrt{\log(n)} \big\}\Big]+\frac{1}{n}
    \]
    Set the universal constant $D$ in the formulation of the theorem as $D= 3C_{11}$.Once Equations \eqref{bound_on_I_1} and \eqref{bound_I_{2,1}} are justified, the proof will be complete. 

\bigskip
\bigskip

{\em Justification of Equation \eqref{bound_I_{2,1}}}

To bound $I^{\eps_{*}}_{2,1}$, we will apply Theorem \ref{max_CLT_abstract} to the collection of random variables $\{v^{\top}X\}_{v \in \mathcal{V}_{\eps_{*},t}}$. Let $C_{13} = \Big[(C_{4,1}\vee C_{3,1})\sqrt{C_{\psi_{1/2}}}\Big] \vee 1$, where $C_{4,1}$ and $C_{3,1}$ are the universal constants from Lemma \ref{relationship_norms} and $C_{\psi_{1/2}}$ is a universal constant from Lemma \ref{triangle_inequality}. As a first step we show that $(C_{13}H_n)^3$ satisfies the constraints \eqref{eq:Bn1} --\eqref{eq:Bn3}. Observe that 
    \begin{equation}\label{needed_upper_bound_H_n}
    \max_{i=1}^{n}\max_{k=1}^{K}\|\|\pi_k X_i\|_{2}\|_{\psi_1} \leq \sqrt{C_{\psi_{1/2}}}\max_{i=1}^{n}\max_{k=1}^{K} \sqrt{\sum_{j=1}^{M}\|[\pi_{k}X_{i}]_j\|^2_{\Psi_{1}}} \leq \sqrt{C_{\psi_{1/2}}}H_n,
    \end{equation}
    where we have used Lemmas \ref{relationship_norms} and \ref{triangle_inequality}. By Lemma \ref{lem:product_spheres_approximation}, if $v \in \mathcal{V}_{\eps,t}$, there exists a unique integer $k$ (satisfying $1 \leq k \leq K$) such that $\pi_{k}\neq 0$.  Given a vector $v \in  \mathcal{V}_{\eps,t}$, let $l(v)$ be this unique integer, so that $l(v) \in \{1, \dots, K\}$. Note that for all $v$, $\pi_{l(v)} v \in \R^M$ is a unit vector.
Equation \eqref{needed_upper_bound_H_n} and Lemmas \ref{relationship_norms} (ii) and \ref{triangle_inequality} show that for any $\eps<t$ (we must have $\eps<t$ so that $\mathcal{V}_{\eps,t}$ is defined),
\begin{equation}
\begin{split}
\max_{v \in \mathcal{V}_{\eps,t} } \frac{1}{n}\sum_{i=1}^{n}\E[|v^TX_i|^{4}]
=& \max_{k=1}^{K}\max_{(v \in \mathcal{V}_{\eps,t})\wedge (l(v)=k) } \frac{1}{n}\sum_{i=1}^{n}\E[|(\pi_kv)^{\top}(\pi_k X_i)|^{4}]
\\
\leq &\max_{k=1}^{K}\max_{(v \in \mathcal{V}_{\eps,t})\wedge (l(v)=k) } \max_{i=1}^{n}\E[|(\pi_kv)^{\top}(\pi_k X_i)|^{4}]
\\
\leq&\max_{k=1}^{K}\max_{(v \in \mathcal{V}_{\eps,t})\wedge (l(v)=k) } \max_{i=1}^{n}\Big(C_{4,1}\|(\pi_kv)^{\top}(\pi_k X_i)\|_{\Psi_1}\Big)^4 
\\
\leq& \max_{k=1}^{K}\max_{i=1}^{n}\Big(C_{4,1}\|\|\pi_k X_i\|_{2}\|_{\Psi_1}\Big)^4
\\
\leq& \Big(C_{4,1}\sqrt{C_{\psi_{1/2}}}H_n\Big)^4 \leq (C_{13}H_n)^6.
\end{split}
\end{equation}
In the last line we used \eqref{needed_upper_bound_H_n} and in the third line we used Lemma~\ref{relationship_norms}(ii). We further use the property that if $0 \leq X \leq Y$ almost surely then also $\|X\|_{\psi}\leq \|Y\|_\psi$, which follows directly from the definition of $\|\cdot\|_\psi$, in the fourth line. Likewise we have,
\begin{equation}
\begin{split}
&\max_{v \in \mathcal{V}_{\eps,t} } \frac{1}{n}\sum_{i=1}^{n}\E[|v^TX_i|^{3}] \leq \Big(C_{3,1}\sqrt{C_{\psi_{1/2}}}H_n\Big)^3 \leq (C_{13}H_n)^3  \\
&\max_{i=1}^{n}\max_{k=1}^{K}\|\|\pi_k X_i\|_{2}\|_{\psi_1} \leq \sqrt{C_{\psi_{1/2}}}H_n \leq  (C_{13}H_n)^3.
\end{split}
\end{equation}

By Theorem \ref{max_CLT_abstract}  there is a universal constant $C_2$ such that  
\begin{align*}
I^{\eps_{*}}_{2,1} &= \sup_{t \in[s, t_{*}]}\sup_{w \geq t-\eps_{*}}\Big|\Pr\Big(\max_{v \in \mathcal{V}_{\eps_{*},t}}v^\top S_n^Y \leq w\Big) - \Pr\Big(\max_{v \in \mathcal{V}_{\eps_*,t}}v^\top S_n^X \leq w\Big)\Big|
\\
&\leq \sup_{t \in [s,t_{*}]}C_2 \Big[\Big\{(C_{13}H_n)^3\mA^{t}_n\Big(s-\frac{1}{n}-C_2\log_2(n)\phi_{*}, C_2\phi_{*}\Big) \Big\}\vee \big\{\phi_{*} \sqrt{\log(n)} \big\}\Big]
\end{align*}
for
\[
\phi_{*} = \Big(\frac{\log^{4}|V_{\eps_{*},t}|(C_{13}H_n)^6}{n}\Big)^{1/6}.
\]
A straightforward calculation shows that there is a universal constant $C_{14}$ such that for all $t \leq t_{*}$, 
\[
\log(|\mathcal{V}_{\eps_{*}, t}|) \leq \log(|\mathcal{V}_{\eps_{*}, t_{*}}|)  \leq C_{14}M\log(KnH_n).
\]
Let $C_{15} = \big[C_{13}C_{14}\big]^{4/6}$. Then
\[
C_{15}\delta_{*} =C_{15}\Big(\dfrac{H_n^6M^4 \log^{4}(H_nKn)}{n}\Big)^{1/6} \geq \phi_{*}.
\]
It follows that
\begin{align*}
&\sup_{t \in [s,t_{*}]}\Big[\Big\{(C_{13}H_n)^3\mA^{t}_n(s-n^{-1}-C_2\log_2(n)\phi_{*}, C_2\phi_{*})\Big\}\vee\big\{ \phi_{*}\sqrt{\log(n)} \big\}\Big]
\\
&\leq C_3\Big[\Big\{H_n^3\sup_{t \in [s, t_{*}]}\mA^{t}_n(s-C_3\log(n)\delta_{*},C_3\delta_{*}) \Big\}\vee \big\{\delta_{*}\sqrt{\log(n)}\big\}\Big],
\end{align*}
for sufficiently large universal constant $C_3$, where in the last line we used $1/n\leq \delta_{*}$, $n \geq 3$ and $\log_2(n) \leq 2\log(n)$. This shows the bound in \eqref{bound_I_{2,1}}.
\bigskip
\newline 
{\em Justification of Equation \eqref{bound_on_I_1}}
\bigskip
\newline
\bigskip
By the triangle inequality,
\begin{equation}\label{I_1_bound}
I_{1} \leq \Pr\Big(\max_{k=1}^{K}\|\pi_{k}S^{X}_{n}\|_{2} > t_{*}\Big)+\Pr\Big(\max_{k=1}^{K}\|\pi_{k}S^{Y}_{n}\|_{2} > t_{*}\Big).
\end{equation}
By Lemma~\ref{lim_max_sum} (iv), 
\begin{align*}
&\Pr\Big(\max_{k=1}^{K}\|\pi_{k}S^{X}_{n}\|_{2} > t_{*}\Big)+\Pr\Big(\max_{k=1}^{K}\|\pi_{k}S^{Y}_{n}\|_{2} > t_{*}\Big)\\
&\leq 2\exp\Big(-\frac{t_{*}}{\|\max_{k=1}^{K}\|\pi_{k}S^{X}_{n}\|_{2}\|_{\psi_1}}\Big)+2\exp\Big(-\frac{t_{*}}{\|\max_{k=1}^{K}\|\pi_{k}S^{Y}_{n}\|_{2}\|_{\psi_1}}\Big).
\end{align*}
Therefore it suffices to show that 
\begin{equation}\label{condition_t_*}
t_{*} \geq \Big(\|\max_{k=1}^{K}\|\pi_{k}S^{Y}_{n}\|_{2}\|_{\psi_1}\log(4n)\Big)\vee\Big(\|\max_{k=1}^{K}\|\pi_{k}S^{X}_{n}\|_{2}\|_{\psi_1}\log(4n)\Big).
\end{equation}
First, note that 
\begin{align*}
\Big\|\max_{k=1}^{K}\|\pi_{k}S^{X}_{n}\|_{2}\Big\|_{\Psi_{1}} &\leq K_{\psi_1}\log(K+1) \max_{k=1}^{K} \Bigg\|\sqrt{\sum_{j=1}^{M}([\pi_{k}S^{X}_{n}]_j)^{2}}\Bigg\|_{\Psi_{1}} \tag{Lemma~\ref{lim_max_sum}(ii)}
\\
&= K_{\psi_1}\log(K+1) \max_{k=1}^{K}\sqrt{\Big\|\sum_{j=1}^{M}([\pi_{k}S^{X}_{n}]_j)^{2}\Big\|_{\Psi_{1/2}}} \tag{Lemma~\ref{relationship_norms}(i)}
\\
&\leq K_{\psi_1}\log(K+1)\max_{k=1}^{K}\sqrt{\sum_{j=1}^{M}C_{\psi_{1/2}}\Big\|([\pi_{k}S^{X}_{n}]_j)^{2}\Big\|_{\Psi_{1/2}}} 
\tag{Lemma~\ref{triangle_inequality}}
\\
&= K_{\psi_1}\sqrt{C_{\psi_{1/2}}}\log(K+1) \max_{k=1}^{K}\sqrt{\sum_{j=1}^{M}\Big\|[\pi_{k}S^{X}_{n}]_j\Big\|^2_{\Psi_{1}}}
\tag{Lemma~\ref{relationship_norms}(i)}
\\
&= K_{\psi_1}\sqrt{C_{\psi_{1/2}}}\log(K+1) \max_{k=1}^{K}\sqrt{\sum_{j=1}^{M}\big(K_1\max_{i=1}^{n}\Big\|[\pi_{k}X_i]_j\Big\|_{\Psi_{1}}\big)^2}
\tag{Lemma~\ref{lim_max_sum}(iii)}
\\
&\leq K^2_1K_{\psi_1}\sqrt{C_{\psi_{1/2}}}\log(K+1)H_n.
\end{align*}
In the final line, we used the definition of $H_n$ and the bound 
\[
\sum_{j=1}^{M} \Big\{ \big\|[\pi_{k}S^{X}_{n}]_j\big\|_{\Psi_{1}} \Big\}^2
\leq  \sum_{j=1}^{M} \Big\{ K_1\max_{i=1,\dots,n}\big\|[\pi_{k} X_i]_j\big\|_{\Psi_{1}}\Big\}^2 = K_1^2 \sum_{j=1}^{M} \max_{i=1,\dots,n}\big\|[\pi_{k} X_i]_j\big\|_{\Psi_{1}}^2
\]
which follows by Lemma~\ref{lim_max_sum}(iii) since the $X_i$ are centered by assumption.

Then note that $\|[\pi_{k}Y_{i}]_j\|^2_{\Psi_{1}} \leq C_4 \|[\pi_{k}X_{i}]_j\|^2_{\Psi_{1}}$ for a universal constant $C_4$ since
\begin{align*}
\|[\pi_{k}Y_{i}]_j\|_{\Psi_{1}} &\leq \|1\|_{\psi_2} \|[\pi_{k}Y_{i}]_j\|_{\Psi_{2}} =  \|1\|_{\psi_2} \sqrt{\Var([\pi_{k}Y_{i}]_j)}\sqrt{8/3} = \|1\|_{\psi_2} \sqrt{[\Var([\pi_{k}X_{i}]_j)}\sqrt{8/3}
\\ 
&\lesssim \big\|[\pi_{k}X_{i}]_j\big\|_{\Psi_{1}}. 
\end{align*}
The first inequality follows from Lemma \ref{relationship_norms} (i) and the first inequality because each $Y_i$ is Gaussian.  This bound implies that 
\[\Big\|\max_{k=1}^{K}\|\pi_{k}S^{X}_{n}\|_{2}\Big\|_{\Psi_{1}} \leq C_5\log(K+1)H_n
\]
by the same reasoning used to bound $\Big\|\max_{k=1}^{K}\|\pi_{k}S^{X}_{n}\|_{2}\Big\|_{\Psi_1}$ and a universal constant $C_5$. Then set $D_1$ in the definition of $t_{*}$ so that \eqref{condition_t_*} is satisfied. \hfill $\Box$

\bigskip

\noindent\textbf{Proof of Lemma~\ref{lem:A1_A3_Delta_M_MG}} 

Recall the notation introduced in Section \ref{sub-section:additional_notation}.
\begin{align*}
&\sup_{t\geq s} \Big|\Pr(T^{M,G} \leq t) - \Pr(T^{M} \leq t)\Big|
\\
&= \sup_{t \geq s}\Big|\Pr\Big(\max_{k=1}^{K}\|\pi_k g\|_2^{2} \leq t\Big) - \Pr\Big(\max_{k=1}^{K}\Big\|\frac{1}{\sqrt{n}}\sum_{i=1}^{n}\Phi_{i,k}\Big\|_2^{2} \leq t\Big)\Big|
\\
&\leq \sup_{t \geq s} \Big|\Pr\Big(\max_{k=1}^{K}\|\pi_k g\|_2 \leq \sqrt{t}\Big) - \Pr\Big(\max_{k=1}^{K}\Big\|\frac{1}{\sqrt{n}}\sum_{i=1}^{n}\Phi_{i,k}\Big\|_2\leq \sqrt{t}\Big)\Big|
\\
&=\sup_{t \geq \sqrt{s}} \Big|\Pr\Big(\max_{k=1}^{K}\|\pi_k g\|_2 \leq t\Big) - \Pr\Big(\max_{k=1}^{K}\Big\|\frac{1}{\sqrt{n}}\sum_{i=1}^{n}\Phi_{i,k}\Big\|_2\leq t\Big)\Big|
\\
&=\sup_{t \geq \sqrt{s}} \Big|\Pr(g \in E_{t}) - \Pr\Big(\frac{1}{\sqrt{n}}\sum_{i=1}^{n}\Phi_{i}\in E_t\Big)\Big|
\end{align*}
where $\Phi_{i,k}$ is defined through $\pi_k \Phi =: \Phi_{i,k}$, as previously noted in the second paragraph of section \ref{sub-section:additional_notation}.

Let $D_2 = D \vee 2$ for the constant $D$ of Theorem~ \ref{clt_l2_abstract}, and let $D_1$ be the constant by the same name of Theorem~ \ref{clt_l2_abstract}. Under (A2) we have by 
Lemma~\ref{lem:tail_sum} 
\begin{equation}
\max_{k=1}^{K} \sqrt{\sum_{j=1}^{M}\max_{i=1}^{n}\|[\pi_{k}\Phi_{i}]_j\|^2_{\Psi_{1}}}\leq C_{\phi}\sqrt{\frac{2\alpha}{2\alpha-1}}.
\end{equation}
Therefore the factor $C_{\phi}\sqrt{\frac{2\alpha}{2\alpha-1}}$ satisfies \eqref{eq:Hn3}, and we may set $H_n$ in the statement of Theorem~\ref{clt_l2_abstract} as $H_n = \big( C_{\phi} \vee 1\big)\sqrt{\frac{2\alpha}{2\alpha-1}}$. Note that $H_n>1$. Let $\delta_{*} = \big(\dfrac{H_n^6M^4 \log^{4}(H_nKn)}{n}\big)^{1/6}$. There are now two cases.  If $\sqrt{s} \leq D_2\log_2(n)\delta_{*}/2$ then,
\begin{align*}
&\sup_{t \geq \sqrt{s}}\Big|\Pr(g \in E_{t}) - \Pr\Big(\frac{1}{\sqrt{n}}\sum_{i=1}^{n}\Phi_{i}\in E_t\Big)\Big|
\\
&\leq 2\frac{D_2\log_2(n)\delta_{*}}{\sqrt{s}}
\\
&\leq H_n\frac{4D_2}{\sqrt{s}}\log(n)\Big(\dfrac{M^4 \log^{4}(H_nKn)}{n}\Big)^{1/6}
\\
& \leq 4D_2H_n[1+\log(H_n)]^{4/6}\frac{1}{\sqrt{s}}\Big(\dfrac{M^4 [\log(n)]^6\log^{4}(Kn)}{n}\Big)^{1/6}
\\
&\leq 4D_2H_n[1+\log(H_n)]^{4/6}\frac{1}{\sqrt{s \wedge 1}}\Big(\dfrac{M^{10}\log^{10}(Kn)}{n}\Big)^{1/6},
\end{align*}
and so Equation \eqref{eq:lem:A1_A3_Delta_M_MG} holds with $C=C^{(1)}=4D_2H_n[1+\log(H_n)]^{4/6}$. If $\sqrt{s} >D_2\log_2(n)\delta_{*}/2$, a straightforward computation shows that $\sqrt{s} >1/n$ for all $n \geq 3$: 
\begin{align*}
\sqrt{s} >D_2\log_2(n)\delta_{*}/2 > \log_2(n)n^{-1/6} = n^{-1} \log_2(n)n^{5/6} >n^{-1}
\end{align*}
where in the last line we used that $\log_2(n)n^{5/6} \geq 1$ if $n \geq 3$.

By Theorem~ \ref{clt_l2_abstract}, for $t_{*} = D_1H_n\log(K+1)\log(n)$, where $H_n$ is any number that satisfies \eqref{eq:Hn3} and $H_n \geq 1$, 
\begin{multline*}
\sup_{t \geq \sqrt{s}}\Big|\Pr(g \in E_{t}) - \Pr\Big(\frac{1}{\sqrt{n}}\sum_{i=1}^{n}\Phi_{i}\in E_t\Big)\Big| \\
\leq 
D_2\Big[\Big\{\sup_{t \in [\sqrt{s},t_{*}]}H_n^3\mA^{t}_n(\sqrt{s}-D_2\log(n)\delta_{*}, D_2\delta_{*}) \Big\}\vee \big\{\delta_{*}\sqrt{\log(n)}\big\}\Big],
\end{multline*}
where
\[
\mA_n^{t}(w, \delta):=\sup_{x \geq w} \Pr\Big(\Big|\max_{v \in \mathcal{V}_{\frac{1}{n}, t}}v^\top g - x\Big|\leq \delta\Big).
\]
 We used that $\mA^{t}_n$ is non-increasing in its first argument, and non-decreasing in its second argument when replacing $D$ by $D_2$ in the above bound.  
Apply Lemma~ \ref{lem: Belloni anti-concentration} to conclude that 
\begin{align*}
&\sup_{t\in [\sqrt{s},t_{*}]}\mA_n^{t}(\sqrt{s}-D_2\log(n)\delta_{*}, D_2\delta_{*}) \leq \sup_{t\in [\sqrt{s},t_{*}]}\mA_n^{t}(\sqrt{s}/2, D_2\delta_{*})\leq 34 \frac{D_2\delta_{*}}{\sqrt{s}}\sup_{t\in [\sqrt{s},t_{*}]} \log(|\mathcal{V}_{1/n,t}|).
\end{align*}
By Lemma \ref{lem:product_spheres_approximation}(a),
\[
\sup_{t\in [\sqrt{s},t_{*}]} \log(|\mathcal{V}_{1/n,t}|) \leq 3M\log(K(1 + t_{*}n)) \leq C_{\phi, \alpha}M\log(Kn),
\]
where in the last inequality we plugged in the value of $t_{*}$ and noted that for a constant $C_{\phi, \alpha}$ that is universal for fixed values of $\phi$ and $\alpha$ we have $3M\log(K(1 + t_{*}n)) \leq C_{\phi, \alpha}M\log(Kn)$. The result now follows from plugging in the definition of $\delta_{*}$, and noting that the constant $H_n$ inside of the logarithmic term can be pulled out of the argument of the logarithm and absorbed into $C_{\phi, \alpha}$. \hfill $\Box$

\bigskip

\noindent\textbf{Proof of Lemma~\ref{lem:A1_A4_Delta_M_MG}} 

We follow the proof of Lemma~\ref{lem:A1_A3_Delta_M_MG}, which shows that it suffices to bound
\[
\sup_{t \geq \sqrt{s}} \Big|\Pr(g \in E_{t}) - \Pr\Big(\frac{1}{\sqrt{n}}\sum_{i=1}^{n}\Phi_{i}\in E_t\Big)\Big|,
\]
and defines the sequence $H_n$ and the constant $D_2$ and $\delta_*$. Repeating the argument made in the proof of Lemma~\ref{lem:A1_A3_Delta_M_MG}, if  $\sqrt{s} \leq D_2\log_2(n)\delta_{*}/2$ then,
\begin{align*}
&\sup_{t \geq \sqrt{s}}\Big|\Pr(g \in E_{t}) - \Pr\Big(\frac{1}{\sqrt{n}}\sum_{i=1}^{n}\Phi_{i}\in E_t\Big)\Big|
\\
&\leq 2\frac{D_2\log_2(n)\delta_{*}}{\sqrt{s}}
\\
&\leq H_n\frac{4D_2}{\sqrt{s}}\log(n)\Big(\dfrac{M^4 \log^{4}(H_nKn)}{n}\Big)^{1/6}
\\
& \leq 4D_2H_n[1+\log(H_n)]^{4/6}\frac{1}{\sqrt{s}}\Big(\dfrac{M^4 [\log(n)]^6\log^{4}(Kn)}{n}\Big)^{1/6}.
\end{align*}
and so Equation \eqref{lem:A1_A4_Delta_M_MG_main_bound} holds with $C=C^{(1)}=4D_2H_n[1+\log(H_n)]^{4/6}$. If $\sqrt{s} >D_2\log_2(n)\delta_{*}/2$, a straightforward computation shows that $\sqrt{s} >1/n$ for all $n \geq 3$. 

By Theorem~ \ref{clt_l2_abstract}, for $t_{*} = D_1H_n\log(K+1)\log(n)$, where $H_n$ is any number that satisfies \eqref{eq:Hn3} and $H_n \geq 1$, 
\begin{multline*}
\sup_{t \geq \sqrt{s}}\Big|\Pr(g \in E_{t}) - \Pr\Big(\frac{1}{\sqrt{n}}\sum_{i=1}^{n}\Phi_{i}\in E_t\Big)\Big|
\\
\leq 
D_2\Big[\Big\{\sup_{t \in [\sqrt{s},t_{*}]}H_n^3\mA^{t}_n(\sqrt{s}-D_2\log(n)\delta_{*}, D_2\delta_{*}) \Big\}\vee \big\{\delta_{*}\sqrt{\log(n)}\big\}\Big],
\end{multline*}
where
\[
\mA_n^{t}(w, \delta):=\sup_{x \geq w} \Pr\Big(\Big|\max_{v \in \mathcal{V}_{\frac{1}{n}, t}}v^\top g - x\Big|\leq \delta\Big).
\]
We used that $\mA^{t}_n$ is non-increasing in its first argument, and non-decreasing in its second argument when replacing $D$ by $D_2$ in the above bound.  Apply Lemma~ \ref{lem:ac_ep_net} (with $\eps=1/n$, noting that $\delta_*$ is (up to a numeric constant) larger than $1/n$) to conclude that 
\begin{align*}
	&\sup_{t\in [\sqrt{s},t_{*}]}\mA_n^{t}(\sqrt{s}-D_2\log(n)\delta_{*}, D_2\delta_{*}) \leq \sup_{t\in [\sqrt{s},t_{*}]}\mA_n^{t}(\sqrt{s}/2, D_2\delta_{*})\\
    &\leq F_{\alpha, \beta, \sqrt{s}, C_\phi, C_2}\Big[D_2\Big(\frac{M^{4}\log^{4}(Kn)}{n}\Big)^{1/6}\vee 1/n\Big]\log^{1+\frac{3\beta}{2\alpha-1}}(K+1)\log^{\frac{1}{2}+\frac{3\beta}{2\alpha-1}}(n)\\
    &\leq D_2F_{\alpha, \beta, \sqrt{s}, C_\phi, C_2}\Big(\frac{M^{4}\log^{4}(Kn)}{n}\Big)^{1/6}\log^{1+\frac{3\beta}{2\alpha-1}}(K+1)\log^{\frac{1}{2}+\frac{3\beta}{2\alpha-1}}(n),
\end{align*}
and so Equation \eqref{lem:A1_A4_Delta_M_MG_main_bound} holds for sufficiently large constant $C$. We have set the parameter $m$ equal to $n$ in our application of Lemma~ \ref{lem:ac_ep_net}. \hfill $\Box$

\subsubsection{Gaussian Comparison Bounds for $L^{\infty}-L^{2}$ Mixed Norms}\label{sec:gausscomparison}

We begin the section by proving two Lemmas that are used in the proof of Theorem~\ref{main:thm}. Recall the definition
\[
\Delta_{Me, MG}(s;f) := \sup_{t \geq s} \Big|\Pr_e(T^{M,e}(\Phi,e) \leq t) - \Pr(T^{M,G}(g)\leq t) \Big|. 
\]
from the proof of Theorem~\ref{main:thm}. We also introduce some additional notation used below. Let $\Phi_i \in \R^{KM}$ have entries $\phi_j^{f_{k,i}}, j=1,\dots,M$ in the $k$'th copy of $\R^M$. Let $\Sigma_i^\Phi$ denote the covariance matrix of $\Phi_i$ and let $\Sigma^\Phi = n^{-1}\sum_{i=1}^n \Sigma^{\Phi}_i$. Let $\hat \Sigma^\Phi$ denote the empirical covariance matrix of $\Phi_1,\dots,\Phi_n$. The main result of this section is the following. It's proof relies on intermediate technical results that are stated and proved further below.  

\begin{lemma}\label{lem:bootstrap}
	Suppose conditions (A1)--(A3) hold. There exists a constant $C_{gc}$ that may depend on $\alpha$ and $C_{\phi}$ but is universal for fixed values of those parameters, such that for all $n$ satisfying $n \geq 3$  and all $K \geq 1$ and all $M \geq 3$ we have
	\begin{equation}\label{eq:mainlemgauscomp}
		\Pr\Big(\Delta_{Me,MG}(s;f) > \frac{C_{gc}}{\sqrt{s} \wedge 1}\Big[M^{4/3} \log(Kn)\Big( \frac{\log(KMn)\log^{2}(n)}{n}\Big)^{1/6}\Big]\Big) \leq 1/n.
	\end{equation}
	
\end{lemma}

\begin{lemma}\label{lem:bootstrap-2}
	Suppose conditions (A1)--(A4) hold. There exists a constant $C_1$ that may depend on $C_{\phi}$, $C_2$, $\alpha$, $\beta$, and $s$ such that for all  $K \geq 1$, and all $n$ satisfying $n\geq 3$ and  and all $M \geq 3$, we have
	\begin{multline}\label{eq:mainlemgauscomp-2}
		\Pr\Big(\Delta_{Me,MG}(s;f) > C_1\Big[M^{2/3} \log^{1/3}(Kn)\Big( \frac{\log(KMn)\log^{2}(n)}{n}\Big)^{1/6}\Big]\log^{\frac{1}{2}+\frac{3\beta}{2\alpha-1}}(K+1)\log^{\frac{3\beta}{2\alpha-1}}(n)\Big)
        \\
        \leq 1/n. 
	\end{multline}
	
\end{lemma}
The following two results appear in \cite{CCK17}. 

\begin{lemma}\label{max_norm_approx_2}
	Let $X_1,\dots,X_n \in \R^p$ be independent with covariance matrices $\Sigma_1, \dots, \Sigma_n$, and let $\Sigma = n^{-1}(\sum_{i=1}^{n}\Sigma_{i})$. Let $\hat \Sigma$ denote the sample covariance matrix of $X_1,\dots,X_n$. Assume that $B_n$ satisfies~\eqref{eq:Bn1}--\eqref{eq:Bn3}. There exists a universal constant $C$ (not depending on $\tau$) such that for all $\tau \in (0, e^{-1})$,
	\[
	\Pr\Big(\|\Sigma-\hat \Sigma\|_{\infty}> C[n^{-1}B_{n}^{2}\log(pn)\log^{2}(1/\tau)]^{1/2}\Big)\leq  \tau.
	\]
\end{lemma}
The proof is given within the proof of Proposition 4.1, case (E.1), in~\cite{CCK17}. Let\begin{equation}\label{eq_bound:smoothmax}
\tilde F_\beta(w,x) :=\beta^{-1} \log\left( {\textstyle \sum}_{j=1}^p\exp \left( \beta (w_{j}-x_j) \right) \right), \ w \in \R^{p}, x \in \R^p.
\end{equation}

The next result is a slight modification of Theorem 1 in \cite{CCK12c} which is also mentioned in the proof of Proposition 4.1 in \cite{CCK17}.

\begin{lemma}[Modification of \cite{CCK12c}, Theorem 1]\label{lem:smooth_ev_bound}
	Let $g \in \mathcal{C}^2(\R)$ with uniformly bounded first and second derivatives, and let $F_{\beta}$ be the smooth-max function in~\eqref{eq_bound:smoothmax}. Suppose that $p>2$. Fix $\beta>0$. Let $X, Y \in \R^{p}$ with $X \sim N(0, \Sigma_{1})$, and $Y \sim N(0, \Sigma_{2})$. Then 
	\[
	\sup_{w \in \R^p} |\E[g(\tilde F_{\beta}(X,w))-\E[g(\tilde F_{\beta}(Y,w))]|\leq (\|g''\|_\infty/2+\beta\|g'\|_{\infty}) \|\Sigma_{1}-\Sigma_{2}\|_{\infty}.
	\]
\end{lemma}
The following Lemma follows from the previous one and an argument similar to the the one made leading up to \eqref{eq:approxbound}, and the proof is included below. 
\begin{lemma}\label{lem:gaussian_comparison_abstract}
Let $X, Y \in \R^{p} (p \geq 3)$ with $X \sim N(0, \Sigma_{1})$, and $Y \sim N(0, \Sigma_{2})$. Let $\Delta=\|\Sigma_{1} - \Sigma_{2}\|_{\infty}$.	Then for any $\phi>0$ and any $x \in \R$, there is a universal constant $C$  which does depend on $\Sigma_1$, $\Sigma_2$, $\phi$, or $p$ or $x$ such that
\begin{align*}
		&\Big|\Pr\Big(\max_{1 \leq j \leq p}X_j \leq x\Big)-\Pr\Big(\max_{1 \leq j \leq p}Y_j \leq x\Big)\Big|
        \\
		& \leq \Pr\Big(\Big|\max_{j=1}^{p} Y_j -x\Big|\leq \phi^{-1}\Big)\vee \Pr\Big(\Big|\max_{j=1}^{p} Y_j -x-2\phi^{-1}\Big|\leq \phi^{-1}\Big) +C\phi^{2}\log(p)\Delta.\\
	\end{align*}
    
\end{lemma}

\begin{lemma}\label{lem:gaussian_comparison_max_stat}
Let $X, Y \in \R^{p}$ with $X \sim N(0, \Sigma_{1})$, and $Y \sim N(0, \Sigma_{2})$. Let $\Delta=\|\Sigma_{1} - \Sigma_{2}\|_{\infty}$. Suppose there is a factor $D = D_m(\Sigma_1,t,m)$ such that for any $m \in \N, m \geq 3$ and all $\tau\geq 0$,
\begin{equation}\label{def:D}
\sup_{x \geq t/2}\Pr(|\max_{j=1}^{p}X_j-x| \leq \tau) \leq D\big(\tau \vee 1/m\big).
\end{equation}
There is a numeric (universal) constant $C$ (independent of $D$ and all other parameters) such that for any $m\geq 3$
\begin{equation}
\sup_{x \geq t}\Big|\Pr\Big(\max_{1 \leq j \leq p}X_j \leq x\Big)-\Pr\Big(\max_{1 \leq j \leq p}Y_j \leq x\Big)\Big| \leq C\Big(D^{2/3} \vee \frac{1}{t \wedge 1}\Big)\big( \{\Delta\log(p)\} \vee m^{-2}\big)^{1/3}.
\end{equation}
\end{lemma}
\begin{proposition}\label{prop:gaussbootsphere}
Let $X$ and $Y$ be centered Gaussians on $(\R^M)^K$ $(M \geq 3, K \geq 1)$. Let $\Sigma_1 = \Var(X)$, and $\Sigma_2 = \Var(Y)$. Suppose $s >0$ and suppose $0<\eps <1/\exp(1)$. Further, let
\[
	U = \big(\max_{k=1}^{K}\sqrt{\Tr\{\Var(\pi_{k}Y)\}}\big)\vee \big(\max_{k=1}^{K}\sqrt{\Tr\{\Var(\pi_{k}X)\}}\big)
	\]	
 There is a universal constant $C$ (independent of all parameters) such that 
\begin{multline*}
\sup_{t \geq s}\Big|\Pr\Big(\max_{k=1}^K \|\pi_k Y\|_2 \leq t\Big) - \Pr\Big(\max_{k=1}^K \|\pi_k X\|_2\leq t\Big)\Big|
\\
\leq C\frac{1+\log(U\vee 1)}{(s \wedge 1)}M\Big[ M^{1/3}\|{\Sigma}_1- \Sigma_2 \|_{\infty}^{1/3}+\eps\Big]\log\Big(\frac{K}{\eps}\Big).
\end{multline*}
	
\end{proposition}

\begin{proposition}\label{prop:gaussbootsphere-new}
Let $X$ and $Y$ be centered Gaussians on $(\R^M)^K$ $(M \geq 3, K \geq 1)$, and let $m \in \N, m \geq 3$.  Let $\Sigma_1 = \Var(X)$, and $\Sigma_2 = \Var(Y)$, and suppose that $\Sigma_2$ has the properties indicated in Lemma~ \ref{lem:acnew}. Suppose $s >0$ and suppose $0<\eps <1/\exp(1)$. Let 
    \[
    U = \big(\max_{k=1}^{K}\sqrt{\Tr\{\Var(\pi_{k}Y)\}}\big)\vee \big(\max_{k=1}^{K}\sqrt{\Tr\{\Var(\pi_{k}X)\}}\big)
    \]
    Then there is a constant $D_{\alpha, \beta, s, C_1, C_2}$ that is universal for fixed values of the parameters and independent of $\Sigma_1$ and $\Sigma_2$ in its subscript such that 
	\begin{align*}
		&\sup_{t\geq s}\Big|\Pr\Big(\max_{k=1}^K \|\pi_k X\|_2 \leq t\Big) - \Pr\Big(\max_{k=1}^K \|\pi_k Y\|_2\leq t\Big)\Big|\\
		& \leq D_{\alpha, \beta, s, C_1, C_2}(1+\log(U \vee 1))^{\frac{1}{3}}\Big[M^{2/3}\|\Sigma_{1}-\Sigma_{2}\|_{\infty}^{1/3}\log^{1/3}\Big(\frac{K}{\eps}\Big)+(\eps\vee m^{-2/3})\Big]
        \\
        & \quad \quad\times \log^{\frac12+\frac{3\beta}{2\alpha-1}}(K+1)\log^{\frac{3\beta}{2\alpha-1}}(m).\\
		\end{align*}
\end{proposition}

\subsubsection{Proof for section~\ref{sec:gausscomparison}}

\noindent\textbf{Proof of Lemma~\ref{lem:gaussian_comparison_max_stat}}  

Note that if $D$ satisfies \eqref{def:D}, then $D_2= D \wedge {m}$ also satisfies \eqref{def:D}.   Let $D_3 = 1 \vee D_2$. Let $\tau_{*} = (\log(p)\Delta/D_3)^{1/3}$. If $\tau_{*} \geq t/4$, then
\begin{align*}
&\sup_{x \geq t}\Big|\Pr\Big(\max_{1 \leq j \leq p}X_j \leq x\Big)-\Pr\Big(\max_{1 \leq j \leq p}Y_j \leq x\Big)\Big| \leq \frac{4\tau_{*}}{t} \leq 4\frac{(\log(p)\Delta)^{1/3}}{tD_3^{1/3}} \leq 4\frac{\big(\log(p)\Delta\big)^{1/3}}{t}\\
& \leq 4\frac{\big(\log(p)\Delta\big)^{1/3}}{1 \wedge t} \leq 4\Big(D^{2/3}\vee \frac{1}{t \wedge 1}\Big)\big(\{\Delta\log(p)\} \vee m^{-2}\big)^{1/3}.
\end{align*}
Let $C^{(1)} = 4$. From now on we assume $\tau_{*} <t/4$. By Lemma \ref{lem:gaussian_comparison_abstract} applied with $\phi = \tau_*^{-1}$ we have for a numeric constant $C_1$, 
\begin{equation}\label{eq:110}
\sup_{x \geq t}\Big|\Pr\Big(\max_{1 \leq j \leq p}X_j \leq x\Big)-\Pr\Big(\max_{1 \leq j \leq p}Y_j \leq x\Big)\Big| \leq \sup_{x\geq t/2}\Pr\Big(\Big|\max_{j=1}^{p}[S_n^Y]_j -x\Big|\leq \tau_{*}\Big) +C_1\tau_{*}^{-2}\Delta\log(p).
\end{equation}
First consider the case $\tau_{*} <1/{m}$. Then by \eqref{eq:110} %Lemma \ref{lem:gaussian_comparison_abstract},
\begin{align*}
&\sup_{x \geq t}\Big|\Pr\Big(\max_{1 \leq j \leq p}X_j \leq x\Big)-\Pr\Big(\max_{1 \leq j \leq p}Y_j \leq x\Big)\Big|
\\
&\leq \sup_{x\geq t/2}\Pr\Big(\Big|\max_{j=1}^{p}[S_n^Y]_j -x\Big|\leq 1/{m}\Big) +C_1\tau_{*}^{-2}\log(p)\Delta
\\
&\leq \frac{D_3}{{m}}+C_1[\Delta\log(p)]^{1/3}[D_3]^{2/3}
\\
&= D_3^{2/3}\Big[C_1\big(\log(p)\Delta\big)^{1/3}+D_3^{1/3}/{m}\Big]
\\
&\leq D_3^{2/3}\Big[C_1\big(\log(p)\Delta\big)^{1/3}+{m}^{-2/3}\Big]
\\
& \leq 2(C_1 \vee 1)[D_3]^{2/3}\Big[\big(\log(p)\Delta\big)^{1/3}+{m}^{-2/3}\Big]
\\ 
&\leq 2(C_1 \vee 1)[D \vee 1]^{2/3}\Big[\big(\log(p)\Delta\big)^{1/3}+{m}^{-2/3}\Big]
\\ 
&\leq2(C_1 \vee 1)\Big(D^{2/3} \vee \frac{1}{t \wedge1}\Big)\Big[\big(\log(p)\Delta\big)^{1/3}+{m}^{-2/3}\Big]
\end{align*}
where we used the definition of $\tau_*$ and the assumption in~\eqref{def:D} in the second inequality. Let $C^{(2)} = 4(C_1 \vee 1)$. Now consider the case $\tau_{*} \geq 1/{m}$. By \eqref{eq:110} %Lemma \ref{lem:gaussian_comparison_abstract},
\begin{align*}
&\sup_{x \geq t}\Big|\Pr\Big(\max_{1 \leq j \leq p}X_j \leq x\Big)-\Pr\Big(\max_{1 \leq j \leq p}Y_j \leq x\Big)\Big| \leq \sup_{x\geq t/2}\Pr\Big(\Big|\max_{j=1}^{p}[S_n^Y]_j -x\Big|\leq \tau^{*}\Big) +C_1\tau_{*}^{-2}\Delta\log(p)
\\
&\leq [D_3]\tau_{*}+C_1[\Delta\log(p)]^{1/3}[D_3]^{2/3} \leq (1+C_1)[D_3]^{2/3}[\Delta\log(p)]^{1/3}\\
&\leq (1+C_1)\Big(D^{2/3} \vee \frac{1}{t\wedge 1}\Big)\Big(\{\Delta\log(p)\}\vee {m}^{-2}\Big)^{1/3},
\end{align*}
where we used the definition of $\tau_*$ and the assumption in~\eqref{def:D} in the second inequality and the definition of $\tau_*$ again in the third inequality.
Set $C^{(3)} = (1+C_1)$, and set $C$ in the formulation of the Lemma as $C=C^{(1)} \vee C^{(2)} \vee C^{(3)}$. This completes the proof of Lemma~\ref{lem:gaussian_comparison_max_stat}.
\hfill $\Box$

\bigskip

\noindent\textbf{Proof of Lemma~\ref{lem:bootstrap} and Lemma~\ref{lem:bootstrap-2}} 

By Lemma~\ref{max_norm_approx_2}, for a universal constant $C$ and any $\tau \in (0,e^{-1})$ 
\begin{equation}\label{eq:proofAnticonhelp1}
	\Pr\Big(\|\Sigma^\Phi-\hat \Sigma^\Phi\|_{\infty}> C[n^{-1}\log(KMn)\log^{2}(1/\tau)]^{1/2}\Big)\leq \tau,
\end{equation}
where we plugged in $KM$ in place of $p$ in the bound of  Lemma~\ref{max_norm_approx_2}.
We also used the fact that under (A2) $B_n$ in Lemma~\ref{max_norm_approx_2} can bounded from above by a constant that is independent of $n$ and depends only on the constants $C_\phi, \alpha$ appearing in (A2). Next, let $(\hat \Sigma^\Phi)_k$ denote the $k'th$ diagonal block of size $M\times M$. We may assume without loss of generality assume that 
\[
\sqrt{M}C[n^{-1}\log(KMn)\log^{2}(n)]^{1/2} \leq 1
\]
since otherwise the bound in~\eqref{eq:mainlemgauscomp} can be made to hold by picking $C_{gc}$ sufficiently large. Then on the event $\Omega_1 := \{\|\Sigma^\Phi-\hat \Sigma^\Phi\|_{\infty} \leq C[n^{-1}\log(KMn)\log^{2}(n)]^{1/2} \}$, which holds with probability greater than $1-1/n$ by~\eqref{eq:proofAnticonhelp1},
\begin{align*}
	\max_{k=1}^{K}\sqrt{\sum_{j=1}^{M}[(\hat \Sigma^\Phi)_k]_{jj}} 
	&\leq \sqrt{M}\|\Sigma^{\Phi}-\hat \Sigma^\Phi\|_{\infty}+\max_{k=1}^{K}\sqrt{\sum_{j=1}^{M}[(\Sigma)_k]_{jj}}
	\\ 
	&\leq \sqrt{M}C[n^{-1}\log(KMn)\log^{2}(n)]^{1/2} + U \leq 2U
\end{align*}
where 
\[
U := 1 \vee\max_{k=1}^{K}\sqrt{\sum_{j=1}^{M}[(\Sigma^\Phi)_k]_{jj}} 
\]
is bounded by a constant that depends on $C_\phi, \alpha$ only by (A2). 

Now let $X$ denote a centered Gaussian vector in $\R^{KM}$ with covariance matrix $\Sigma^\Phi$ and let $Y = [\hat\Sigma^\Phi]^{1/2} Z$ for a isotropic Gaussian $Z$ on $\R^{KM}$. The distribution of $T^{M,e}(\Phi,e)$,  conditional on $\mathcal{D}$ is equal to the conditional distribution of $\max_{k=1}^K \|\pi_k Y\|_2^2$ given $[\hat\Sigma^\Phi]^{1/2}$ while $T^{M,G}$ has the same distribution as $\max_{k=1}^K \|\pi_k X\|_2^2$. Let 
\[
U_n := 1 \vee \max_{k=1}^{K}\sqrt{\Tr\{(\Var(\pi_{k}\hat\Sigma^\Phi)\}}\vee \max_{k=1}^{K}\sqrt{\Tr\{(\Var(\pi_{k}\Sigma^\Phi)\}}.
\]
Given the earlier discussion, $U_n$ is bounded by a constant that depends on $C_\phi, \alpha$ on the event $\Omega_1$. 

To prove Lemma ~\ref{lem:bootstrap}, apply Proposition~\ref{prop:gaussbootsphere} with $\eps = 1/n$ to obtain the existence of a $C$ that may depend only on $\alpha$ and $C_\phi$ such that almost surely for any $n \geq 3$, 
\[
\Delta_{Me, MG}(f;s) \leq  C \frac{1}{(\sqrt{s} \wedge 1)}\Big[ M^{4/3}\|\hat \Sigma^\Phi - \Sigma^\Phi \|_{\infty}^{1/3}+\frac{M}{n}\Big]\log(Kn),
\]
where we absorbed the term $1+\log(U_n)$ into $C$ based on the previously indicated bound on $B_n$. The claim follows by combining this bound with the bound $\|\hat \Sigma^\Phi - \Sigma^\Phi \|_{\infty}$ derived earlier and picking $C_{gc}$ to be sufficiently large. 

To prove Lemma~\ref{lem:bootstrap-2}, apply Proposition~\ref{prop:gaussbootsphere-new} with $\eps=1/n$, noting that under (A2) and (A4), each $\Phi_i$ has the properties indicated in Lemma \ref{lem:acnew}. It follows that
\begin{align*}
\Delta_{Me, MG}(X;s) &\leq D_{\alpha, \beta, s, C_1, C_2}(1+\log(U \vee 1))^{\frac{1}{3}}\Big[M^{2/3}\|\Sigma_{1}-\Sigma_{1}\|_{\infty}^{1/3}\log^{1/3}(Kn)+\frac{1}{n^{2/3}}\Big]
\\
& \quad\quad \times \log^{\frac{1}{2}+\frac{3\beta}{2\alpha-1}}(K+1)\log^{\frac{3\beta}{2\alpha-1}}(n)
\end{align*}
on $\Omega_1$.
The claim follows by combining this bound with the bounds on $U_n$ and $\|\hat \Sigma^\Phi - \Sigma^\Phi \|_{\infty}$ derived earlier, and picking $C_1$ sufficiently large. 
\hfill $\Box$

\bigskip

\noindent\textbf{Proof of Lemma~\ref{lem:gaussian_comparison_abstract}} 

Recall the notation in the proof leading up to Equation \eqref{eq:approxbound}. Arguing as was done to obtain Equation \eqref{eq:approxbound} and applying Lemma~\ref{lem:smooth_ev_bound} with $g(x) = g_0(\phi x), \beta = \phi \log p$ for a fixed, three times continuously differentiable $g_0$ and with $\beta = \phi \log p$ gives the following bound:
\begin{align*}
&\Big|\Pr\Big(\max_{1\leq j \leq p}X_j \leq x-\phi^{-1}\Big)-\Pr\Big(\max_{1\leq j \leq p}Y_j\leq x-\phi^{-1}\Big)\Big|
\\
&\leq \Pr\Big(\Big|\max_{1\leq j \leq p}Y_{j} -x\Big| \leq \phi^{-1}\Big) \vee  \Pr\Big(\Big|\max_{1\leq j \leq p}Y_{j} -x-2\phi^{-1}\Big| \leq \phi^{-1}\Big)
\\
& \quad\quad + \sup_{x \in \R} \Big|\E[g( F_{\beta}(X,x))-\E[g(F_{\beta}(Y,x))]\Big| 
\\
&= \Pr\Big(\Big|\max_{1\leq j \leq p}Y_{j} -x\Big| \leq \phi^{-1}\Big) \vee  \Pr\Big(\Big|\max_{1\leq j \leq p}Y_{j} -x-2\phi^{-1}\Big| \leq \phi^{-1}\Big)
\\
& \quad\quad + \sup_{x \in \R} \Big|\E[g(\tilde F_{\beta}(X,\boldsymbol{1}x))-\E[g(\tilde F_{\beta}(Y,\boldsymbol{1}x))]\Big| 
\\
&\leq \Pr\Big(\Big|\max_{1\leq j \leq p}Y_{j} -x\Big| \leq \phi^{-1}\Big) \vee  \Pr\Big(\Big|\max_{1\leq j \leq p}Y_{j} -x-2\phi^{-1}\Big| \leq \phi^{-1}\Big)
\\
&\quad\quad+\Big( \frac{\|g_0''\|_\infty}{2}\phi^{2}+\|g_0'\|_\infty\phi^{2}\log(p)\Big)\Delta
    \\
&\leq \Pr\Big(\Big|\max_{1\leq j \leq p}Y_{j} -x\Big| \leq \phi^{-1}\Big) \vee  \Pr\Big(\Big|\max_{1\leq j \leq p}Y_{j} -x-2\phi^{-1}\Big| \leq \phi^{-1}\Big)  + C \phi^2 \Delta \log p  
\end{align*}
where the constant $C$ in the last line depends on the derivatives of $g_0$ and we defined $\boldsymbol{1} \in \R^p$ as a vector with all entries equal to $1$. Here, we applied arguments similar to those that were used to derive Equation \eqref{eq:approxbound} to obtain the first inequality, and Lemma~\ref{lem:smooth_ev_bound} to obtain the second inequality. \hfill $\Box$

\bigskip

\textbf{Proof of Proposition~\ref{prop:gaussbootsphere} and Proposition~\ref{prop:gaussbootsphere-new}}

If $\eps\geq s/2$,
\begin{align*}
	&\sup_{t \geq s}\Big|\Pr\Big(\max_{k=1}^K \|\pi_k Y\|_2 \leq t\Big) - \Pr\Big(\max_{k=1}^K \|\pi_k X\|_2\leq t\Big)\Big|\leq 2\frac{\eps}{s}\leq 2\frac{\eps}{s\wedge 1},
\end{align*}
and so the bound of Proposition ~\ref{prop:gaussbootsphere} holds with $C=2$. For the remainder of the proof, assume $\eps <s/2$.
For any $t_{*} \geq s$,
\begin{align*}
&\sup_{t\geq s}\Big|\Pr\Big(\max_{k=1}^K \|\pi_k Y\|_2 \leq t\Big) - \Pr\Big(\max_{k=1}^K \|\pi_k X\|_2\leq t\Big)\Big|
\\
&\leq \sup_{t\geq t_{*}}\Big|\Pr\Big(\max_{k=1}^K \|\pi_k Y\|_2 \leq t\Big) - \Pr\Big(\max_{k=1}^K \|\pi_k X\|_2\leq t\Big)\Big|\\
&\hspace{1cm}+\sup_{t\in [s,t_{*}]}\Big|\Pr\Big(\max_{k=1}^K \|\pi_k Y\|_2 \leq t\Big) - \Pr\Big(\max_{k=1}^K \|\pi_k X\|_2\leq t\Big)\Big|.
\end{align*}
By Lemma \ref{lem:tail-bound-from-trace-bound}, there is a universal constant $D_1$ such that if $t_{*}=D_1\sqrt{\log(K+1)}U\sqrt{\log(\frac{4}{\eps})}$ and $t \geq t_{*}$, 
\[
\Big|\Pr\Big(\max_{k=1}^K \|\pi_k Y\|_2 \leq t\Big) - \Pr\Big(\max_{k=1}^K \|\pi_k X\|_2\leq t\Big)\Big| \leq \eps.
\]
By Lemma~\ref{lem:V_approx}, 
\begin{equation}\label{eq:bound_in_terms_of_I_1_I_2}
\sup_{t \in [s, t_{*}]}\Big|\Pr\Big(\max_{k=1}^K \|\pi_k Y\|_2 \leq t\Big) - \Pr\Big(\max_{k=1}^K \|\pi_k X\|_2\leq t\Big)\Big| \leq \sup_{t \in [s, t_{*}]}I_{2}(t,\eps)+\sup_{t \in [s, t_{*}]}I_{1}(t,\eps),
\end{equation}
where
\begin{align*}
	& I_{1}(t, \eps) =  \max_{w \in \{t-\eps, t\}}\Big|\Pr\Big(\max_{v \in \mathcal{V}_{\eps, t}}v^{\top}X \leq w\Big)- \Pr\Big(\max_{v \in \mathcal{V}_{\eps, t}}v^{\top}Y \leq w\Big)\Big|,
	\\
	&I_{2}(t,\eps) = \Big|\Pr\Big(\max_{v\in V_{\eps,t}}v^{\top}Y \leq t\Big)-\Pr\Big(\max_{v\in V_{\eps,t}}v^{\top}Y \leq t-\eps\Big)\Big|.
\end{align*}
Therefore 
\[
\sup_{t\geq s}\Big|\Pr\Big(\max_{k=1}^K \|\pi_k Y\|_2 \leq t\Big) - \Pr\Big(\max_{k=1}^K \|\pi_k X\|_2\leq t\Big)\Big| \leq \sup_{t \in [s, t_{*}]}I_{2}(t,\eps)+\sup_{t \in [s, t_{*}]}I_{1}(t,\eps) + \eps.
\]

We now introduce some additional notation. Let $f:\mathcal{V}_{\eps, t}\rightarrow \{1, \dots,|\mathcal{V}_{\eps, t}|\} \subset \mathbb{N}$ be a bijection and define $\tilde{Y},\tilde{X} \in \mathbb{R}^{|\mathcal{V}_{\eps, t}|}$ component-wise as follows:  for each $1 \leq \ell \leq |\mathcal{V}_{\eps, t}|$ let $[\tilde{Y}]_{\ell}=(f^{-1}(\ell))^\top Y$ and $[\tilde{X}]_{\ell}=(f^{-1}(\ell))^\top X$. 
Note that 
\[
I_{1}(t, \eps) = \max_{w \in \{t-\eps, t\}}\Big|\Pr\Big(\max_{\ell=1}^{ |\mathcal{V}_{\eps, t}|}[\tilde{X}]_{\ell} \leq w\Big)- \Pr\Big(\max_{\ell=1}^{ |\mathcal{V}_{\eps, t}|}[\tilde{Y}]_{\ell}  \leq w\Big)\Big|,
\]
and 
\[
I_{2}(t,\eps) = \Pr\Big(\max_{\ell=1}^{ |\mathcal{V}_{\eps, t}|}[\tilde{Y}]_{\ell}  \leq t\Big)-\Pr\Big(\max_{\ell=1}^{ |\mathcal{V}_{\eps, t}|}[\tilde{Y}]_{\ell}  \leq t-\eps\Big) \leq \Pr\Big( \Big|\max_{\ell=1}^{ |\mathcal{V}_{\eps, t}|}[\tilde{Y}]_{\ell} - t\Big| \leq \eps\Big).
\]
Assume that 
\begin{align*}
\sup_{w  \geq s/4}\Pr\Big(\big|\max_{\ell=1}^{ |\mathcal{V}_{\eps, t}|}[\tilde{Y}]_{\ell} -w\big| \leq \tau\Big) \leq D_{m}(\tau \vee 1/{m}), 
\end{align*}
Under this assumption it follows from the definition of $I_{2}(t,\eps)$ that 
\[
\sup_{t \in [s, t_{*}]}I_{2}(t,\eps) \leq D_{m}(\eps \vee {m}^{-1}).
\]
To deal with $I_1$, apply Lemma \ref{lem:gaussian_comparison_max_stat} to obtain
\begin{align*}
\sup_{t \in [s, t_{*}]}I_{1}(t, \eps) &\leq \sup_{t \in [s/2,t_*]} \Big|\Pr\Big(\max_{\ell=1}^{ |\mathcal{V}_{\eps, t}|}[\tilde{X}]_{\ell} \leq w\Big)- \Pr\Big(\max_{\ell=1}^{ |\mathcal{V}_{\eps, t}|}[\tilde{Y}]_{\ell}  \leq w\Big)\Big|
\\
&\leq D_8\sup_{t \in [s/2, t_{*}]}\Big(D_{m} ^{2/3}\vee\frac{1}{s\wedge 1}\Big)\big({\big\{}\log\big(|V_{\eps,t}|\big)\|\Var(\tilde Y) - \Var(\tilde Z)\|_{\infty}{\big\}} {\vee {m}^{-2}}\big)^{1/3}.
\end{align*}
To bound the difference of the variances, we argue as follows. Since each $v \in \mathcal{V}_{\eps, t}$ has at most $M$ non-zero entries and satisfies $\|v\|_2 =1$, we have for $v,w \in \mathcal{V}_{\eps, t}$
\begin{equation}\label{covariance_operator_bound}
|v^\top (\Sigma_1-\Sigma_2) w| \leq \|v\|_1 \|(\Sigma_1-\Sigma_2) w\|_\infty \leq \|v\|_1 \|\Sigma_1- \Sigma_2\|_\infty \|w\|_1 \leq M \|\Sigma_1- \Sigma_2\|_\infty,
\end{equation}
(we used the inequality $\|v\|_1 \leq \sqrt{M}\|v\|_{2}$ which follows by H\"older's inequality), we obtain	
\[
\|\Var(\tilde Y) - \Var(\tilde Z)\|_{\infty} \leq M \|\Sigma_1- \Sigma_2\|_\infty.
\]
Thus 
\[
\sup_{t \in [s, t_{*}]}I_{1}(t, \eps) \leq D_8\Big(D_{m} ^{2/3}\vee\frac{1}{s\wedge 1}\Big) \sup_{t \in [s/2, t_{*}]}\Big(\big\{M\log\big(|V_{\eps,t}|\big)\|\Sigma_1 - \Sigma_2\|_{\infty}\big\}\vee {m}^{-2}\Big)^{1/3}.
\]
By Lemma \ref{lem:product_spheres_approximation}
\[
\sup_{t \in [s/2, t_{*}]} \log\big(|V_{\eps,t}|\big) \leq D_{10} M \log(K(1+t_*\eps^{-1}))
\]
for a universal constant $D_{10}$. Then note that 
\begin{align*}
&\log\Big(K\big(1+t_{*}/\eps\big)\Big) = \log\Big(K\big(1+D_1\sqrt{\log(K+1)}U\big[\log(\frac{4}{\eps})\big]^{1/2}/\eps\big)\Big)\\
&\leq \log\Big(K\big(D_5K(U\vee1)\eps^{-2}\big)\Big) \leq D_6(1+\log(U\vee1))\log\big(K\eps^{-1}\big).
\end{align*}
for a universal constants $D_5$ and $D_6$ (we used $K/\eps >\exp(1)$) giving the bound
\[
\sup_{t \in [s, t_{*}]}I_{1}(t, \eps) \leq D_{11}\Big(D_{m}^{2/3}\vee\frac{1}{s\wedge 1}\Big) \Big(\big\{M^2 (1+\log(U\vee1))\log\big(K\eps^{-1}\big)\|\Sigma_1 - \Sigma_2\|_{\infty}\big\} \vee {m}^{-2}\Big)^{1/3}.
\]
To complete the proof of Proposition~\ref{prop:gaussbootsphere}, apply Lemma~ \ref{lem: Belloni anti-concentration} to obtain for any $\tau>0$ and any ${m} \geq 3$ and for any $s$ and $\eps$ satisfying $\eps < s$ (so that $\mathcal{V}_{\eps, t}$ is defined for all $t \geq s$)
\begin{equation}\label{eq:drop_max_n}
\sup_{w  \geq s/4}\Pr\Big(\big|\max_{\ell=1}^{ |\mathcal{V}_{\eps, t}|}[\tilde{Y}]_{\ell} -w\big| \leq \tau\Big) \leq \frac{17}{s/4}\log(|\mathcal{V}_{\eps, t}|)\tau \leq \frac{68}{s \wedge 1}\log(|\mathcal{V}_{\eps, t}|)\tau \leq D_{m}(\tau \vee 1/{m}), 
\end{equation}
for 
\[
D_{m} = \frac{68}{s \wedge 1}\log(|\mathcal{V}_{\eps, t}|) \leq \frac{D_{12}}{s \wedge 1} M (1+\log(U\vee1))\log\big(K\eps^{-1}\big).
\]
Thus
\[
\sup_{t \in [s, t_{*}]}I_{1}(t, \eps) \leq \frac{D_{14}}{s\wedge 1} M^{4/3} (1+\log(U\vee1))\log\big(K\eps^{-1}\big) \|\Sigma_1 - \Sigma_2\|_{\infty}^{1/3},
\]
where the term with $\vee m^{-2}$ can be dropped because $n$ can be chosen arbitrarily large in light of \eqref{eq:drop_max_n}.
Combined with the earlier bounds established in this proof
\begin{align*}
&\sup_{t\geq s}\Big|\Pr\Big(\max_{k=1}^K \|\pi_k Y\|_2 \leq t\Big) - \Pr\Big(\max_{k=1}^K \|\pi_k X\|_2
\leq t\Big)\Big|
\leq \sup_{t \in [s, t_{*}]}I_{2}(t,\eps)+\sup_{t \in [s, t_{*}]}I_{1}(t,\eps) + \eps 
\\
&\leq \frac{D_{15}(1+\log(U\vee1))}{s\wedge 1} M \log\big(K\eps^{-1}\big) \Big(M^{1/3}\|\Sigma_1 - \Sigma_2\|_{\infty}^{1/3} + \eps \Big) + \eps.
\end{align*}
The final $+\eps$ can be dropped by enlarging the constant since $M \log(K/\eps) \geq 1$ and this completes the proof of Proposition~\ref{prop:gaussbootsphere}.

We now move to the proof of Proposition~\ref{prop:gaussbootsphere-new}. Here, we apply Lemma \ref{lem:ac_ep_net} to obtain 
\begin{align*}
&\sup_{w  \geq s/4}\Pr\Big(\big|\max_{\ell=1}^{ |\mathcal{V}_{\eps, t}|}[\tilde{Y}]_{\ell} -w\big| \leq \tau\Big) 
= \sup_{w  \geq s/4} \Pr\Big(\max_{v \in \mathcal{V}_{\eps, t}} v^\top Y \in [w-\tau,w+\tau]\Big)
\\
&\leq D_{\alpha, \beta, s/4, C_1, C_2}(\tau \vee {m}^{-1})\log^{\frac{1}{2}+\frac{3\beta}{2\alpha-1}}(K+1)\log^{\frac{3\beta}{2\alpha-1}}(n) = D_{{m}} (\tau \vee {m}^{-1})
\end{align*}
for $D_{m} = D_{\alpha, \beta, s/4, C_1, C_2}\log^{\frac{1}{2}+\frac{3\beta}{2\alpha-1}}(K+1)\log^{\frac{3\beta}{2\alpha-1}}({m})$. Combining this with the bounds obtained earlier yields 
\begin{align*}
&\sup_{t\geq s}\Big|\Pr\Big(\max_{k=1}^K \|\pi_k Y\|_2 \leq t\Big) - \Pr\Big(\max_{k=1}^K \|\pi_k X\|_2
\leq t\Big)\Big|
\\
& \leq \sup_{t \in [s, t_{*}]}I_{2}(t,\eps)+\sup_{t \in [s, t_{*}]}I_{1}(t,\eps) + \eps  
\\
& \leq  \frac{D_{16}(1+\log(U\vee1))^{1/3}}{s\wedge 1}\log^{\frac{1}{2}+\frac{3\beta}{2\alpha-1}}(K+1)\log^{\frac{3\beta}{2\alpha-1}}({m}) 
\\
&\quad \quad \times \Big( M^{2/3}\|\Sigma_1 - \Sigma_2\|_{\infty}^{1/3} \log^{1/3}\big(K\eps^{-1}\big) + (\eps \vee {m}^{-2/3})\Big),
\end{align*}
for sufficiently large constant $D_{16}$, noting the bound 
\[
\big\{M^{2/3}\|\Sigma_1 - \Sigma_2\|_{\infty}^{1/3} \log^{1/3}\big(K\eps^{-1}\big)\}\vee {m}^{-2/3} \leq M^{2/3}\|\Sigma_1 - \Sigma_2\|_{\infty}^{1/3} \log^{1/3}\big(K\eps^{-1}\big) +{m}^{-2/3}.
\]
\hfill $\Box$

\subsubsection{Remainder Bounds}\label{Appendix:Remainder}
We begin this section by stating Lemmas that are used in the proof of Theorems~ \ref{main:thm} and \ref{main:thm-power}. Recall the notation introduced in Assumptions \ref{assumptions}, and Section \ref{sub-section:additional_notation}, and let $K_{\phi, \alpha} = C_{\phi}\sqrt{\frac{2\alpha}{2\alpha-1}}$, and recall the definition of the conditional Orlicz norm given in Definition \ref{def:condorlicz}.

\begin{lemma}\label{lem:|T^d-T^M|}
Recall the definition of $T^M$ and the definition of $T^d$ made in Section \ref{sub-section:additional_notation}. Assume that (A1) and (A2) hold. There exists a universal constant $C$ such that for any $M \geq 1$ and $K \geq 1$,
\begin{equation}
\|T^{M} - T^{d}\|_{\Psi_{\frac{1}{2}}} \leq C(q^{-1}C_H^2+C_HK_{\phi, \alpha}+K_{\phi, \alpha}^{2})[\log(K+1)]^{2}[T^{-q}+M^{1-2\alpha}].
\end{equation}
\end{lemma}

\begin{lemma}\label{lem:|T^{d,e}-T^{M,e}|}
Recall the definition of $T^{M,e}$ and the definition of $T^{d,e}$ made in Section \ref{sub-section:additional_notation}. Assume that (A1) and (A2) hold.There exists a universal constant $C$ and a random variable $Y \in \sigma(f)$ such that for all $M \geq 1$ and $K \geq 1$,
\begin{equation}
\|T^{M,e} - T^{d,e}\|_{\Psi_{1}|\sigma(f)} \leq Y,
\end{equation}
and
\begin{equation}
\|Y\|_{\Psi_{\frac{1}{2}}} \overset{a.s}{\leq} C(q^{-1}C_H^2+C_H K_{\phi, \alpha}+K_{\phi, \alpha}^{2}) [\log(K+1)]^{3}[T^{-q}+M^{1-2\alpha}].
\end{equation}
\end{lemma}
The rest of this section contains results that are used to prove Lemmas~ \ref{lem:|T^d-T^M|} and \ref{lem:|T^{d,e}-T^{M,e}|}. 

Note that Lemma~\ref{lem:|T^d-T^M|} follows straightforwardly from Lemmas~\ref{lem:norm_bound:T_M-T} and Lemmas~\ref{lem:smooth_norm_bound}, using Lemma~\ref{triangle_inequality} to combine the two bounds. Likewise, Lemma~\ref{lem:|T^{d,e}-T^{M,e}|} follows straightforwardly from Lemma~\ref{lem:boot_smooth_norm_bound} and Lemmas~\ref{lem: gaussian_tail_bound}.

\begin{lemma}\label{lem:norm_bound:T_M-T}
	Recall the definition of $T^M$ in Section \ref{sub-section:additional_notation}. Assume that (A2) holds. There exists a universal constant $C$ such that for any $M\geq 1$, and any $K \geq 1$,
	\begin{equation}\label{norm_bound:T_M-T}
		\Big\|\max_{k=1}^{K}\int_{I}\Big(\frac{\sum_{i=1}^{n}f_{k,i}}{\sqrt{n}}\Big)^{2}-T^{M}\Big\|_{\Psi_{\frac{1}{2}}} \leq C K^2_{\phi, \alpha} \log^{2}(K+1)M^{1-2\alpha}.
	\end{equation}
	Additionally, 
	
\[
\Big\|\max_{k=1}^{K}\int_{I}\Big(\frac{\sum_{i=1}^{n}f_{k,i}}{\sqrt{n}}\Big)^{2}\Big\|_{\Psi_{\frac{1}{2}}} \leq C K^2_{\phi, \alpha} \log^{2}(K+1).
\]
\end{lemma}

\begin{lemma}\label{lem:smooth_norm_bound}
Recall the definition of $T^d$ in Section \ref{sub-section:additional_notation}. Assume (A1), (A2) hold. There exists a constant $C$ that does not depend on $n$, $M$, $K$, $T$, $\alpha$, $q$, $C_{H}$ or $C_{\phi}$ such that for all $K \geq 1$,
\begin{equation}
\Big\|\max_{k=1}^{K}\int_{I}\Big(\frac{\sum_{i=1}^{n}f_{k,i}}{\sqrt{n}}\Big)^{2}-T^{d}\Big\|_{\Psi_{\frac{1}{2}}} \leq C(q^{-1}C_H^2+C_H K_{\phi, \alpha})\frac{\log^{2}(K+1)}{T^{q}}.
\end{equation}
\end{lemma}

\begin{lemma}\label{lem:boot_smooth_norm_bound}
Recall the definition of $T^{d,e}$ in Section \ref{sub-section:additional_notation}. There exists a constant $C$ that does not depend on $n$, $M$, $K$, $T$, $\alpha$, $q$, $C_{H}$ or $C_{\phi}$, and a random variable  $Y \in \sigma(f)$  such that for all $K\geq 1$
\[
\Big\|\max_{k=1}^{K}\int_{I}\Big(\frac{\sum_{i=1}^{n}e_{i}[f_{k,i}-\bar{f}_{k,\cdot}]}{\sqrt{n}}\Big)^{2} - T^{e,d}\Big\|_{\Psi_{1}|\sigma(f)} \leq Y,
\] 
and 
\[
\|Y\|_{\Psi_{\frac{1}{2}}} \overset{a.s}{\leq}C(q^{-1}C_H^2+C_H K_{\phi, \alpha})\frac{\log^{3}(K+1)}{T^{q}}.
\]
\end{lemma}

\begin{lemma}\label{lem: gaussian_tail_bound}
	Recall the definition of $T^{M,e}$ in Section \ref{sub-section:additional_notation}. There exists a constant $C$ that does not depend on $n$, $M$, $K$, $T$, $\alpha$, $q$, $C_{H}$ or $C_{\phi}$, and a random variable $W \in \sigma(f)$ such that for all $K \geq 1$,
	
	 \[
	 \Big\|T^{e,M}-\max_{k=1}^{K}\int_{I}\Big(\frac{\sum_{i=1}^{n}e_{i}[f_{k,i}-\bar{f}_{k,\cdot}]}{\sqrt{n}}\Big)^{2}\Big\|_{\Psi_{1}|\sigma(f)}\leq W\]
	 
	 almost surely, and

	\[
	 \|W\|_{\Psi_{\frac{1}{2}}} \leq C K^2_{\phi, \alpha}M^{1-2\alpha}\log^{3}(K+1).
	\]
	 
\end{lemma}

\textbf{Proof of Lemma \ref{lem:norm_bound:T_M-T}}

The definition of $T^M$ and the $L^{\infty}$ triangle inequality give the point-wise bound
\begin{align*}
\Big|\max_{k=1}^{K}\int_{I}\Big(\frac{\sum_{i=1}^{n}f_{k,i}}{\sqrt{n}}\Big)^{2}- T^{M}\Big| 
&= \Big|\max_{k=1}^{K}\int_{I}\Big(\frac{\sum_{i=1}^{n}f_{k,i}}{\sqrt{n}}\Big)^{2}- \max_{k=1}^{K}\sum_{j=1}^{M}\Big(\frac{\sum_{i=1}^{n}\phi^{f_{k,i}}_{j}}{\sqrt{n}}\Big)^{2}\Big|
\\
&\leq \max_{k=1}^{K}\Big|\int_{I}\Big(\frac{\sum_{i=1}^{n}f_{k,i}}{\sqrt{n}}\Big)^{2}-\sum_{j=1}^{M}\Big(\frac{\sum_{i=1}^{n}\phi^{f_{k,i}}_{j}}{\sqrt{n}}\Big)^{2}\Big|.
\end{align*}
Let $K_{f_{k,i}}$ be the covariance kernel of $f_{k,i}$. The identity 

\[
K_{\frac{\sum_{i=1}^{n}f_{k,i}}{\sqrt{n}}} = \frac{1}{n}\sum_{i=1}^{n}K_{f_{k,i}}
\]
holds and so the operator inequality 

\[ 
\mathcal{K}_{\frac{\sum_{i=1}^{n}f_{k,i}}{\sqrt{n}}} = \frac{\sum_{i=1}^{n}\mathcal{K}_{f_{k,i}}}{n}
\]
(see Theorem \ref{mercer} for the definition of $\mathcal{K}_X$ for a stochastic process $X$) is seen to hold by linearity.  

Thus we have the equality in $L^{2}$ (see Theorem \ref{kl-theorem} for the definition of $Z^{X}_{j}$ in terms of $\mathcal{K}_{X}$)
\[
Z^{\frac{\sum_{i=1}^{n}f_{k,i}}{\sqrt{n}}}_{j} =  \frac{\sum_{i=1}^{n}Z^{f_{k,i}}_{j}}{n}.
\]  
 
Hence by linearity, for  each $1 \leq j < \infty$ (see Theorem \ref{kl-theorem} for a definition of the linear map $X \xrightarrow{L_{Z^{X}_{j}}} \phi^{X}_{j}$.),
\[
\phi^{\frac{\sum_{i=1}^{n}f_{k,i}}{\sqrt{n}}}_{j} \overset{a.s}{=} \frac{\sum_{i=1}^{n}\phi^{f_{k,i}}_{j}}{\sqrt{n}}.
\]
By Lemma \ref{kl-integral}, part (i),
\[
\int_{I}\Big(\frac{\sum_{i=1}^{n}f_{k,i}}{\sqrt{n}}\Big)^{2}d\mu-\sum_{j=1}^{M}\Big(\phi^{\frac{\sum_{i=1}^{n}f_{k,i}}{\sqrt{n}}}_{j}\Big)^{2}= \sum_{j=M+1}^{\infty}\Big(\phi^{\frac{\sum_{i=1}^{n}f_{k,i}}{\sqrt{n}}}_{j}\Big)^{2},
\]
and so almost surely,
\[
\int_{I}\Big(\frac{\sum_{i=1}^{n}f_{k,i}}{\sqrt{n}}\Big)^{2}d\mu-\sum_{j=1}^{M}\Big(\frac{\sum_{i=1}^{n}\phi^{f_{k,i}}_{j}}{\sqrt{n}}\Big)^{2} = \sum_{j=M+1}^{\infty}\Big(\frac{\sum_{i=1}^{n}\phi^{f_{k,i}}_{j}}{\sqrt{n}}\Big)^{2}.
\]
We have
\begin{align*}
&\Big\|\int_{I}\Big(\frac{\sum_{i=1}^{n}f_{k,i}}{\sqrt{n}}\Big)^{2}d\mu-\sum_{j=1}^{M}\Big(\frac{\sum_{i=1}^{n}\phi^{f_{k,i}}_{j}}{\sqrt{n}}\Big)^{2}\Big\|_{\Psi_{\frac{1}{2}}} \leq C_{\Psi_{\frac{1}{2}}}\sum_{j=M+1}^{\infty}\Big\|\Big(\frac{\sum_{i=1}^{n}\phi^{f_{k,i}}_{j}}{\sqrt{n}}\Big)^{2}\Big\|_{\Psi_{\frac{1}{2}}}.
\\
&\leq K_{\frac{1}{2}}C_{\Psi_{\frac{1}{2}}}\sum_{j=M+1}^{\infty}\max_{i=1}^{n}\Big\|(\phi^{f_{k,i}}_{j})^{2}\Big\|_{\Psi_{\frac{1}{2}}} \leq  K_{p}C_{\Psi_{\frac{1}{2}}}\sum_{j=M+1}^{\infty}\max_{i=1}^{n}\|\phi^{f_{k,i}}_{j}\|_{\Psi_{1}}^{2}&
\\
&\leq K_{\frac{1}{2}}C_{\Psi_{\frac{1}{2}}}C_{\phi}^{2}\sum_{j=M+1}^{\infty}j^{-2\alpha}\leq  K_{\frac{1}{2}}C_{\Psi_{\frac{1}{2}}}C_{\phi}^{2}\frac{M^{1-2\alpha}}{2\alpha-1},&
\end{align*}
where the first inequality is by 
Lemma \ref{kl-integral}, part (ii), the second by  Lemma \ref{lim_max_sum} (iii), the third by Lemma \ref{relationship_norms}, the fourth by condition (A2) of Assumptions \ref{assumptions}, and the last by the first statement of Lemma \ref{lem:tail_sum}. Then by Lemma \ref{lim_max_sum} (ii), recalling the first inequality in the proof,
\[
\Big\|\max_{k=1}^{K}\int_{I}\Big(\frac{\sum_{i=1}^{n}f_{k,i}}{\sqrt{n}}\Big)^{2}d\mu- T^{M}\Big\|_{\Psi_{\frac{1}{2}}} 
\leq 
K_{\Psi_{1}}^2 K_{\frac{1}{2}}C_{\Psi_{\frac{1}{2}}}\log^{2}(K+1)C_{\phi}^{2}\frac{M^{1-2\alpha}}{2\alpha-1}.
\]
Set $C= K_{\frac{1}{2}} C_{\Psi_{\frac{1}{2}}} K_{\Psi_{1}}^2$. This completes the proof of the first bound in the Lemma. 

The second bound is proved similarly after observing that
\[
\int_{I}\Big(\frac{\sum_{i=1}^{n}f_{k,i}}{\sqrt{n}}\Big)^{2}d\mu 
= \sum_{j=1}^{\infty}\Big(\phi^{\frac{\sum_{i=1}^{n}f_{k,i}}{\sqrt{n}}}_{j}\Big)^{2}.
\] 
Now proceed similarly to the proof of the first statement of the Lemma, using the second statement of Lemma \ref{lem:tail_sum} to obtain
\begin{equation}\label{eq_for_application_of_Lemma_33}
\Bigg \|\sum_{j=1}^{\infty}\Big(\phi^{\frac{\sum_{i=1}^{n}f_{k,i}}{\sqrt{n}}}_{j}\Big)^{2} \Bigg\|_{\psi_1} \leq (K_{\frac12}C_{\Psi_{\frac12}})^{1/2}C_{\phi}\sqrt{\frac{2\alpha}{2\alpha-1}}.
\end{equation}
The second statement in the Lemma then follows from an application of Lemma~ \ref{relationship_norms} (i) and Lemma~\ref{lim_max_sum} (ii).
\hfill $\Box$

\bigskip 
\noindent
\textbf{Proof of Lemma \ref{lem:smooth_norm_bound}}

Define $S_{k}(x) = \frac{\sum_{i=1}^{n}f_{k,i}(x)}{\sqrt{n}}$ to lighten notation. By definition of  $T^d$ and the $L^{\infty}$ triangle inequality,
\begin{align*}
&\Big|\max_{k=1}^{K}\int_{I}\Big(\frac{\sum_{i=1}^{n}f_{k,i}}{\sqrt{n}}\Big)^{2}- T^d\Big| = \Big|\max_{k=1}^{K}\sum_{\ell=1}^{T}\int_{t_{\ell-1}}^{t_{\ell}}S_{k}^{2}(x)dx -\max_{k=1}^{K}\sum_{\ell=1}^{T}S_{k}(t_{\ell})^{2}\Big|
		\\
		&\leq \max_{k=1}^{K}\Big|\sum_{\ell=1}^{T}\int_{t_{\ell-1}}^{t_{\ell}}S_{k}^{2}(x) dx -\sum_{\ell=1}^{T}S_{k}(t_{\ell})^{2}\Big|&
		\\
		&\leq \max_{k=1}^{K}\sum_{\ell=1}^{T}\int_{t_{\ell-1}}^{t_{\ell}}|S_{k}(x) -S_{k}(t_{\ell})||S_{k}(x) +S_{k}(t_{\ell})|dx,&
	\end{align*}
For each $\ell$ and $k$, and each $x \in [t_{\ell-1},t_\ell]$ 
\begin{align*}
&\Big\||S_{k}(x) -S_{k}(t_{\ell})||S_{k}(x) +S_{k}(t_{\ell})|\Big\|_{\tilde \Psi_{\frac{1}{2}}}&
\\
&\leq \Big\||S_{k}(x) -S_{k}(t_{\ell})||S_{k}(x) +S_{k}(t_{\ell})|\Big\|_{\Psi_{\frac{1}{2}}}
		\\
		&\leq \|S_{k}(x) -S_{k}(t_{\ell})\|_{\Psi_{1}}\|S_{k}(x) +S_{k}(t_{\ell})\|_{\Psi_{1}}\tag*{Lemma \ref{relationship_norms}}&
		\\
		&\leq K_1^2 \Big(\max_{i=1}^{n}\|f_{k,i}(x) -f_{k,i}(t_{\ell})\|_{\Psi_{1}}\Big)\Big(\max_{i=1}^{n}\|f_{k,i}(x) +f_{k,i}(t_{\ell})\|_{\Psi_{1}}\Big)\tag*{Lemma \ref{lim_max_sum} (iii)}&
		\\
		&\leq K_1^2C_{H}T^{-q} \max_{i=1}^{n}\|f_{k,i}(x) +f_{k,i}(t_{\ell})\|_{\Psi_{1}}\tag*{Assumption (A1)}&
		\\
		&\leq 2K_1^2C_{H}T^{-q}\max_{i=1}^{n}\Big\|\sup_{s \in [t_{\ell-1},t_{\ell}]} f_{k,i}(s) \Big\|_{\Psi_{1}}\tag*{Triangle ineq. and monotonicity}&
		\\
		&\leq 2K_1^2C_{H}T^{-(q+1)}\max_{i=1}^{n}\Big\|\sup_{s \in [0,1]} f_{k,i}(s) \Big\|_{\Psi_{1}}\tag*{Supremum over larger set}&
		\\
		&\leq CK_1^2 K_{*}C_{H}T^{-q} \Big(\frac{C_{H}}{q}+ C_{\phi}\sqrt{\frac{2\alpha}{2\alpha-1}}\Big)\tag*{Lemma \ref{lem:sup_process}}
	\end{align*}
for a universal constant $C$. The use of Lemma~\ref{lem:sup_process} is justified by \eqref{eq_for_application_of_Lemma_33}. Thus we have
\begin{align*}
		&\Big\|\int_{t_{\ell-1}}^{t_{\ell}}|S_{k}(x) -S_{k}(t_{\ell})||S_{k}(x) +S_{k}(t_{\ell})|dx\Big\|_{\Psi_{\frac{1}{2}}}&
		\\
		&\leq C_{\Psi_{\frac{1}{2}}}\Big\|\int_{t_{\ell-1}}^{t_{\ell}}|S_{k}(x) -S_{k}(t_{\ell})||S_{k}(x) +S_{k}(t_{\ell})|dx\Big\|_{\tilde \Psi_{\frac{1}{2}}}\tag*{Lemma \ref{orlicz-equiv}}&
		\\
		&\leq C_{\Psi_{\frac{1}{2}}}\int_{t_{\ell-1}}^{t_{\ell}}2K_1^2K_*C_{H}T^{-q} \Big(\frac{C_{H}}{q}+ C_{\phi}\sqrt{\frac{2\alpha}{2\alpha-1}}\Big) dx\tag*{bound above and Lemma \ref{lem:riemann_sample_paths}}&
		\\
		&\leq C_{T^{d}}T^{-q-1}\Big(q^{-1}C_H^2+C_HC_{\phi}\sqrt{\frac{2\alpha}{2\alpha-1}}\Big)&
%		\\
	\end{align*}
	where $C_{T^{d}}$ is a universal constant. We may now apply the triangle inequality up to the constant $C_{\Psi_{\frac{1}{2}}}$ to yield, for each $k$, 
	
\begin{equation}
\Big\|\sum_{\ell=1}^{T}\int_{t_{\ell-1}}^{t_{\ell}}S_{k}^{2}(x) dx -\sum_{\ell=1}^{T}S_{k}(t_{\ell})^{2}\Big\|_{\Psi_{\frac{1}{2}}}\leq C_{\Psi_{\frac{1}{2}}}C_{T^{d}}T^{-q}\Big(q^{-1}C_H^2+C_HC_{\phi}\sqrt{\frac{2\alpha}{2\alpha-1}}\Big).
\end{equation}
Thus by Lemma \ref{lim_max_sum} part (ii)
\begin{flalign*}
&\Big\|\max_{k=1}^{K}\int_{I}\Big(\frac{\sum_{i=1}^{n}f_{k,i}}{\sqrt{n}}\Big)^{2}d\mu- T^d\Big\|_{\Psi_{\frac{1}{2}}} \leq K_{\Psi_{1}}^{2}\log^2(K+1)C_{\Psi_{\frac{1}{2}}}C_{T^{d}}T^{-q}\Big(q^{-1}C_H^2+C_HC_{\phi}\sqrt{\frac{2\alpha}{2\alpha-1}}\Big).&
\end{flalign*}
Setting $C$ equal to $K_{\Psi_{1}}^{2}C_{T^{d}}C_{\Psi_{\frac{1}{2}}}$ finishes the proof. \hfill $\Box$

\bigskip

\noindent
\textbf{Proof of Lemma \ref{lem:boot_smooth_norm_bound}}

Recalling the definition of $T^{d,e}$, we need to bound
	\begin{align*}
		 &\Big|\max_{k=1}^{K}\int_{0}^{1}\Big(\frac{\sum_{i=1}^{n}e_i[f_{k,i}(x)-\bar f_{k,\cdot}(x)]}{\sqrt{n}}\Big)^{2}dx - \max_{k=1}^{K}\frac{1}{T}\sum_{\ell=1}^{T}\Big(\frac{\sum_{i=1}^{n}e_i[f_{k,i}(t_{\ell})-\bar f_{k,\cdot}(t_{\ell})]}{\sqrt{n}}\Big)^2\Big|&\\
		&\leq \max_{k=1}^{K}\Big|\int_{0}^{1}\Big(\frac{\sum_{i=1}^{n}e_i[f_{k,i}(x)-\bar f_{k,\cdot}(x)]}{\sqrt{n}}\Big)^{2}dx - \frac{1}{T}\sum_{\ell=1}^{T}\Big(\frac{\sum_{i=1}^{n}e_i[f_{k,i}(t_{\ell})-\bar f_{k,\cdot}(t_{\ell})]}{\sqrt{n}}\Big)^2\Big| &\\
		&\leq \max_{k=1}^{K}\Big|\sum_{\ell=1}^{T}\Big\{\int_{t_{\ell-1}}^{t_{\ell}}\Big(\frac{\sum_{i=1}^{n}e_i[f_{k,i}(x)-\bar f_{k,\cdot}(x)]}{\sqrt{n}}\Big)^{2}dx - \frac{1}{T}\Big(\frac{\sum_{i=1}^{n}e_i[f_{k,i}(t_{\ell})-\bar f_{k,\cdot}(t_{\ell})]}{\sqrt{n}}\Big)^2\Big\}\Big| &\\
		&\leq \frac{1}{n}\max_{k=1}^{K}\Big|\sum_{\ell=1}^{T}\int_{t_{\ell-1}}^{t_{\ell}}\Big\{\Big(\sum_{i=1}^{n}e_i[f_{k,i}(x)-\bar f_{k,\cdot}(x)]\Big)^{2} - \Big(\sum_{i=1}^{n}e_i[f_{k,i}(t_{\ell})-\bar f_{k,\cdot}(t_{\ell})]\Big)^2\Big\}dx\Big| \\
        & =:g(f,e)
	\end{align*}
    where $f$ stands for the collection $(f_{k,i})_{i=1,\dots,n, k=1,\dots,K}$ and $e \in \R^n$ has entries $e_i$.
We introduce the following notation to decrease the typographical length of the remaining argument. Let $A_{\ell,f_{k}}(x) \in \mathbb{R}^{n}$, $B_{\ell,f_{k}}(x) \in \mathbb{R}^{n}$ be defined coordinate wise for $x \in [t_{\ell-1}, t_{\ell}]$ as follows:
	\begin{align*}
		&[A_{\ell,f_{k}}]_{i}(x) = [(f_{k,i}(x)-f_{k,i}(t_{\ell}))-(\bar f_{k,\cdot}(x)-\bar f_{k,\cdot}(t_{\ell}))].&\\
		&[B_{\ell,f_{k}}]_{i}(x) = [(f_{k,i}(x)+f_{k,i}(t_{\ell}))-(\bar f_{k,\cdot}(x)+\bar f_{k,\cdot}(t_{\ell}))]. &\\
	\end{align*}
	Applying the difference of squares formula now gives:
\begin{align*}&\Big|\max_{k=1}^{K}\int_{I}\Big(\frac{\sum_{i=1}^{n}e_{i}[f_{k,i}(x)-\bar{f}_{k,\cdot}(x)]}{\sqrt{n}}\Big)^{2}dx - T^{d,e}\Big|\leq  \frac{1}{n}\max_{k=1}^{K}\Big|\sum_{\ell=1}^{T}\int_{t_{\ell-1}}^{t_{\ell}}\langle e, A_{\ell,f_{k}}(x) \rangle\langle e, B_{\ell,f_{k}}(x) \rangle dx\Big|&
\end{align*}
Hence it suffices to estimate $\|\sum_{\ell=1}^{T}\int_{t_{\ell-1}}^{t_{\ell}}\langle e, A_{\ell,f_{k}}(x) \rangle\langle e, B_{\ell,f_{k}}(x) \rangle dx\|_{\Psi_{1}|\sigma(f)}$, using monotonicity of the conditional norm. We will now use Definition \ref{def:condorlicz} with $X = f, Y = e$ and $g$ as defined earlier in this proof. Note that  the definition applies in our setting because the functions $f_{k,i}$ are independent of the Gaussian variables $e$.
\begin{align*}
		&\Big[\Big\|\sum_{\ell=1}^{T}\int_{t_{\ell-1}}^{t_{\ell}}\langle e, A_{\ell,f_{k}}(x) \rangle\langle e,B_{\ell,f_{k}}(x) \rangle dx\Big\|_{\Psi_{1}|\sigma(f)}\Big](\omega)
		\\ &\overset{a.s}{=}\Big\|\sum_{\ell=1}^{T}\int_{t_{\ell-1}}^{t_{\ell}}\langle e, A_{\ell,f^{\omega}_{k}}(x) \rangle\langle e,B_{\ell,f^{\omega}_{k}}(x) \rangle dx\Big\|_{\Psi_{1}}\\
		&\leq\sum_{\ell=1}^{T}\Big\|\int_{t_{\ell-1}}^{t_{\ell}}\langle e, A_{\ell,f^{\omega}_{k}}(x) \rangle\langle e,B_{\ell,f^{\omega}_{k}}(x) \rangle dx\Big\|_{\Psi_{1}}&
	\end{align*}
where the notation $f^{\omega}_{k}$ refers to a fixed realization of the random functions $\{f_{k,i}\}_{i=1,k=1}^{n,K}$. This implies that for any $\omega$ the variables  $\langle e, A_{\ell,f^{\omega}_{k}}(x) \rangle$ and $\langle e,B_{\ell,f^{\omega}_{k}}(x) \rangle$ are Gaussian random variables with standard deviations $\|A_{\ell,f^{\omega}_{k}}(x)\|_{2}$ and $\|B_{\ell,f^{\omega}_{k}}(x)\|_{2}$ respectively. Elementary computations show that if $Z \sim N(0, \sigma^{2})$, then $\|Z\|_{\psi_{2}} = \sqrt{\frac{8}{3}}\sigma$. Thus,
\begin{align*}
&\Big\|\langle e, A_{\ell,f^{\omega}_{k}}(x) \rangle\langle e,B_{\ell,f^{\omega}_{k}}(x) \rangle \Big\|_{\Psi_{1}}
\\
&\leq \Big\|\langle e, A_{\ell,f^{\omega}_{k}}(x)\rangle \Big\|_{\Psi_{2}}\Big\|\langle e,B_{\ell,f^{\omega}_{k}}(x) \rangle \Big\|_{\Psi_{2}}
\\
&\leq \frac{8}{3} \|A_{\ell,f^{\omega}_{k}}(x)\|_{2}\|B_{\ell,f^{\omega}_{k}}(x)\|_{2}.
\end{align*}
By continuity of each $f^{\omega}_{k,i}$, the above is a continuous function of $x$, thus Riemann integrable. An application of Lemma~\ref{lem:riemann_sample_paths} yields 
\[
\sum_{\ell=1}^{T}\Big\|\int_{t_{\ell-1}}^{t_{\ell}}\langle e, A_{\ell,f^{\omega}_{k}}(x) \rangle\langle e,B_{\ell,f^{\omega}_{k}}(x) \rangle dx\Big\|_{\Psi_{1}}
\leq 
\frac{8}{3}\sum_{\ell=1}^{T}\int_{t_{\ell-1}}^{t_{\ell}} \|A_{\ell,f^{\omega}_{k}}(x)\|_{2}\|B_{\ell,f^{\omega}_{k}}(x)\|_{2} dx
\]
and so
\begin{align*}
&\Big\|\max_{k=1}^{K}\int_{I}\Big(\frac{\sum_{i=1}^{n}e_{i}[f_{k,i}-\bar{f}_{k,\cdot}]}{\sqrt{n}}\Big)^{2} - T^{d,e}\Big\|_{\Psi_{1}|\sigma(f)}
\\
& \leq \frac{8}{3}K_{\Psi_{1}}\frac{\log(K+1)}{n}\max_{k=1}^{K}\sum_{\ell=1}^{T}\int_{t_{\ell-1}}^{t_{\ell}}\|A_{\ell,f_{k}}(x)\|_{2}\|B_{\ell,f_{k}}(x)\|_{2} dx.\\
\end{align*}

We define $Y$ to be equal to the right-hand-side of this inequality, i.e.
\begin{align*}\label{sigma_1}
		&Y :=  \frac{8}{3} K_{\Psi_{1}}\frac{\log(K+1)}{n}\max_{k=1}^{K}\sum_{\ell=1}^{T}\int_{t_{\ell-1}}^{t_{\ell}}\|A_{\ell,f_{k}}(x)\|_{2}\|B_{\ell,f_{k}}(x)\|_{2} dx,&\\
	\end{align*}
	and proceed to compute the $\Psi_{\frac{1}{2}}$ norm of $Y$. To this end we first bound the $\tilde{\Psi}_{1/2}$ norm of the integrand. By Lemma \ref{relationship_norms} (ii) applied for $p=1/2$,
\[
\Big\|\|A_{\ell,f_{k}}(x)\|_{2}\|B_{\ell,f_{k}}(x)\|_{2} \Big\|_{\tilde\Psi_{1/2}} \leq \Big\|\|A_{\ell,f_{k}}(x)\|_{2}\|B_{\ell,f_{k}}(x)\|_{2} \Big\|_{\Psi_{1/2}} \leq \Big\|\|A_{\ell,f_{k}}(x)\|_{2}\Big\|_{\Psi_{1}}\Big\|\|B_{\ell,f_{k}}(x)\|_{2}\Big\|_{\Psi_{1}}.
\]

We will now estimate $\|\|A_{\ell,f_{k}}(x)\|_{2}\|_{\Psi_{1}}$ and $\|\|B_{\ell,f_{k}}(x)\|_{2}\|_{\Psi_{1}}$ separately. Note that by the Euclidean triangle inequality,

	\begin{align*}
		&\|A_{\ell,f_{k}}(x)\|_{2} \leq \sqrt{\sum_{i=1}^{n}(f_{k,i}(x)-f_{k,i}(t_{\ell}))^2}+\sqrt{\sum_{i=1}^{n}(\bar f_{k,\cdot}(x)-\bar f_{k,\cdot}(t_{\ell}))^2},&\\
	\end{align*}
	and by Lemma~\ref{triangle_inequality} (triangle inequality for the $\Psi_1$ norm),
	
	\begin{align*}
		&\|\|A_{\ell,f_{k}}(x)\|_{2}\|_{\Psi_{1}} \leq \Big\|\sqrt{\sum_{i=1}^{n}(f_{k,i}(x)-f_{k,i}(t_{\ell}))^2}\Big\|_{\Psi_{1}}+\Big\|\sqrt{\sum_{i=1}^{n}(\bar f_{k,\cdot}(x)-\bar f_{k,\cdot}(t_{\ell}))^2}\Big\|_{\Psi_{1}}.&\\
	\end{align*}
	To even further lighten notation, let $D_{k, i, \ell}(x) = f_{k,i}(x)-f_{k,i}(t_{\ell})$ be defined on $[t_{\ell-1}, t_{\ell}]$. To shorten notation we drop the argument $x$ in $D_{k, i, \ell}$ below. We have,
	
	\begin{align*}
		&\Big\|\sqrt{\sum_{i=1}^{n}(f_{k,i}(x)-f_{k,i}(t_{\ell}))^2}\Big\|_{\Psi_{1}} =\Big\|\sqrt{\sum_{i=1}^{n}(D_{k, i, \ell})^2}\Big\|_{\Psi_{1}}  =\sqrt{ \|\sum_{i=1}^{n}(D_{k, i, \ell})^2\|_{\Psi_{\frac{1}{2}}} } \tag*{Lemma \ref{relationship_norms}(ii)}&\\
		&=\sqrt{ \|\sum_{i=1}^{n}[(D_{k, i, \ell})^2-\E[(D_{k, i, \ell})^2]+\E[(D_{k, i, \ell})^2]\|_{\Psi_{\frac{1}{2}}} }&\\
		& \lesssim \sqrt{ \|\sum_{i=1}^{n}[(D_{k, i, \ell})^2-\E[(D_{k, i, \ell})^2]\|_{\Psi_{\frac{1}{2}}} }+\sqrt{ \|\sum_{i=1}^{n}\E[(D_{k, i, \ell})^2]\|_{\Psi_{\frac{1}{2}}} }\tag*{Lemma \ref{triangle_inequality}}&\\
		& \lesssim \sqrt{ \sqrt{n}\max_{i}\|(D_{k, i, \ell})^2-\E[(D_{k, i, \ell})^2]\|_{\Psi_{\frac{1}{2}}} }+\sqrt{n}\max_{i} \sqrt{\E[(D_{k, i, \ell})^2]}\tag*{Lemma \ref{lim_max_sum} (iii) and Lemma \ref{triangle_inequality}.}&\\
		& \lesssim \sqrt{ \sqrt{n}\max_{i}\|(D_{k, i, \ell})^2\|_{\Psi_{\frac{1}{2}}} +\sqrt{n}\max_{i}\E[(D_{k, i, \ell})^2] }+\sqrt{n}\max_{i} \sqrt{\E[(D_{k, i, \ell})^2]}\tag*{ Lemma \ref{triangle_inequality}.}&\\
		& \lesssim n^{\frac{1}{4}}\sqrt{\max_{i}\|(D_{k, i, \ell})^2\|_{\Psi_{\frac{1}{2}}}} +\sqrt{n}\max_{i} \sqrt{\E[(D_{k, i, \ell})^2]}\tag*{sub-additivity of square root} &\\
		& \lesssim n^{\frac{1}{4}}\max_{i}\|(D_{k, i, \ell})\|_{\Psi_{1}} +\sqrt{n}\max_{i}\|(D_{k, i, \ell})\|_{\Psi_{1}}\tag*{ Lemma \ref{relationship_norms} (i) and (ii).} &\\
		& \lesssim \sqrt{n}C_{H}T^{-q},&\tag*{(A1) of Assumption~\ref{assumptions}}\\
	\end{align*}
	where all bounds are uniform
    in $x \in [t_{\ell-1}, t_{\ell}]$ and all constants are universal. 
    The second piece in the bound on $\|\|A_{\ell,f_{k}}(x)\|_{2}\|_{\Psi_{1}} $ is of lower order:
\begin{align*}
		\Big\|\sqrt{\sum_{i=1}^{n}(\bar f_{k,\cdot}(x)-\bar f_{k,\cdot}(t_{\ell}))^2}\Big\|_{\Psi_{1}} &=\sqrt{n}\|\bar f_{k,\cdot}(x)-\bar f_{k,\cdot}(t_{\ell})\|_{\Psi_{1}}=\frac{1}{\sqrt{n}}\Big\|\sum_{i=1}^{n}D_{k,i,\ell}\Big\|_{\Psi_{1}}&
		\\
		&\lesssim \max_{i}\|D_{k,i,\ell}\|_{\Psi_{1}} \leq C_{H}T^{-q},&
	\end{align*}
	where the first inequality follows from Lemma \ref{lim_max_sum} (iii) and the final bound from Assumption (A1). Therefore by the triangle inequality,
	\[
	\|\|A_{\ell,f_{k}}(x)\|_{2}\|_{\Psi_{1}} \lesssim  \sqrt{n}C_{H}T^{-q}.
	\]	
	Moving to the next factor note that
	\begin{align*}
		&\|B_{\ell,f_{k}}(x)\|_{2} \leq \sqrt{\sum_{i=1}^{n}(f_{k,i}(x)+f_{k,i}(t_{\ell}))^2}+\sqrt{\sum_{i=1}^{n}(\bar f_{k,\cdot}(x)+\bar f_{k,\cdot}(t_{\ell}))^2}.&
	\end{align*}
	
	We will estimate the norm of the first piece in this bound; the second has lower order as can be seen by an argument similar to the one made when bounding the second piece in the upper bound to $\|\|A_{\ell,f_{k}}(x)\|_{2}\|_{\Psi_{1}}$. Define $S_{k,i,\ell}(x) = f_{k,i}(x)+f_{k,i}(t_{\ell})$. By similar arguments as in the analysis of $A$, 
	\begin{align*}
		&\Big\|\sqrt{\sum_{i=1}^{n}(f_{k,i}(x)+f_{k,i}(t_{\ell}))^2}\Big\|_{\Psi_{1}} = \Big\|\sqrt{\sum_{i=1}^{n}S_{k,i,\ell}^2(x)}\Big\|_{\Psi_{1}}
		\lesssim \sqrt{n}\max_{i}\|S_{k,i,\ell}\|_{\Psi_{1}},&
	\end{align*}
	where the last inequality follows from an argument similar to the one made when bounding the first piece of the upper bound to $\|\|A_{\ell,f_{k}}(x)\|_{2}\|_{\Psi_{1}}$. Finally, for each $i$, by Lemma \ref{lem:sup_process},
		\begin{align*}
		&\|S_{k,i,\ell}\|_{\Psi_{1}} \lesssim \|\sup_{s \in [0,1]}f_{k,i}(s)\|_{\Psi_{1}} \lesssim C_{\phi}\sqrt{\frac{2\alpha}{2\alpha-1}}+\frac{C_{H}}{q}.&
	\end{align*}
Thus we have proved, uniformly in $x \in [t_{\ell-1},t_\ell]$
\[
\Big\|\|A_{\ell,f_{k}}(x)\|_{2}\|B_{\ell,f_{k}}(x)\|_{2} \Big\|_{\Psi_{\frac{1}{2}}}
\lesssim n T^{-q}C_H\Big(C_{\phi}\sqrt{\frac{2\alpha}{2\alpha-1}}+\frac{C_{H}}{q}\Big).
\]
This upper bound is a constant and thus Riemann integrable function of $x$.
Thus
\begin{align*}
		&\Big\|\sum_{\ell=1}^{T}\int_{t_{\ell-1}}^{t_{\ell}}\|A_{\ell,f_{k}}(x)\|_{2}\|B_{\ell,f_{k}}(x)\|_{2} dx\Big\|_{\Psi_{\frac{1}{2}}} &\\
		&\leq C_{\Psi_{\frac{1}{2}}}\sum_{\ell=1}^{T}\Big\|\int_{t_{\ell-1}}^{t_{\ell}}\|A_{\ell,f_{k}}(x)\|_{2}\|B_{\ell,f_{k}}(x)\|_{2} dx\Big\|_{\Psi_{\frac{1}{2}}}&\\
		&\lesssim \sum_{\ell=1}^{T}\int_{t_{\ell-1}}^{t_{\ell}}n T^{-q}C_H\Big(C_{\phi}\sqrt{\frac{2\alpha}{2\alpha-1}}+\frac{C_{H}}{q}\Big)dx&\tag{Lemma~\ref{lem:riemann_sample_paths}}\\
		&= n T^{-q}C_H\Big(C_{\phi}\sqrt{\frac{2\alpha}{2\alpha-1}}+\frac{C_{H}}{q}\Big).
	\end{align*}
    Thus, by the above bound and Lemma~\ref{lim_max_sum} part (ii)
	\begin{align*}
		\|Y \|_{\Psi_{\frac{1}{2}}} &=  \Big\|\frac{8}{3}K_{\Psi_{1}}\frac{\log(K+1)}{n}\max_{k=1}^{K}\sum_{\ell=1}^{T}\int_{t_{\ell-1}}^{t_{\ell}}\|A_{\ell,f_{k}}(x)\|_{2}\|B_{\ell,f_{k}}(x)\|_{2} dx\Big\|_{\Psi_{\frac{1}{2}}}&
		\\
		&\leq CT^{-q}\log^{3}(K+1)C_H\Big(C_{\phi}\sqrt{\frac{2\alpha}{2\alpha-1}}+\frac{C_{H}}{q}\Big).&
	\end{align*}
	for a sufficiently large universal constant $C$.
\hfill $\Box$

\bigskip

\noindent
\textbf{Proof of Lemma \ref{lem: gaussian_tail_bound}}

Define $T^{e,1}$ and $S^{e}_{k}$ as follows:
	
	\[
	T^{e,1}= \max_{k}\int_{I}\Big(\frac{\sum_{i=1}^{n}e_{i}[f_{k,i}-\bar{f_{k,\cdot}}]}{\sqrt{n}}\Big)^{2}dx,
	\]
	and 
	\[
	S^{e}_{k}= \frac{\sum_{i=1}^{n}e_{i}[f_{k,i}-\bar{f_{k,\cdot}}]}{\sqrt{n}}.
	\]
An elementary calculation shows that the process $S^{e}_{k}$ admits Karhunen–Loève basis vectors $Z_{j}^{S^{e}_{k}}$ equal to $Z^{f_{1,k}}_{j}$. Then, 
	\begin{align*}
		&\phi^{S^{e}_{k}}_{j} = \int_{I}Z^{S^{e}_{k}}_{j}(t)S^{e}_{k}dt = \frac{1}{\sqrt{n}}\sum_{i=1}^{n}\int_{I}Z^{f_{1,k}}_{j}(t)e_i[f_{k,i}(t)-\bar f_{k,\cdot}]dt = \frac{\sum_{i=1}^{n}e_{i}[\phi^{f_{k,i}}_{j}-\bar\phi^{f_{k,\cdot}}_{j}]}{\sqrt{n}} &
	\end{align*}
	where in the last equality we used the definition of each $\phi^{f_{k,i}}$ and linearity of the integral with respect to the sum on $i$. We are now ready to compute the norm bounds indicated in the statement of the lemma.  Note that
\begin{align*}
		|T^{e} - T^{M,e}| &=\Big|\max_{k}\int_I\Big(\frac{\sum_{i=1}^{n}e_i [f_{k,i}-\bar f_{k,\cdot}]}{\sqrt{n}}\Big)^2dx - \max_{k}\sum_{j=1}^{M}\Big(\sum_{i=1}^{n}\frac{e_i [\phi_j^{f_{k,i}} - \bar\phi_j^{f_{k,\cdot}}]}{\sqrt{n}}\Big)^2\Big|&
		\\
		&\leq \max_{k}\Big|\int_I\Big(\frac{\sum_{i=1}^{n}e_i [f_{k,i}-\bar f_{k,\cdot}]}{\sqrt{n}}\Big)^2dx - \sum_{j=1}^{M}\Big(\frac{\sum_{i=1}^{n}e_i [\phi_j^{f_{k,i}} - \bar\phi_j^{f_{k,\cdot}}]}{\sqrt{n}}\Big)^2\Big|&
		\\
		&= \max_{k}\sum_{j=M+1}^{\infty}\Big(\frac{\sum_{i=1}^{n}e_i [\phi_j^{f_{k,i}} - \bar\phi_j^{f_{k,\cdot}}]}{\sqrt{n}}\Big)^2&
        \\
        & =: g(f,e)  &
	\end{align*}
	where the second to last equality follows from Lemma \ref{kl-integral}, while the inequality follows from the properties of the $\max$ function. We will now bound $\|T^{e} - T^{M,e}\|_{\Psi_{1}|\sigma(f)}$.
 Recall that (see Definition \ref{def:condorlicz}), 
 
 \begin{align*}
    \big[\|T^{e} - T^{M,e}\|_{\Psi_{1}|\sigma(f)}\big](\omega) =\Bigg\|\max_{k}\sum_{j=M+1}^{\infty}\Big(\frac{\sum_{i=1}^{n}e_i [\phi_j^{f^{\omega}_{k,i}} - \bar\phi_j^{f^{\omega}_{k,\cdot}}]}{\sqrt{n}}\Big)^2\Bigg\|_{\Psi_1},
    \end{align*}
where $\phi_{j}^{f^{\omega}_{k,i}} := \int_I f_{k,i}(t;\omega) Z_j^{f_{1,k}}(t)dt$ refer to the $j^{th}$ Karhuen-Loeve score of a fixed realization of the $f_{k,i}$. Let 
	\[
	U_{jk}(\omega):= \Big[(1/n)\sum_{i=1}^{n}(\phi_j^{f^{\omega}_{k,i}} - \bar \phi_j^{f^{\omega}_{k,\cdot}})^{2} \Big]^{1/2}.
	\]
For appropriate dependent standard normal variables $Z_{jk}$ which are independent of $\sigma(f)$ and therefore of $\sigma(\cup_{k=1}^{K}\cup_{j=M+1}^{\infty}U_{jk})$, since the $U_{jk}$ are derived from the $f_{k,i}$ deterministically, 

\begin{equation}
\big[\|T^{e} - T^{M,e}\|_{\Psi_{1}|\sigma(f)}\big](\omega)  = \Big\|\max_{k=1}^{K}\sum_{j=M+1}^{\infty} (U_{jk}(\omega) Z_{jk})^{2}\Big\|_{\psi_1}
\end{equation}

Recall that if $Z$ is standard normal, then $\|Z\|_{\psi_{2}}$ exists and $\|Z^{2}\|_{\Psi_{1}} = (\|Z\|_{\psi_{2}})^{2} = 8 \Var(Z)/3$. The first inequality follows from Lemma \ref{relationship_norms} (i) and the second from a standard computation. It follows that from the above derivations, for a universal constant $C$, we have a.s.
	\[
	\Big\|\sum_{j=M+1}^\infty \Big( \frac{1}{\sqrt{n}} \sum_{i=1}^n e_i (\phi_j^{f_{k,i}} - \bar \phi_j^{f_{k,\cdot}})  \Big)^2\Big\|_{\Psi_{1}|\sigma(f)}(\omega)
	\leq \sum_{j=M+1}^{\infty} (C/n)\sum_{i=1}^{n}(\phi_j^{f_{k,i}^\omega}  - \bar \phi_j^{f_{k,\cdot}^\omega})^{2}.
	\]
	Thus, for a universal constant $C$, by Lemma \ref{lim_max_sum}(ii)
	\begin{equation}\label{sigma_2}
		\|T^{e} - T^{e,M}\|_{\Psi_{1}|\sigma(f)} \leq \frac{C}{n} \log(K+1)  \max_{k=1}^K \sum_{j=M+1}^{\infty} \sum_{i=1}^{n}(\phi_j^{f_{k,i}} - \bar \phi_j^{f_{k,\cdot}})^{2} \quad a.s.
	\end{equation}
	Thus for a universal constant $C$ whose values differ from one to line, recalling the definition of $\|\cdot\|_{\tilde{\psi}_{\frac{1}{2}}}$ from Lemma \ref{orlicz-equiv},
	\begin{align*}
		\Big\|\sum_{i=1}^{n}(\phi_j^{f_{k,i}}  -\bar \phi_j^{f_{k,\cdot}})^{2}\Big\|_{\tilde \Psi_{\frac{1}{2}}} 
		&= \Big\| \Big[\sum_{i=1}^{n}(\phi_j^{f_{k,i}})^2\Big] - n(\bar \phi_j^{f_{k,\cdot}})^{2}\Big\|_{\tilde \Psi_{\frac{1}{2}}}
		\\
		&\leq n\max_{i=1}^n \|(\phi_j^{f_{k,i}})^2\|_{\tilde \Psi_{\frac{1}{2}}}  + \Big\|(\sqrt{n}\bar \phi_j^{f_{k,\cdot}})^{2}\Big\|_{\tilde \Psi_{\frac{1}{2}}} \tag{Lemma~ \ref{triangle_inequality}}
		\\
		&\leq Cn\max_{i=1}^n \|\phi_j^{f_{k,i}}\|_{\Psi_{1}}^2 +C \|\sqrt{n}\bar \phi_j^{f_{k,\cdot}}\|_{\Psi_{1}}^2 \tag{Lemma~\ref{orlicz-equiv},  Lemma~\ref{relationship_norms}(i)}
		\\
		&\leq Cn\max_{i=1}^n \|\phi_j^{f_{k,i}}\|_{\Psi_{1}}^2 + C \max_{i=1}^n \|\phi_j^{f_{k,i}}\|_{\Psi_{1}}^2 \tag{Lemma~\ref{lim_max_sum}(iii)}
		\\
		&\leq n C \max_{i=1}^n \|\phi_j^{f_{k,i}}\|_{\Psi_{1}}^2,
	\end{align*}
 where the value of the constant $C$ changed from line to line. Therefore, we find by Lemma \ref{orlicz-equiv}
	\[
	\Big\| \|T^{e} -T^{e,M}\|_{\Psi_{1}|\sigma(f)}\Big\|_{\tilde \Psi_{\frac{1}{2}}} \leq C \log^3 (K+1) \max_{k=1}^K \sum_{j=M+1}^\infty \max_{i=1}^n \|\phi_i^{k,j}\|_{\Psi_{1}}^2.
	\]
	Let $R_2(M) := \sum_{j=M+1}^\infty C_{\phi}^{2}j^{-2\alpha}$. By Assumption (A2) stated in Assumption~\ref{assumptions},
	\[
	\Big\| \|T^{e}- T^{e,M}\|_{\Psi_{1}|\sigma(f)}\Big\|_{ \Psi_{\frac{1}{2}}} \leq  C \log^3 (K+1) R_2(M) \lesssim \frac{C_{\phi}^{2}}{2\alpha-1}\log^3 (K+1) M^{1-2\alpha}.
	\]
	\hfill $\Box$

\subsubsection{Results related to stochastic processes}
The random elements defined in \ref{assumptions} are also measurable stochastic processes:
\begin{lemma}\label{measurability}
	Suppose $f$ is a Borel measurable random element of the space $(C[I])^K$, defined on a measure space $\Omega$. Then $X(\omega, t):=f(\omega)(t)$ is a  measurable $K$-dimensional stochastic process defined on $\Omega \times I$.
\end{lemma}
The following result must be known to the community, but we were unable to find a reference and so provide a proof for completeness.

\begin{lemma}\label{lem:gaussian_process}
	Let $f = \oplus_{k=1}^{K}f_k$ be a $K$ dimensional stochastic process with index set $I$. Suppose for all $1 \leq k \leq K$ the real valued process $f_k$ is centered, second order, and mean-square continuous, so that it satisfies the hypotheses of the Karhunen–Loève Theorem (Theorem \ref{kl-theorem}). Then,

	\begin{enumerate}
		\item There exists a $K$ dimensional Gaussian process $G$ with real valued marginals $G_{1}, \dots, G_{K}$ such that for all $1 \leq k_{1} \leq k_{2} \leq K$ and all $t_{1}, t_{2}\in I$, 
		
		\[\Cov\Big(G_{k_{1}}(t_1), G_{k_{2}}(t_2)\Big) = \Cov \Big(f_{k_{1}}(t_1), f_{k_{2}}(t_2)\Big).\]
		\item For $1 \leq k \leq K$ let $\{\zeta^{k}_1,\zeta^{k}_2, \dots \}$ be the Karhunen–Loève scores of $G_k$, and let $\{\phi^{f_{k}}_1,\phi^{f_{k}}_2, \dots \}$ be the Karhunen–Loève scores of $f_k$. For any finite collection of natural numbers $\{n_i\}_{i=1}^{N}$, $\oplus_{k=1}^{K}\oplus_{i=1}^{N}\zeta^{k}_{n_i}$ is a Gaussian random vector, and $\Var(\oplus_{k=1}^{K}\oplus_{i=1}^{N}\zeta^{k}_{n_i}) = \Var(\oplus_{k=1}^{K}\oplus_{i=1}^{N}\phi^{f_{k}}_{n_i})$. 
	\end{enumerate}

\end{lemma}
We recall some classical results which we state here for the reader's convenience.

\begin{theorem}(Mercer's Theorem)\label{mercer}
	Let $K(s,t)$ be a continuous symmetric and non-negative definite function defined on $I \times I \subset \R^2$. Consider the linear operator $\mathcal{K}$ defined for each $f \in L^2(I)$ by $\mathcal{K}(f)=\int_{I}K(s, \cdot)f(s)ds$. Each eigen-value of $\mathcal{K}$, $\lambda_{j}$, is real and non-negative, and the corresponding eigen-vectors can be chosen to be an orthonormal set. Further, if $(\lambda_{j}, e_{j})$ are the eigen-value and eigen-vector pairs derived from $\mathcal{K}$,
	
	\[
	K(s,t)=\sum_{j=1}^{\infty}\lambda_{j}e_{j}(s)e_{j}(t),
	\]
	for all $s, t$, with the sum converging absolutely and uniformly. 
\end{theorem}

For a proof, observe that the operator $\mathcal{K}$ is compact by Theorem~4.6.2 in \cite{TE15}, and this implies that the pairs $(\lambda_j, e_j)$ exist. The decomposition of $K$ follows from Theorem~4.6.5 in \cite{TE15}. The following result follows from Theorem~7.3.5 in \cite{TE15} and its proof.

\begin{theorem}\label{kl-theorem}(Karhunen–Loève Theorem)
	Let $I$ be compact, let $\Omega$ be a probability space, and let $X(\omega, t): \Omega \times I \rightarrow \mathbb{R}$  be an integrable stochastic process. Suppose that
	
	\begin{enumerate}
		\item (centered, second order) For all $s, t \in I$, $\E[X_t]=0$ and $Cov(X_t, X_s) <\infty$.
		\item (mean square continuous) If $t_{n} \rightarrow t$, then $\lim_{n\rightarrow \infty}\E[(X_{t_{n}}- X_t)^{2}]=0$. 
	\end{enumerate}
	Let $K(s,t)=\Cov(X_s, X_t)$. For each $j$, let $Z_{j} \in L^{2}(I)$ be the eigen-vector of $\mathcal{K}$ corresponding to the $j^{th}$ eigen-value $\lambda_{j}$ of $\mathcal{K}$ (where this operator is defined in the statement of Theorem \ref{mercer}), so that the functions $Z_{j}$ are orthonormal in $L^{2}[0,1]$. Further, let $\phi_{j}=\int_{I}Z_{j}(t)X_t$. Then the random variables $\phi_{j}$ are pair-wise uncorrelated, satisfy $\Var(\phi_j)=\lambda_j$, and for all $M$,
	
	\[\sup_{t \in I}\E[|X_{t}-\sum_{j=1}^{M}\phi_{j}Z_{j}(t)|^{2}]=\sup_{t\in I}[K(t,t)-\sum_{j=1}^{M}\lambda_{j}e_{j}^2(t)]\]
	Furthermore, 
	\[ 
	\lim_{M \rightarrow \infty}\sup_{t\in I}[K(t,t)-\sum_{j=1}^{M}\lambda_{j}e_{j}^2(t)]=0.
	\]
	
\end{theorem}

\begin{lemma}\label{kl-integral}
Let $f_t$ be a stochastic process that satisfies the assumptions of Theorem \ref{kl-theorem}, and recall the notation introduced in the statement of that theorem. Let $K$ denote the covariance kernel of $f_t$.
%Let $R_{M+1} = \sum_{j=M+1}^{\infty}\phi_{j}^{2}$, and let $K$ be the covariance kernel of $f_t$. 
Then, $\E[\int f_t^2 dt] < \infty$. Moreover, for all $M \geq 0$, $R_{M+1} := \sum_{j=M+1}^{\infty}\phi_{j}^{2}$ is almost surely finite, 
\[
(i) 0\leq \int_{I}f_{t}^{2}dt - \sum_{j=1}^{M}\phi_{j}^{2} =R_{M+1},
\]
and 
	
	\[
	(ii) \|R_{M+1}\|_{\Psi_{\frac{1}{2}}} \leq C_{\Psi_{\frac{1}{2}}} \sum_{j=M+1}^{\infty}\|\phi_{j}^{2} \|_{\Psi_{\frac{1}{2}}},
	\]
	where the right-hand side is defined to be $+\infty$ if for any $j\geq M+1$, $\|\phi_{j}\|_{\psi_{\frac{1}{2}}} =\infty$ (recall Definition \ref{def:orlicz-norm}.)
	
\end{lemma}

The next few results are elementary and are included for the reader's convenience:

\begin{definition}\label{def:subExp}
	Let $d$ be a metric on $I$, and let $\{f_t; t \in I\}$ be a real-valued stochastic process.  If for each $(s,t) \in I^{2}$ it holds that $\|f_s-f_t \|_{\Psi_{1}} \leq d(s,t)$, we say that $f \in \textrm{subExp}(d)$. 
\end{definition}

The following proof is elementary and can be found in~\cite{Decker24} (see Lemma E.0.9).

\begin{lemma} \label{lem:dismetric}
	Let $0<q<1$. Then function $d_q(s,t):=|s-t|^{q}$ is a metric on $\mathbb{R}.$ 
\end{lemma}

We next recall that sub-exponentiality implies a.s. continuity of sample paths, see Exercise 17 on p.~106 in \cite{VW96}. 

\begin{lemma}\label{lem:cts_sample_paths}
	Let $0<q \leq 1$, $C_{H}>0$ be real constants and let $d_{q}(x,y)=|x-y|^q$ (by Lemma \ref{lem:dismetric} it is a metric). If $f \in \textrm{subExp}(C_{H}d_{q})$ then almost every sample path of $f$ is a continuous function on $I$.
\end{lemma}

\begin{lemma}\label{sup_lemma}
	Let $0<q \leq 1$, $C_{H}>0$, and recall that $d_{q}(s,t) = |s-t|^q$ is a metric (Lemma \ref{lem:dismetric}). There is a constant $K_*$ that does not depend on $C_{H}$ or $q$ such that for all $f \in \textrm{subExp}(C_{H}d_{q})$,
	\begin{equation}
		\Big\|\sup_{(\theta_{1}, \theta_{2})\in I^2}\{|f_{\theta_{1}}-f_{\theta_{2}}|\}\Big\|_{\Psi_{1}} \leq K_* \frac{C_{H}}{q}.
	\end{equation}
\end{lemma}

\begin{lemma}\label{lem:sup_process}
Let $0<q \leq 1$, $C_{H}>0$, and recall that $d_{q}(s,t) = |s-t|^q$ is a metric (Lemma \ref{lem:dismetric}). 
Then for the constant $K_*$ from Lemma~\ref{sup_lemma}, for all $f \in \textrm{subExp}(C_{H}d_{q})$ there
    \begin{equation}
		\Big\|\sup_{t \in I}f_t\Big\|_{\Psi_{1}} \leq \frac{K_{*}C_{H}}{q} + \Bigg\|\sqrt{\int_{I}f_t^2 dt}\Bigg\|_{\Psi_{1}} 
	\end{equation}
\end{lemma}

\textbf{Proof of Lemma \ref{measurability}} 

The random element $f$ is assumed to be a measurable map from $(\Omega, \mathcal{F}, \mathbb{P})$ into $((C[I])^K, \mathcal{B})$, where $\mathcal{B}$ is the Borel sigma algebra generated by the canonical topology (i.e. the topology induced by the norm $f \rightarrow \max_{k=1}^{K}\sup_{t\in I}|\pi_k f(t)|$.)  We must show that $X(\omega, t) := f(\omega)(t)$ is a jointly measurable map defined on $\Omega \times I$ and taking values in $\R^K$. It suffices that for almost every $\omega$ the map $t\rightarrow X(\omega, t): I \rightarrow \R^K$ is continuous with respect to the Euclidean norm and also that for each $t$ the map $\omega \rightarrow X(\omega, t):\Omega \rightarrow \R^K$ is measurable. These two separate conditions imply that $X(\omega, t)$ is jointly measurable by Lemma 4.51 of \cite{CAB}. 

First, note that for fixed $\omega$, $t\rightarrow X(\omega, t) = t\rightarrow \oplus_{k=1}^{K}\pi_{k}[f(\omega)(t)]$, and we have assumed that for every $\omega$ it holds for every $k$ that $\pi_{k}[f(\omega)(\cdot)] \in C[I]$. Continuity of each component of a map into $K$ dimensional Euclidean space implies continuity of the  map. 

Second, note that for each fixed $t$, $\omega \rightarrow X(\omega, t) = e_{t}\circ f(\omega)$, where $e_{t}: (C[I])^K \rightarrow \R^K$ is defined by $e_t(f) = f(t)$. It is a well known result of point set topology that $e_t$ is continuous with respect to the canonical supremum topology. Since $f$ is assumed to be measureable, the composition is measureable. \hfill $\Box$

\bigskip

\textbf{Proof of Lemma \ref{lem:gaussian_process}}

Recall that $X$ is a $K$ dimensional Gaussian process if and only if $X$ is a measureable $\R^{K}$ valued stochastic process and for every finite collection $\{t_{1}, \cdots, t_{L}\} \subset I$ the vector $\oplus_{\ell=1}^{L}\oplus_{k=1}^{K}X_k(t_{\ell}) \in (\R^K)^L$ has a multivariate normal distribution. We will construct a Gaussian process $G$ with the properties indicated in (a) by invoking the Kolmogorov Extension Theorem (see \cite{oksendal2013stochastic}, Theorem 2.1.5).

Given a collection $t_1, \dots t_L \subset I$, let $\Sigma_{t_1, \dots t_L}$ be the $(KL) \times (KL) $ positive semi-definite matrix defined by $\Sigma_{t_1, \dots t_L}:=\Var(\oplus_{l=1}^{L} \oplus_{k=1}^{K}f_{k}(t_l))$. Let $g_{t_1, \dots t_L}$ be a centered Gaussian vector with covariance matrix $\Sigma_{t_1, \dots t_L}$, and let $\nu_{t_{1}, \dots t_L}$ be the measure on $(\R^K)^L$ that is defined on each Borel $F \subset (\R^K)^L$ by $\nu_{t_{1}, \dots t_L}(F) = \Pr(g_{t_{1}, \dots t_L}\in F)$. 

Let $e_{1}, \dots, e_{KL}$ be the standard basis of $(\R^{K})^L$. To facilitate computations, we use an alternate labeling of the basis vectors. For $1 \leq l \leq L$, and $1 \leq k \leq K$,  let $f_{l,k} = {e_{(l-1)L+k}}$. Let $\sigma$ be a permutation of $\{1, \dots, L\}$, and define the invertible linear operator $L_{\sigma}: (\R^{K})^{L} \rightarrow (\R^{K})^{L}$ through its action on each basis vector $f_{l,k}$:

\begin{equation}
	L_{\sigma}(f_{l,k}) = f_{\sigma(l),k}.
\end{equation}
It is straightforward to see that if $F_{1}, F_{2}, \dots F_{L}$ are Borel sets in $\R^{K}$,

\[
L_{\sigma}(F_{1} \times F_{2} \times \dots F_{L}) = F_{\sigma(1)} \times F_{\sigma(2)} \times \dots F_{\sigma(L)}.
\]
Thus,
\begin{align*}
&\nu_{t_{1}, \dots t_L}(F_{1} \times F_{2} \times \dots F_{L})\\
&= \Pr(g_{t_1, \dots t_L}\in F_{1} \times F_{2} \times \dots F_{L})\\
&=\Pr(L_{\sigma}g_{t_1, \dots t_L}\in F_{\sigma(1)} \times F_{\sigma(2)} \times \dots F_{\sigma(L)})\\
& = \Pr(Y \in F_{\sigma(1)} \times F_{\sigma(2)} \times \dots F_{\sigma(L)}),\\
\end{align*}
for $Y$ a centered gaussian with covariance matrix $L_{\sigma}\Big(\Sigma_{t_1, \dots t_L}\Big)L_{\sigma}^{\top}$. Since 
\begin{align*}
& L_{\sigma}\Big(\Sigma_{t_1, \dots t_L}\Big)L_{\sigma}^{\top}= L_{\sigma}\Var(\oplus_{l=1}^{L} \oplus_{k=1}^{K}f_{k}(t_l))L_{\sigma}^{\top}\\
&=\Var(L_{\sigma}\oplus_{l=1}^{L} \oplus_{k=1}^{K}f_{k}(t_l))\\
&=\Var(\oplus_{l=1}^{L} \oplus_{k=1}^{K}f_{k}(t_{\sigma(l)}))\\
&=\Sigma_{t_{\sigma(1)}, \dots t_{\sigma(L)}},
\end{align*}
it follows that 
\begin{align*}
	&\nu_{t_{1}, \dots t_L}(F_{1} \times F_{2} \times \dots \times F_{L}) = \Pr(Y \in F_{\sigma(1)} \times F_{\sigma(2)} \times \dots \times F_{\sigma(L)})\\
	&=\Pr(g_{t_{\sigma(1)}, \dots, \sigma(L)}\in  F_{\sigma(1)} \times F_{\sigma(2)} \times \dots \times F_{\sigma(L)} )\\
	&=\nu_{t_{\sigma(1)}, \dots, \sigma(L)}( F_{\sigma(1)} \times F_{\sigma(2)} \times \dots \times F_{\sigma(L)}).
\end{align*}
Thus the family of jointly gaussian measures $\nu_{t_{1}, \dots t_L}$ satisfies the first consistency condition in the Kolomogorov Extension Theorem. The second consistency condition holds trivially. Then by the Kolmogorov Extension Theorem there exists a probability space $(\Omega, \mathcal{F}, P)$ and a stochastic process $G_t=\oplus_{k=1}^{K}G_{k}(t)$ on $\Omega$, such that for every finite sub-collection $\{t_{1}, \dots, t_{L}\}$ $\oplus_{\ell=1}^{T}\oplus_{k=1}^{K}G_k(t_\ell)$ is a Gausssian random vector with covariance matrix $\Var(\oplus_{l=1}^{L} \oplus_{k=1}^{K}f_{k}(t_l))$. This implies in particular that  for all $1 \leq k_{1} \leq k_{2} \leq K$ and all $t_{1}, t_{2}\in I$, $\Cov(G_{k_{1}}(t_1), G_{k_{2}}(t_2)) = \Cov(f_{k_{1}}(t_1), f_{k_{2}}(t_2))$.

To prove part (b), recall that we assumed that for each $1 \leq k \leq K$, the stochastic process $f_k$ is centered, second order, and mean square continuous. These conditions imply that $(s,t) \mapsto \Cov(f_{k}(s),f_{k}(t)) =\Cov(G_{k}(s),G_{k}(t)) $ is a continuous function defined on $I \times I$. Let $Z^{f_k}_j$ denote the eigenfunctions of $(s,t) \mapsto \Cov(f_{k}(s),f_{k}(t))$. By Theorem~\ref{kl-theorem} the corresponding KL scores are given by $\phi^{f_{k}}_{j} = \int_I Z^{f_k}_j(t) f_k(t) dt$. Since $\Cov(f_{k}(s),f_{k}(t)) =\Cov(G_{k}(s),G_{k}(t))$ for all $s,t,k$, the covariance operator of $G_{k}$ has the same eingenfunctions and eigenvalues. The corresponding KL scores satisfy $\zeta^{k}_{j} = \int_I Z^{f_k}_j(t) f_k(t) dt$ by Theorem~\ref{kl-theorem}. 

If $Z^{f_k}_j$ corresponds to a zero eigenvalue, then almost surely $\phi^{f_{k}}_{j} = \zeta^{k}_{j}=0$, since $f_k$ and $G_k$ concentrate in sub-space that is orthogonal to $Z^{f_k}_j$ as discussed on page 182 of \cite{TE15}. Hence, in what follows it suffices to prove joint gaussianity of the collections $(\zeta^{k}_{j})_{(k,j) \in S}$ for finite sets $S \subset \{1,\dots,K\}\times \N$ such that each $Z^{f_k}_j$ corresponds to a non-zero eigenvalues.

Each such $Z^{f_k}_j$ is continuous. To see this, denote by $\lambda$ the corresponding eigenvalue and note that
\[
\lambda(Z(t) - Z(t')) = \int_I Z(s)(K(s,t)-K(s',t)) ds \to 0 \quad \mbox{as } t \to t'
\]
by continuity and boundedness of the covariance kernel $K$ combined with square integrability of $Z$ and the dominated convergence theorem. By Theorem \ref{kl-theorem},
\begin{align*}
\Cov(\phi^{f^{k_1}}_{j_1},\phi^{f^{k_2}}_{j_2}) &= \E\Big[\Big(\int_{I}f_{k_1}(s) Z^{f_{k_1}}_{j_1}(s)ds\Big)\Big(\int_{I}f_{k_2}(t)Z^{f_{k_2}}_{j_2}(t)dt\Big)\Big]\\
%	& =\int_{I}\int_{I}\E[f_{k_1}(s)f_{k_2}(t)] Z^{f_{k_1}}_{j_1}(s)Z^{f_{k_2}}_{j_2}(t)dsdt\\
	& =\int_{I}\int_{I}\E[f_{k_1}(s)f_{k_2}(t)] Z^{f_{k_1}}_{j_1}(s)Z^{f_{k_2}}_{j_2}(t)dsdt\\
	& =\int_{I}\int_{I}\Cov(f_{k_1}(s),f_{k_2}(t)) Z^{f_{k_1}}_{j_1}(s)Z^{f_{k_2}}_{j_2}(t)dsdt\\
	& =\int_{I}\int_{I}\Cov(G_{k_1}(s),G_{k_2}(t)) Z^{G_{k_1}}_{j_1}(s)Z^{G_{k_2}}_{j_2}(t)dsdt\\
	& =\Cov(\zeta^{k_1}_{j_1},\zeta^{k_2}_{j_2}),\\
\end{align*}
where we used Fubini's Theorem to justify interchange of the expectation and integral, and the result of part (a) in the second last equality. Further, interpreting $(   \zeta^{k}_{j})_{(k,j) \in S}$ as vector in $\R^S$  
\begin{align*}
&(\zeta^{k}_{j})_{(k,j) \in S} = \Big(\int_{I}G_k(t)Z^{f_k}_{j}(t)dt\Big)_{(k,j) \in S}=\Big(\lim_{M \rightarrow \infty}M^{-1}\sum_{m=1}^{M}G_k(m/M)Z^{f_k}_{j}(m/M)\Big)_{(k,j) \in S}\\
&=\lim_{M \rightarrow \infty}\Big(M^{-1}\sum_{m=1}^{M}G_k(m/M)Z^{f_k}_{j}(m/M)\Big)_{(k,j) \in S} = \lim_{M \rightarrow \infty}\sum_{m=1}^{M}\Big(M^{-1}G_k(m/M)Z^{f_k}_{j}(m/M)\Big)_{(k,j) \in S}.
\end{align*}
We have used that each $Z^{f_kk}_{j}$ is a continuous function, and that $G_{k}$ has an almost surely uniformly continuous modification; it satisfies $\|G_{k}(s) - G_{k}(t)\|_{\Psi_{1}}  \lesssim \sqrt{\Var(G_{k}(s) - G_{k}(t))} = \sqrt{\Var(f_{k}(s) - f_{k}(t))} \lesssim \|f_{k}(s) - f_{k}(t))\|_{\Psi_{1}} \leq C_{H}|s-t|^q$
and then we can apply the conclusion of Exercise 17, page 106 of \cite{VW96} Therefore the product $G_k Z_{j}^{k}$ is continuous, and so Riemann integrable, and we may therefore apply Lemma \ref{lem:riemann}, which justifies the above display.

Note that $\oplus_{m=1}^M\big(M^{-1}G_k(m/M)Z^{k}_{j}(m/M)\Big)_{(k,j) \in S}$ is a multi-variate centered gaussian because $G$ is a gaussian process. Thus for each $M$, $\sum_{m=1}^{M}\big(M^{-1}G_k(m/M)Z^{f_k}_{j}(m/M)\big)_{(k,j) \in S}$ is a centered Gaussian. The sequence converges almost surely by Lemma \ref{lem:riemann}, hence converges in distribution. It is well known that the distributional limit, when it exists, of a sequence of Gaussians is gaussian. 

\hfill $\Box$

 \bigskip

\textbf{Proof of Lemma \ref{kl-integral}}

For a second order and mean-square continuous process $g_t$ with $\E[g_t^2] < \infty$, Fubini's Theorem gives $\int_{I\times \Omega}g(\omega, t)d(\mu \times \omega) = \E[\int_{I}g_{t} d \mu_t] = \int_{I}\E[g_{t}] d \mu_t$, here $\mu$ corresponds to Lebesgue measure on $I = [0,1]$. Apply this with $g_t = f_t$ to conclude that $\E\big[\int_{I}f_{t}^2dt\big] < \infty$. This implies that $\int_{I}f_{t}^2dt$ is almost surely finite. Next, note that 
\begin{align*}
0 &\leq \int_{I}\Big(f_{t}-\sum_{j=1}^{M}\phi_{j}Z_{j}(t)\Big)^{2}dt
\\
& = \int_{I}f_{t}^2 dt+\int_{I}\Big(\sum_{j=1}^{M}\phi_{j}Z_{j}(t)\Big)^{2}dt - 2\Big(\int_{I}f_{t}\sum_{j=1}^{M}\phi_{j}Z_{j}(t)\Big)dt 
\\
&= \int_{I}f_{t}^{2} dt - \sum_{j=1}^{M}\phi_{j}^{2},
\end{align*}
where the last equality follows by Theorem~\ref{kl-theorem}, using the fact that $\int_I Z_j(t)Z_k(t)dt = \delta_{kj}$ and $\int_I f(t)Z_j(t)dt = \phi_j$. Let $S_M=\sum_{j=1}^{M}\phi_{j}^{2}$, and note that by the bound just proved and since $S_M$ is non-decreasing,  $S_M$ converges point-wise to an almost surely finite limit. Note that 

\begin{align*}
\E\Big|\sum_{j=1}^{M}\phi_{j}^{2} - \int_{I}f_{t}^{2}dt\Big| 
&= \E\Big[ \int_{I}\Big(f_{t}-\sum_{j=1}^{M}\phi_{j}Z_{j}(t)\Big)^{2}dt \Big]
\\
&= \int_{I}\E\Big(f_{t}-\sum_{j=1}^{M}\phi_{j}Z_{j}(t)\Big)^{2}dt
\\
&\leq \sup_{t \in I}\E\Big(f_{t}-\sum_{j=1}^{M}\phi_{j}Z_{j}(t)\Big)^{2}
\\
&\leq \sup_{t \in I}\Big|K(t,t)-\sum_{j=1}^{M}\lambda_{j}Z_{j}^2(t)\Big|
\end{align*}
where the first equality was established in the beginning of this proof, the second equality follows from Fubini's Theorem, and the last inequality is from Theorem~\ref{kl-theorem}, which also shows the the final term tends to zero as $M$ goes to $\infty$. It follows that $S_M$ converges almost surely to  $\int_{I}f_{t}^{2}dt$. Therefore for any $M\geq 0$ (in case $M=0$ the sum $\sum_{j=1}^{M}\phi_{j}^{2}$ is defined to be $0$) ,
\[
0\leq \int_{I}f_{t}^{2}dt - \sum_{j=1}^{M}\phi_{j}^{2} = \lim_{N \to \infty}\big(S_N-\sum_{j=1}^{M}\phi_{j}^{2}\big)=R_{M+1}\leq \int_{I}f_{t}^{2}dt,
\]
and this shows that $R_{M+1}$ is almost surely finite and also (i).
To complete the proof of (ii), apply Lemma~\ref{triangle_inequality} with $p=1/2$, setting $X_j$ in the statement of the Lemma as $X_j = \phi_{j+M}^2$. Note that we may assume that $\sum_{j=M+1}^\infty \|\phi_j^2\|_{\psi_{1/2}}<\infty$ since otherwise the bound holds trivially.
 \hfill $\Box$

\noindent

\bigskip

\noindent
\textbf{Proof of Lemma \ref{sup_lemma}} 
By Corollary 2.2.5 in \cite{VW96}, there exists a constant $C$ depending only on $C_{H}$ such that 
\begin{equation}
\Big\|\sup_{\theta_{1}, \theta_{2}} |f_{\theta_{1}}-f_{\theta_{2}}| \Big\|_{\Psi_{1}} \leq C\int_{0}^{1}\Psi_{1}^{-1}(D(\eps, d_{q}))d\eps.
\end{equation}
Inspection of the proof of Theorem~2.2.4 in \cite{VW96} shows that $C \leq C_{1}C_{H}$ for a universal constant $C_{1}$. Recall that $\Psi_{1}^{-1}(x)=\log(x+1)$. By elementary computations for every $1 \geq \eps>0$, $D(\eps, d_q, [0,1])\leq \frac{2^{1/q}}{2}\eps^{-1/q}$. Thus, using $x \geq 1 \rightarrow \log(x+1)\leq \log(2x)$,
\begin{align*}
& \Big\|\sup_{\theta_{1}, \theta_{2}}\{|f_{\theta_{1}}-f_{\theta_{2}}|\}\Big\|_{\Psi_{1}} 
\leq C\Big[\int_{0}^{1}\log\Big(\frac{2^{1/q}}{2}\eps^{-1/q}+1\Big)d\eps\Big]&\\
		& \leq C\Big[\frac{1}{q}\int_{0}^{1}\log\Big(\frac{2}{\eps}\Big) d\eps\Big]\\
		& = \frac{C}{q}\Big(\log(2) - \int_{0}^{1}\log(\eps)d\eps\Big) =  \frac{C}{q}\Big(\log(2) - [\lim_{\delta \downarrow 0^{+}} \eps\log(\eps)-\eps]|^{1}_{\delta}\Big) =\frac{C}{q}(\log(2)+1) \leq \frac{2C_1C_H}{q}.&\\
	\end{align*}
	Let $K_{*}:=2C_1$. \hfill $\Box$

\bigskip
\noindent
\textbf{Proof of Lemma \ref{lem:sup_process}}  We have
\begin{align*}
		|f_{t}| &\leq \inf_{s \in[0,1]}\Big\{ |f_{t}-f_{s}|+|f_{s}| \Big\} &\\
		& \leq \inf_{s \in I} \Big\{\sup_{\theta_{1}, \theta_{2} \in I}|f_{\theta_{1}}-f_{\theta_{2}}|+|f_{s}|\Big\}&\\
		&\leq \sup_{\theta_{1}, \theta_{2} \in I}\Big\{|f_{\theta_{1}}-f_{\theta_{2}}|\Big\}+\inf_{s \in I} |f_{s}|&\\
		&\leq \sup_{\theta_{1}, \theta_{2} \in I}\Big\{|f_{\theta_{1}}-f_{\theta_{2}}|\Big\}+ \int_{0}^{1}|f_{s}| ds\\
		&\leq \sup_{\theta_{1}, \theta_{2} \in I}\Big\{|f_{\theta_{1}}-f_{\theta_{2}}|\Big\} + \sqrt{\int_{0}^{1}f_{s}^{2}ds}.
\end{align*}
By monotonicity, i.e. $0 \leq X \leq Y$ a.s. implies $\|X\|_{\Psi_1}\leq\|Y\|_{\Psi_1}$, and the triangle inequality, 
\begin{equation}
		\Big\|\sup_{t}|f_{t}|\Big\|_{\Psi_{1}} \leq \Big\|\sup_{\theta_{1}, \theta_{2} \in I}\{|f_{\theta_{1}}-f_{\theta_{2}}|\}\Big\|_{\Psi_{1}}+\Big\|\sqrt{\int_{0}^{1}f_{s}^{2}}\Big\|_{\Psi_{1}}.
	\end{equation}
{The claim follows from Lemma \ref{sup_lemma}.} \hfill $\Box$

\subsubsection{Miscellaneous Technical Results}
The following bound is standard:
\begin{lemma}\label{lem:tail_sum}
	For $M \geq 1$ and $\alpha>\frac{1}{2}$
	\[
	\sum_{j=M+1}^{\infty}j^{-2\alpha} \leq \int_{M}^{\infty}x^{-2\alpha} dx=\frac{M^{1-2\alpha}}{2\alpha-1}.
	\]
	Further, 
	\[
	\sum_{j=1}^{\infty}j^{-2\alpha} \leq \Big(1+\int_{1}^{\infty}x^{-2\alpha} dx\Big) =\frac{2\alpha}{2\alpha-1}.
	\]
\end{lemma}

\begin{lemma}\label{lem:tail_bound_TMG}
{Assume that (A2) holds.} The random variable $T^{M,G}$ defined in section \ref{sub-section:additional_notation} {is absolutely continuous with respect to Lebesgue measure}.  Furthermore under the conditions of \ref{assumptions}, there is a universal constant $C$ such that for all $M \geq 1$,
\begin{equation}
\|T^{M,G}\|_{\Psi_{1}} \leq C\log(K+1)C_{\phi}^{2}\frac{2\alpha}{2\alpha-1}.
\end{equation}
This implies that for each $t \geq 0$,
\begin{equation}
\Pr(T^{M,G} >t) \leq 2\exp\Big(-\frac{2\alpha-1}{{2\alpha}C{C_\phi^2}}\cdot\frac{t}{\log(K+1)}\Big).
\end{equation}
Further, for each $\epsilon >0$,
\begin{equation}
\Pr\Big(T^{M,G} \geq \log(K+1)\log\big(\tfrac{2}{\eps}\big)\frac{2\alpha C{C_\phi^2}}{2\alpha-1}\Big) \leq \eps.
\end{equation}
\end{lemma}

\noindent
\textbf{Proof of Lemma~\ref{lem:tail_bound_TMG}}

Let $E$ be a null set, and recall that $T^{M,G} = \max_{k=1}^{K}\sum_{j=1}^{M}g_{k,j}^2$, where $\oplus_{k=1}^{K}\oplus_{j=1}^{M}g_{k,j}$ is Gaussian and $\Var(g_{k,j}) \leq C_{1,2}^2C_{\phi}^2j^{-2\alpha}$ where $C_{1,2}$ is the constant from Lemma~ \ref{relationship_norms}. We have
\[
\Pr(T^{M,G} \in E)  = \Pr\Big(\max_{k=1}^{K}\sum_{j=1}^{M}g_{jk}^{2} \in E\Big)  \leq \Pr\Big(\exists k: \sum_{j=1}^{M}g_{jk}^{2} \in E\Big) {\leq K \max_{k=1}^K\Pr\Big(\sum_{j=1}^{M}g_{jk}^{2} \in E\Big)}
\]
by the union bound. 

By construction, for fixed $k$, $g_{jk}$ are uncorrelated and Gaussian. Each $g_{jk}^{2}$ is absolutely continuous, being equal in distribution to a scaled $\chi^{2}(1)$ random variable. Convolution preserves absolutely continuity. The norm bound, {follows from (A2) by routine calculations after noting that chi-square random variables have finite sub-Gaussian norms and applying Lemma~\ref{lim_max_sum} (ii) and Lemma~\ref{lem:tail_sum}. The tail bound follows from Lemma~\ref{lim_max_sum} (iv) and the final bound by rearranging the tail bound}. 
\hfill $\Box$

\bigskip

\begin{lemma}[A Kolmogorov Bound implies a bound on quantiles]
	\label{lem:koltoquant}
	Let $F$ and $G$ be cumulative distribution functions, and  let $q_{F}$ and $q_{G}$ be their respective quantile functions. Let $t_*\in \R, \delta > 0$, and suppose
	\begin{equation}\label{eq:helpsupbound}
	\sup_{t\ge t_{*}}|F(t)-G(t)|\leq \delta.
	\end{equation}
	If $q_{F}(\alpha)>t_{*}$ then %or if $G(t_*) + \delta < \alpha$ 
	\[
	q_{G}(\alpha-\delta) \leq q_{F}(\alpha) \leq q_{G}(\alpha+\delta).
	\]
Further, $G(t_*) + \delta < \alpha$ implies $q_{F}(\alpha)>t_{*}$.
\end{lemma}

\noindent
\textbf{Proof of Lemma~\ref{lem:koltoquant}}

Let $H: \R \to \R$ be a non-decreasing right continuous function, and recall that $q_{H}(\alpha):=\inf A_{H}(\alpha)$, where
\[
A_{H}(\alpha) := \{t: H(t) \geq \alpha\},
\]
and $\inf \varnothing = \infty$ by convention. We will first prove that $G(t_*)+\delta < \alpha$ implies $q_{F}(\alpha)>t_{*}$ and then complete the proof assuming the latter. By assumption $G(t_*)+\delta < \alpha$ implies $F(t_*) \leq G(t_*) + \delta < \alpha$. By right-continuity of $F$ there exists $\eps >0$ such that $F(t_*+\eps) < \alpha$ and thus by definition $q_F(\alpha) \ge t_*+\eps > t_*.$ From now on assume $q_{F}(\alpha)>t_{*}$.

Assume that $x \in A_{F}(\alpha)$. Then $F(x)\geq \alpha$ and
$t_{*}< q_{F}(\alpha) = \inf A_{F}(\alpha)$ implies $t_{*}<x$. Since $t_{*}<x$, \eqref{eq:helpsupbound} implies $G(x)+\delta \geq F(x) \geq \alpha$ where the last bound follows from $x \in A_{F}(\alpha)$

This inequality means that $x \in A_{G+\delta}(\alpha)$. Hence $A_{F}(\alpha) \subseteq A_{G+\delta}(\alpha)$. Therefore,
\[
q_{F}(\alpha)=\inf A_{F}(\alpha) \geq \inf A_{G+\delta}(\alpha)=\inf A_{G}(\alpha-\delta) = q_{G}(\alpha-\delta).
\]
Likewise, $x \in A_{F}(\alpha)$ implies $ G(x)-\delta \leq F(x)$ so that $ A_{G-\delta}(\alpha) \subseteq A_{F}(\alpha)$, and therefore,
\[
q_{F}(\alpha)=\inf A_{F}(\alpha) \leq \inf A_{G-\delta}(\alpha)=\inf A_{G}(\alpha+\delta) = q_{G}(\alpha+\delta).
\]
Combining the two inequalities yields the result. \hfill $\Box$

\bigskip

\begin{lemma}\label{lem:tail-bound-from-trace-bound}
Let $X \in (\mathbb{R}^{M})^K$ be a centered Gaussian vector with positive semi definite covariance matrix $\Sigma$. Let $\Sigma_k:=\Var(\pi_k X)$, and suppose $\max_{k=1}^{K}\sqrt{\sum_{i=1}^{M}[\Sigma_k]_{jj}} \leq U$. Then for all $0<\eps \leq 1$,
\[
\Big\|\max_{k=1}^{K}\|\pi_{k}X\|_{2}\Big\|_{\psi_{2}} \leq K_{\Psi_{1}}^{\frac{1}{2}}\sqrt{\frac{8}{3}} \log^{\frac{1}{2}}(K+1)U,
\]
and 
\[	\Pr\Big(\max_{k=1}^{K}\|\pi_{k}X\|_{2} \geq \sqrt{\frac{8}{3}}K_{\Psi_{1}}^{\frac{1}{2}}\sqrt{\log(K+1)}U\sqrt{\log\big(\tfrac{2}{\eps}\big)}\Big) \leq \eps,
\]
where $K_{\Psi_{1}}$ is the universal constant of Lemma~ \ref{lim_max_sum} (ii).
\end{lemma}
\noindent
\textbf{Proof of Lemma~\ref{lem:tail-bound-from-trace-bound}} 
We have 
\begin{align*}
	&\Big\|\max_{k=1}^{K}\|\pi_{k}X\|_{2}\Big\|_{\psi_{2}} = \Big(\Big\|\max_{k=1}^{K}\|\pi_{k}X\|^{2}_{2}\Big\|_{\Psi_{1}}\Big)^{\frac{1}{2}}\tag*{Lemma~ \ref{relationship_norms}(i)}\\ 
	& \leq K_{\Psi_{1}}^{\frac{1}{2}} \log^{\frac{1}{2}}(K+1)\Big(\max_{k=1}^{K}\sum_{j=1}^{M} \|[\pi_{k}X]^{2}_j\|_{\Psi_{1}}\Big)^{\frac{1}{2}} \tag*{Lemma~ \ref{lim_max_sum} (ii) } \\
	& \leq K_{\Psi_{1}}^{\frac{1}{2}} \log^{\frac{1}{2}}(K+1)\Big(\max_{k=1}^{K}\sum_{j=1}^{M} \|[\pi_{k}X]_j\|_{\psi_{2}}^{2}\Big)^{\frac{1}{2}} \tag*{Lemma~ \ref{relationship_norms}(i)}.
	\end{align*}

	Since $[\pi_{k}X]_j$ is a centered Gaussian, $\|[\pi_{k}X]_j\|_{\psi_{2}}=\sqrt{\frac{8}{3}}\sqrt{\Var([\pi_{k}X]_j)}=\sqrt{\frac{8}{3}[\Sigma_k]_{jj}}$ (this follows by homogeneity of the norm; the exact constant follows by calculating $\|N(0,1)\|_{\psi_{2}}$ from the definition). Thus, 
	
	\[
	\Big\|\max_{k=1}^{K}\|\pi_{k}X\Big\|_{2}\|_{\psi_{2}} \leq K_{\Psi_{1}}^{\frac{1}{2}}\sqrt{\frac{8}{3}} \log^{\frac{1}{2}}(K+1)U.
	\]
 	By Lemma ~\ref{lim_max_sum} (iv),for any $t$, 
	
	\[
	\Pr\Big(\max_{k=1}^{K}\|\pi_{k}X\|_{2} \geq t\Big) \leq 2 \exp\Bigg(-\Bigg(\frac{t}{K_{\Psi_{1}}^{\frac{1}{2}}\sqrt{\frac{8}{3}}\log^{\frac{1}{2}}(K+1)U} \Bigg)^2\Bigg).
	\]
	Rearranging the bound yields the result. \hfill $\Box$
\subsubsection{Useful Properties of Orlicz spaces}\label{section:orlicz}

The material in this section is known in the community, but scattered. We provide a collection of the results we will use along with precise statements for the reader's convenience. We also provide short proofs for results we could not find in the literature in the exact form we need them. Recall the following definitions, see also Section 2.7.1 in \cite{V18} and Section~2.2.1 in \cite{VW96}. If $X$ is a topological space, we define $\mathcal{B}_{X}$ to be the generated Borel $\sigma$-field on $X$.

\begin{definition}\label{def:orlicz-norm}
	Let $(\Omega, \mathcal{F}, \mathbb{P})$ be a probability space. If $X:\Omega \rightarrow \mathbb{R}$ is $\mathcal{B}_{\mathbb{R}}/\mathcal{F}$ measureable  and $\psi$ is a measureable scalar function,
	\begin{equation}
		\|X\|_{\psi} := \inf\Big\{B>0: \E\psi\big(\tfrac{|X|}{B}\big) \leq 1\Big\},
	\end{equation}
	with the convention that $\inf \varnothing =\infty$. We then define
	\begin{equation}
		L^{\psi}= \{X, \|X\|_{\psi} < \infty\}.
	\end{equation}
	A non-zero function $\psi:[0,\infty) \rightarrow [0,\infty)$ is an Orlicz function if and only if $\psi$ is convex, non-decreasing, $\psi(0)=0$ and $\psi$ is not identically zero. 
\end{definition}

\begin{lemma}\label{orlicz-banach}
If $\psi$  is an Orlicz function, $L^{\psi}$ is a Banach space.
\end{lemma}

This is mentioned in Section~2.7.1 of \cite{V18}, see \cite{Decker24} for a proof (see Lemma H.0.1). Of special interest in this article are the functions $\psi_{p}$,
\[
\psi_{p}(x) := \exp(|x|^{p})-1.
\]
If $p\geq 1$, $\psi_{p}$ is an Orlicz function. Otherwise, $\|\cdot\|_{\psi_p}$ is only a quasi-norm (the triangle inequality fails). Nonetheless we have:

\begin{lemma}[The $\psi_{p}$ quasi-norm is equivalent to an Orlicz norm]\label{orlicz-equiv}
	If $1 \leq p$, $\psi_{p}$ is an Orlicz function. If $0<p$, there exists an Orlicz function $\tilde{\psi}_{p}$ such that $\|X\|_{\tilde{\psi}_{p}} \leq \|X\|_{{\psi}_{p}} \leq C_{\psi_{p}}\|X\|_{\tilde{\psi}_{p}}$, for $C_{\psi_{p}}:=e^{-1}[\frac{e}{\log(2)p}]^{\frac{1}{p}}.$
\end{lemma}

This result can be found in Exercise~2.14.1 in~\cite{VW96}, the exact form of the constants $C_{\psi_{p}}$is worked out it \cite{Decker24} (see Lemma H.0.2). This equivalence implies that the triangle inequality holds up to a constant for infinite sums in $L^{\psi_{p}}$.

\begin{lemma}[Triangle inequality for infinite sums, up to constant]\label{triangle_inequality}
Whenever $p>0$, if $\sum_{j=1}^{\infty}\|X_{j}\|_{\psi_{p}} < \infty$, there exists $L\in L^{\psi_p}$ such that
\[
\lim_{N\to\infty} \Big\|\sum_{j=1}^{N}X_{j} - L \Big\|_{\psi_p} = 0.
\]
Further $\|L\|_{\psi_p} \leq \tilde C_{\psi_{p}} \sum_{j=1}^{\infty}\|X_{j}\|_{\psi_{p}}$ where $\tilde C_{\psi_{p}} = 1$ for $p\geq 1$ and is the constant $C_{\psi_{p}}$ from Lemma~\ref{orlicz-equiv} otherwise. When $X_i \geq 0$ a.s., the sum $\sum_{j=1}^{N}X_{j}$ also converges to $L$ almost surely and in this case we will write $\sum_{j=1}^\infty X_j$ instead of $L$.
\end{lemma}

For completeness, the proof is provided below. We now present some technical results on limits and tail bounds that will are used extensively throughout the proofs.

\begin{lemma}[Limits, Maxima, Independent sums, Tail bounds]\label{lim_max_sum}	For all $i$, let $X_{i}$ be a random variable.
\begin{itemize}
\item [(i)] If $\psi$ is an Orlicz function,
		\[
		\|\liminf_{n \rightarrow \infty}X_{n}\|_{\psi} \leq \liminf_{n \rightarrow \infty}\|X_{n}\|_{\psi}.
		\]
		\item [(ii)]   There exists a universal constant $K_{\Psi_{1}}$ such that for $p>0$
		\[
		\Big\|\max_{1\leq i \leq K}X_{i}\Big\|_{\psi_{p}} \leq K_{\Psi_{1}}^{\frac{1}{p}}\log^{\frac{1}{p}}(K+1)\max_{i}\|X_{i}\|_{\psi_{p}}.
		\]
		\item [(iii)]  Let $S_{n}=\sum_{i=1}^{n}X_{i}$ and suppose the $X_{i}$ are independent and centered. For each $0<p\leq 2$,
		\[
		\|S_{n}/\sqrt{n}\|_{\psi_{p}} \leq K_{p}\max_{i}\|X_{i}\|_{ \psi_{p}},
		\]
		for a universal constant $K_p$.		
		\item[(iv)] For all $t \geq 0, p > 0$,
		\[
		\Pr(|X| \geq t) \leq 2\exp\Big(-\Big(\frac{t}{\|X\|_{\psi_{p}}}\Big)^{p}\Big).
		\]
	\end{itemize}
\end{lemma}
 
The proofs of part (i) and (iii) of the above Lemma are given below for completeness. Part (iv) follows from the Markov inequality, part (ii) with an unspecified constant is given in Lemma 2.2.2 of~\cite{VW96}, a derivation with this form of constant can be found in \cite{Decker24} (see Lemma H.0.4).

\begin{lemma}[Relationships between quasi-norms]\label{relationship_norms}

Let $p, q >0$ be real numbers, and let $X$ and $Y$ be a random variables.
	\begin{itemize}
		\item [(i)]$\|X\|_{\psi_{p}} = \||X|^{p/q}\|_{\psi_{q}}^{q/p}$ and $\|XY\|_{\psi_{p}} \leq \|X\|_{\psi_{2p}}\|Y\|_{\psi_{2p}}$.
		\item [(ii)] For any $p,q > 0$ there exists a universal constant $C_{p,q}$ such that  $\|X\|_{L_{q}} \leq C_{p,q}\|X\|_{\psi_{p}}$.
	\end{itemize}
\end{lemma}
The proofs of both statements of part (i) in case $p=1$ and $q=2$ are given in Lemma 2.7.7 and 2.7.6 of \cite{V18}. A proof in the general case is given below.

\begin{lemma}\label{lem:riemann_sample_paths}
Suppose $X(\omega, t)$ is a measurable stochastic process with sample paths that are a.s. Riemann integrable. Suppose that $\|X(\cdot, t)\|_{\psi} \leq h(t)$ for a Riemann integrable function $h$. Let $f(\omega)=\int_{I}X(\omega, t)dt$. Then for any Orlicz function $\psi$, $\|f(\omega)\|_{\psi} \leq \int_{I}h(t)dt$. 
\end{lemma}

A more general version of this results involving Bochner integrals instead of Riemann integrals can be found in~\cite{Decker24}(see Lemma H.0.6).

All results discussed above are immediately generalized to conditional versions under the simple dependence structure assumed in the following definition. A more general definition of conditional Orlicz norm in terms of conditional expectation is possible, but not required in this article. 

\begin{definition}(Conditional Orlicz Norm)\label{def:condorlicz}
Let $X$ and $Y$ be independent random elements taking values in separable metric spaces $\mathcal{X}$ and $\mathcal{Y}$ respectively Let $g:\mathcal{X}\times \mathcal{Y}\rightarrow \R$ be measurable. Then 
\begin{equation}\label{eq:cond_orlicz_norm_def}
\big[\|g(X,Y)\|_{\psi |\sigma(Y)}\big](\omega) := h(Y(\omega)) \quad \mbox{for} \quad h(y) := \|g(X,y)\|_{\psi}.		
\end{equation}
\end{definition}

\begin{remark}\label{remark:conditional_tail_bound} 
Definition \ref{def:condorlicz} allows the results of Lemma \ref{lim_max_sum} and Lemma \ref{relationship_norms} to be generalized to the conditional case. For example, for each $y \in \mathcal{Y}$, let $
f(y,t):=\Pr\Big(g\big(X,y\big) \geq t\Big)$.
By Lemma \ref{lim_max_sum} (iv) we have for all $t \geq 0$ and any $p>0$, $
f(y,t) \leq 2\exp\Big(-\Big(\frac{t}{\|g(X,y)\|_{\psi_{p}}}\Big)^{p}\Big)$
and so 
\[
\Pr\Big(g\big(X,Y\big) \geq t\mid Y\Big) = f(Y,t)  \leq 2\exp\Big(-\Big(\frac{t}{\|g(X,Y)\|_{\psi |\sigma(Y)}}\Big)^{p}\Big),
\]
where all inequalities are almost sure and we used \eqref{eq:cond_orlicz_norm_def}.
\end{remark}

\newpage

\subsubsection{Proofs for section~\ref{section:orlicz}}

\textbf{Proof of Lemma \ref{triangle_inequality}} 
If $\sum_{j=1}^{\infty}\|X_{j}\|_{\psi_{p}} < \infty$, by Lemma \ref{orlicz-equiv}, $S_{M} = \sum_{j=1}^{M}X_{j}$ is a Cauchy sequence in $L^{\tilde{\psi}_{p}}$. By the same lemma, $\tilde{\psi}_{p}$ is an Orlicz function, and so by Lemma \ref{orlicz-banach}, $S_{M}$ converges in $L^{\tilde{\psi}_{p}}$ to a limit $L$ satisfying $L \in L^{\tilde{\psi}_{p}}$. By Lemma \ref{orlicz-equiv}, $L \in L^{{\psi}_{p}}$ and the convergence also takes place in $L^{{\psi}_{p}}$. This proves the first claim. For the second claim, observe that 
\[
\Big|\|S_{M}\|_{\tilde{\psi}_{p}} - \| L \|_{\tilde{\psi}_{p}} \Big| \leq \Big\|S_{M} - L\Big\|_{\tilde{\psi}_{p}}
\]
by the reverse triangle inequality applied in the normed vector space $L^{\tilde{\psi}_{p}}$. Fix $\eps >0$. Since $S_{M} \xrightarrow{\tilde{\psi}_{p}} L$,  $\|S_{M} - L\|_{\tilde{\psi}_{p}}\leq \eps$ for all sufficiently large $M$. Thus, 
\[
\| L\|_{\tilde{\psi}_{p}} \leq \| S_M\|_{\tilde{\psi}_{p}} + \eps \leq  \sum_{j=1}^{\infty}\|X_{j}\|_{\tilde{\psi}_{p}} +\eps.
\]
Since $\eps$ was arbitrary, $\| L\|_{\tilde{\psi}_{p}} \leq \sum_{j=1}^{\infty}\|X_{j}\|_{\tilde{\psi}_{p}}$. By Lemma \ref{orlicz-equiv} again, this implies the bound in the statement of the lemma we are proving. When $X_j \geq 0$ a.s., the sum $\sum_{j=1}^N X_j(\omega)$ is non-decreasing for almost all $\omega$ and hence converges almost surely. The limit is a.s. finite and equal to $L$ by uniqueness of limits in law. \hfill $\Box$

\bigskip

\noindent
\textbf{Proof of Lemma \ref{lim_max_sum} (i)}

Without loss of generality, $\lim\inf_{n \rightarrow \infty}\|X_{n}\|_{\psi}<\infty$, so let $K$ be any strict upper bound. By Definition \ref{def:orlicz-norm}, $\E[\psi(\frac{X_{n}}{K})] \leq 1$ for all large $n$.
By continuity of the convex function $\psi$, Fatou's Lemma, monotonicity of $\psi$, and the previous observation,
\[
\E\Big[\psi\Big(\lim_{n\rightarrow \infty}\inf_{k \geq n}\frac{X_{n}}{K}\Big)\Big] =  \E\Big[\lim_{n\rightarrow \infty}\psi\Big(\inf_{k \geq n}\frac{X_{n}}{K}\Big)\Big] \leq \liminf_{n\rightarrow \infty}\E\Big[\psi\Big(\inf_{k \geq n}\frac{X_{n}}{K}\Big)\Big]  \leq \liminf_{n\rightarrow \infty}\E\Big[\psi\Big(\frac{X_{n}}{K}\Big)\Big]\leq 1.
\]
Thus by Definition \ref{def:orlicz-norm}, 
\[
\Big\|\liminf_{n \rightarrow \infty}X_{n}\Big\|_{\psi} \leq \inf_{K}\Big\{K>\liminf_{n \rightarrow \infty}\|X_{n}\|_{\psi}\Big\} = \liminf_{n \rightarrow \infty}\|X_{n}\|_{\psi}.
\]
\hfill $\Box$

\bigskip

\textbf{Proof of Lemma \ref{lim_max_sum}, (iii)} 

First consider the case $0<p \leq 1$. By Proposition A.1.6 in \cite{VW96}, originally proved in \cite{LT19} as Theorem 6.21, there exists a constant $K_{p}$ such that 
	\[
	\|S_{n}\|_{\psi_{p}} \leq K_{p}\Big(\E|S_{n}|+ \|\max_{1 \leq i \leq n} |X_{i}|\|_{\psi_{p}}\Big).
	\]
 
	By Lemma \ref{lim_max_sum} (ii),
	\[
	\Big\|\max_{1 \leq i \leq n}X_{i}\Big\|_{\psi_{p}} \leq K_{\Psi_{1}}^{\frac{1}{p}}\log^{1/p}(n+1) \max_{1 \leq i \leq n} \|X_{i}\|_{\psi_{p}}.
	\]
	By the Cauchy-Schwarz inequality, 
	\begin{align*}
		\|S_{n}\|_{\psi_{p}} &\leq K_{p}\Big(\{\E[|S_{n}|^{2}]\}^{1/2}+ K_{\Psi_{1}}^{\frac{1}{p}} \log^{1/p}(n+1)\max_{i}\|X_{i}\|_{\psi_{p}}\Big)
		\\ 
		&\leq  C_{p}\Big(\sqrt{n}\max_{i}\|X_{i}\|_{L_{2}}+\sqrt{n}\max_{i}\|X_{i}\|_{\psi_{p}}\Big)
		\\ 
		&\leq C_p\max_{i}\sqrt{n}\|X_{i}\|_{\psi_{p}},
	\end{align*}
	where the value of the constant $C_p$ can change from line to line. In the last line we applied Lemma \ref{relationship_norms}(ii). The proof in case case $p>1$ follows from A.1.6 in \cite{VW96} in a similar way. \hfill $\Box$

\bigskip

\textbf{Proof of Lemma \ref{relationship_norms}, (i)}

To show the first statement in (i), we will first show that $\| |X|^{p}\|_{\Psi_{1}} = \|X\|_{\psi_{p}} ^{p}$. The general case will follow from this. Pick $B >\|X\|_{\psi_{p}}$. It follows that
\begin{equation}
\E\Big[\exp\Big(\Big|\frac{|X|^{p}}{B^{p}}\Big|\Big)\Big] = \E\Big[\exp\Big(\Big|\frac{|X|}{B}\Big|^{p}\Big)\Big] \leq 2.
\end{equation}
Thus $B^{p} \geq \||X|^{p}\|_{\Psi_{1}}$. Let $B_{n}\downarrow \|X\|_{\psi_{p}}$. By Fatou's Lemma, and continuity of both $\exp \circ |\cdot |$ and $w \rightarrow w^{p}$,
\begin{equation}
\E\Big[\exp\Big(\Big|\frac{|X|^{p}}{ \|X\|_{\psi_{p}} ^{p}}\Big|\Big)\Big]  = \E\Big[\lim_{n\rightarrow \infty}\exp\Big(\Big|\frac{|X|^{p}}{B_{n}^{p}}\Big|\Big)\Big] \leq \liminf_{n\rightarrow \infty}\E\Big[\exp\Big(\Big|\frac{|X|^{p}}{B_{n}^{p}}\Big|\Big)\Big] \leq 2. 
\end{equation}
Thus, $\||X|^{p}\|_{\Psi_{1}} \leq  \|X\|_{\psi_{p}} ^{p}$. The reverse inequality is obtained similarly: Let $B >\||X|^p\|_{\Psi_{1}}$, and note that
\begin{equation}
\E\Big[\exp\Big(\Big|\Big(\frac{X}{B^{\frac{1}{p}}}\Big)\Big|^{p}\Big)\Big] = \E\Big[\exp\Big(\Big|\frac{|X|^p}{B}\Big|\Big)\Big] \leq 2.
\end{equation}
An application of Fatou's Lemma then shows that $\|X\|_{\psi_{p}}^{p} \leq \||X|^{p}\|_{\Psi_{1}}$.
The general result follows by noting that
\begin{equation}
\|X\|_{\psi_{p}} = \||X|^{p}\|_{\Psi_{1}}^{\frac{1}{p}} =\||X|^{q\frac{p}{q}}\|_{\Psi_{1}}^{\frac{1}{p}}=\||X|^{\frac{p}{q}}\|_{\psi_{q}}^{\frac{q}{p}}.
\end{equation}
To show the second statement in item (i),we will first show that $\|XY\|_{\Psi_{1}} \leq \|Y\|_{\psi_{2}}\|X\|_{\psi_{2}}$. The general result will follow from since by the first half of part (i) 
\[
\|XY\|_{\psi_{p}} = \||XY|^p\|^{\frac{1}{p}}_{\Psi_{1}} \leq   \||X|^p\|_{\psi_{2}}^{\frac{1}{p}}\||Y|^p\|_{\psi_{2}}^{\frac{1}{p}} =  \|X\|_{\psi_{2p}}\|Y\|_{\psi_{2p}}.
\]
To show $\|XY\|_{\Psi_{1}} \leq \|Y\|_{\psi_{2}}\|X\|_{\psi_{2}}$ we may assume that $\|Y\|_{\psi_{2}}=\|X\|_{\psi_{2}}=1$ since both sides are absolutely homogeneous with respect to multiplication by positive constants.

Let $B_{1}>1$ and let $B_{2}>1$. Then,
\begin{align*}
&\E\Big[\exp\Big(\frac{|XY|}{(B_{1}B_{2})^{2}}\Big)\Big] \leq \E\Big[\exp\Big(\frac{|X|^{2}+|Y|^{2}}{2(B_{1}B_{2})^{2}}\Big)\Big] =\E\Big[\exp\Big(\Big|\frac{Y}{\sqrt{2}B_{1}B_{2}}\Big|^{2}\Big)\exp\Big(\Big|\frac{X}{\sqrt{2}B_{1}B_{2}}\Big|^{2}\Big)\Big]
\\
& \leq \Big(\E\Big[\exp\Big(\Big|\frac{Y}{B_{1}B_{2}}\Big|^{2}\Big)\Big]\Big)^{\frac{1}{2}}\Big(\E\Big[\exp\Big(\Big|\frac{X}{B_{1}B_{2}}\Big|^{2}\Big)\Big]\Big)^{\frac{1}{2}} \leq 2.
\end{align*}
Choosing sequences $B_{1,n}$ and $B_{2,n}$ that each converge downwards to 1, along with Fatou's Lemma, finishes the proof. \hfill $\Box$

\bigskip
\noindent
\textbf{Proof of Lemma \ref{relationship_norms}, (ii)} 
Let $B>\|X\|_{\psi_{p}}$. Then $\sum_{j=1}^{\infty}\frac{\E[ (|\frac{X}{B}|^{p})^{j}]}{j!} \leq 1$, and so for all natural numbers $j$,
\[
\E[|X|^{pj}]\leq B^{pj}j!.
\]
Let $j = \textrm{ceil}(\frac{q}{p})$, so that $pj \geq q$. By Jensen's inequality,
\[
\E[|X|^{q}]^{pj/q} \leq \E[|X|^{pj}].
\]
These two inequalities combine to give
\[
\E[|X|^{q}]^{1/q} \leq \E[|X|^{pj}]^{\frac{1}{pj}} \leq B (j!)^{\frac{1}{pj}}.
\]
Set $C_{p,q} = (j!)^{1/pj}$. Note that  $\textrm{ceil}(\frac{q}{p}) \leq \frac{q}{p}+1$, and so $C_{p,q} \leq \Gamma(\frac{q}{p}+2)^{\frac{1}{q}}$. \hfill $\Box$

\bigskip

\noindent 
\textbf{Proof of Lemma \ref{lem:riemann_sample_paths}} 
The following result is proved in \cite{Sp94}:

\begin{lemma}\label{lem:riemann}
	If $f$ is Riemann integrable on $[a,b]$, then for every $\eps>0$ there is some $\delta_{f,\eps}>0$ such that if $P = \{0, t_{1}, \dots, 1\}$ is any partition of $[a,b]$ with all lengths $t_{i}-t_{i-1} <\delta$, then
	\[
	\Big|\sum_{i=1}^{n}f(x_{i})(t_{i}-t_{i-1})-\int_{[0,1]}f(x)dx\Big| \leq \eps,
	\]
	for any Riemann sum formed by choosing $x_{i}\in [t_{i-1}, t_{i}]$. 
\end{lemma}

	We are now ready to prove the main statement. Let 
	\[
	f_{N}(\omega) = N^{-1}\sum_{i=1}^{N}X(\omega, i/N).
	\] 
	Then $f_{N}(\omega) \rightarrow f(\omega)$ whenever the sample path corresponding to $\omega$ is Riemann integrable.  Since this is assumed to be the case almost surely (by Lemma \ref{lem:riemann}), $f_{N} \rightarrow f$ a.s. By the triangle inequality for the $\psi$ norm and Lemma \ref{lim_max_sum}(i),
	\begin{align*}
		\Big\|\int_{[0,1]}X(\cdot, t)dt\Big\|_{\psi} &\leq \liminf_{N}\Big\|N^{-1}\sum_{i=1}^{N}X(\cdot, i/N)\Big\|_{\psi}
		\\
		&\leq \liminf_{N} N^{-1}\sum_{i=1}^{N}\|X(\cdot, i/N)\|_{\psi} 
		\\
		&\leq \liminf_{N} N^{-1}\sum_{i=1}^{N} h(i/N) 
		\\
		&=\int_{[0,1]} h(t)dt.
	\end{align*}
	The first inequality follows from Lemma \ref{lim_max_sum}(i), the second inequality from the triangle inequality, and the final equality from the fact that, $\sum_{i=1}^{N} h(i/N)$ is a right Riemann sum for the function $h(t)$ and so (like any other Riemann sum) converges to $\int_{[0,1]}h(t) dt$. 
\hfill $\Box$

\begin{lemma}\label{maximizer_is_unique}
Let $Z =[X, Y]^{\top}$ be a centered multivariate Gaussian such that $X \in \R^{M}$ and $Y \in \R^{M}$ and $X$ and $Y$ are absolutely continuous with respect to Lebesgue measure. If for some orthogonal linear map $O$, $X=OY$, $\Pr(\|X\|_2 = \|Y\|_2)=1$. Otherwise, $\Pr(\|X\|_2 = \|Y\|_2)=0$.
\end{lemma}

\textbf{Proof of Lemma \ref{maximizer_is_unique}} 

It is well known that for each $y \in \R^M$, we have $\textrm{law} \{X|Y=y\}=\mu(y) + N(0,\Sigma_Z)$, where $\Sigma_Z :=\Sigma_{XX} - \Sigma_{XY}\Sigma_{YY}^{-1} \Sigma_{YX}$, for $\Sigma_{XX}: = \Var(X)$, $\Sigma_{YY}: = \Var(Y)$, and $\Sigma_{XY}: = \Cov(X,Y)$, and $\mu(y):=\Sigma_{XY}\Sigma_{YY}^{-1}y$. Therefore, 
\[
\Pr(\|X\| = \|Y\|) = \int_{\R^d} \Pr\Big(\|X\| = \|y\|~\Big|~Y=y\Big) d\Pr^Y(y) = \int_{\R^d} \Pr\Big(\|\mu(y) + N(0,\Sigma_Z)\| = \|y\|\Big) d\Pr^Y(y).
\] 
If $\Sigma_Z$ is not the zero-matrix, we have for every $y \in \R^M$, 
\[
\Pr\Big(\|\mu(y) + N(0,\Sigma_Z)\| = \|y\|\Big)=0.
\]
If $\Sigma_Z$ is the zero-matrix, we have for all $y$,  $\textrm{law} \{X|Y=y\}=\Sigma_{XY}\Sigma_{YY}^{-1}y$, and so $X=AY$, almost surely, for $A=\Sigma_{XY}[\Sigma_{YY}]^{-1}$. If  $A$ is orthogonal, we have 
\[
\Pr(\|X\|_2 = \|Y\|_2)=\Pr(\|AY\|_2 = \|Y\|_2) =\Pr(\|Y\|_2 = \|Y\|_2)=1.
\]
Now assume that $A$ is not orthogonal. Note that $\|Ay\|^2 = y^\top A^\top A y$. The matrix $A^\top A$ is symmetric, thus there exists an orthonormal basis, say $v_1,\dots,v_d$, of $\R^d$ of eigenvectors of $A^\top A$. Let $\lambda_1,\dots,\lambda_d$ denote the corresponding eigenvalues. Then for $y = \sum \beta_i v_i$ we have $\|Ay\|^2 = \|y\|^2$ if and only if $\sum_i \lambda_i \beta_i^2 = \sum \beta_i^2$. The set of $\beta_i$ which satisfy this is a Lebesgue null set unless all $\lambda_i = 1$. Indeed, if say $\lambda_1 \neq 1$ then $\beta_1^2 = - \frac{1}{\lambda_1-1}\sum_{i \neq 1}(\lambda_i-1)\beta_i^2$. Therefore, the set $\{\|Ay\|^2 = y^\top A^\top A y\}$ is the countable union of the graphs  smooth functions and so has Lebesgue measure zero. \hfill $\Box$

\end{document}